\documentclass[11pt,a4paper,american]{extarticle}
\usepackage[T1]{fontenc}
\usepackage[latin1]{inputenc}
\usepackage{mathrsfs}
\usepackage{dsfont}
\usepackage{amsmath}
\usepackage{amssymb}
\usepackage{stmaryrd}
\usepackage{setspace}
\usepackage{wasysym}
\usepackage{esint}
\makeatletter

\usepackage{babel}
\usepackage{bbm}
\usepackage{hyperref}
\usepackage{authblk}
\usepackage{tikz}
\usetikzlibrary{matrix,decorations.pathmorphing,arrows.meta,positioning}
\tikzset{
  bgdiag/.style = {line width=0.7pt, baseline=-0.6ex},
  offshell/.style = {dashed, dash pattern=on 4pt off 3pt},
}
\usepackage{cancel}
\usepackage{wrapfig}
\usepackage{comment}
\date{}
\allowdisplaybreaks
\numberwithin{equation}{section}
\usepackage{svg}
\usepackage{shuffle}
\svgsetup{inkscapelatex=true}
\svgsetup{
  inkscapeexe={C:/Program Files/Inkscape/bin/inkscape.exe}
}

\author{Renann Lipinski Jusinskas\thanks{renannlj@fzu.cz}}
\affil{Institute of Physics of the Czech Academy of Sciences \\ CEICO - Central European Institute for Cosmology and Fundamental Physics
\authorcr  Na Slovance 2, 182 21, Prague - Czech Republic}

\usepackage[style=numeric,sorting=none,sortcites=true]{biblatex}

\makeatother

\usepackage{babel}
\begin{document}
\title{Perturbiner methods in scattering amplitudes}
\maketitle
\begin{abstract}
Berends--Giele currents have recently attracted renewed attention across the amplitude community, appearing in new contexts ranging from gravitational amplitudes to cosmological correlators and string theory. They can be derived from first principles using the perturbiner method, a thirty-year-old framework that organizes tree-level scattering data in the form of classical multi-particle solutions to field theory equations of motion. After its rediscovery about a decade ago, it has been systematically studied and applied in many different domains. This review aims to provide a pedagogical introduction to the method, covering its basic formulation across a broad but certainly not exhaustive range of models and progressing to some of the most recent findings in the literature. This includes several unpublished results, many of which represent tacit knowledge of the community that has not previously appeared in print.
\tableofcontents{}
\end{abstract}

\section{Origin and rediscovery\label{sec:origin}}

The computation of scattering amplitudes has always been a central
challenge in quantum field theory. In gauge theories, the problem
is particularly acute: the number of Feynman diagrams grows factorially
with the number of external particles, and the individual diagrams
are plagued by gauge-dependent intermediate expressions that largely
cancel in the final result. By the mid-1980s, the state of the art
for the covariant computation of gluon scattering was the six-point
amplitude, using the traditional diagrammatic approach. Around then,
Parke and Taylor published their celebrated formula for arbitrary
multiplicity, maximally helicity violating (MHV) amplitudes \cite{Parke:1986gb}.
The situation changed dramatically in 1987, when Berends and Giele
introduced their famous currents \cite{Berends:1987me}. Via a partially
off-shell recursion, these objects provided an efficient algorithm
for computing gluon amplitudes to arbitrary multiplicity. The Parke--Taylor
formula had already revealed that the final answers are strikingly
simple; the Berends-Giele recursion explained, at least operationally,
how to reach them systematically. These developments marked the beginning
of a period of sustained progress in amplitude methods, with tree-level
results being extended and organized in increasingly powerful ways
\cite{Mangano:1987xk,Mangano:1990by} (see also \cite{Dixon:1996wi}).

About a decade later, Rosly and Selivanov introduced an alternative
perspective on the recursive structure of scattering amplitudes \cite{Rosly:1996cp,Rosly:1996vr,Rosly:1997ap,Selivanov:1996gw,Selivanov:1998hn}.
Rather than working directly with the recursion relations, they observed
that the off-shell currents could be understood as coefficients in
a formal multi-particle expansion of the classical field\footnote{The connection between solutions to classical field equations and
tree-level scattering amplitudes had already been known for a long
time \cite{Boulware:1968zz} (see also \cite{Arefeva:1974jv}). For the self-dual Yang--Mills theory,
a construction similar to the perturbiner but predating it was already
known at tree level \cite{Bardeen:1995gk} and even at one-loop \cite{Cangemi:1996rx}.}. This effectively established a generating function for all tree-level
scattering data, constructed directly from the classical equations
of motion. They called this object the \emph{perturbiner}. The name
deserves a brief remark: it was coined in reference to Alexander Turbiner,
a mathematical physicist known for his work on quasi-exactly solvable
systems, and it should be pronounced accordingly (toor-bee-NEHR)\footnote{This observation was raised by Andrei Mikhailov in a private communication.
In the first paper of the perturbiner series \cite{Rosly:1996cp},
the origin of the name is mentioned in a footnote. The preprint has
been since withdrawn, but the first version can still be found on
\href{https://arxiv.org/abs/hep-th/9610070v1}{arXiv}.}. Despite the simplicity of the construction, the perturbiner attracted
little attention at the time of its introduction. The original papers
accumulated about 25 citations over the following two decades, a
modest number for a method with such broad applicability. In hindsight,
this is perhaps unsurprising, as the amplitude community was focused
on the powerful on-shell methods.

The landscape of amplitude methods was transformed in the years that
followed by a series of conceptual breakthroughs. The discovery of
Britto--Cachazo--Feng--Witten (BCFW) recursion \cite{Britto:2004ap,Britto:2005fq}
showed that tree-level amplitudes in gauge theories and gravity could
be reconstructed from on-shell data alone, bypassing the need for
off-shell intermediates entirely. The Cachazo--He--Yuan (CHY) formalism
\cite{Cachazo:2013gna,Cachazo:2013hca} provided a strikingly compact
representation of tree-level amplitudes as integrals over the moduli
space of punctured Riemann spheres, connecting scattering amplitudes
to the geometry of the scattering equations. Perhaps most consequentially
for the perturbiner story, the discovery of color-kinematics duality
and the Bern--Carrasco--Johansson (BCJ) relations \cite{Bern:2008qj}
revealed a deep algebraic structure underlying gauge theory amplitudes.
This structure led to the formulation of the field-theory double copy
\cite{Bern:2010ue}, heavily inspired by the known Kawai--Lewellen--Tye
(KLT) formula in string theory \cite{Kawai:1985xq}, establishing
that graviton amplitudes could be expressed as squares of gauge theory
building blocks. More broadly, the amplituhedron \cite{Arkani-Hamed:2013jha},
which reformulates $\mathcal{N}=4$ super Yang--Mills amplitudes
as the volume of a geometric object, and the associahedron \cite{Mizera:2017cqs,Arkani-Hamed:2017mur},
which encodes bi-adjoint scalar amplitudes in the combinatorics of
a polytope, are striking examples of how geometric thinking has reshaped
the field. These developments collectively shifted the focus of the
amplitude community toward the algebraic and geometric (on-shell) structures
underlying scattering data.

The rediscovery of the perturbiner came in this context. In 2015,
Mafra and Schlotterer, working on the pure spinor superspace formulation
of ten-dimensional super Yang--Mills theory, independently discovered
new multi-particle expansions similar to those that Rosly and Selivanov
had introduced \cite{Mafra:2015vca,Mafra:2016mcc} (see also \cite{Lee:2015upy,Mafra:2016ltu}).
Once the connection was recognized, the original works began to attract
significantly more attention (the citation count has since grown to
over 110) and the perturbiner was established as a standard tool in
the modern amplitude toolkit.

Since then, the perturbiner method has found applications in many
different contexts. In \cite{Mizera:2018jbh}, the perturbiner was
applied to the non-linear sigma model (NLSM), the special Galileon,
and Born-Infeld theory, where KLT double-copy relations were shown
to hold at the level of Berends--Giele currents, which is a stronger
statement than a double copy of amplitudes alone. Similar double-copy
constructions were later found in \cite{Escudero:2022zdz,Correa:2024mub}.
In \cite{Garozzo:2018uzj}, string-inspired (mass-deformed) corrections
of the Yang--Mills action were studied using off-shell currents.
With suitable non-linear gauge transformations, these currents were
used in the construction of double-copy representations of gravitational
amplitudes with $\alpha'$ corrections. The double copy at the level
of classical solutions (called classical double copy) has also been
connected to perturbiner methods \cite{Raeymaekers:2025akr}.

Given that the perturbiner is based on classical equations of motion,
it is natural to try to understand it in terms of $L_{\infty}$ algebras.
This algebraic structure underlying field theory interactions was
explored in \cite{Lopez-Arcos:2019hvg,Gomez:2020vat,Escudero:2022zdz}
in different models, placing the recursive construction in a more
abstract algebraic framework. A world-line perspective on the perturbiner
has been developed in \cite{Ben-Shahar:2021doh,Ahmadiniaz:2021ayd},
connecting the recursive structure to first-quantized methods.

The extension of the perturbiner to gravity represents one of the
most significant recent developments. A general recursive treatment
of Einstein gravity, circumventing the infinite tower of interaction
vertices through a careful expansion of the inverse metric, was achieved
in \cite{Gomez:2021shh}. A complementary formulation using double
field theory was later developed in \cite{Cho:2021nim}. These constructions
made the gravitational Berends--Giele currents available for the
first time in a fully systematic form, and opened the door to the
coupling of gravity to arbitrary matter content. Applications in string
theory have been explored in chiral string models \cite{Guillen:2021mwp,Carabine:2023yxv},
and more recently in the open bosonic string \cite{Garozzo:2024myw}.
In these cases, the perturbiner was used as a validation check to
scattering amplitudes computed using string theory methods.

Beyond flat space, the perturbiner has been formulated in anti-de
Sitter space \cite{Armstrong:2022mfr}, where the natural observables
are boundary correlators rather than S-matrix elements. A recent result
\cite{Gomez:2026yno}, using the perturbiner construction on a flat
space with a boundary, has shown that tree-level gluon correlators
in $\textrm{AdS}_{4}$ admit an exact decomposition in terms of flat-space
amplitudes at all multiplicities, with the recursive structure of
the perturbiner making this decomposition manifest. At loop level,
the perturbiner has been extended to one-loop integrands through both
an algebraic recursion \cite{Gomez:2022dzk,Gomez:2024xec}, and a
more systematic quantum effective action formalism \cite{Lee:2022aiu}
for higher loops, with the two approaches being complementary in their
scope and ease of implementation. Other applications of the perturbiner include \cite{Chen:2023bji,Tao:2023wls,Miller:2024oza,Tao:2025fch,Miller:2025wpq,Lopez-Arcos:2025cly}.

This review aims to provide a pedagogical and self-contained account
of the perturbiner method and its applications, accessible to readers
at different levels of familiarity with the subject. While the earlier
sections cover material that is by now standard in the literature,
the review contains several results, computational details, and perspectives
that have not appeared elsewhere, as detailed in the outline below.

\paragraph*{The organization of the review:}

Section \ref{sec:multiparticle solutions} establishes the construction
of classical multi-particle solutions for different types of fields.
The reader more familiar with the perturbiner may find useful the
discussion on spinors and different gauge choices for the vector field,
something that is usually not presented in the literature. Section
\ref{sec:gauge-theories} introduces colored theories and the color-stripped
perturbiner. Again, this is fairly standard material,
but an important ingredient for the later sections. A minor novelty
is the construction of the perturbiner for the $\mathcal{N}=4$ super
Yang--Mills theory, though there is really nothing new about the
methodology, just a mixture of different fields (vector, scalars,
and spinors) that is not really discussed elsewhere. In section \ref{sec:NLSM}
the NLSM is presented, both the coordinate-valued and group-valued
formulations. Particular attention is devoted to the treatment of
the infinite tower of interaction vertices. This section also presents
a new cubic recursion for the NLSM, without the need for auxiliary
fields, which should be interesting even to more experienced readers.
Section \ref{sec:gravity} introduces the perturbiner framework for
gravitons coupled to matter. Since these developments are relatively
recent, the construction is presented in detail, including the vielbein
formulation and coupling to spinors. In section \ref{sec:AdS} we
explore the perturbiner construction beyond Minkowski space. The first
part reviews a more recent work on the perturbiner construction in
flat space with a boundary in one spatial direction, and some interesting
new ingredients (such as the concatenated polarizations). The second
part presents the classical multi-particle solutions of scalars, gluons,
and gravitons in anti de Sitter, including the discussion of a new
democratic gauge choice that puts the different polarization directions
on equal footing. This is likely the most demanding section of the
review. For readers less familiar with free solutions of field equations
in anti-de Sitter spacetime, appendix \ref{sec:Free-solutions-AdS}
provides a self-contained introduction. Section \ref{sec:loop} introduces
the construction of a fully off-shell formulation of the perturbiner,
and its application to the construction of one-loop integrands. A
general proof of the algebraic recursive construction of the one-loop
integrands is presented. The construction for pure gravity assumes
some familiarity with the BV-BRST formalism. Finally, section \ref{sec:summary}
surveys selected applications of the perturbiner that have appeared
in the recent literature, and concludes with an assessment of the
current state of the perturbiner framework, together with a discussion
of several open problems and promising directions for future research.

\paragraph*{A note on conventions:}

We will generally work in $D$ spacetime dimensions. Vector indices
are denoted by $m,n,\ldots$, unless otherwise stated (e.g. four-dimensional
compactification from $D=10$, boundary directions in anti de Sitter).
The Minkowski metric will be denoted by $\eta_{mn}$, with inverse
$\eta^{mn}$, and mostly plus signature. The d'Alembertian operator
will be denoted by $\Box=\eta^{mn}\partial_{m}\partial_{n}$, with
$\partial_{m}=\partial/\partial x^{m}$. For Grassmann variables $\psi$
and $\chi$, we will work with the Hermitian conjugation $(\psi\chi)^{\dagger}=\chi^{\dagger}\psi^{\dagger}=-\psi^{\dagger}\chi^{\dagger}$,
though this is mostly used to verify the reality of the fermionic
actions. Other specific notations will be introduced
along the way, but they are mostly uniform along the whole review.

\section{Classical multi-particle solutions\label{sec:multiparticle solutions}}

Classical multi-particle solutions are formal solutions to the non-linear
field equations of a given theory. In this section we are going to
describe them and develop the basic setup to be used in the rest of
the review.

These multi-particle solutions use the solutions to the free equation
of motion as building blocks, i.e., when the coupling constants vanish,
which in flat space are usually given in terms of plane waves\footnote{In (anti) de Sitter, the free solutions can be expanded as plane waves
in the boundary directions times a function of the radial coordinates
(see appendix \ref{sec:Free-solutions-AdS}). They will be used in
section \ref{sec:AdS}.}. In a scattering process, free particle solutions are simply the
asymptotic states, sufficiently far from the interaction region. For
instance, the Klein--Gordon equation
\begin{equation}
(\Box-\mathrm{m}^{2})\phi=0,\label{eq:massless-free-scalar}
\end{equation}
with mass squared $\mathrm{m}^{2}$, admits solutions of the form
\begin{equation}
\phi(x)=\phi_{p}e^{\mathrm{i}k_{p}\cdot x},\label{eq:free scalar}
\end{equation}
with mass-shell condition $k_{p}^{2}+\mathrm{m}^{2}=0$. The \emph{letter}
$p$ is a particle label. The field amplitude $\phi_{p}$ can be seen
as a (single-particle) scalar ``polarization''. In more general
cases, like with spinors or gauge vectors, the polarization carries
more physical information.

In deriving the multi-particle solutions, there is one condition that
will be imposed in all theories to be considered here: \emph{single-particle
polarizations are always nilpotent}. This implies that equation \eqref{eq:free scalar}
is also a solution of the massive scalar equation with polynomial
interactions! Nilpotency is an unusual requirement, a priori difficult
to justify. After all, there is no physical reason why squaring a
scalar field should give zero. The key point is that these multi-particle
solutions are generators of scattering trees, not individual physical
configurations for the fields. In this context, each particle label
$p$ represents a distinct external leg of the amplitude. Even if
two particles have identical quantum numbers (same momentum and polarization),
they remain distinguishable by their labels. Nilpotency simply enforces
this distinguishability algebraically, and this will be clear soon
when interactions are turned on.

This section lays the groundwork for the rest of the review, introducing
the basic notation for the construction of classical multi-particle
solutions.

\subsection{Scalars\label{subsec:Scalars}}

Let us now consider a simple cubic interaction, with field equation
given by
\begin{equation}
(\Box-\mathrm{m}^{2})\phi=-\frac{\lambda}{2}\phi^{2},\label{eq:scalar-cubic}
\end{equation}
and coupling constant $\lambda$. We have already discussed that \eqref{eq:free scalar}
is formally a solution when the nilpotency condition is enforced.
Now consider the addition of another single-particle solution, such
that
\begin{equation}
\phi(x)=\phi_{1}e^{\mathrm{i}k_{1}\cdot x}+\phi_{2}e^{\mathrm{i}k_{2}\cdot x}.\label{eq:scalar-2particle}
\end{equation}
While the left hand side of \eqref{eq:scalar-cubic} still vanishes,
the right hand side yields
\begin{equation}
-\frac{\lambda}{2}\phi^{2}=-\lambda\phi_{1}\phi_{2}e^{\mathrm{i}k_{12}\cdot x},
\end{equation}
with $k^m_{12}=k^m_{1}+k^m_{2}$. Thus, equation \eqref{eq:scalar-2particle}
is not a solution of the interacting theory. However, this can be corrected if we consider
\begin{equation}
\phi(x)=\phi_{1}e^{\mathrm{i}k_{1}\cdot x}+\phi_{2}e^{\mathrm{i}k_{2}\cdot x}+\frac{\lambda}{(s_{12}+\mathrm{m}^{2})}\phi_{1}\phi_{2}e^{\mathrm{i}k_{12}\cdot x},\label{eq:scalar-2particle-solution}
\end{equation}
with $s_{12}=(k_{1}+k_{2})^{2}$ and $s_{12}+\mathrm{m}^{2}\neq 0$.

As we increase the number of single-particle states, we have to consider
higher order corrections characterized by multi-particle momenta,
\begin{equation}
\phi(x)=\sum_{p}\phi_{p}e^{\mathrm{i}k_{p}\cdot x}+\sum_{p_{1}<p_{2}}\Phi_{p_{1}p_{2}}e^{\mathrm{i}(k_{p_{1}}+k_{p_{2}})\cdot x}+\sum_{p_{1}<p_{2}<p_{3}}\Phi_{p_{1}p_{2}p_{3}}e^{\mathrm{i}(k_{p_{1}}+k_{p_{2}}+k_{p_{3}})\cdot x}+\ldots,
\end{equation}
in which we have simple sums, double sums, etc, over single-particle
labels $p_{i}$. Note that these sums are \emph{ordered}, i.e., $p_{1}<\ldots<p_{n}$,
because there is no way to distinguish different orderings of the
labels. The multi-particle ansatz above can be compactly written as
\begin{equation}
\phi(x)=\sum_{P}\Phi_{P}e^{\mathrm{i}k_{P}\cdot x}.\label{eq:scalar-multiparticle}
\end{equation}
In this case, the sum is over all possible ordered \emph{words} of
any non-zero length, such that $P=p_{1}p_{2}\ldots p_{n}$, with $|P|=n$ denoting
the length of a given word. In addition, we have $k_{P}=k_{p_{1}}+\ldots+k_{p_{n}}$,
and $\Phi_{P}$ denotes the multi-particle coefficients of the ansatz.
Their relevant property is that they can be recursively determined
by plugging \eqref{eq:scalar-multiparticle} back into the field equation
\eqref{eq:scalar-cubic}, and comparing the multi-particle plane waves.
We have, for instance,
\begin{equation}
(\Box-\mathrm{m}^{2})\phi=-\sum_{P}(s_{P}+\mathrm{m}^{2})\Phi_{P}e^{\mathrm{i}k_{P}\cdot x},\label{eq:scalar-multiparticle-kinetic}
\end{equation}
where the generalized Mandelstam variable is given by $s_{P}=k_{P}^{2}$.
The interaction term can be cast as
\begin{align}
\frac{\lambda}{2}\phi^{2} & =\frac{\lambda}{2}\left(\sum_{Q}\Phi_{Q}e^{\mathrm{i}k_{Q}\cdot x}\right)\left(\sum_{R}\Phi_{R}e^{\mathrm{i}k_{R}\cdot x}\right),\nonumber \\
 & =\frac{\lambda}{2}\sum_{P}e^{\mathrm{i}k_{P}\cdot x}\left[\sum_{{Q,R\atop Q\cup R=P}}\Phi_{Q}\Phi_{R}\right].\label{eq:scalar-multiparticle-interaction}
\end{align}
From the first to the second line, we are simply collecting the terms
with non-empty words $Q$ and $R$ such that $k_{Q}+k_{R}=k_{P}$,
which explains the sum inside the square brackets\footnote{In the rare occasions empty words appear, they will be denoted by
$\emptyset$.}. That is, we sum over all non-empty ordered sub-words $Q$ and $R$
that form the word $P$. For example, for the word $P=123$, we have
the set of pairs
\begin{equation}
\{(Q,R)\}=\{(12,3),(13,2),(23,1),(1,23),(2,13),(3,12)\}.
\end{equation}
This construction can be easily generalized to $i\geq2$ subwords,
and we will use the shorthand notation 
\begin{equation}
\sum_{P=P_{1}\cup\ldots\cup P_{i}}=\sum_{{P_{1},P_{2},\ldots,P_{i}\atop P_{1}\cup\ldots\cup P_{i}=P}}.\label{eq:shorthand-ordered}
\end{equation}

A direct comparison between \eqref{eq:scalar-multiparticle-kinetic}
and \eqref{eq:scalar-multiparticle-interaction} finally leads to
the recursive definition of the multi-particle coefficients,
\begin{equation}
\Phi_{P}=\frac{\lambda}{2}\frac{1}{(s_{P}+\mathrm{m}^{2})}\sum_{P=Q\cup R}\Phi_{Q}\Phi_{R}.\label{eq:cubic-scalar-recursion}
\end{equation}
Single-letter coefficients are of course the single-particle polarizations,
$\Phi_{p}=\phi_{p}$. A quick check for the word $P=12$, i.e., $\{(Q,R)\}=\{(1,2),(2,1)\}$,
yields
\begin{align}
\Phi_{12} & =\frac{\lambda}{2}\frac{1}{(s_{12}+\mathrm{m}^{2})}(\Phi_{1}\Phi_{2}+\Phi_{2}\Phi_{1}),\nonumber \\
 & =\frac{\lambda}{(s_{12}+\mathrm{m}^{2})}\phi_{1}\phi_{2},
\end{align}
 matching the corresponding term in equation \eqref{eq:scalar-2particle-solution}.

\subsection{Spinors\label{subsec:WZmodel}}

The procedure to obtain multi-particle solutions of field equations
involving fermions is similar. A minor bother is the fact that spinorial
representations depend on the spacetime dimension, so in this section
we will work in $D=4$.

We have to introduce some additional notation. Dirac spinors, $\xi_{a}$,
with $a=1,2,3,4$, can be cast according to their chirality,
\begin{equation}
\xi=\left(\begin{array}{c}
\psi_{\alpha}\\
\bar{\chi}^{\dot{\alpha}}
\end{array}\right),
\end{equation}
such that the index $\alpha=1,2$ is associated to a chiral spinor,
and $\dot{\alpha}=1,2$ to an anti-chiral spinor. Complex conjugation
in our standard metric signature maps chiral and anti-chiral representations,
i.e., $(\psi_{\alpha})^{\dagger}=\bar{\psi}_{\dot{\alpha}}$. Spinor
indices can be raised and lowered using $\epsilon^{\alpha\beta}$
and $\epsilon_{\alpha\beta}$, respectively, e.g. $\psi^{\alpha}=\epsilon^{\alpha\beta}\psi_{\beta}$,
with $\epsilon^{12}=\epsilon_{21}=+1$, $\epsilon^{21}=\epsilon_{12}=-1$,
and $\epsilon_{\alpha\gamma}\epsilon^{\gamma\beta}=\delta_{\alpha}^{\beta}$.
Finally, Dirac matrices, $\Gamma_{ab}^{m}$, are recast in terms of
Pauli matrices, $\sigma_{\alpha\dot{\alpha}}^{m}$ and $\bar{\sigma}^{m\dot{\alpha}\alpha}=\epsilon^{\alpha\beta}\epsilon^{\dot{\alpha}\dot{\beta}}\sigma_{\beta\dot{\beta}}^{m}$,
with
\begin{equation}
\Gamma^{m}=\left(\begin{array}{cc}
0 & \sigma^{m}\\
\bar{\sigma}^{m} & 0
\end{array}\right).
\end{equation}
The Clifford algebra, $\{\Gamma^{m},\Gamma^{n}\}=2\eta^{mn}$, then
translates to
\begin{align}
(\sigma_{\alpha\dot{\alpha}}^{m}\bar{\sigma}^{n\dot{\alpha}\beta}+\sigma_{\alpha\dot{\alpha}}^{n}\bar{\sigma}^{m\dot{\alpha}\beta}) & =2\eta^{mn}\delta_{\alpha}^{\beta},\\
(\bar{\sigma}^{m\dot{\alpha}\alpha}\sigma_{\alpha\dot{\beta}}^{n}+\bar{\sigma}^{n\dot{\alpha}\alpha}\sigma_{\alpha\dot{\beta}}^{m}) & =2\eta^{mn}\delta_{\dot{\beta}}^{\dot{\alpha}}.
\end{align}
In addition, the Pauli matrices satisfy
\begin{equation}
\sigma_{\alpha\dot{\alpha}}^{m}\bar{\sigma}^{n\dot{\beta}\beta}\eta_{mn}=2\delta_{\alpha}^{\beta}\delta_{\dot{\alpha}}^{\dot{\beta}}.
\end{equation}

As a first example of a field theory involving spinors, we will look
at the massless Wess-Zumino model \cite{Wess:1973kz,Wess:1974tw}, which is arguably the simplest
interacting supersymmetric model in four dimensions. Its field content
comprises a scalar field $\phi$, a chiral spinor $\psi_{\alpha}$,
and their Hermitian conjugates, denoted by $\bar{\phi}$ and $\bar{\psi}_{\dot{\alpha}}$.
The action is given by
\begin{equation}
\mathcal{S}=\int d^{4}x\bigg\{-\partial_{m}\bar{\phi}\partial^{m}\phi+\mathrm{i}\psi^{\alpha}\sigma_{\alpha\dot{\alpha}}^{m}\partial_{m}\bar{\psi}^{\dot{\alpha}}-\frac{\lambda}{2}\phi\psi_{\alpha}\psi^{\alpha}+\frac{\lambda}{2}\bar{\phi}\bar{\psi}_{\dot{\alpha}}\bar{\psi}^{\dot{\alpha}}+\frac{\lambda^{2}}{4}\phi^{2}\bar{\phi}^{2}\bigg\},\label{eq:WZ-action}
\end{equation}
in which $\lambda$ is the coupling constant. Note that under the
global supersymmetry transformations
\begin{equation}
\begin{array}{ccc}
\delta_{\textrm{susy}}\phi=\mathrm{i}\kappa_{\alpha}\psi^{\alpha}, &  & \delta_{\textrm{susy}}\psi^{\alpha}=\bar{\kappa}_{\dot{\alpha}}\bar{\sigma}^{m\dot{\alpha}\alpha}\partial_{m}\phi+\mathrm{i}\frac{\lambda}{2}\bar{\phi}^{2}\kappa^{\alpha},\\
\delta_{\textrm{susy}}\bar{\phi}=\mathrm{i}\bar{\kappa}_{\dot{\alpha}}\bar{\psi}^{\dot{\alpha}}, &  & \delta_{\textrm{susy}}\bar{\psi}^{\dot{\alpha}}=\bar{\sigma}^{m\dot{\alpha}\alpha}\kappa_{\alpha}\partial_{m}\bar{\phi}-\mathrm{i}\frac{\lambda}{2}\phi^{2}\bar{\kappa}^{\dot{\alpha}},
\end{array}
\end{equation}
with parameters $\{\kappa_{\alpha},\bar{\kappa}_{\dot{\alpha}}\}$,
the action \eqref{eq:WZ-action} changes by a boundary term.

The classical equations of motion associated to \eqref{eq:WZ-action}
are\begin{subequations}
\begin{align}
\Box\phi & =-\frac{\lambda}{2}\bar{\psi}_{\dot{\alpha}}\bar{\psi}^{\dot{\alpha}}-\frac{\lambda^{2}}{2}\phi^{2}\bar{\phi},\\
\Box\bar{\phi} & =+\frac{\lambda}{2}\psi_{\alpha}\psi^{\alpha}-\frac{\lambda^{2}}{2}\bar{\phi}^{2}\phi,\\
\mathrm{i}\sigma_{\alpha\dot{\alpha}}^{m}\partial_{m}\bar{\psi}^{\dot{\alpha}} & =-\lambda\phi\psi_{\alpha},\\
\mathrm{i}\partial_{m}\psi^{\alpha}\sigma_{\alpha\dot{\alpha}}^{m} & =\lambda\bar{\phi}\bar{\psi}_{\dot{\alpha}},
\end{align}
\end{subequations}and it is straightforward to analyze their multi-particle
solutions. Analogously to \eqref{eq:scalar-multiparticle}, we start
with the ansatze
\begin{equation}
\begin{array}{ccc}
\phi(x)=\sum_{P}\Phi_{P}e^{\mathrm{i}k_{P}\cdot x}, &  & \psi^{\alpha}(x)=\sum_{P}\Psi_{P}^{\alpha}e^{\mathrm{i}k_{P}\cdot x},\\
\bar{\phi}(x)=\sum_{P}\bar{\Phi}_{P}e^{\mathrm{i}k_{P}\cdot x}, &  & \bar{\psi}^{\dot{\alpha}}(x)=\sum_{P}\bar{\Psi}_{P}^{\dot{\alpha}}e^{\mathrm{i}k_{P}\cdot x},
\end{array}\label{eq:WZ-multi-ansatze}
\end{equation}
and plug them back into the equations of motion. Here $\Phi_{P}$,
$\bar{\Phi}_{P}$, $\Psi_{P}^{\alpha}$, and $\bar{\Psi}_{P}^{\dot{\alpha}}$
denote the multi-particle coefficients. We already have some experience with
the scalar equations, and it is easy to derive their recursions, 
\begin{align}
\Phi_{P} & =\frac{\lambda}{2s_{P}}\sum_{P=Q\cup R}\bar{\Psi}_{Q\dot{\alpha}}\bar{\Psi}_{R}^{\dot{\alpha}}+\frac{\lambda^{2}}{2s_{P}}\sum_{P=Q\cup R\cup S}\Phi_{Q}\Phi_{R}\bar{\Phi}_{S},\\
\bar{\Phi}_{P} & =-\frac{\lambda}{2s_{P}}\sum_{P=Q\cup R}\Psi_{Q\alpha}\Psi_{R}^{\alpha}+\frac{\lambda^{2}}{2s_{P}}\sum_{P=Q\cup R\cup S}\Phi_{Q}\bar{\Phi}_{R}\bar{\Phi}_{S}.
\end{align}
The only new ingredient is the sum over three nonempty subwords $Q$,
$R$, and $S$ that form the word $P=Q\cup R\cup S$. This piece comes
from the quartic vertex in the action \eqref{eq:WZ-action}. For the
spinors, we obtain
\begin{align}
k_{Pm}\sigma_{\alpha\dot{\alpha}}^{m}\bar{\Psi}_{P}^{\dot{\alpha}} & =\lambda\sum_{P=Q\cup R}\Psi_{Q\alpha}\Phi_{R},\\
k_{Pm}\sigma_{\alpha\dot{\alpha}}^{m}\Psi_{P}^{\alpha} & =-\lambda\sum_{P=Q\cup R}\bar{\Psi}_{Q\dot{\alpha}}\bar{\Phi}_{R}.
\end{align}
In order to get them in a similar form to the scalars, we can simply
multiply both sides of these equations by $\bar{\sigma}^{m}k_{Pm}$, which leads to
\begin{align}
\bar{\Psi}_{P}^{\dot{\alpha}} & =\frac{\lambda}{s_{P}}\bar{\sigma}^{m\dot{\alpha}\alpha}k_{Pm}\sum_{P=Q\cup R}\Psi_{Q\alpha}\Phi_{R},\\
\Psi_{P}^{\alpha} & =-\frac{\lambda}{s_{P}}\bar{\sigma}^{m\dot{\alpha}\alpha}k_{Pm}\sum_{P=Q\cup R}\bar{\Psi}_{Q\dot{\alpha}}\bar{\Phi}_{R}.
\end{align}
Once more, one letter words correspond to the single particle polarizations,
i.e., $\Psi_{p}^{\alpha}=\psi_{p}^{\alpha}$ and $\bar{\Psi}_{p}^{\dot{\alpha}}=\bar{\psi}_{p}^{\dot{\alpha}}$.

Observe that whenever particle $i$ has an assigned polarization,
e.g. $\phi_{i}$, we do not assign the same label to a different polarization,
e.g. $\psi_{i}^{\alpha}$ or $\bar{\phi}_{i}$. Each single particle
label belongs to a unique polarization. As an example, let us work
out some specific sets of polarizations. First, consider a particle
configuration without fermions, with scalars $\phi_{1}$, $\phi_{2}$,
$\bar{\phi}_{3}$, and $\bar{\phi}_{4}$. In this case, the multi-particle
coefficients $\Psi_{P}^{\alpha}$ and $\bar{\Psi}_{P}^{\dot{\alpha}}$
are always zero, and the only non-trivial coefficients are
\begin{equation}
\begin{array}{ccc}
\Phi_{123}=\frac{\lambda^{2}}{s_{123}}\phi_{1}\phi_{2}\bar{\phi}_{3}, &  & \bar{\Phi}_{134}=\frac{\lambda^{2}}{s_{134}}\phi_{1}\bar{\phi}_{3}\bar{\phi}_{4},\\
\Phi_{124}=\frac{\lambda^{2}}{s_{124}}\phi_{1}\phi_{2}\bar{\phi}_{4}, &  & \bar{\Phi}_{234}=\frac{\lambda^{2}}{s_{234}}\phi_{2}\bar{\phi}_{3}\bar{\phi}_{4}.
\end{array}
\end{equation}
Second, consider the case with one scalar, $\phi_{1}$, and two spinors,
$\psi_{2}^{\alpha}$ and $\psi_{3}^{\alpha}$. The only non-trivial
coefficients are
\begin{equation}
\begin{array}{ccccc}
\bar{\Phi}_{23}=\frac{\lambda}{s_{23}}\psi_{2}^{\alpha}\psi_{3\alpha}, &  & \bar{\Psi}_{12}^{\dot{\alpha}}=\frac{\lambda}{s_{12}}\bar{\sigma}^{m\dot{\alpha}\alpha}k_{1m}\psi_{2\alpha}\phi_{1}, &  & \bar{\Psi}_{13}^{\dot{\alpha}}=\frac{\lambda}{s_{13}}\bar{\sigma}^{m\dot{\alpha}\alpha}k_{1m}\psi_{3\alpha}\phi_{1},\end{array}
\end{equation}
where we have already used the on-shell conditions $k_{2m}\bar{\sigma}^{m\dot{\alpha}\alpha}\psi_{2\alpha}=k_{3m}\bar{\sigma}^{m\dot{\alpha}\alpha}\psi_{3\alpha}=0$.

\subsection{Gauge vectors}

As the final case study in this section, we will look at the field
equations of quantum electrodynamics. They are now standard textbook material and can be cast as\begin{subequations}\label{eq:QED-eom}
\begin{align}
\mathrm{i}\partial_{m}\psi^{\alpha}\sigma_{\alpha\dot{\alpha}}^{m} & =\mathrm{m}\bar{\psi}_{\dot{\alpha}}-e\sigma_{\alpha\dot{\alpha}}^{m}A_{m}\psi^{\alpha},\\
\mathrm{i}\sigma_{\alpha\dot{\alpha}}^{m}\partial_{m}\bar{\psi}^{\dot{\alpha}} & =\mathrm{m}\psi_{\alpha}+e\sigma_{\alpha\dot{\alpha}}^{m}A_{m}\bar{\psi}^{\dot{\alpha}}\\
\partial_{n}(\partial^{n}A^{m}-\partial^{m}A^{n}) & =e\psi^{\alpha}\sigma_{\alpha\dot{\alpha}}^{m}\bar{\psi}^{\dot{\alpha}}.\label{eq:Maxwell}
\end{align}
\end{subequations}Here we have $\mathrm{m}$ as the mass of the spinor,
and $e$ its electric charge. Note that these equations imply the
conservation of the electric charge,
\begin{equation}
\partial_{m}(e\psi^{\alpha}\sigma_{\alpha\dot{\alpha}}^{m}\bar{\psi}^{\dot{\alpha}})=0,\label{eq:QED-chargeconservation}
\end{equation}
which directly follows from the divergence of \eqref{eq:Maxwell}
or using the spinor equations of motion. Naturally, we have to consider
the gauge redundancy of the theory, and it is easy to check that these
equations are preserved under the $U(1)$ transformations
\begin{equation}
\begin{array}{ccccc}
A_{m}\to A_{m}+\partial_{m}\lambda, &  & \psi^{\alpha}\to e^{+\mathrm{i}e\lambda}\psi^{\alpha}, &  & \bar{\psi}^{\dot{\alpha}}\to e^{-\mathrm{i}e\lambda}\bar{\psi}^{\dot{\alpha}},\end{array}
\end{equation}
where $\lambda=\lambda(x)$ is the gauge parameter.

In order to obtain the multi-particle solutions of \eqref{eq:QED-eom},
we start with the ansatze
\begin{equation}
\begin{array}{ccccc}
\psi^{\alpha}(x)=\sum_{P}\Psi_{P}^{\alpha}e^{\mathrm{i}k_{P}\cdot x}, &  & \bar{\psi}^{\dot{\alpha}}(x)=\sum_{P}\bar{\Psi}_{P}^{\dot{\alpha}}e^{\mathrm{i}k_{P}\cdot x}, &  & A_{m}(x)=\sum_{P}\mathcal{A}_{Pm}e^{\mathrm{i}k_{P}\cdot x}.\end{array}\label{eq:QEDansatz}
\end{equation}
The procedure to find the recursions for the respective multi-particle
coefficients is straightforward. We first obtain
\begin{align}
-\sigma_{\alpha\dot{\alpha}}^{m}k_{Pm}\Psi_{P}^{\alpha} & =\mathrm{m}\bar{\Psi}_{P\dot{\alpha}}-e\sigma_{\alpha\dot{\alpha}}^{m}\sum_{P=Q\cup R}\mathcal{A}_{Qm}\Psi_{R}^{\alpha},\\
-\sigma_{\alpha\dot{\alpha}}^{m}k_{Pm}\bar{\Psi}_{P}^{\dot{\alpha}} & =\mathrm{m}\Psi_{P\alpha}+e\sigma_{\alpha\dot{\alpha}}^{m}\sum_{P=Q\cup R}\mathcal{A}_{Qm}\bar{\Psi}_{R}^{\dot{\alpha}},\\
(s_{P}\eta^{mn}-k_{P}^{m}k_{P}^{n})\mathcal{A}_{Pn} & =-e\sigma_{\alpha\dot{\alpha}}^{m}\sum_{P=Q\cup R}\Psi_{Q}^{\alpha}\bar{\Psi}_{R}^{\dot{\alpha}},\label{eq:QED-vectorrecursion}
\end{align}
which are not yet primed for a recursion. The charge conservation
equation \eqref{eq:QED-chargeconservation} simply reads
\begin{equation}
\sigma_{\alpha\dot{\alpha}}^{m}k_{Pm}\sum_{P=Q\cup R}\Psi_{Q}^{\alpha}\bar{\Psi}_{R}^{\dot{\alpha}}=0.\label{eq:multi-particle-chargeconservation}
\end{equation}

The spinor equations can be easily manipulated and lead to
\begin{align}
\Psi_{P}^{\alpha} & =+\frac{e}{(s_{P}+\mathrm{m}^{2})}\bar{\sigma}^{m\dot{\alpha}\alpha}\sum_{P=Q\cup R}\left(\mathrm{m}\mathcal{A}_{Qm}\bar{\Psi}_{R\dot{\alpha}}+k_{Pm}\sigma_{\beta\dot{\alpha}}^{n}\mathcal{A}_{Qn}\Psi_{R}^{\beta}\right),\label{eq:QEDpsirecursion}\\
\bar{\Psi}_{P}^{\dot{\alpha}} & =-\frac{e}{(s_{P}+\mathrm{m}^{2})}\bar{\sigma}^{m\dot{\alpha}\alpha}\sum_{P=Q\cup R}\left(\mathrm{m}\mathcal{A}_{Qm}\Psi_{R\alpha}+k_{Pm}\sigma_{\alpha\dot{\beta}}^{n}\mathcal{A}_{Qn}\bar{\Psi}_{R}^{\dot{\beta}}\right).\label{eq:QEDpsibarrecursion}
\end{align}
For the vector equation, however, the left hand side involves a projector,
$(s_{P}\eta^{mn}-k_{P}^{m}k_{P}^{n})$, therefore it cannot be unambiguously
inverted. This is related to the gauge redundancy of the theory, so
we need to choose a gauge in order to define the recursion.

\subsubsection{Temporal gauge\label{subsec:Temporal-gauge}}

In the temporal gauge, we choose $A^{0}(x)=0$, which translates to
$\mathcal{A}_{P}^{0}=0$ in the multi-particle ansatz. The remaining
components, $\mathcal{A}_{P}^{i}$, with $i=1,2,3$ should be amenable
to a recursive definition. Indeed, we obtain
\begin{align}
k_{Pi}\mathcal{A}_{P}^{i} & =\frac{e}{k_{P}^{0}}\sigma_{\alpha\dot{\alpha}}^{0}\sum_{P=Q\cup R}\Psi_{Q}^{\alpha}\bar{\Psi}_{R}^{\dot{\alpha}},\label{eq:temp-gauge-div}\\
\mathcal{A}_{P}^{i} & =\frac{e}{s_{P}}\left(\frac{k_{P}^{i}}{k_{P}^{0}}\sigma_{\alpha\dot{\alpha}}^{0}-\sigma_{\alpha\dot{\alpha}}^{i}\right)\sum_{P=Q\cup R}\Psi_{Q}^{\alpha}\bar{\Psi}_{R}^{\dot{\alpha}}.
\end{align}
The first equation comes from the component $m=0$ in \eqref{eq:QED-vectorrecursion},
while the second equation comes from the components $m=i$, already
taking into account the result for $k_{Pi}\mathcal{A}_{P}^{i}$ from
the first equation. In terms of the single-particle solutions, the
linearized equations of motion lead to $k_{pi}\varepsilon_{p}^{i}=0$,
with $\varepsilon_{p}^{i}$ denoting the single-particle polarization.
This is consistent with equation \eqref{eq:temp-gauge-div} when $P$
is a one-letter word.

\subsubsection{Coulomb gauge}

In the Coulomb gauge, we impose that the spatial divergence of the
gauge vector vanishes, i.e., $\partial_{i}A^{i}(x)=0$, which translates
to $k_{Pi}\mathcal{A}_{P}^{i}=0$. Once more, the different components
of equation \eqref{eq:QED-vectorrecursion} lead to
\begin{align}
\mathcal{A}_{P}^{0} & =-\frac{e}{\tilde{s}_{P}}\sigma_{\alpha\dot{\alpha}}^{0}\sum_{P=Q\cup R}\Psi_{Q}^{\alpha}\bar{\Psi}_{R}^{\dot{\alpha}},\\
\mathcal{A}_{P}^{i} & =\frac{e}{s_{P}}\left(\frac{k_{P}^{i}k_{P}^{0}}{\tilde{s}_{P}}\sigma_{\alpha\dot{\alpha}}^{0}-\sigma_{\alpha\dot{\alpha}}^{i}\right)\sum_{P=Q\cup R}\Psi_{Q}^{\alpha}\bar{\Psi}_{R}^{\dot{\alpha}},
\end{align}
with $\tilde{s}_{P}=(k_{P}^{1})^{2}+(k_{P}^{2})^{2}+(k_{P}^{3})^{2}$
. It is trivial to check the consistency of the gauge choice using
\eqref{eq:QED-chargeconservation}. For the single-particle solutions,
the polarizations have the same behavior as in the temporal gauge.

\subsubsection{Lorenz gauge}

The multi-particle solution takes its most elegant form in the Lorenz
gauge, $\partial_{m}A^{m}(x)=0$. In this case, we have the transversality
condition
\begin{equation}
k_{P}\cdot\mathcal{A}_{P}=0,\label{eq:Lorenz-multi-particle-QED}
\end{equation}
with the recursion simply defined by
\begin{equation}
\mathcal{A}_{P}^{m}=-\frac{e}{s_{P}}\sigma_{\alpha\dot{\alpha}}^{m}\sum_{P=Q\cup R}\Psi_{Q}^{\alpha}\bar{\Psi}_{R}^{\dot{\alpha}}.\label{eq:QEDgaugerecursion}
\end{equation}
The gauge condition applied to this relation leads to the charge conservation
equation \eqref{eq:multi-particle-chargeconservation}. We will often
refer to the coefficient $\mathcal{A}_{P}^{m}$ as a multi-particle
current, which is motivated by \eqref{eq:Lorenz-multi-particle-QED}.
More generally, it is common to find in the literature different multi-particle
coefficients named as currents, i.e., $\Phi_{P}$ in \eqref{eq:cubic-scalar-recursion},
even though they are associated to different kinds of fields. In terms
of single-particle solutions, we have transversal polarizations, $k_{p}\cdot\varepsilon_{p}=0$,
with a residual gauge invariance given by $\delta\varepsilon_{p}=k_{p}\lambda$.

\subsection{Tree level scattering amplitudes}

These multi-particle solutions have a very special physical meaning.
They encode the complete set of tree level scattering amplitudes of
the respective theory.

In order to see this, it is helpful to have a diagrammatic realization,
so we will choose the simple scalar model of subsection \ref{subsec:Scalars}.
The two-particle current, $\Phi_{pq}$, is assigned to a ``cubic
tree''.
\begin{figure}[htbp]
  \centering
  $\Phi_{pq}=\dfrac{\lambda}{(s_{pq}-\mathrm{m}^{2})}\,\phi_{p}\phi_{q}
   \qquad\longrightarrow\qquad$
  \begin{tikzpicture}[bgdiag]
    \coordinate (v) at (0,0);
    \draw[offshell,{Stealth}-] (-1.8,0) -- (v)
          node[midway,below=1pt] {$\tfrac{1}{(s_{pq}-\mathrm{m}^{2})}$};
    \draw[-{Stealth[reversed]}] (v) -- (38:1.5)  node[right] {$\phi_{p}$};
    \draw[-{Stealth[reversed]}] (v) -- (-38:1.5) node[right] {$\phi_{q}$};
    \node at (-0.10,0.46) {$\lambda$};
  \end{tikzpicture}
  \caption{Graphic depiction of the scalar two-particle current.}
  \label{fig:scalar3pt}
\end{figure}
The dashed line in figure \ref{fig:scalar3pt} can be seen as an
off-shell leg. If we compute a three-particle current, e.g. $\Phi_{123}$,
we obtain
\begin{equation}
\Phi_{123}=\frac{\lambda}{(s_{123}+\mathrm{m}^{2})}\Phi_{12}\phi_{3}+\textrm{permutations }(1,2,3).
\end{equation}
\begin{figure}[htbp]
  \centering
  $\Phi_{123}\qquad\longrightarrow\qquad$
  \begin{tikzpicture}[bgdiag]
    \coordinate (v1) at (0,0);
    \coordinate (v2) at (40:1.5);
    \path (v2) ++(90:1.2)  coordinate (p1);
    \path (v2) ++(-10:1.5) coordinate (p2);
    \path (v1) ++(-40:1.5) coordinate (p3);
    \draw[offshell,-{Stealth}] (v1) -- (-1.8,0)
          node[midway,below=3pt] {$\tfrac{1}{(s_{123}-\mathrm{m}^{2})}$};
    \draw[{Stealth[reversed]}-] (p3) node[below right=-1pt] {$\phi_{3}$} -- (v1);
    \draw (v1) -- (v2)
          node[midway,below right=2pt and 1pt]
               {$\tfrac{1}{(s_{12}-\mathrm{m}^{2})}$};
    \draw[{Stealth[reversed]}-] (p1) node[above] {$\phi_{1}$} -- (v2);
    \draw[{Stealth[reversed]}-] (p2) node[right] {$\phi_{2}$} -- (v2);
    \node at (-0.1,0.5) {$\lambda$};
    \node at (0.70,1.24) {$\lambda$};
  \end{tikzpicture}
  \caption{Graphic depiction of one of the terms in the scalar
           three-particle current.}
  \label{fig:scalar4pt}
\end{figure}
The first term is diagrammatically expressed in figure \ref{fig:scalar4pt}.
The recursive definition of the currents is intuitive, generating
single-particle trees with the possible permutations.
In these diagrams, if we attach to the dashed lines a single-particle
state by canceling the respective propagator, we end up with expressions
that look like $n$-point tree level amplitudes, $A_{n}$. Indeed,
\begin{align}
\lim_{s_{12}\to\mathrm{m}^{2}}\phi_{3}(s_{12}+\mathrm{m}^{2})\Phi_{12} & =\lambda\phi_{1}\phi_{2}\phi_{3},\nonumber \\
 & \propto A_{3},\label{eq:scalar3pt}
\end{align}
and
\begin{align}
\lim_{s_{123}\to\mathrm{m}^{2}}\phi_{4}(s_{123}+\mathrm{m}^{2})\Phi_{123} & =\lambda^{2}\phi_{1}\phi_{2}\phi_{3}\phi_{4}\left(\frac{1}{(s_{12}+\mathrm{m}^{2})}+\frac{1}{(s_{13}+\mathrm{m}^{2})}+\frac{1}{(s_{23}+\mathrm{m}^{2})}\right),\nonumber \\
 & \propto A_{4}.\label{eq:scalar4pt}
\end{align}
Note that the limits in these equations are compatible with momentum
conservation and the on-shell condition on the attached state. In
$A_{3}$, we have $s_{12}=-\mathrm{m}^{2}=k_{3}^{2}$, and in $A_{4}$
we have $s_{123}=-\mathrm{m}^{2}=k_{4}^{2}$. The relative coefficient
between \eqref{eq:scalar3pt} and \eqref{eq:scalar4pt} is also the
same that one would obtain using Feynman rules. More generally, we
can define the $(n+1)$-point tree level amplitude as
\begin{equation}
A_{n+1}=\mathcal{N}\lim_{s_{N}\to\mathrm{m}^{2}}\phi_{n+1}(s_{N}+\mathrm{m}^{2})\Phi_{N},\label{eq:(n+1)-pointscalar}
\end{equation}
where $N=1\ldots n$. Although not explicitly, the amplitude is invariant
under arbitrary permutations of the single-particle labels. The overall
normalization $\mathcal{N}$, which does not depend on $n$, cannot
be fixed by the multi-particle solution. The reason is clear, since
the classical solutions are obtained from the equations of motion.
They are blind to an overall numerical factor in front of the action,
whereas the scattering amplitudes are proportional to it. Here and
in the rest of the review the usual $D$-dimensional momentum conservation
deltas, $\delta^{D}(k_{1}+\ldots+k_{n+1})$, are left implicit.

The construction of tree level scattering amplitudes is similar when
different particle flavors are involved. The key difference is the
canceled propagator when attaching the on-shell state. In the massless
Wess-Zumino model of subsection \ref{subsec:WZmodel}, the tree level
amplitudes defined via the multi-particle coefficients can be computed
in different ways, depending on the particle content involved. They
can be summarized as\begin{subequations}\label{eq:WZNpoint}
\begin{align}
A_{n+1}^{\textrm{WZ}} & =\mathcal{N}\lim_{s_{N}\to0}\phi_{n+1}s_{N}\bar{\Phi}_{N},\\
A_{n+1}^{\textrm{WZ}} & =\mathcal{N}\lim_{s_{N}\to0}\bar{\phi}_{n+1}s_{N}\Phi_{N},\\
A_{n+1}^{\textrm{WZ}} & =\mathcal{N}\lim_{s_{N}\to0}\psi_{n+1}^{\alpha}k_{Nm}\sigma_{\alpha\dot{\alpha}}^{m}\bar{\Psi}_{N}^{\dot{\alpha}},\\
A_{n+1}^{\textrm{WZ}} & =\mathcal{N}\lim_{s_{N}\to0}\bar{\psi}_{n+1}^{\dot{\alpha}}k_{Nm}\sigma_{\alpha\dot{\alpha}}^{m}\Psi_{N}^{\alpha}.
\end{align}
\end{subequations}Let us take, as an example, the following polarization
set, $\{\phi_{1},\bar{\phi}_{2},\psi_{3},\bar{\psi}_{4}\}$. It is
straightforward to build the respective three-particle coefficients.
The only non-trivial ones are\begin{subequations}
\begin{align}
\Phi_{134} & =-\frac{\lambda^{2}}{s_{134}s_{13}}k_{1m}(\psi_{3}\sigma^{m}\bar{\psi}_{4})\phi_{1}\\
\bar{\Phi}_{234} & =+\frac{\lambda^{2}}{s_{234}s_{24}}k_{2m}(\psi_{3}\sigma^{m}\bar{\psi}_{4})\bar{\phi}_{2}\\
\bar{\Psi}_{124}^{\dot{\alpha}} & =+\frac{\lambda^{2}}{s_{124}s_{24}}\bar{\sigma}^{m\dot{\alpha}\beta}k_{124m}(\sigma_{\beta\dot{\beta}}^{n}\bar{\psi}_{4}^{\dot{\beta}})k_{2n}\phi_{1}\bar{\phi}_{2},\\
\Psi_{123}^{\alpha} & =+\frac{\lambda^{2}}{s_{123}s_{13}}\bar{\sigma}^{m\dot{\beta}\alpha}k_{123m}(\sigma_{\beta\dot{\beta}}^{n}\psi_{3}^{\beta})k_{1n}\phi_{1}\bar{\phi}_{2}.
\end{align}
\end{subequations}Then we can directly check that the amplitude definitions
in \eqref{eq:WZNpoint} agree with one another. In order to show this
we need momentum conservation (e.g. $s_{13}=s_{24}$) and the linearized
equations of motion for the spinor polarizations. Like in the scalar
case, the overall normalization $\mathcal{N}$ is independent of the
number of points and can be fixed from first principles from the action
\eqref{eq:WZ-action}.

Finally, we can work out the QED tree level amplitudes using the multi-particle
currents in \eqref{eq:QEDansatz}. In analogy with \eqref{eq:(n+1)-pointscalar}
and \eqref{eq:WZNpoint}, they are given by\begin{subequations}
\begin{align}
A_{n+1}^{\textrm{QED}} & =\mathcal{N}\lim_{s_{N}\to\mathrm{m}^{2}}\psi_{n+1}^{\alpha}(k_{Nm}\sigma_{\alpha\dot{\alpha}}^{m}\bar{\Psi}_{N}^{\dot{\alpha}}+\mathrm{m}\Psi_{N\alpha}),\\
A_{n+1}^{\textrm{QED}} & =\mathcal{N}\lim_{s_{N}\to\mathrm{m}^{2}}\bar{\psi}_{n+1}^{\dot{\alpha}}(k_{Nm}\sigma_{\alpha\dot{\alpha}}^{m}\Psi_{N}^{\alpha}+\mathrm{m}\bar{\Psi}_{P\dot{\alpha}}),\\
A_{n+1}^{\textrm{QED}} & =\mathcal{N}\lim_{s_{N}\to0}\varepsilon_{n+1}^{m}s_{N}\mathcal{A}_{Nm},\label{eq:QEDvectoramplitude}
\end{align}
\end{subequations}evaluated with the currents defined in \eqref{eq:QEDpsirecursion},
\eqref{eq:QEDpsibarrecursion}, and \eqref{eq:QEDgaugerecursion}.
As an example, we can compute the three-point amplitude with polarizations
$\{\varepsilon_{1},\psi_{2},\bar{\psi}_{3}\}$ and the four-point
amplitude with polarizations $\{\psi_{1},\psi_{2},\bar{\psi}_{3},\bar{\psi}_{4}\}$.
They are given by
\begin{equation}
A_{3}^{\textrm{QED}}(\varepsilon_{1},\psi_{2},\bar{\psi}_{3})=-\mathcal{N}e\varepsilon_{1m}(\psi_{2}\sigma^{m}\bar{\psi}_{3}),
\end{equation}
and
\begin{equation}
A_{4}^{\textrm{QED}}(\psi_{1},\psi_{2},\bar{\psi}_{3},\bar{\psi}_{4})=\mathcal{N}\left(\frac{e^{2}}{s_{13}}(\psi_{1}\sigma^{m}\bar{\psi}_{3})(\psi_{2}\sigma^{m}\bar{\psi}_{4})+\frac{e^{2}}{s_{23}}(\psi_{2}\sigma^{m}\bar{\psi}_{3})(\psi_{1}\sigma^{m}\bar{\psi}_{4})\right),\label{eq:4pointQED}
\end{equation}
which match known QED amplitudes up to the normalization $\mathcal{N}$.
Note, in particular, that we cannot use equation \eqref{eq:QEDvectoramplitude}
to compute \eqref{eq:4pointQED}, since there are no photons in the
external legs, i.e., $\varepsilon_{p}=0$. Regarding the residual
gauge transformations of the single-particle polarizations in the
Lorenz gauge, $A_{n+1}^{\textrm{QED}}$ should be invariant under
$\delta\varepsilon_{p}=\mathrm{i}k_{p}\lambda$. In \eqref{eq:QEDvectoramplitude}
we can directly check the gauge invariance of the amplitude with respect
to the residual transformation of the polarization $\varepsilon_{n+1}^{m}$,
\begin{align}
\delta A_{n+1}^{\textrm{QED}} & =\mathcal{N}\lim_{s_{N}\to0}\overbrace{k_{n+1}^{m}\lambda_{n+1}}^{\delta\varepsilon_{n+1}^{m}}s_{N}\mathcal{A}_{Nm},\nonumber \\
 & =-\mathcal{N}\lim_{s_{N}\to0}\lambda_{n+1}s_{N}(k_{N}^{m}\mathcal{A}_{Nm}),\nonumber \\
 & =0.
\end{align}
From the first to the second line we have used momentum conservation,
precisely yielding the transversality condition of the multi-particle
current $\mathcal{A}_{Nm}$.

The construction of the classical multi-particle solutions together
with the definition of the tree level scattering amplitudes constitutes
the perturbiner method. In the following sections we are going to
explore the perturbiner in increasingly complex field-theoretic models.

\section{Perturbiner in gauge theories and relation to Berends--Giele currents\label{sec:gauge-theories}}

In the theories discussed in section \ref{sec:multiparticle solutions},
the common ingredient of all multi-particle solutions was the plane
wave $e^{\mathrm{i}k_{P}\cdot x}$. And it is easy to understand why
they fit in a recursive construction involving polynomial interactions,
since the product of plane waves is also a plane wave. We could speculate
on the existence of a similar ingredient in field theories of interest,
embodied by a global group structure. We will call it $T[\lambda]$,
in which $\lambda$ is some parameter. In order to make sense inside
a recursive relation, such a structure has to satisfy a closure property
that can be generically cast as
\[
T[\lambda]T[\tilde{\lambda}]\sim T[f(\lambda,\tilde{\lambda})],
\]
in which $f$ is some function that combines the two parameters. For
the plane wave, $T$ is the exponential function, $\lambda$ ($\tilde{\lambda}$)
is the momentum $k_{m}$ ($\tilde{k}_{m}$), and $f$ is simply $\lambda+\tilde{\lambda}$.

This is all very suggestive, and we are going to consider in this
section Lie-valued fields. We will assume a group $G$, with structure
constants $f_{ab}^{\phantom{ab}c}$ and generators $T_{a}$, with
$a=1,\ldots,\dim(G)$, satisfying
\begin{equation}
[T_{a},T_{b}]=\mathrm{i}f_{ab}^{\phantom{ab}c}T_{c}.
\end{equation}
As a consequence of the group algebra, we can expand any nested product
of structure constants in terms of products of group generators. For
example,
\begin{align}
\mathrm{i}f_{ab}^{\phantom{ab}c}T_{c} & =[T_{a},T_{b}],\\
\mathrm{i}^{2}f_{ab}^{\phantom{ab}e}f_{ec}^{\phantom{ab}d}T_{d} & =[[T_{a},T_{b}],T_{c}],\\
\mathrm{i}^{3}f_{ab}^{\phantom{ab}f}f_{fc}^{\phantom{ab}g}f_{gd}^{\phantom{ab}e}T_{e} & =[[[T_{a},T_{b}],T_{c}],T_{d}],\\
\mathrm{i}^{3}f_{ab}^{\phantom{ab}f}f_{cd}^{\phantom{ab}g}f_{fg}^{\phantom{ab}e}T_{e} & =[[T_{a},T_{b}],[T_{c},T_{d}]].
\end{align}
This feature is connected to the decomposition of scattering amplitudes
in terms of partial amplitudes of colored field theories, which is
a color decomposition \cite{Dixon:1996wi}. We are going to explore the
perturbiner in this context.

Beyond Yang--Mills theory and the bi-adjoint scalar, we will extend
the color-stripped perturbiner construction to $\mathcal{N}=4$ super
Yang--Mills, which has a non-trivial mixture of gauge bosons, scalars,
and fermions. 

\subsection{Yang--Mills theory}

We start with non-Abelian gauge theories, which are at the heart of
the standard model of particle physics. We will work with a Lie-valued
gauge vector $A_{m}=A_{m}^{a}T_{a}$, which will be often referred
to as a gluon, and the Yang--Mills action
\begin{equation}
\mathcal{S}=\frac{1}{g_{\textrm{YM}}^{2}}\int d^{D}x\,\textrm{Tr}\left(-\frac{1}{4}F_{mn}F^{mn}\right).\label{eq:YM-action}
\end{equation}
Here we have the field-strength
\begin{equation}
F_{mn}=\partial_{m}A_{n}-\partial_{n}A_{m}-\mathrm{i}[A_{m},A_{n}],\label{eq:YM-fieldstrength}
\end{equation}
and the coupling constant, $g_{\textrm{YM}}$.

\subsubsection{The color-dressed perturbiner}

The classical dynamics of the gauge field is governed by the Yang--Mills
equation of motion,
\begin{equation}
\partial^{n}F_{mn}=\mathrm{i}[A^{n},F_{mn}],\label{eq:YMeom}
\end{equation}
which can be recast as
\begin{align}
\Box A_{m}^{a} & =f_{bc}^{\phantom{ab}a}A^{nb}\partial_{m}A_{n}^{c}-2f_{bc}^{\phantom{ab}a}A^{nb}\partial_{n}A_{m}^{c}-f_{be}^{\phantom{ab}a}f_{cd}^{\phantom{ab}e}A^{nb}A_{n}^{c}A_{m}^{d}\nonumber \\
 & +\partial_{m}(\partial^{n}A_{n}^{a})+f_{bc}^{\phantom{ab}a}A_{m}^{b}(\partial^{n}A_{n}^{c}).\label{eq:YMeom-expanded}
\end{align}

We would like to investigate the multi-particle solutions of this
equation, so we follow the procedure described in section \ref{sec:multiparticle solutions}.
First, we choose the Lorenz gauge $\partial^{m}A_{m}^{a}=0$, so the
second line of \eqref{eq:YMeom-expanded} vanishes. We then plug in
the multi-particle ansatz
\begin{equation}
A_{m}^{a}(x)=\sum_{P}\mathcal{A}_{Pm}^{a}e^{\mathrm{i}k_{P}\cdot x},\label{eq:YM-perturbiner-color-dressed}
\end{equation}
and obtain the recursion relation for the multi-particle current,
\begin{align}
\mathcal{A}_{Pm}^{a} & =\frac{\mathrm{i}}{s_{P}}f_{bc}^{\phantom{ab}a}\sum_{P=Q\cup R}[2(k_{R}\cdot\mathcal{A}_{Q}^{b})\mathcal{A}_{Rm}^{c}-k_{Rm}(\mathcal{A}_{Q}^{b}\cdot\mathcal{A}_{R}^{c})]\nonumber \\
 & +\frac{1}{s_{P}}f_{be}^{\phantom{ab}a}f_{cd}^{\phantom{ab}e}\sum_{P=Q\cup R\cup S}(\mathcal{A}_{Q}^{b}\cdot\mathcal{A}_{R}^{c})\mathcal{A}_{Sm}^{d}.\label{eq:YMcolordressedrecursion}
\end{align}
We will use the notation $a^{m}b_{m}=a\cdot b$ whenever convenient.
The currents are conserved in the sense they satisfy $k_{P}\cdot\mathcal{A}_{P}^{a}=0$.

Since there is a residual gauge redundancy for the single-particle
polarizations, we can analyze its multi-particle realization. So let
us consider a multi-particle expansion for the gauge parameter,
\begin{equation}
\lambda^{a}=\sum_{P}\Lambda_{P}^{a}e^{\mathrm{i}k_{P}\cdot x}.\label{eq:multi-particleparameterYM}
\end{equation}
We know that the gauge field transforms as
\begin{equation}
\delta A_{m}=\partial_{m}\lambda-\mathrm{i}[A_{m},\lambda],\label{eq:YM-gaugetransf}
\end{equation}
and we would like to find a solution for the coefficients $\Lambda_{P}$
such that the Lorenz gauge is preserved, i.e.,
\begin{equation}
\partial^{m}\delta A_{m}=0.
\end{equation}
After plugging in the expansion \eqref{eq:multi-particle-chargeconservation}
in this equation, we obtain
\begin{equation}
\Lambda_{P}^{a}=\frac{\mathrm{i}}{s_{P}}f_{bc}^{\phantom{ab}a}\sum_{P=Q\cup R}(k_{R}\cdot\mathcal{A}_{Q}^{b})\Lambda_{R}^{c}.\label{eq:multigaugeYM}
\end{equation}
This recursion determines the multi-particle gauge parameters in terms
of the single-particle ones, such that the Lorenz gauge is preserved.

Finally, we can define tree level scattering amplitudes for gluons.
Similarly to what was done for the photons in \eqref{eq:QEDvectoramplitude},
we introduce
\begin{equation}
A_{n+1}^{\textrm{YM}}=\gamma_{ab}\lim_{s_{N}\to0}s_{N}(\varepsilon_{n+1}^{a}\cdot\mathcal{A}_{N}^{b}),\label{eq:Nptgluonamplitude}
\end{equation}
with $\gamma_{ab}=\textrm{Tr}(T_{a}T_{b})$, which is often normalized
to $\delta_{ab}$. For the sake of simplicity, we are taking the overall
normalization in \eqref{eq:Nptgluonamplitude} to be $1$. In section
\ref{sec:gravity} we will come back to this discussion. Given the
multi-particle gauge parameters of equation \eqref{eq:multigaugeYM},
it is almost trivial to verify the residual gauge invariance of the
full amplitude. Given that the single-particle polarizations transform
as $\delta\varepsilon_{p}^{m}=\mathrm{i}k_{p}^{m}\lambda_{p}$, we
have
\begin{equation}
\delta\mathcal{A}_{Pm}^{a}=\mathrm{i}k_{Pm}\Lambda_{P}+f_{bc}^{\phantom{bc}a}\sum_{P=Q\cup R}\mathcal{A}_{Qm}^{b}\Lambda_{R}^{c}.
\end{equation}
By replacing this expression in the transformation of the amplitude,
we obtain
\begin{align}
\delta A_{n+1}^{\textrm{YM}} & =\mathrm{i}\gamma_{ab}\lim_{s_{N}\to0}s_{N}[\lambda_{n+1}^{a}(k_{n+1}\cdot\mathcal{A}_{1\ldots n}^{b})+(\varepsilon_{n+1}^{a}\cdot k_{1\ldots n})\Lambda_{1\ldots n}]\nonumber \\
 & +\gamma_{ab}\lim_{s_{N}\to0}s_{N}\sum_{1\ldots n=Q\cup R}f_{cd}^{\phantom{cd}b}(\varepsilon_{n+1}^{a}\cdot\mathcal{A}_{Q}^{c})\Lambda_{R}^{d}.
\end{align}
The first line vanishes using momentum conservation and the transversality
of $\varepsilon_{n+1}$ and $\mathcal{A}_{1\ldots n}$. The second
line vanishes in the limit $s_{N}\to0$. Therefore, the prescription
\eqref{eq:Nptgluonamplitude} is invariant under the residual gauge
transformations of the Lorenz gauge.

\subsubsection{Berends--Giele currents}

Until the mid 80's, tree level computations for gluon amplitudes relied
on Feynman diagrams, and it is easy to imagine how cumbersome they
become as we increase the number of external legs. Beyond the compact
formula for MHV amplitudes discovered by Parke and Taylor \cite{Parke:1986gb},
the state of the art by 1987 in a covariant formulation was the six-gluon
amplitude \cite{Mangano:1987vj,Berends:1987cv}. It was then observed
that gluon amplitudes admit a color decomposition, which is given
by
\begin{equation}
A_{n+1}^{\textrm{YM}}=\sum_{\textrm{perm}\{1,\ldots,n\}}\textrm{Tr}(T_{a_{1}}\ldots T_{a_{n}}T_{a_{n+1}})\mathscr{C}(1,\ldots,n,n+1).\label{eq:YM-colordecomposition}
\end{equation}
The so-called \emph{partial} amplitudes, $\mathscr{C}$, are a function
of the gluon momenta and polarizations. As it turns out, they can
be expressed as
\begin{equation}
\mathscr{C}(1,\ldots,n,n+1)=s_{N}J(1,\ldots,n)\cdot\varepsilon_{n+1},\label{eq:BGpartialamplitude}
\end{equation}
where the objects $J_{m}(1,\ldots,n)$ are recursively defined and
satisfy the conservation equation
\begin{equation}
k_{1\ldots n}\cdot J(1,\ldots,n)=0.
\end{equation}
These are the famous Berends--Giele currents \cite{Berends:1987me},
and we will see their expression soon.

Apart from the color structure, the partial amplitude \eqref{eq:BGpartialamplitude}
clearly resembles the gluon amplitude of equation \eqref{eq:Nptgluonamplitude}.
In this case, the Berends-Giele currents $J_{m}(1,\ldots,n)$ should
be related to the multi-particle currents $\mathcal{A}_{Pm}^{a}$.
In order to see this, it is useful to start with a couple of examples.

First, let us analyze the two-particle current
\begin{equation}
\mathcal{A}_{12m}^{a}=\frac{\mathrm{i}}{s_{12}}f_{bc}^{\phantom{ab}a}[2(k_{2}\cdot\varepsilon_{1}^{b})\varepsilon_{2m}^{c}-2(k_{1}\cdot\varepsilon_{2}^{c})\varepsilon_{1m}^{b}+(k_{1m}-k_{2m})(\varepsilon_{1}^{b}\cdot\varepsilon_{2}^{c})].
\end{equation}
By reintroducing the group generators, it can be reexpressed as
\begin{align}
\mathcal{A}_{12m} & =  \tilde{\mathcal{A}}_{12m}+\tilde{\mathcal{A}}_{21m},\\
\tilde{\mathcal{A}}_{ij}^{m} & =  \frac{1}{s_{ij}}[2(k_{j}\cdot\varepsilon_{i})\varepsilon_{j}^{m}-2\varepsilon_{i}^{m}(k_{i}\cdot\varepsilon_{j})+(k_{i}^{m}-k_{j}^{m})(\varepsilon_{i}\cdot\varepsilon_{j})]\label{eq:YM-2particle-Lievalued}
\end{align}
 with $\mathcal{A}_{12m}=\mathcal{A}_{12m}^{a}T_{a}$ and $\varepsilon_{i}=\varepsilon_{i}^{a}T_{a}$.
A similar decomposition can be done for the three--particle current,
\begin{align}
\mathcal{A}_{123m}^{a}= & +\frac{\mathrm{i}}{s_{123}}f_{bc}^{\phantom{ab}a}\left[2(k_{3}\cdot\mathcal{A}_{12}^{b})\varepsilon_{3m}^{c}-2(k_{12}\cdot\varepsilon_{3}^{c})\mathcal{A}_{12m}^{b}+(k_{12m}-k_{3m})(\mathcal{A}_{12}^{b}\cdot\varepsilon_{3}^{c})\right]\nonumber \\
 & +\frac{\mathrm{i}}{s_{123}}f_{bc}^{\phantom{ab}a}\left[2(k_{2}\cdot\mathcal{A}_{13}^{b})\varepsilon_{2m}^{c}-2(k_{13}\cdot\varepsilon_{2}^{c})\mathcal{A}_{13m}^{b}+(k_{13m}-k_{2m})(\mathcal{A}_{13}^{b}\cdot\varepsilon_{2}^{c})\right]\nonumber \\
 & +\frac{\mathrm{i}}{s_{123}}f_{bc}^{\phantom{ab}a}\left[2(k_{1}\cdot\mathcal{A}_{23}^{b})\varepsilon_{1m}^{c}-2(k_{23}\cdot\varepsilon_{1}^{c})\mathcal{A}_{23m}^{b}+(k_{23m}-k_{1m})(\mathcal{A}_{23}^{b}\cdot\varepsilon_{1}^{c})\right]\nonumber \\
 & +\frac{1}{s_{123}}f_{be}^{\phantom{ab}a}f_{cd}^{\phantom{ab}e}[(\varepsilon_{1}^{b}\cdot\varepsilon_{2}^{c})\varepsilon_{3m}^{d}+(\varepsilon_{1}^{b}\cdot\varepsilon_{3}^{c})\varepsilon_{2m}^{d}]\nonumber \\
 & +\frac{1}{s_{123}}f_{be}^{\phantom{ab}a}f_{cd}^{\phantom{ab}e}[(\varepsilon_{2}^{b}\cdot\varepsilon_{1}^{c})\varepsilon_{3m}^{d}+(\varepsilon_{2}^{b}\cdot\varepsilon_{3}^{c})\varepsilon_{1m}^{d}]\nonumber \\
 & +\frac{1}{s_{123}}f_{be}^{\phantom{ab}a}f_{cd}^{\phantom{ab}e}[(\varepsilon_{3}^{b}\cdot\varepsilon_{1}^{c})\varepsilon_{2m}^{d}+(\varepsilon_{3}^{b}\cdot\varepsilon_{2}^{c})\varepsilon_{1m}^{d}],
\end{align}
which can be recast as
\begin{equation}
\mathcal{A}_{123}^{m}=\tilde{\mathcal{A}}_{123}^{m}+\tilde{\mathcal{A}}_{231}^{m}+\tilde{\mathcal{A}}_{312}^{m}+\tilde{\mathcal{A}}_{132}^{m}+\tilde{\mathcal{A}}_{321}^{m}+\tilde{\mathcal{A}}_{213}^{m},
\end{equation}
where $\mathcal{A}_{123m}=\mathcal{A}_{123m}^{a}T_{a}$, and 
\begin{multline}
\tilde{\mathcal{A}}_{ijk}^{m}=+\frac{1}{s_{ijk}}[2(k_{k}\cdot\tilde{\mathcal{A}}_{ij})\varepsilon_{k}^{m}-2\tilde{\mathcal{A}}_{ij}^{m}(k_{ij}\cdot\varepsilon_{k})+(k_{ij}^{m}-k_{k}^{m})(\tilde{\mathcal{A}}_{ij}\cdot\varepsilon_{k})]\\
+\frac{1}{s_{ijk}}[2(k_{jk}\cdot\varepsilon_{i})\tilde{\mathcal{A}}_{jk}^{m}-2\varepsilon_{i}^{m}(k_{i}\cdot\tilde{\mathcal{A}}_{jk})+(k_{i}^{m}-k_{jk}^{m})(\varepsilon_{i}\cdot\tilde{\mathcal{A}}_{jk})]\\
+\frac{1}{s_{ijk}}(\varepsilon_{i}^{n}\varepsilon_{j}^{m}\varepsilon_{kn}-\varepsilon_{in}\varepsilon_{j}^{n}\varepsilon_{k}^{m}+\varepsilon_{in}\varepsilon_{j}^{m}\varepsilon_{k}^{n}-\varepsilon_{im}\varepsilon_{j}^{n}\varepsilon_{k}^{m}).\label{eq:YM-3particle-Lievalued}
\end{multline}
Note that while the current $\mathcal{A}_{P}^{m}$ is expressed in
terms of ordered words $P=p_{1}\ldots p_{n}$, with $p_{1}<\ldots<p_{n}$,
the currents $\tilde{\mathcal{A}}_{m}$ can have different orderings,
e.g. $\tilde{\mathcal{A}}_{21}^{m}$ and $\tilde{\mathcal{A}}_{312}^{m}$.
This has to be the case, since the order of the single-particle labels
determines the order in which the group generators appear. The combinations
of different orderings precisely yield the commutators of the group
generators.

The identification of the Berends--Giele currents $J_{m}(1,\ldots,n)$
with $\tilde{\mathcal{A}}_{m}$ is strongly suggested, though we have
to strip the color structure of the latter. As we will see next, this
is elegantly accomplished with the perturbiner method.

\subsubsection{The color-stripped perturbiner}

The basic idea behind the color-stripped perturbiner is to remove
the group structure from the multi-particle currents \cite{Mafra:2015vca}. Inspired by
the discussion above, let us consider the following ansatz,
\begin{equation}
A^{m}(x)=\sum_{P}\boldsymbol{A}_{P}^{m}e^{\mathrm{i}k_{P}\cdot x}T^{a_{P}},\label{eq:CS-YM-perturbiner}
\end{equation}
where $P,Q,\ldots$ denote unordered words\footnote{Whether a given word is ordered or unordered will be clear from the
context, so there will be no distinction in their abstract representation. }, and
\begin{equation}
T^{a_{P}}=T^{a_{p_{1}}}\ldots T^{a_{p_{n}}}.
\end{equation}
The information about the order in which the group generators appear
is now contained in $P$. The momenta in the expansion are of course
blind to this ordering, so we have, for instance, $k_{12}=k_{21}$.

Using \eqref{eq:CS-YM-perturbiner}, commutators of Lie-valued objects
are expressed in a simple form. For example,
\begin{align}
[A^{m}(x),B^{n}(x)] & =\sum_{Q,R}\boldsymbol{A}_{Q}^{m}\boldsymbol{B}_{R}^{n}e^{\mathrm{i}k_{QR}\cdot x}[T^{a_{Q}},T^{a_{R}}],\nonumber \\
 & =\sum_{Q,R}(\boldsymbol{A}_{Q}^{m}\boldsymbol{B}_{R}^{n}-\boldsymbol{B}_{Q}^{n}\boldsymbol{A}_{R}^{m})e^{\mathrm{i}k_{QR}\cdot x}T^{a_{Q}}T^{a_{R}},\nonumber \\
 & =\sum_{P}e^{\mathrm{i}k_{P}\cdot x}T^{a_{P}}\sum_{{Q,R\atop QR=P}}(\boldsymbol{A}_{Q}^{m}\boldsymbol{B}_{R}^{n}-\boldsymbol{B}_{Q}^{n}\boldsymbol{A}_{R}^{m}).
\end{align}
From the second to the third line, we are introducing a new constrained
sum. It goes over all pairs $(Q,R)$ of non-empty words that can be
concatenated to form the word $P$. Like in equation \eqref{eq:shorthand-ordered},
we will introduce the shorthand notation
\begin{equation}
\sum_{P=P_{1}\ldots P_{n}}=\sum_{{P_{1},P_{2},\ldots,P_{n}\atop P_{1}\ldots P_{n}=P}},
\end{equation}
where the sum goes over all possible deconcatenations of the word
$P$ into $n$ non-empty sub-words $P_{i}$.

In order to derive the recursion for the current $\boldsymbol{A}_{P}^{m}$,
we simply substitute the ansatz \eqref{eq:CS-YM-perturbiner} into
the Yang--Mills equation of motion. For convenience, we will work
in the Lorenz gauge, $\partial^{m}A_{m}=0$, such that
\begin{multline}
\boldsymbol{A}_{Pm}=\frac{1}{s_{P}}\sum_{P=QR}[2(k_{R}\cdot\boldsymbol{A}_{Q})\boldsymbol{A}_{Rm}-(\boldsymbol{A}_{Q}\cdot\boldsymbol{A}_{R})k_{Rm}-(Q\leftrightarrow R)]\\
+\frac{1}{s_{P}}\sum_{P=QRS}[(\boldsymbol{A}_{Q}\cdot\boldsymbol{A}_{S})\mathbf{A}_{Rm}-(\boldsymbol{A}_{R}\cdot\boldsymbol{A}_{S})\mathbf{A}_{Qm}+(Q\leftrightarrow S)].\label{eq:BG-YM}
\end{multline}
The multi-particle coefficients $\boldsymbol{A}_{Pm}$ match the Berends--Giele
currents $J_{m}$, with color ordering given by the word $P$. The
conservation of the current follows directly from the gauge choice,
$k_{P}\cdot\boldsymbol{A}_{P}=0$, which is usually a good consistency
check when performing computations.

Let us look at some examples. Starting with $P=12$, we get
\begin{equation}
\boldsymbol{A}_{12m}=\frac{1}{s_{12}}[2(k_{2}\cdot\varepsilon_{1})\varepsilon_{2m}-2(k_{1}\cdot\epsilon_{2})\varepsilon_{1m}+(\varepsilon_{1}\cdot\varepsilon_{2})(k_{1m}-k_{2m})],\label{eq:CS-YM-2particle}
\end{equation}
which matches the \emph{color-stripped} version of equation \eqref{eq:YM-2particle-Lievalued}.
It is easy to check that $k_{12}\cdot\boldsymbol{A}_{12}=0$ using
the transversality of the single-particle polarizations. For $P=123$,
we obtain
\begin{multline}
\boldsymbol{A}_{123m}=+\frac{1}{s_{123}}[2(k_{23}\cdot\varepsilon_{1})\boldsymbol{A}_{23m}-2(k_{1}\cdot\boldsymbol{A}_{23})\varepsilon_{1m}+(\varepsilon_{1}\cdot\boldsymbol{A}_{23})(k_{1m}-k_{23m})]\\
+\frac{1}{s_{123}}[2(k_{3}\cdot\boldsymbol{A}_{12})\varepsilon_{3m}-2(k_{12}\cdot\varepsilon_{3})\boldsymbol{A}_{12m}+(\boldsymbol{A}_{12}\cdot\varepsilon_{3})(k_{12m}-k_{3m})]\\
+\frac{1}{s_{123}}[2(\varepsilon_{1}\cdot\varepsilon_{3})\varepsilon_{2m}-(\varepsilon_{2}\cdot\varepsilon_{3})\varepsilon_{1m}-(\varepsilon_{1}\cdot\varepsilon_{2})\varepsilon_{3m}],\label{eq:CS-YM-3particle}
\end{multline}
which satisfies $k_{123}\cdot\boldsymbol{A}_{123}=0$ and matches
the color-stripped version of equation \eqref{eq:YM-3particle-Lievalued}.
The path to equation \eqref{eq:BG-YM} is a first principles derivation
of the Berends--Giele recursion relations \cite{Berends:1987me}.

There is one last point to address before discussing amplitudes in
the color-stripped framework. The left hand side of equation \eqref{eq:CS-YM-perturbiner}
is Lie algebra valued. On the right hand side this is less obvious,
since we have products of generators $T^{a}$ rather than commutators.
The latter, however, are in-built in the recursive definition of the
multi-particle currents, cf. equation \eqref{eq:BG-YM}. For example,
consider the two-particle currents with labels $i$ and $j$. Because
$\boldsymbol{A}_{ji}^{m}=-\boldsymbol{A}_{ij}^{m}$, their contribution
to $A^{m}(x)$ comes with $e^{\mathrm{i}k_{ij}\cdot x}$ multiplied
by
\begin{equation}
\boldsymbol{A}_{ij}^{m}T^{a_{i}}T^{a_{j}}+\boldsymbol{A}_{ji}^{m}T^{a_{j}}T^{a_{i}}=\boldsymbol{A}_{ij}^{m}[T^{a_{i}},T^{a_{j}}],
\end{equation}
similar to equation \eqref{eq:YM-2particle-Lievalued}. In the case
of three-particle currents, we have
\begin{multline}
\boldsymbol{A}_{ijk}^{m}T^{a_{i}}T^{a_{j}}T^{a_{k}}+\boldsymbol{A}_{jki}^{m}T^{a_{j}}T^{a_{k}}T^{a_{i}}+\boldsymbol{A}_{kij}^{m}T^{a_{k}}T^{a_{i}}T^{a_{j}}\\
+\boldsymbol{A}_{kji}^{m}T^{a_{k}}T^{a_{j}}T^{a_{i}}+\boldsymbol{A}_{jik}^{m}T^{a_{j}}T^{a_{i}}T^{a_{k}}+\boldsymbol{A}_{ikj}^{m}T^{a_{i}}T^{a_{k}}T^{a_{j}}\\
=\frac{1}{3}\left(\boldsymbol{A}_{ijk}^{m}-\boldsymbol{A}_{kij}^{m}\right)[[T^{a_{i}},T^{a_{j}}],T^{a_{k}}]+\textrm{cyclic}(i,j,k),
\end{multline}
which follows from the identities\begin{subequations}\label{eq:3pshuffle}
\begin{align}
\boldsymbol{A}_{ijk}^{m}+\boldsymbol{A}_{jik}^{m}+\boldsymbol{A}_{jki}^{m} & =0,\\
\boldsymbol{A}_{ijk}^{m}+\boldsymbol{A}_{ikj}^{m}+\boldsymbol{A}_{kij}^{m} & =0.
\end{align}
\end{subequations}More generally, the multi-particle currents of
\eqref{eq:BG-YM} satisfy the shuffle identities, denoted by
\begin{equation}
\boldsymbol{A}_{Q\shuffle R}^{m}=0.\label{eq:shuffle-identity}
\end{equation}
There is a small abuse of notation here. The operation $Q\shuffle R$
is defined as the sum of the possible shuffles of the two words, $Q=q_{1}\ldots q_{|Q|}$
and $R=r_{1}\ldots r_{|R|}$. It satisfies
\begin{equation}
P\shuffle\emptyset=\emptyset\shuffle P=P,
\end{equation}
and
\begin{equation}
q_{1}\ldots q_{|Q|}\shuffle r_{1}\ldots r_{|R|}=q_{1}(q_{2}\ldots q_{|Q|}\shuffle r_{1}\ldots r_{|R|})+r_{1}(r_{2}\ldots r_{|R|}\shuffle q_{1}\ldots q_{|Q|}).
\end{equation}
Therefore, the identity \eqref{eq:shuffle-identity} is defined as
the sum over multi-particle currents labeled by the unordered words
generated in the shuffle (see \cite{Mafra:2016ltu} and references
therein). For example,
\begin{equation}
\boldsymbol{A}_{i\shuffle jk}^{m}=\boldsymbol{A}_{ijk}^{m}+\boldsymbol{A}_{jik}^{m}+\boldsymbol{A}_{jki}^{m}.
\end{equation}
The shuffle identities manifest the fact that the color-stripped ansatz
\eqref{eq:CS-YM-perturbiner} is Lie algebra valued and, therefore,
well defined.

\subsubsection{Partial amplitudes and Kleiss--Kuijf relations}

As we have already discussed, the scattering amplitudes in the Yang--Mills
theory can be decomposed according to the color structure, cf. equation
\eqref{eq:YM-colordecomposition}. The partial amplitudes are defined
as
\begin{equation}
\mathscr{C}(1,\ldots,n,n+1)=\lim_{s_{1\ldots n}\to0}s_{1\ldots n}(\varepsilon_{n+1}\cdot\boldsymbol{A}_{1\ldots n}).\label{eq:YMpartialamplitude}
\end{equation}
Besides being cyclic,
\begin{equation}
\mathscr{C}(1,\ldots,n,n+1)=\mathscr{C}(n+1,1,\ldots,n),
\end{equation}
they satisfy a reflection property,
\begin{equation}
\mathscr{C}(1,\ldots,n,n+1)=-(-1)^{n}\mathscr{C}(n+1,n,\ldots,1).
\end{equation}
These properties follow directly from the group structure, via the
trace of a product of the respective group generators \cite{DelDuca:1999rs}.

Let us have a look at some examples. The three-point partial amplitude
is given by
\begin{equation}
\mathscr{C}(1,2,3)=\varepsilon_{1}^{m}\varepsilon_{2}^{n}\varepsilon_{3}^{p}V_{mnp},
\end{equation}
in which we used momentum conservation to make the three-point gluon
vertex manifest, with
\begin{equation}
V_{mnp}=\eta_{mn}(k_{1}-k_{2})_{p}+\eta_{np}(k_{2}-k_{3})_{m}+\eta_{mp}(k_{3}-k_{1})_{n}.\label{eq:three-gluon-vertex}
\end{equation}
Note that $\mathscr{C}(1,2,3)$ is antisymmetric in the exchange of
any pair of single-particle labels. The full amplitude is simply
\begin{align}
A_{3}^{\textrm{YM}} & =\textrm{Tr}(T_{a_{1}}T_{a_{2}}T_{a_{3}})\mathscr{C}(1,2,3)+\textrm{Tr}(T_{a_{2}}T_{a_{1}}T_{a_{3}})\mathscr{C}(2,1,3),\nonumber \\
 & =\textrm{Tr}([T_{a_{1}},T_{a_{2}}]T_{a_{3}})\mathscr{C}(1,2,3),
\end{align}
which agrees with equation \eqref{eq:Nptgluonamplitude} once we restore
the color index of the polarizations.

The four-point partial amplitude can be simply expressed as
\begin{equation}
\mathscr{C}(1,2,3,4)=s_{12}\boldsymbol{A}_{12}\cdot\boldsymbol{A}_{34}+(\varepsilon_{1}\cdot\varepsilon_{3})(\varepsilon_{2}\cdot\varepsilon_{4})-(\varepsilon_{2}\cdot\varepsilon_{3})(\varepsilon_{1}\cdot\varepsilon_{4})-\{(123)\to(231)\}.
\end{equation}
From the shuffle identities \eqref{eq:3pshuffle}, it follows that
$\mathscr{C}(1,2,3,4)$ satisfies\begin{subequations}
\begin{align}
\mathscr{C}(1,2,3,4)+\mathscr{C}(2,1,3,4)+\mathscr{C}(2,3,1,4) & =0,\\
\mathscr{C}(1,2,3,4)+\mathscr{C}(1,3,2,4)+\mathscr{C}(3,1,2,4) & =0.
\end{align}
\end{subequations}These additional identities were presented in \cite{Kleiss:1988ne},
and became known as Kleiss--Kuijf (KK) relations. They arise when
the Jacobi identities among color factors are translated into relations
among color-ordered tree amplitudes. From the perturbiner construction,
they are a trivial consequence of the shuffle identities. More generally,
we have
\begin{equation}
\lim_{s_{QR}\to0}s_{QR}\left(\varepsilon_{n+1}\cdot\boldsymbol{A}_{Q\shuffle R}\right)=0,
\end{equation}
which directly relate different orderings of the partial amplitudes
to the shuffle operation.

In reference \cite{Mafra:2022wml} the reader can find a deep and
detailed construction of the color-stripped multi-particle currents
of super Yang--Mills theory, with a thorough discussion of many properties
that go beyond our scope here.

\subsection{Bi-adjoint scalar}

We will now analyze a special class of fields, the bi-adjoint scalars.
Their relevance in modern scattering amplitudes can hardly be overstated,
as they are ubiquitous across modern amplitude methods and properties.
To name a few: (1) bi-adjoint scalars are the backbone of the CHY
formalism \cite{Cachazo:2013iea}; (2) in the field theory double copy, the inverse of the
KLT kernel is precisely the matrix of bi-adjoint scalar amplitudes;
(3) more broadly in the double copy context, bi-adjoint scalars play
the role of the identity; and (4) bi-adjoint scalars have simple soft
limits, often used as templates in theories with more intricate dynamics.
The list goes on, but our focus here is more modest. We are interested
in classical multi-particle solutions.

As the name suggests, a bi-adjoint scalar $\phi$ is a spacetime scalar
that transforms in the adjoint representations of two distinct groups.
We have
\begin{equation}
\phi=\phi_{a\tilde{a}}T^{a}\otimes\tilde{T}^{\tilde{a}},
\end{equation}
where $T$ $(\tilde{T}$) is a group generator in the adjoint representation.
Their equation of motion is simply
\begin{align}
\Box\phi & +\frac{1}{2}\llbracket\phi,\phi\rrbracket=0,\label{eq:BA-eom}
\end{align}
where we define
\begin{equation}
\llbracket T^{a}\otimes\tilde{T}^{\tilde{a}},T^{b}\otimes\tilde{T}^{\tilde{b}}\rrbracket=[T^{a},T^{b}]\otimes[\tilde{T}^{\tilde{a}},\tilde{T}^{\tilde{b}}].
\end{equation}

The multi-particle solutions of \eqref{eq:BA-eom} can be studied
in a color-dressed or in a color-stripped form. We will present the
latter, being the more relevant in modern literature \cite{Mafra:2016ltu}. We start with
the ansatz
\begin{equation}
\phi=\sum_{P,Q}\Phi_{P|Q}e^{\mathrm{i}k_{P}\cdot x}T^{a_{P}}\otimes\tilde{T}^{\tilde{a}_{Q}},\label{eq:BA-multiansatz}
\end{equation}
where the single-particle states are just massless scalars with polarization
$\Phi_{p|p}=\phi_{p}$. Note now that we have two orderings, associated to the words $P$ and
$Q$. In addition, \eqref{eq:BA-multiansatz} only makes sense if
$P$ and $Q$ are composed of the same single-particle labels, i.e.,
they are related through permutations of the letters. For example,
we cannot have a non-vanishing coefficient $\Phi_{12|13}$. This is
ultimately explained by the fact that single-particle polarizations
are identified with a unique letter.

The recursive definition of $\Phi_{P|Q}$ is straightforward to obtain
from the equation of motion. We have
\begin{align}
\Box\phi & =-\sum_{P,Q}s_{P}\Phi_{P|Q}e^{\mathrm{i}k_{P}\cdot x}T^{a_{P}}\otimes\tilde{T}^{\tilde{a}_{Q}},\\
\llbracket\phi,\phi\rrbracket & =2\sum_{P,Q}e^{\mathrm{i}k_{P}\cdot x}T^{a_{P}}\otimes\tilde{T}^{\tilde{a}_{Q}}\left(\sum_{P=RS}\sum_{Q=TU}\Phi_{R|T}\Phi_{S|U}-\Phi_{S|T}\Phi_{R|U}\right),
\end{align}
which leads to
\begin{equation}
\Phi_{P|Q}=\frac{1}{s_{P}}\sum_{P=RS}\sum_{Q=TU}\Phi_{R|T}\Phi_{S|U}-(R\leftrightarrow S).\label{eq:BA-recursion}
\end{equation}

Finally, following the usual perturbiner prescription, we can define the
 double-partial amplitude,
\begin{equation}
m(Pn|Qn)=\lim_{s_{P}\to0}\phi_{n|n}s_{P}\Phi_{P|Q},\label{eq:BA-amplitudes}
\end{equation}
which satisfies the usual cyclicity, reflection, and Kleiss--Kuijf
relations for the two arguments, $Pn$ and $Qn$, separately.

For simplicity, we can set $\phi_{p|p}=1$. The three-point and four-point
double-partial amplitudes are given by
\begin{align}
m(123|123) & =1,\\
m(1234|1234) & =\frac{1}{s_{12}}+\frac{1}{s_{23}},\\
m(1234|2134) & =-\frac{1}{s_{12}},\\
m(1234|1324) & =-\frac{1}{s_{23}}.
\end{align}
Different orderings can be obtained via label permutation or using
the color ordering properties. For example, the KK identity
\begin{equation}
m(123|123)+m(123|213)=0,
\end{equation}
leads to $m(123|213)=-1$.

We refer the reader to reference \cite{Mafra:2016ltu} for additional
details on the bi-adjoint model in the perturbiner framework, in particular
regarding its connection to color-kinematics duality and the double
copy.

\subsection{$\mathcal{N}=4$ super Yang--Mills}

We will discuss now the (maximally supersymmetric) $\mathcal{N}=4$
super Yang--Mills theory, which is the four-dimensional theory with
the highest number of supercharges without gravity. This property
leads to highly constrained interactions and a vanishing beta function.
The theory is finite to all orders in perturbation theory for any
gauge group. There is no need for coupling constant renormalization,
making it an ideal laboratory for testing computational techniques.
This scale invariance follows from the ``bigger'' superconformal
symmetry, and enables powerful symmetry-based methods such as dual
superconformal symmetry, Yangian structures (integrability), twistor
geometry, and other modern amplitude techniques (BCFW recursion, amplituhedron,
etc.). More broadly, $\mathcal{N}=4$ super Yang--Mills is dual to
type II-B string theory on $AdS_{5}\times S^{5}$, the core representative
of the AdS/CFT correspondence, linking strongly coupled gauge theory
to classical gravity.

The study of supersymmetric Yang--Mills (SYM) theory appeared early
on in the perturbiner context \cite{Selivanov:1998hn}. Within our
modest approach, it is a simple example of a color ordered field theory
that involves a diverse field content: gauge field, spinors, and scalars.
Since we are not worried about non-perturbative aspects, masslessness
and exact conformal invariance are not really an obstruction regarding
the asymptotic states of the free theory, i.e., the building blocks
of the classical multi-particle solutions. This subsection involves
some extra technical steps, but the notation and logical flow should
be compatible with what we have already seen.

Maybe the easiest way to build the $\mathcal{N}=4$ SYM action is
through the dimensional reduction of the $\mathcal{N}=1$ ten-dimensional
SYM theory \cite{Brink:1976bc}, which appears in the field-theory limit of the open superstring.
The corresponding action is given by
\begin{equation}
S=\frac{1}{g_{\textrm{YM}}^{2}}\int d^{10}x\,\textrm{Tr}\left(-\frac{1}{4}F_{MN}F^{MN}+\frac{\mathrm{i}}{2}\lambda^{a}\gamma_{ab}^{M}\nabla_{M}\lambda^{b}\right),\label{eq:SYM10Daction}
\end{equation}
where the field-strength $F_{MN}$ is defined in analogously to equation
\eqref{eq:YM-fieldstrength} ($M,N=0,\ldots,9$ denote the spacetime
directions), and $\lambda^{a}$ is a Majorana--Weyl spinor, with
$a,b=1,\ldots,16$. The $\gamma$ matrices are the ten-dimensional
analogue of the Pauli matrices, satisfying
\begin{align}
\gamma_{bc}^{M}\gamma^{Nca}+\gamma_{bc}^{N}\gamma^{Mca} & =\eta^{MN}\delta_{b}^{a},\\
\gamma_{ab}^{M}\gamma_{Mcd}+\gamma_{bc}^{M}\gamma_{Mad}+\gamma_{ca}^{M}\gamma_{Mbd} & =0.
\end{align}
The covariant derivative on the spinor is defined as
\begin{equation}
\nabla_{M}\lambda^{a}=\partial_{M}\lambda^{a}-\mathrm{i}[A_{M},\lambda^{a}].
\end{equation}

The action \eqref{eq:SYM10Daction} is invariant under the global
transformations
\begin{align}
\delta_{\textrm{susy}}A_{M} & =\mathrm{i}(\xi\gamma_{M}\lambda),\\
\delta_{\textrm{susy}}\lambda^{a} & =\frac{1}{2}F_{MN}(\gamma^{MN}\xi)^{a},
\end{align}
where $\xi^{a}$ is the supersymmetry parameter.

The dimensional reduction of \eqref{eq:SYM10Daction} is straightforward.
We will decompose the vector indices as $M=\{m,I\}$, with $m=0,1,2,3$
denoting the four spacetime directions, and $I=1,\ldots,6$ denoting
the extra dimensions. The more involved part involves the spinorial
representation. The $16$ components of the spinor $\lambda^{a}$
are mapped to four Dirac spinors with conjugate chiral 
components\footnote{The four-dimensional notation in discussed in subsection \ref{subsec:WZmodel}.}
$\lambda^{i\alpha}$ ($\bar{\lambda}_{i}^{\dot{\alpha}}$). They are
in the (anti) fundamental representation of the so-called $R$-symmetry
group, $SU(4)$, with $i=1,2,3,4$ . This mapping can be systematically
described through the projectors
\[
P_{a}^{i\alpha},P_{ai}^{\dot{\alpha}},P_{i\alpha}^{a},P_{\dot{\alpha}}^{ai},
\]
such that\begin{subequations}
\begin{align}
\lambda^{i\alpha} & =\lambda^{a}P_{a}^{i\alpha},\\
\bar{\lambda}_{i}^{\dot{\alpha}} & =\lambda^{a}P_{ai}^{\dot{\alpha}},\\
\lambda_{\alpha}^{i} & =\epsilon_{\alpha\beta}\lambda^{i\beta},\\
\bar{\lambda}_{i\dot{\alpha}} & =\epsilon_{\dot{\alpha}\dot{\beta}}\bar{\lambda}_{i}^{\dot{\beta}},
\end{align}
\end{subequations}The projectors satisfy\begin{subequations}
\begin{align}
P_{i\alpha}^{a}P_{b}^{i\alpha}+P_{\dot{\alpha}}^{ai}P_{bi}^{\dot{\alpha}} & =\delta_{b}^{a},\\
P_{a}^{i\alpha}P_{j\beta}^{a} & =\delta_{j}^{i}\delta_{\beta}^{\alpha},\\
P_{aj}^{\dot{\alpha}}P_{\dot{\beta}}^{ai} & =\delta_{j}^{i}\delta_{\dot{\beta}}^{\dot{\alpha}},\\
P_{a}^{i\alpha}P_{\dot{\alpha}}^{aj} & =0,\\
P_{ai}^{\dot{\alpha}}P_{j\alpha}^{a} & =0.
\end{align}
\end{subequations}The first three equations simply state the completeness
of the decomposition, while the remaining two are orthogonality conditions.
The $\gamma$ matrices, $\gamma_{ab}^{M}=\{\gamma_{ab}^{m},\gamma_{ab}^{I}\}$
and $\gamma_{M}^{ab}=\{\gamma^{mab},\gamma^{Iab}\}$, can be decomposed
as
\begin{eqnarray}
\gamma_{ab}^{I} & = & \gamma_{ij}^{I}P_{a}^{i\alpha}P_{b}^{j\beta}\epsilon_{\alpha\beta}+\bar{\gamma}^{Iij}P_{ai}^{\dot{\alpha}}P_{bj}^{\dot{\beta}}\epsilon_{\dot{\alpha}\dot{\beta}},\\
\gamma_{ab}^{m} & = & \sigma_{\alpha\dot{\alpha}}^{m}(P_{a}^{i\alpha}P_{bi}^{\dot{\alpha}}+P_{b}^{i\alpha}P_{ai}^{\dot{\alpha}}).
\end{eqnarray}
Besides the four-dimensional Pauli matrices $\sigma^{m}$, we have
introduced here the Clebsch--Gordon coefficients $\gamma_{ij}^{I}$
and $\bar{\gamma}^{Iij}$, which satisfy\begin{subequations}
\begin{align}
\gamma_{ij}^{I} & =-\gamma_{ji}^{I},\\
\bar{\gamma}^{Iij} & =(\gamma_{ij}^{I})^{\dagger},\\
\gamma_{ik}^{I}\bar{\gamma}^{Jkj}+\gamma_{ik}^{J}\bar{\gamma}^{Ikj} & =2\delta^{IJ}\delta_{i}^{j},\\
\bar{\gamma}^{Iij} & =\frac{1}{2}\epsilon^{ijkl}\gamma_{kl}^{I},\\
\gamma_{ij}^{I} & =\frac{1}{2}\epsilon_{ijkl}\bar{\gamma}^{Ikl}.
\end{align}
\end{subequations}The Levi--Civita symbol $\epsilon^{ijkl}$ ($\epsilon_{ijkl}$)
is normalized as $\epsilon^{ijkl}\epsilon_{ijkl}=24$. It is then
straightforward to show that\begin{subequations}
\begin{eqnarray}
\gamma_{ac}^{m}\gamma^{ncb}+\gamma_{ac}^{n}\gamma^{mcb} & = & +2\eta^{mn}\delta_{a}^{b},\\
\gamma_{ac}^{m}\gamma^{Icb}+\gamma_{ac}^{I}\gamma^{mcb} & = & 0,\\
\gamma_{ac}^{I}\gamma^{Jcb}+\gamma_{ac}^{J}\gamma^{Icb} & = & +2\delta^{IJ}\delta_{a}^{b},
\end{eqnarray}
\end{subequations}which demonstrates the consistency of the decomposition
with respect to the $\gamma$ matrices algebra.

Finally, the dimensionally reduced action (up to boundary terms) can
be cast as
\begin{multline}
S=\frac{1}{g_{\textrm{YM}}^{2}}\int d^{4}x\,\textrm{Tr}\bigg(-\frac{1}{4}F_{mn}F^{mn}+\mathrm{i}\lambda^{i\alpha}\sigma_{\alpha\dot{\alpha}}^{m}\nabla_{m}\bar{\lambda}_{i}^{\dot{\alpha}}-\frac{1}{2}\nabla_{m}\phi_{I}\nabla^{m}\phi_{I}\\
+\frac{1}{2}\gamma_{ij}^{I}\lambda^{i\alpha}[\phi_{I},\lambda_{\alpha}^{j}]+\frac{1}{2}\bar{\gamma}^{Iij}\bar{\lambda}_{i}^{\dot{\alpha}}[\phi_{I},\bar{\lambda}_{j\dot{\alpha}}]+\frac{1}{4}[\phi_{I},\phi_{J}]^{2}\bigg),
\end{multline}
where $\phi_{I}\equiv A_{I}$. The coefficients $\gamma_{ij}^{I}$
are usually absorbed in the definition of the scalar fields,\begin{subequations}
\begin{align}
\phi_{ij} & \equiv\frac{1}{2}\gamma_{ij}^{I}\phi_{I},\\
\phi^{I} & =-\frac{1}{2}\bar{\gamma}^{Iij}\phi_{ij}\\
\phi^{ij} & =\frac{1}{2}\epsilon^{ijkl}\phi_{kl},
\end{align}
\end{subequations}so the $\mathcal{N}=4$ SYM action is recast as
\begin{multline}
S=\frac{1}{g_{\textrm{YM}}^{2}}\int d^{4}x\,\textrm{Tr}\bigg(-\frac{1}{4}F_{mn}F^{mn}+\mathrm{i}\lambda^{\alpha i}\sigma_{\alpha\dot{\alpha}}^{m}\nabla_{m}\bar{\lambda}_{i}^{\dot{\alpha}}+\frac{1}{2}\nabla_{m}\phi_{ij}\nabla^{m}\phi^{ij}\\
+\lambda^{i\alpha}[\lambda_{\alpha}^{j},\phi_{ij}]+\bar{\lambda}_{i}^{\dot{\alpha}}[\bar{\lambda}_{j\dot{\alpha}},\phi^{ij}]+\frac{1}{4}[\phi^{ij},\phi^{kl}][\phi_{ij},\phi_{kl}]\bigg).\label{eq:N=00003D4SYMaction}
\end{multline}

The extended supersymmetry is now labeled with $SU(4)$ indices, with
parameters $\xi^{i\alpha}$ ($\xi_{i}^{\dot{\alpha}}$). Using the
transformations
\begin{align}
\delta_{\textrm{susy}}A_{m} & =\mathrm{i}\sigma_{m\alpha\dot{\alpha}}(\xi^{i\alpha}\lambda_{i}^{\dot{\alpha}}+\xi_{i}^{\dot{\alpha}}\lambda^{i\alpha}),\\
\delta_{\textrm{susy}}\phi_{ij} & =\mathrm{i}\epsilon_{ijkl}\xi_{\alpha}^{k}\lambda^{l\alpha}-2\mathrm{i}\xi_{[i}^{\dot{\alpha}}\lambda_{j]\dot{\alpha}},\\
\delta_{\textrm{susy}}\lambda^{i\alpha} & =\frac{1}{2}F_{mn}\sigma^{m\dot{\alpha}\alpha}\sigma_{\beta\dot{\alpha}}^{n}\xi^{i\beta}+2\nabla_{m}\phi^{ij}\sigma^{m\dot{\alpha}\alpha}\xi_{j\dot{\alpha}},\\
\delta_{\textrm{susy}}\lambda_{i}^{\dot{\alpha}} & =\frac{1}{2}F_{mn}\sigma^{m\dot{\alpha}\alpha}\sigma_{\alpha\dot{\beta}}^{n}\xi_{i}^{\dot{\beta}}+2\nabla_{m}\phi_{ij}\sigma^{m\dot{\alpha}\alpha}\xi_{\alpha}^{j},
\end{align}
it is possible to demonstrate the invariance of the Lagrangian up
to total derivatives.

The equations of motion derived from \eqref{eq:N=00003D4SYMaction}
are given by\begin{subequations}\label{eq:N=00003D4SYMeom}
\begin{align}
\nabla^{n}F_{mn} & =\mathrm{i}[\nabla_{m}\phi_{ij},\phi^{ij}]-\{\lambda^{\alpha i},\bar{\lambda}_{i}^{\dot{\alpha}}\}\sigma_{m\alpha\dot{\alpha}},\\
\nabla_{m}\nabla^{m}\phi^{ij} & =\{\lambda^{i\alpha},\lambda_{\alpha}^{j}\}+\frac{1}{2}\epsilon^{ijkl}\{\bar{\lambda}_{k}^{\dot{\alpha}},\bar{\lambda}_{l\dot{\alpha}}\}+[\phi_{kl},[\phi^{ij},\phi^{kl}]],\\
\sigma_{\alpha\dot{\alpha}}^{m}\nabla_{m}\lambda^{\alpha i} & =2\mathrm{i}[\phi^{ij},\bar{\lambda}_{j\dot{\alpha}}],\\
\sigma_{\alpha\dot{\alpha}}^{m}\nabla_{m}\bar{\lambda}_{i}^{\dot{\alpha}} & =2\mathrm{i}[\phi_{ij},\lambda_{\alpha}^{j}].
\end{align}
\end{subequations}In addition to \eqref{eq:CS-YM-perturbiner}, we
are going to consider here color-stripped multi-particle expansions
of the form\begin{subequations}
\begin{align}
\lambda^{\alpha i} & =\sum_{P}\Lambda_{P}^{\alpha i}e^{\mathrm{i}k_{P}\cdot x}T^{a_{P}},\\
\bar{\lambda}_{i}^{\dot{\alpha}} & =\sum_{P}\bar{\Lambda}_{Pi}^{\dot{\alpha}}e^{\mathrm{i}k_{P}\cdot x}T^{a_{P}},\\
\phi^{ij} & =\sum_{P}\Phi_{P}^{ij}e^{\mathrm{i}k_{P}\cdot x}T^{a_{P}},
\end{align}
\end{subequations}where the multi-particle coefficients have the
same statistics of the corresponding field. In the Lorenz gauge, $\partial^{m}A_{m}=0$,
the equations of motion \eqref{eq:N=00003D4SYMeom} can be recast
as
\begin{align}
\Box A_{m} & =2\mathrm{i}[A^{n},\partial_{n}A_{m}]-\mathrm{i}[A^{n},\partial_{m}A_{n}]+\mathrm{i}[\phi^{ij},\partial_{m}\phi_{ij}]+\{\lambda^{\alpha i},\bar{\lambda}_{i}^{\dot{\alpha}}\}\sigma_{m\alpha\dot{\alpha}}\nonumber \\
 & +[A^{n},[A_{n},A_{m}]]-[\phi^{ij},[\phi_{ij},A_{m}]]\\
\Box\phi^{ij} & =2\mathrm{i}[A^{m},\partial_{m}\phi^{ij}]+\{\lambda^{i\alpha},\lambda_{\alpha}^{j}\}+\frac{1}{2}\epsilon^{ijkl}\{\bar{\lambda}_{k}^{\dot{\alpha}},\bar{\lambda}_{l\dot{\alpha}}\}\nonumber \\
 & +[A_{m},[A^{m},\phi^{ij}]]+[\phi_{kl},[\phi^{ij},\phi^{kl}]],\\
\sigma_{\alpha\dot{\alpha}}^{m}\partial_{m}\lambda^{\alpha i} & =\mathrm{i}\sigma_{\alpha\dot{\alpha}}^{m}[A_{m},\lambda^{\alpha i}]+2\mathrm{i}[\phi^{ij},\bar{\lambda}_{j\dot{\alpha}}],\\
\sigma_{\alpha\dot{\alpha}}^{m}\partial_{m}\bar{\lambda}_{i}^{\dot{\alpha}} & =\mathrm{i}\sigma_{\alpha\dot{\alpha}}^{m}[A_{m},\bar{\lambda}_{i}^{\dot{\alpha}}]+2\mathrm{i}[\phi_{ij},\lambda_{\alpha}^{j}],
\end{align}
already prepared for the computation of the multi-particle coefficients.
Using the ansatze above, we obtain
\begin{multline}
s_{P}\boldsymbol{A}_{P}^{m}=\sum_{P=QR}[2(k_{R}\cdot\boldsymbol{A}_{Q})\boldsymbol{A}_{Rm}-(\boldsymbol{A}_{Q}\cdot\boldsymbol{A}_{R})k_{Rm}+\Phi_{Q}^{ij}\Phi_{Rij}k_{Rm}-\Lambda_{Q}^{\alpha i}\bar{\Lambda}_{Ri}^{\dot{\alpha}}\sigma_{m\alpha\dot{\alpha}}-(Q\leftrightarrow R)]\\
+\sum_{P=QRS}[(\boldsymbol{A}_{Q}\cdot\boldsymbol{A}_{S})\boldsymbol{A}_{Rm}-(\boldsymbol{A}_{Q}\cdot\boldsymbol{A}_{R})\boldsymbol{A}_{Sm}+\Phi_{Q}^{ij}\Phi_{Rij}\boldsymbol{A}_{Sm}-\Phi_{Q}^{ij}\Phi_{Sij}\boldsymbol{A}_{Rm}+(Q\leftrightarrow S)],
\end{multline}
\begin{multline}
s_{P}\Phi_{P}^{ij}=\sum_{P=QR}[2(k_{R}\cdot\boldsymbol{A}_{Q})\Phi_{R}^{ij}-\Lambda_{Q}^{\alpha i}\Lambda_{R\alpha}^{i}-\frac{1}{2}\epsilon^{ijkl}\bar{\Lambda}_{Qk}^{\dot{\alpha}}\bar{\Lambda}_{R\dot{\alpha}l}-(Q\leftrightarrow R)]\\
+\sum_{P=QRS}[(\boldsymbol{A}_{Q}\cdot\boldsymbol{A}_{S})\Phi_{R}^{ij}-(\boldsymbol{A}_{Q}\cdot\boldsymbol{A}_{R})\Phi_{S}^{ij}+\Phi_{Q}^{kl}\Phi_{Rkl}\Phi_{S}^{ij}-\Phi_{Q}^{kl}\Phi_{R}^{ij}\Phi_{Skl}+(Q\leftrightarrow S)],
\end{multline}
and

\begin{align}
\sigma_{\alpha\dot{\alpha}}^{m}k_{Pm}\Lambda_{P}^{\alpha i} & =\sum_{P=QR}[\sigma_{\alpha\dot{\alpha}}^{m}\boldsymbol{A}_{Qm}\Lambda_{R}^{\alpha i}+2\Phi_{Q}^{ij}\bar{\Lambda}_{R\dot{\alpha}j}-(Q\leftrightarrow R)],\\
\sigma_{\alpha\dot{\alpha}}^{m}k_{Pm}\bar{\Lambda}_{Pi}^{\dot{\alpha}} & =\sum_{P=QR}[\sigma_{\alpha\dot{\alpha}}^{m}\boldsymbol{A}_{Qm}\bar{\Lambda}_{Ri}^{\dot{\alpha}}+2\Phi_{Qij}\Lambda_{R}^{\alpha j}-(Q\leftrightarrow R)].
\end{align}

Now we can introduce the tree level partial amplitudes in $\mathcal{N}=4$
super Yang--Mills. Since we have several types of fields involved,
they can be defined in four different forms, depending on the reference
leg to which we attach the scattering tree. In addition to equation
\eqref{eq:YMpartialamplitude}, we have\begin{subequations}\label{eq:N=4SYMamplitudes}
\begin{align}
\mathscr{C}(1,\ldots,n,n+1) & =\lim_{s_{N}\to0}\phi_{ij,n+1}s_{N}\Phi_{N}^{ij},\\
\mathscr{C}(1,\ldots,n,n+1) & =\sigma_{m\alpha\dot{\alpha}}\lim_{s_{N}\to0}\lambda_{n+1}^{\alpha i}k_{N}^{m}\bar{\Lambda}_{Ni}^{\dot{\alpha}},\\
\mathscr{C}(1,\ldots,n,n+1) & =\sigma_{m\alpha\dot{\alpha}}\lim_{s_{N}\to0}\bar{\lambda}_{n+1,i}^{\dot{\alpha}}k_{N}^{m}\Lambda_{N}^{\alpha i},
\end{align}
\end{subequations}where we have the word $N=1\ldots n$.

As a quick check, let us look at the three-point scattering with one
gauge vector $\varepsilon_{1m}$ and two spinors, $\lambda_{2}^{\alpha i}$
and $\bar{\lambda}_{3i}^{\dot{\alpha}}$. The relevant two-particle
currents are given by
\begin{eqnarray}
s_{23}\boldsymbol{A}_{23}^{m} & = & -\sigma_{\alpha\dot{\alpha}}^{m}\lambda_{2}^{\alpha i}\bar{\lambda}_{3i}^{\dot{\alpha}},\\
\sigma_{\alpha\dot{\alpha}}^{m}k_{12m}\Lambda_{12}^{\alpha i} & = & \sigma_{\alpha\dot{\alpha}}^{m}\varepsilon_{1m}\lambda_{2}^{\alpha i},\\
\sigma_{\alpha\dot{\alpha}}^{m}k_{13m}\bar{\Lambda}_{31i}^{\dot{\alpha}} & = & -\sigma_{\alpha\dot{\alpha}}^{m}\varepsilon_{1m}\bar{\lambda}_{3i}^{\dot{\alpha}}.
\end{eqnarray}
It is then straightforward to check that\begin{subequations}
\begin{align}
\mathscr{C}(2,3,1) & =\lim_{s_{23}\to0}s_{23}(\varepsilon_{1}\cdot\boldsymbol{A}_{23}),\\
\mathscr{C}(3,1,2) & =\sigma_{m\alpha\dot{\alpha}}\lim_{s_{13}\to0}\lambda_{2}^{\alpha i}k_{31}^{m}\bar{\Lambda}_{31i}^{\dot{\alpha}},\\
\mathscr{C}(1,2,3) & =\sigma_{m\alpha\dot{\alpha}}\lim_{s_{12}\to0}\bar{\lambda}_{3,i}^{\dot{\alpha}}k_{12}^{m}\Lambda_{12}^{\alpha i},
\end{align}
\end{subequations}all coincide, corresponding to cyclic permutations
of the three-point partial amplitude.

In order to check that \eqref{eq:N=4SYMamplitudes} also has
the correct normalization for the scalar prescription, we can compute
the scattering of one scalar $\phi_{1}^{ij}$ and two spinors, $\lambda_{2}^{\alpha i}$
and $\lambda_{3}^{\alpha i}$. The relevant two-particle currents
are
\begin{eqnarray}
s_{23}\Phi_{23}^{ij} & = & -2\lambda_{2}^{\alpha i}\lambda_{3\alpha}^{j},\\
\sigma_{\alpha\dot{\alpha}}^{m}k_{12m}\bar{\Lambda}_{12i}^{\dot{\alpha}} & = & +2\phi_{1ij}\lambda_{2}^{\alpha j},\\
\sigma_{\alpha\dot{\alpha}}^{m}k_{13m}\bar{\Lambda}_{31i}^{\dot{\alpha}} & = & -2\phi_{1ij}\lambda_{3}^{\alpha j},
\end{eqnarray}
and it is easy to show that the different prescriptions in \eqref{eq:N=4SYMamplitudes} all
agree, finally leading to $\mathscr{C}(1,2,3)=-2\phi_{1ij}\lambda_{2}^{\alpha i}\lambda_{3\alpha}^{j}.$
Higher-point examples of course follow the same logic.

In the next section we are going to investigate the non-linear sigma
model, which is also ubiquitous in physics. The interesting feature
is that it is a field theory with an infinite number of interaction
vertices. The perturbiner framework effortlessly handles this apparent
obstacle.

\section{Non-Linear Sigma Model\label{sec:NLSM}}

The non-linear sigma model (NLSM) is a field theory describing maps
from spacetime into a curved target manifold, typically a coset space,
with the defining feature that its fields are subject to nonlinear
constraints rather than free to fluctuate independently. It arises
naturally as an effective theory of Goldstone bosons associated with
spontaneous symmetry breaking, most prominently in low-energy Quantum
Chromodynamics where it governs pion dynamics \cite{Weinberg:1968de,Coleman:1969sm,Callan:1969sn,Weinberg:1978kz}, but also in condensed
matter systems such as spin models and in the world-sheet formulation
of string theory. Its Lagrangian is organized as a derivative expansion
reflecting the absence of a mass gap for Goldstone modes, leading
to a rich structure of symmetries, soft limits, and universal low-energy
behavior. In modern amplitude-based approaches, the NLSM provides
a canonical example of a theory with enhanced infrared structure,
which makes it an interesting testing ground for recursive constructions (e.g. \cite{Bartsch:2022pyi,Bartsch:2025mvy}),
including the perturbiner method.

In this section we will discuss the perturbiner framework applied
to the NLSM, culminating with the presentation of a new tree level
recursion for the model that is cubic in the fields but does not require
auxiliary variables.

\subsection{Coordinate-valued formulation}

The action of the non-linear sigma model is given by
\begin{equation}
S[\phi]=-\frac{1}{2}\int d^{d}x\,\tilde{g}_{ab}(\phi)\partial_{m}\phi^{a}\partial^{m}\phi^{b},\label{eq:NLSM-action}
\end{equation}
where $\phi^{a}(x)$ are scalar fields that serve as local coordinates
on a target manifold $\mathcal{M}$, and the indices $a$, $b$ are
target-space coordinate indices. The tensor $\tilde{g}_{ab}(\phi)$
is a non-degenerate, smooth metric on $\mathcal{M}$, which we take
to be positive-definite. With this interpretation, the action is invariant
under field (coordinate) redefinitions $\phi^{a}\to\tilde{\phi}^{a}(\phi)$,
corresponding to changes of coordinates on the target space.

The equation of motion for the field $\phi^{a}$ is straightforward
to obtain. If we use $\delta\tilde{g}_{ab}=\partial_{c}\tilde{g}_{ab}(\phi)\delta\phi^{c}$,
it is possible to show that the variation of the action is given by
\begin{equation}
\delta S=\int d^{d}x\,\delta\phi^{e}\tilde{g}_{ae}\underbrace{\left(\Box\phi^{a}+\tilde{\Gamma}_{bc}^{a}\partial_{m}\phi^{b}\partial^{m}\phi^{c}\right)}_{\textrm{equation of motion}}-\partial^{m}\underbrace{\left(\tilde{g}_{ab}\delta\phi^{a}\partial_{m}\phi^{b}\right)}_{\textrm{boundary}},\label{eq:NLSM-variation}
\end{equation}
where we define the Christoffel symbol\footnote{Note that the symbol $\partial$ is being used to denote both partial
derivatives in spacetime, $\partial_{m}$=$\frac{\partial}{\partial x^{m}}$,
and partial derivatives with respect to the fields $\phi_{a}$, like
in $\partial_{a}g_{bc}=\frac{\partial}{\partial\phi^{a}}g_{bc}(\phi)$.
They have different indices, and it should be clear from the context
which one is being used.},
\begin{equation}
\tilde{\Gamma}_{bc}^{a}\equiv\frac{1}{2}\tilde{g}^{ad}\left(\partial_{b}\tilde{g}_{cd}+\partial_{c}\tilde{g}_{bd}-\partial_{d}\tilde{g}_{bc}\right).
\end{equation}
The boundary term in equation \eqref{eq:NLSM-variation} will be disregarded
here. However, it is important to emphasize that it encodes the physics
of edges, anomalies, and topology. See, for example, \cite{Witten:1983tw,Freed:2006mx}
. We will discuss flat space with a boundary in section \ref{sec:AdS}.

The equation of motion of the NLSM takes a very familiar form,
\begin{equation}
\Box\phi^{a}+\tilde{\Gamma}_{bc}^{a}\partial_{\mu}\phi^{b}\partial^{\mu}\phi^{c}=0.\label{eq:NLSM-eom}
\end{equation}
This equation is analogous to the geodesic equation for a particle
moving in a curved space. In differential geometry, it is known as
the harmonic map equation \cite{Eells:1964,Misner:1978tq}. The difference is that instead of a world-line,
we now have a whole spacetime (in the present case it is simply Minkowski
space), and the target-space metric, $\tilde{g}_{ab}$, is pulled
back to this world-volume via the fields $\phi^{a}$. Thus the dynamics
of the NLSM can in this sense be viewed as a higher-dimensional generalization
of geodesic motion. At the linearized level, we simply have massless
single-particle solutions.

Our first goal is to analyze tree level scattering amplitudes of the
NLSM, so we will define a recursion for the coefficients of the multi-particle
expansion
\begin{equation}
\phi^{a}=\sum_{P}\Phi_{P}^{a}e^{\mathrm{i}k_{P}\cdot x},\label{eq:NLSM-multi}
\end{equation}
which we recognize from previous sections. Since the target-space
metric $\tilde{g}_{ab}(\phi)$ is a more or less arbitrary function
of $\phi^{a}$, we will assume a Taylor expansion of the form
\begin{equation}
\tilde{g}_{ab}=g_{ab}+\sum_{n=1}^{\infty}\frac{1}{n!}H_{abc_{1}\ldots c_{n}}\phi^{c_{1}}\ldots\phi^{c_{n}},\label{eq:NLSM-Taylor}
\end{equation}
with
\begin{equation}
H_{abc_{1}\ldots c_{n}}=\left.\partial_{c_{1}}\ldots\partial_{c_{n}}\tilde{g}_{ab}\right|_{\phi=0}.
\end{equation}
We take $g_{ab}$ to be some fixed background metric, such that the
expansion \eqref{eq:NLSM-Taylor} is well defined. For the inverse
metric $\tilde{g}{}^{ab}$, it is convenient to introduce a similar
expansion, with
\begin{equation}
\tilde{g}^{ab}=g^{ab}-\sum_{n=1}^{\infty}\frac{1}{n!}I_{c_{1}\ldots c_{n}}^{ab}\phi^{c_{1}}\ldots\phi^{c_{n}},\label{eq:NLSMmetricexpansion}
\end{equation}
such that $\tilde{g}_{ac}\tilde{g}^{cb}=g_{ac}g^{cb}=\delta_{a}^{b}$,
and the coefficients $I_{c_{1}\ldots c_{n}}^{ab}$ are solved in terms
of $H_{abc_{1}\ldots c_{n}}$:
\begin{align}
I_{e}^{ab} & =g^{ac}g^{bd}H_{cde},\\
I_{d_{1}\ldots d_{n}}^{ab} & =g^{ac}g^{bd}H_{cdd_{1}\ldots d_{n}}-\tilde{g}^{bc}\sum_{m=1}^{n-1}\binom{n}{m}I_{(d_{1}\ldots d_{n-m}}^{ae}H_{ecd_{n-m+1}\ldots d_{n})}.
\end{align}
More generally, we have to work with the field expansion of $\tilde{\Gamma}_{bc}^{a}$,
which encodes the target-space geometry in \eqref{eq:NLSM-eom}. In
this case, we have
\begin{align}
\tilde{\Gamma}_{bc}^{a} & =\Gamma_{bc}^{a}+\sum_{n=1}^{\infty}\frac{1}{n!}\Gamma_{bcc_{1}\ldots c_{n}}^{a}\phi^{c_{1}}\ldots\phi^{c_{n}},\\
\Gamma_{bcc_{1}\ldots c_{n}}^{a} & =\frac{1}{2}\left(g^{ad}H_{cdbc_{1}\ldots c_{n}}+g^{ad}H_{dbcc_{1}\ldots c_{n}}-g^{ad}H_{bcdc_{1}\ldots c_{n}}\right)-I_{c_{1}\ldots c_{n}}^{ad}\Gamma_{dbc}\nonumber \\
 & +\frac{1}{2}\sum_{m=1}^{n-1}\binom{n}{m}\left(H_{bcd(c_{1}\ldots c_{m}}-H_{cdb(c_{1}\ldots c_{m}}-H_{dbc(c_{1}\ldots c_{m}}\right)I_{c_{m+1}\ldots c_{n})}^{ad},\label{eq:NLSM-Christoffel-expansion}
\end{align}
where $\Gamma_{bc}^{a}=g^{ad}\Gamma_{dbc}$ is the Christoffel symbol
associated to $g_{ab}$, with
\begin{equation}
\Gamma_{abc}=\frac{1}{2}\left(\partial_{b}g_{ac}+\partial_{c}g_{ab}-\partial_{a}g_{bc}\right).
\end{equation}
In terms of the multi-particle expansion, we have
\begin{align}
\tilde{\Gamma}_{bc}^{a} & =\Gamma_{bc}^{a}+\sum_{P}\Gamma_{Pbc}^{a}e^{\mathrm{i}k_{P}\cdot x},\label{eq:NLSM-multiChristoffel}\\
\Gamma_{Pbc}^{a} & =\sum_{n=1}^{|P|}\sum_{P=P_{1}\cup\cdots\cup P_{n}}\frac{1}{n!}\Gamma_{bcc_{1}\ldots c_{n}}^{a}\Phi_{P_{1}}^{c_{1}}\ldots\Phi_{P_{n}}^{c_{n}},
\end{align}
with $\Gamma_{bcc_{1}\ldots c_{n}}^{a}$ given in equation \eqref{eq:NLSM-Christoffel-expansion}.

Now we are in a good position to determine the recursion for $\Phi_{P}^{a}$.
Substituting \eqref{eq:NLSM-multi} and \eqref{eq:NLSM-multiChristoffel}
in the equation of motion leads to
\begin{equation}
\Phi_{P}^{a}=-\frac{1}{s_{P}}\left(\Gamma_{bc}^{a}\sum_{P=Q\cup R}(k_{Q}\cdot k_{R})\Phi_{Q}^{b}\Phi_{R}^{c}+\sum_{P=Q\cup R\cup S}(k_{Q}\cdot k_{R})\Phi_{Q}^{b}\Phi_{R}^{c}\Gamma_{Sbc}^{a}\right).\label{eq:NLSM-recursion}
\end{equation}
All higher-point vertices of the action \eqref{eq:NLSM-action} are
nicely packed in the multi-particle object $\Gamma_{Sbc}^{a}$.

The $N$-point tree level amplitude of the NLSM can then be computed
via the usual perturbiner procedure,
\begin{equation}
A_{N}^{\textrm{NLSM}}=\lim_{s_{2\ldots N}\to0}g_{ab}\phi_{1}^{a}\left(s_{2\ldots N}\Phi_{2\ldots N}^{b}\right).\label{eq:NLSM-amplitude}
\end{equation}
These amplitudes are intrinsically connected to the target-space geometry,
there is no surprise there. Since scattering amplitudes are physical
quantities, invariant under field redefinitions, what can we expect
to find? 

Let us start with the three-point amplitude. Within our framework,
this computation requires the two-particle current, which is given
by
\begin{align}
\Phi_{23}^{a} & =-\frac{2}{s_{23}}\Gamma_{bc}^{a}(k_{2}\cdot k_{3})\phi_{2}^{b}\phi_{3}^{c},\nonumber \\
 & =-\Gamma_{bc}^{a}\phi_{2}^{b}\phi_{3}^{c}.
\end{align}
The pole $s_{23}=0$ disappears from $\Phi_{23}^{a}$, and we have a vanishing three-point amplitude:
\begin{equation}
A_{3}^{\textrm{NLSM}}=\lim_{s_{23}\to0}s_{23}\left(\Gamma_{abc}\phi_{1}^{a}\phi_{2}^{b}\phi_{3}^{c}\right).
\end{equation}
Suppose, for a moment, that
particle $1$ is taken off the mass-shell, so we drop the limit. The
resulting object would be proportional to the Christoffel symbol,
which is not a tensor. The amplitude would not be invariant under
field-redefinitions, which contradicts its role as a physical observable.

For the four-point amplitude, a direct computation leads to
\begin{multline}
A_{4}^{\textrm{NLSM}}= s_{23}\left(\Gamma_{bc}^{e}\Gamma_{eda}-\partial_{d}\Gamma_{abc}\right)\phi_{1}^{a}\phi_{2}^{b}\phi_{3}^{c}\phi_{4}^{d}
+s_{24}\left(\Gamma_{bd}^{e}\Gamma_{eca}-\partial_{c}\Gamma_{abd}\right)\phi_{1}^{a}\phi_{2}^{b}\phi_{3}^{c}\phi_{4}^{d}\\
+s_{34}\left(\Gamma_{cd}^{e}\Gamma_{eba}-\partial_{b}\Gamma_{acd}\right)\phi_{1}^{a}\phi_{2}^{b}\phi_{3}^{c}\phi_{4}^{d},\label{eq:NLSM4pt}
\end{multline}
which is non-vanishing. The only covariant object built from the metric
with two derivatives is the Riemann tensor and its contractions (Ricci
tensor and scalar curvature),
\begin{equation}
R_{abcd}=\partial_{c}\Gamma_{abd}-\partial_{d}\Gamma_{abc}+\Gamma_{bc}^{e}\Gamma_{eda}-\Gamma_{bd}^{e}\Gamma_{eca}.
\end{equation}
Equation \eqref{eq:NLSM4pt} is not immediately in this form. However,
we can make this manifest using momentum conservation. More concretely,
we can use
\begin{equation}
s_{ij}=\frac{2}{3}s_{ij}-\frac{1}{3}s_{ik}-\frac{1}{3}s_{jk},
\end{equation}
where $\{i,j,k\}$ can be any permutation of $\{2,3,4\}$, and substitute
this rearrangement in $A_{4}^{\textrm{NLSM}}$ to finally obtain 
\begin{equation}
A_{4}^{\textrm{NLSM}}=\frac{1}{3}[s_{23}(R_{abcd}+R_{acbd})+s_{24}(R_{abdc}+R_{adbc})+s_{34}(R_{acdb}+R_{adcb})]\phi_{1}^{a}\phi_{2}^{b}\phi_{3}^{c}\phi_{4}^{d},
\end{equation}
where the Riemann tensor explicitly appears.

By following this pattern, we can have an educated guess for $A_{5}^{\textrm{NLSM}}$.
It can be schematically written as
\begin{equation}
A_{5}^{\textrm{NLSM}}\sim\sum s_{ij}(\nabla R)_{abcde}\phi_{1}^{a}\phi_{2}^{b}\phi_{3}^{c}\phi_{4}^{d}\phi_{5}^{e},
\end{equation}
where $\nabla R$ denotes the covariant derivative of the Riemann
tensor (it is the only allowed object at this order). The sum is over
different permutations of indices and single-particle labels. The
full computation of $A_{5}^{\textrm{NLSM}}$ is more or less straightforward,
though a bit lengthy. In general, the five-point amplitude vanishes
only if
\begin{equation}
\nabla_{a}R_{bcde}=0,
\end{equation}
i.e., for locally symmetric manifolds $\mathcal{M}$ (essentially,
the Riemann tensor is parallel with respect to the Levi--Civita connection).
Roughly speaking, $\phi\to-\phi$ is a symmetry of the action \eqref{eq:NLSM-action},
which forces the metric expansion \eqref{eq:NLSMmetricexpansion}
to contain only even powers of $\phi$. As a consequence, all odd-point
amplitudes of the NLSM vanish. We will return to this point in the
next subsection, where the group-valued formulation makes this structure
particularly transparent.

\subsection{Group-valued formulation}

The group-valued formulation of the NLSM \cite{Coleman:1969sm,Callan:1969sn} is probably the most popular
one in the scattering amplitudes community. The associated action
can be simply cast as
\begin{equation}
S[\phi]=\frac{1}{2}\int d^{d}x\,\textrm{Tr}(\partial_{m}U^{-1}\partial^{m}U),\label{eq:NLSM-group-action}
\end{equation}
where $U=e^{\mathrm{i}\phi^{a}T_{a}}$, and the group generators $T_{a}$
are in the adjoint representation. Note that it is common to find
$S[\phi]$ written as
\begin{equation}
S[\phi]=-\frac{1}{2}\int d^{d}x\,\textrm{Tr}(\eta^{mn}J_{m}J_{n}),
\end{equation}
which explicitly uses the Maurer--Cartan form, $J_{m}=U^{-1}\partial_{m}U$.

As we have already seen, a characteristic feature of the NLSM is the
presence of an infinite number of interaction vertices. We can easily
see this unfolding when making a Taylor expansion of $U$ (and $U^{-1}$)
around $\phi=0$, with\begin{subequations}
\begin{align}
U & =\mathds{1}+\mathrm{i}\phi^{a}T_{a}-\frac{1}{2}(\phi^{a}T_{a})^{2}-\frac{\mathrm{i}}{6}(\phi^{a}T_{a})^{3}+\ldots,\\
U^{-1} & =\mathds{1}-\mathrm{i}\phi^{a}T_{a}-\frac{1}{2}(\phi^{a}T_{a})^{2}+\frac{\mathrm{i}}{6}(\phi^{a}T_{a})^{3}+\ldots,\\
\partial_{m}U & =\mathrm{i}\partial_{m}\phi^{a}T_{a}-\frac{1}{2}(\partial_{m}\phi^{a}T_{a})(\phi^{b}T_{b})-\frac{1}{2}(\phi^{a}T_{a})(\partial_{m}\phi^{b}T_{b})\nonumber \\
 & -\frac{\mathrm{i}}{6}(\partial_{m}\phi^{a}T_{a})(\phi^{b}T_{b})^{2}-\frac{\mathrm{i}}{6}(\phi^{a}T_{a})(\partial_{m}\phi^{b}T_{b})(\phi^{c}T_{c})-\frac{\mathrm{i}}{6}(\phi^{a}T_{a})^{2}(\partial_{m}\phi^{b}T_{b})\ldots,\\
(U^{-1}\partial_{m}U) & =\mathrm{i}\partial_{m}\phi^{a}T_{a}-\frac{1}{2}\phi^{b}\partial_{m}\phi^{a}[T_{a},T_{b}]-\frac{\mathrm{i}}{6}\phi^{c}\phi^{b}\partial_{m}\phi^{a}[[T_{a},T_{b}],T_{c}]\ldots,
\end{align}
\end{subequations}and so on, where the ellipsis denote $\mathcal{O}(\phi^{4})$.
For the Lagrangian, we obtain
\begin{multline}
\textrm{Tr}\left((U^{-1}\partial_{m}U)(U^{-1}\partial^{m}U)\right)=-\textrm{Tr}[(\partial_{m}\phi^{a}T_{a})(\partial^{m}\phi^{b}T_{b})]\\
-\frac{1}{12}\textrm{Tr}([T_{b},T_{d}][T_{a},T_{c}])\phi^{c}\phi^{d}\partial_{m}\phi^{a}\partial^{m}\phi^{b}+\mathcal{O}(\phi^{6}),
\end{multline}
and it is easy to show that it contains only even powers of $\phi$,
after all $U(-\phi)=U^{-1}(\phi)$.

This observation is connected to the discussion at the end of the
previous subsection. Since every interaction vertex has an even number
of legs, any Feynman diagram contributing to a $N$-point amplitude
must be built from even-valent vertices. For a tree diagram with $N$
external legs and $I$ internal propagators, we have
\begin{equation}
N+2I=\sum_{i}n_{i},
\end{equation}
where the sum goes over the vertices $i$ (with valence $n_{i}$,
all even) used to build the diagram. Therefore, $N$ should also be
even and it is not possible to draw any diagrams with an odd number
of external legs. From the perspective of classical multi-particle
solutions, this property simply implies there are no multi-particle
coefficients with words of even length.

The action \eqref{eq:NLSM-group-action} leads to the equation of
motion
\begin{equation}
\partial^{m}(U^{-1}\partial_{m}U)=0.\label{eq:NLSM-group-eom}
\end{equation}
When written in terms of $U=e^{\mathrm{i}\phi^{a}T_{a}}$, we immediately
see a proliferation of interaction terms, even though the underlying
expression is extremely simple. For example, by introducing the color-stripped
multi-particle expansion
\begin{equation}
\phi=\sum_{P}\Phi_{P}T^{a_{P}}e^{\mathrm{i}k_{P}\cdot x},
\end{equation}
we obtain
\begin{align}
U & =\mathds{1}+\sum_{P}\left(\sum_{n=1}^{|P|}\frac{\mathrm{i}^{n}}{n!}\sum_{P=P_{1}\ldots P_{n}}\Phi_{P_{1}}\ldots\Phi_{P_{n}}\right)T^{a_{P}}e^{\mathrm{i}k_{P}\cdot x},\\
U^{-1} & =\mathds{1}+\sum_{P}\left(\sum_{n=1}^{|P|}\frac{(-\mathrm{i})^{n}}{n!}\sum_{P=P_{1}\ldots P_{n}}\Phi_{P_{1}}\ldots\Phi_{P_{n}}\right)T^{a_{P}}e^{\mathrm{i}k_{P}\cdot x},
\end{align}
which leads to an unnecessarily cumbersome recursive definition of
the coefficients $\Phi_{P}$, expressed as
\begin{equation}
s_{P}\Phi_{P}=\sum_{n=2}^{|P|}\frac{\mathrm{i}^{n+1}}{n!}\sum_{P=P_{1}\ldots P_{n}}\left(s_{P}+\sum_{m=1}^{n-1}(-1)^{m}\binom{n}{m}(k_{P}\cdot k_{P_{m+1}\ldots P_{n}})\right)\Phi_{P_{1}}\ldots\Phi_{P_{n}}.
\end{equation}
It is straightforward to show, for instance, that $\Phi_{12}=0$.
Next, it is possible to show that 
\begin{equation}
\Phi_{123}=\left(\frac{1}{6}-\frac{1}{4s_{123}}(s_{12}+s_{23})\right)\phi_{1}\phi_{2}\phi_{3},\label{eq:NLSM-group-3ptv1}
\end{equation}
and also $\Phi_{1234}=0$. As we had already anticipated, the coefficients
with an even number of single-particle labels vanish. Such Berends--Giele-like recursions for the NLSM have already appeared in \cite{Chen:2014dfa}.

We can do a more streamlined construction though. Via a field redefinition,
we propose a multi-particle expansion for the field $U$ instead,
with
\begin{equation}
U=\mathds{1}+\mathrm{i}\sum_{P}U_{P}T^{a_{P}}e^{ik_{P}\cdot x}.\label{eq:NLSM-group-Uexpansion}
\end{equation}
With the adjusted normalization, $U_{p}=\phi_{p}$ denotes the single-particle
polarizations. The question then is how to define $U^{-1}$. Suppose
\begin{equation}
U^{-1}=\mathds{1}-\mathrm{i}\sum_{P}\bar{U}_{P}T^{a_{P}}e^{ik_{P}\cdot x}.\label{eq:NLSM-group-Ubarexpansion}
\end{equation}
We can then compute the product between the ansatze \eqref{eq:NLSM-group-Uexpansion}
and \eqref{eq:NLSM-group-Ubarexpansion}, which leads to
\begin{equation}
U^{-1}U=\mathds{1}+\sum_{P}\left[\mathrm{i}(U_{P}-\bar{U}_{P})+\sum_{P=QR}\bar{U}_{Q}U_{R}\right]T^{a_{P}}e^{ik_{P}\cdot x}.
\end{equation}
However, we know that $U^{-1}U=\mathds{1}$, so the expression inside
the square brackets should vanish. Therefore, we obtain a recursive
expression for $\bar{U}_{P}$, given by
\begin{equation}
\bar{U}_{P}=U_{P}-\mathrm{i}\sum_{P=QR}\bar{U}_{Q}U_{R}.
\end{equation}
We could have computed $UU^{-1}$ instead, which leads to
\begin{equation}
\bar{U}_{P}=U_{P}-\mathrm{i}\sum_{P=QR}U_{Q}\bar{U}_{R}.
\end{equation}
The two recursions for $\bar{U}_{P}$ are equivalent, just not manifestly.
This could serve as a consistency check for the actual computations of scattering
amplitudes later.

Using the equation of motion \eqref{eq:NLSM-group-eom}, we obtain
\begin{equation}
U_{P}=\frac{\mathrm{i}}{s_{P}}\sum_{P=QR}(k_{P}\cdot k_{R})\bar{U}_{Q}U_{R},
\end{equation}
which is then used to compute partial amplitudes in the group-valued
form of the NLSM. They are defined as
\begin{equation}
\mathscr{C}(1,\ldots,n,n+1)=\lim_{s_{N}\to0}\phi_{n+1}s_{N}U_{N},
\end{equation}
where $N$ denotes the canonically ordered word $N=1,\ldots,n$.

Finally we can compute some examples and check the consistency of
our field redefinition. For the two-particle current, we obtain
\begin{align}
U_{12} & =\frac{\mathrm{i}}{s_{12}}(k_{12}\cdot k_{2})\bar{U}_{1}U_{2},\nonumber \\
 & =\frac{\mathrm{i}}{2}\phi_{1}\phi_{2},\\
\bar{U}_{12} & =-U_{12}.
\end{align}
Since $s_{12}$ is canceled because of the numerator contribution,
the partial amplitude $\mathscr{C}(1,2,3)$ vanishes, even though
the two-particle coefficient is nonzero. For the three-particle coefficient,
we obtain
\begin{align}
U_{123} & =\frac{\mathrm{i}}{s_{123}}[(k_{123}\cdot k_{3})\bar{U}_{12}U_{3}+(k_{123}\cdot k_{23})\bar{U}_{1}U_{23}],\nonumber \\
 & =-\frac{1}{4s_{123}}(s_{12}+s_{23})\phi_{1}\phi_{2}\phi_{3},\label{eq:NLSM-group-3ptv2}\\
\bar{U}_{123} & =U_{123},
\end{align}
which leads to the same four-point partial amplitude as the coefficient
$\Phi_{123}$ in \eqref{eq:NLSM-group-3ptv1}. More generally, we
can show that
\begin{equation}
\bar{U}_{1\ldots n}=(-1)^{n+1}U_{1\ldots n},
\end{equation}
such that
\begin{align}
U_{P} & =-\frac{\mathrm{i}}{s_{P}}\sum_{P=QR}(-1)^{|Q|}(k_{P}\cdot k_{R})U_{Q}U_{R},\nonumber \\
 & =-\frac{\mathrm{i}}{2s_{P}}\sum_{P=QR}(-1)^{|Q|}(s_{P}+s_{R}-s_{Q})U_{Q}U_{R}.\label{eq:NLSM-group-cubic}
\end{align}
and the scattering trees of the NLSM can be manifestly described by
a cubic interaction! Note, however, that equation \eqref{eq:NLSM-group-cubic}
cannot be immediately derived from an ordinary equation of motion,
since it involves the length $|Q|$.

Some higher-point examples are
\begin{equation}
U_{1234}=\frac{\mathrm{i}}{8}\left(1-\frac{1}{s_{123}}(s_{12}+s_{23})-\frac{1}{s_{234}}(s_{23}+s_{34})\right)\phi_{1}\phi_{2}\phi_{3}\phi_{4},
\end{equation}
which leads to a vanishing five-point amplitude, and
\begin{align}
U_{12345} & =\frac{1}{16}\frac{1}{s_{12345}}(s_{1234}+s_{45})\frac{1}{s_{123}}(s_{12}+s_{23})\phi_{1}\phi_{2}\phi_{3}\phi_{4}\phi_{5}\nonumber \\
 & +\frac{1}{16}\frac{1}{s_{12345}}(s_{1234}+s_{2345})\frac{1}{s_{234}}(s_{23}+s_{34})\phi_{1}\phi_{2}\phi_{3}\phi_{4}\phi_{5}\nonumber \\
 & +\frac{1}{16}\frac{1}{s_{12345}}(s_{2345}+s_{12})\frac{1}{s_{345}}(s_{34}+s_{45})\phi_{1}\phi_{2}\phi_{3}\phi_{4}\phi_{5}\nonumber \\
 & -\frac{1}{16}\frac{1}{s_{12345}}(s_{12}+s_{23}+s_{34}+s_{45}+s_{1234}+s_{2345})\phi_{1}\phi_{2}\phi_{3}\phi_{4}\phi_{5},\label{eq:NLSM-5ptcurrent}\\
 & =\frac{1}{16}\frac{1}{s_{12345}}(s_{1234}+s_{45})\frac{1}{s_{123}}(s_{12}+s_{23}-s_{123})\phi_{1}\phi_{2}\phi_{3}\phi_{4}\phi_{5}\nonumber \\
 & +\frac{1}{16}\frac{1}{s_{12345}}(s_{1234}+s_{2345}-s_{234})\frac{1}{s_{234}}(s_{23}+s_{34})\phi_{1}\phi_{2}\phi_{3}\phi_{4}\phi_{5}\nonumber \\
 & +\frac{1}{16}\frac{1}{s_{12345}}(s_{2345}+s_{12})\frac{1}{s_{345}}(s_{34}+s_{45}-s_{345})\phi_{1}\phi_{2}\phi_{3}\phi_{4}\phi_{5},
\end{align}
which yields the six-point NLSM partial amplitude. Note that in the
last line of equation \eqref{eq:NLSM-5ptcurrent} the six-point contact
term is manifest, though it can be nicely absorbed by the numerators
on the poles $s_{123}$, $s_{234}$, and $s_{345}$.

More generally, the recursion \eqref{eq:NLSM-group-cubic} can be
suggestively rewritten as
\begin{equation}
U_{P}=\frac{\mathrm{i}}{4s_{P}}\sum_{P=QR}\left([(-1)^{|Q|}-(-1)^{|R|}](s_{Q}-s_{R})-[(-1)^{|Q|}+(-1)^{|R|}]s_{P}\right)U_{Q}U_{R}.
\end{equation}
For $|P|$ even, $(-1)^{|Q|}=(-1)^{|R|}$, and the pole $s_{P}$ disappears.
In turn, this implies that odd-point partial amplitudes vanish, as
expected. For $|P|$ odd, $(-1)^{|Q|}=-(-1)^{|R|}$, and only the
first term in the sum remains. This can be summarized
as
\begin{align}
\left.U_{P}\right|_{|P|\textrm{ even}} & =-\frac{\mathrm{i}}{2}\sum_{P=QR}(-1)^{|Q|}U_{Q}U_{R},\\
\left.U_{P}\right|_{|P|\textrm{ odd}} & =\frac{\mathrm{i}}{2s_{P}}\sum_{P=QR}(-1)^{|Q|}(s_{Q}-s_{R})U_{Q}U_{R},
\end{align}
which are in agreement with the examples already presented.

\subsection{Perturbiner methods and EFTs}

It has been known for a while that the NLSM can be recast as a cubic
theory even at the level of the Lagrangian \cite{Cheung:2016prv},
though with an awkward introduction of extra fields. The cubic character
of the recursion \eqref{eq:NLSM-group-cubic} makes a much clearer
statement in this direction, primed for the investigation of color-kinematics
duality.

Even before the cubic formulation, the NLSM was known to satisfy the
pre-requisites for a double-copy construction \cite{Chen:2013fya},
though the resulting theory was not immediately obvious. It was later
in \cite{Cachazo:2014xea,Cheung:2014dqa} that the so-called special
Galileon theory was identified, which can be seen as two copies of
the NLSM. The special Galileon is a scalar theory invariant under
an extended shift symmetry (the Galileon symmetry) which enforces
an enhanced soft behavior, with amplitudes vanishing faster than those
of the NLSM as any external momentum is taken soft. In fact, several
scalar effective field theories have been found to be connected via
soft limits, double-copy, dimensional reduction \cite{Cheung:2014dqa,Cachazo:2016njl,Cheung:2016yqr}.
The notable ones are the NLSM, Dirac--Born--Infeld theory, and
(special) Galileon.

The perturbiner framework proves to be particularly well-suited for
the exploration of these effective field theories, as systematically
investigated in \cite{Mizera:2018jbh}. There, the NLSM and its relatives
were studied using the cubic formulation of \cite{Cheung:2016prv},
making the recursive structure fully explicit and computationally
efficient. A notable achievement is the implementation of KLT double-copy
relations directly at the level of the currents (multi-particle coefficients),
i.e., before taking the on-shell limit. These relations hold exactly
for fields without gauge redundancies, and up to pure gauge terms
for gauge fields, without the need for further field redefinitions.
This is a stronger statement than a double copy of amplitudes alone,
and a natural home for the perturbiner method.

In the next section we are going to learn about classical multi-particle
solutions involving gravitons, with the construction of \emph{gravitational}
Berends--Giele currents.

\section{Gravity coupled to matter\label{sec:gravity}}

Applying the perturbiner framework to gravity is more subtle than
in gauge theory. While Yang--Mills theory admits a formulation in
terms of finitely many interaction vertices (cubic and quartic), the
Einstein--Hilbert action expanded around flat space contains an endless
number of graviton self-interactions. This is similar to what we have
encountered in section \ref{sec:NLSM} for the NLSM. From the perspective
of conventional perturbation theory, this non-polynomial structure
obscures the recursive organization that underlies the perturbiner
expansions. For this reason, gravity long appeared to be outside the
natural domain of perturbiner methods, despite the existence of remarkable
recursive structures in gravitational scattering amplitudes.

Modern amplitude theory has nevertheless revealed that gravity possesses
an unexpected degree of hidden simplicity. Tree-level graviton amplitudes
can be reconstructed recursively through BCFW recursion relations \cite{Britto:2004ap,Britto:2005fq},
and can often be represented as double copies of gauge-theory amplitudes
through color-kinematics duality. These developments strongly suggest
that the apparent complexity of the Einstein--Hilbert action is largely
redundant from the viewpoint of on-shell observables. The perturbiner
approach provides an alternative realization of this idea directly
at the level of classical field equations.

Historically, perturbiner constructions in gravity were first explored
in the self-dual sector, where the equations dramatically simplify.
A general recursive treatment of Einstein gravity, however, remained
elusive until \cite{Gomez:2021shh}. The main obstacle is clear:
a naive perturbiner expansion would require handling infinitely many
interaction vertices simultaneously. The crucial observation allowing
one to circumvent this difficulty is that the inverse metric must
itself be expanded recursively as a multi-particle field. Once this
is done, Einstein\textquoteright s equations reorganize into well-defined
recursive relations for multi-particle graviton currents, closely
paralleling the Berends--Giele recursion of Yang--Mills theory.

This construction leads to multi-particle solutions of Einstein\textquoteright s
equations that encode tree-level graviton scattering in arbitrary
spacetime dimension. Moreover, the formalism extends naturally to
gravity coupled to matter fields, including scalars, gauge bosons,
fermions, and supersymmetric systems. In this way, perturbiner methods
provide an off-shell recursive framework for gravitational amplitudes
that complements on-shell approaches such as BCFW recursion, CHY representations,
and double-copy constructions.

In this section we are going to describe how gravitons fit in the
perturbiner framework, including their interaction with matter. We
will cover all the relevant material to build the generalization of
the Berends--Giele currents to include gravitons coupled to bosonic
and fermionic matter.

\subsection{Pure Einstein gravity}

We start with Einstein Field Equations, which in the most general
form can be written as
\begin{equation}
R_{mn}-\frac{1}{2}g_{mn}R+\Lambda g_{mn}=\kappa T_{mn}.\label{eq:EFE}
\end{equation}
On the left hand side, we have the spacetime metric, $g_{mn}$ (with
inverse $g^{mn}$), the Ricci tensor, $R_{mn}$, and the Ricci scalar
(or scalar curvature), $R=g^{mn}R_{mn}$. On the right hand side we
have the matter energy-momentum tensor, $T_{mn}$. The cosmological
constant is denoted by $\Lambda$, and $\kappa$ is the gravitational
coupling constant. This is textbook material, see e.g. \cite{Carroll:2004st}.

Equation\eqref{eq:EFE} is covariant under general coordinate
transformations $x^{m}\to\tilde{x}^{m}(x)$. The metric transforms
as
\begin{equation}
\tilde{g}_{mn}(\tilde{x})=\frac{\partial x^{p}}{\partial\tilde{x}^{m}}\frac{\partial x^{q}}{\partial\tilde{x}^{n}}g_{pq}(x),\label{eq:metric-transformation}
\end{equation}
and so do $R_{mn}$ and $T_{mn}$. Obviously, these are local transformations.
For our purposes, the field equations of general relativity will be
seen as a gauge theory of perturbations around a fixed background
described by a metric that satisfies equation \eqref{eq:EFE}. These
perturbations are the gravitons. We assume $D\geq4$. In $D=3$ there
are no dynamical degrees of freedom (global topology only). In $D=2$,
there are no dynamical degrees of freedom either, but there is a rich
topology and conformal structure (e.g. string theory). 

In this section we are going to work with multi-particle solutions
around flat space, so we take $\Lambda=0$. In section \ref{sec:AdS}
we discuss solutions around less trivial spacetimes, including $\Lambda\neq0$. 

First let us consider the case where matter is not present, $T_{mn}=0$.
Equation \eqref{eq:EFE} then reduces to
\begin{equation}
R_{mn}=0.\label{eq:EFE-emptyspace}
\end{equation}
The Ricci tensor is written in terms of the Riemann tensor, $R_{mn}=R_{\phantom{p}mpn}^{p}$,
with
\begin{align}
R_{\phantom{q}mnp}^{q} & =\partial_{n}\Gamma_{mp}^{q}-\partial_{p}\Gamma_{mn}^{q}+\Gamma_{nr}^{q}\Gamma_{mp}^{r}-\Gamma_{pr}^{q}\Gamma_{mn}^{r},\nonumber \\
 & =g^{qr}[\partial_{n}\Gamma_{rmp}-\partial_{p}\Gamma_{rmn}+g^{st}(\Gamma_{spr}\Gamma_{tmn}-\Gamma_{snr}\Gamma_{tmp})].\label{eq:Riemann}
\end{align}
where $\Gamma_{mn}^{p}$ is the metric connection (Christoffel symbols),
written as
\begin{align}
\Gamma_{mn}^{p} & =g^{pq}\Gamma_{qmn},\label{eq:Christoffel}\\
\Gamma_{pmn} & =\frac{1}{2}(\partial_{m}g_{np}+\partial_{n}g_{mp}-\partial_{p}g_{mn}).
\end{align}
The expression for $\Gamma_{mn}^{p}$ follows from the metric compatibility
condition
\begin{align}
\nabla_{p}g_{mn} & =\partial_{p}g_{mn}-\Gamma_{mp}^{q}g_{nq}-\Gamma_{np}^{q}g_{mq},\nonumber \\
 & =0,
\end{align}
as long as we stick to a torsion free space, with $\Gamma_{mn}^{p}=\Gamma_{nm}^{p}$.

\subsubsection{Gravitons}

In terms of the metric, equation \eqref{eq:EFE-emptyspace} can be
suggestively recast as
\begin{multline}
g^{pq}\partial_{p}\partial_{q}g_{mn}=\frac{1}{2}\partial_{m}g^{pq}\partial_{n}g_{pq}-\partial_{m}g^{pq}\partial_{p}g_{nq}-\partial_{n}g^{pq}\partial_{p}g_{mq}+g^{pq}g^{rs}\partial_{p}g_{mr}(\partial_{q}g_{ns}-\partial_{s}g_{nq})\\
+\partial_{m}(g^{pq}\Gamma_{npq})+\partial_{n}(g^{pq}\Gamma_{mpq})-(g^{pq}\Gamma_{rpq})\Gamma_{mn}^{r},\label{eq:EFE-empty-expanded}
\end{multline}
where we are freely using $g^{np}\partial_{q}g_{mn}=-\partial_{q}g^{np}g_{mn}$,
which follows from $\partial_{q}(g_{mn}g^{np})=0$.

Now we can investigate the single-particle solutions of equation \eqref{eq:EFE-emptyspace},
which we identify as the graviton field. Since we are focusing on
flat space, the metric is expanded as
\begin{equation}
g_{mn}(x)=\eta_{mn}+\tilde{H}_{mn}(x).
\end{equation}
At the linearized level, equation \eqref{eq:EFE-empty-expanded} leads
to
\begin{equation}
\Box\tilde{H}_{mn}=\partial_{m}\left(\partial^{p}\tilde{H}_{np}-\frac{1}{2}\eta^{pq}\partial_{n}\tilde{H}_{pq}\right)+\partial_{n}\left(\partial^{p}\tilde{H}_{mp}-\frac{1}{2}\eta^{pq}\partial_{m}\tilde{H}_{pq}\right).\label{eq:graviton-linearized-flat}
\end{equation}
The next step is to factor in the gauge redundancy introduced by the
invariance under general coordinate transformations. Equation \eqref{eq:metric-transformation}
leads to the infinitesimal transformation
\begin{equation}
\delta\tilde{H}_{mn}=\partial_{n}\Lambda_{m}+\partial_{m}\Lambda_{n},\label{eq:graviton-residual}
\end{equation}
where $\delta x^{m}=\Lambda^{m}(x)$. We will work with the harmonic
gauge, which at the linearized level is expressed as
\begin{equation}
\partial^{n}\tilde{H}_{mn}=\frac{1}{2}\eta^{np}\partial_{m}\tilde{H}_{np}.
\end{equation}
This can be considered the analogue of the Lorenz gauge in the Yang--Mills
theory. And just like in the Lorenz gauge, there is a residual gauge
transformation with parameter $\Lambda^{m}$ satisfying the massless
equations,
\begin{equation}
\Box\Lambda_{m}=0.
\end{equation}
Its longitudinal part can be further used to fix $\eta^{mn}\tilde{H}_{mn}=0$,
a traceless condition. It then directly follows from the harmonic
gauge that the graviton is transversal $\partial^{n}\tilde{H}_{mn}=0$,
now with a residual gauge transformation \eqref{eq:graviton-residual}
with transversal parameter $\Lambda^{m}$, (i.e., $\partial_{m}\Lambda^{m}=0$).
In a plane-wave basis, we have
\begin{equation}
g_{mn}(x)=\eta_{mn}+h_{mn}e^{\mathrm{i}k\cdot x},\label{eq:graviton-single-particle}
\end{equation}
with $k^{2}=0$, and graviton polarization, $h_{mn}$, satisfying
\begin{align}
h_{mn} & =h_{nm},\\
k^{n}h_{mn} & =0,\\
\eta^{mn}h_{mn} & =0,
\end{align}
with a residual gauge redundancy given by
\begin{equation}
\delta h_{mn}=k_{m}\lambda_{n}+k_{n}\lambda_{m},\label{eq:graviton-residual-traceless}
\end{equation}
and $k_{m}\lambda^{m}=0$. All together, these conditions describe
a massless spin two single-particle state.

\subsubsection{Multi-graviton expansion}

Next, we would like to investigate the classical multi-particle solutions
of equation \eqref{eq:EFE-empty-expanded}.

We will work with the harmonic gauge
\begin{equation}
g^{np}\Gamma_{np}^{m}=0.\label{eq:harmonic-gauge}
\end{equation}
In this case, the equation of motion is significantly simplified,
and we are left with
\begin{equation}
g^{pq}\partial_{p}\partial_{q}g_{mn}=\frac{1}{2}\partial_{m}g^{pq}\partial_{n}g_{pq}-\partial_{m}g^{pq}\partial_{p}g_{nq}-\partial_{n}g^{pq}\partial_{p}g_{mq}+g^{pq}g^{rs}\partial_{p}g_{mr}(\partial_{q}g_{ns}-\partial_{s}g_{nq}).\label{eq:EFE-gauge-fixed}
\end{equation}
Since $g^{mn}$ is the inverse of $g_{mn}$, this equation encodes
an infinite number of interaction vertices. This is very similar to
what happened in the NLSM model, where we have introduced multi-particle
expansions for $U$ and $U^{-1}$ (cf. equations \eqref{eq:NLSM-group-Uexpansion}
and \eqref{eq:NLSM-group-Ubarexpansion}), as long as the condition
$U^{-1}U=\mathbb{I}$ was imposed. Here we do something similar, starting
with
\begin{equation}
g_{mn}=\eta_{mn}+\sum_{P}H_{Pmn}e^{\mathrm{i}k_{P}\cdot x},\label{eq:multi-graviton}
\end{equation}
where $H_{Pmn}$ denotes the multi-graviton currents. Its inverse
can be expanded as
\begin{equation}
g^{mn}=\eta^{mn}-\sum_{P}I_{P}^{mn}e^{\mathrm{i}k_{P}\cdot x}.\label{eq:multi-graviton-inverse}
\end{equation}
As usual, one-letter words are mapped to single-particle polarizations.
The currents $I_{P}^{mn}$ are then recursively determined in terms
of $H_{Pmn}$ by imposing that $g^{mn}g_{np}=\delta_{p}^{m}$. We
then have
\begin{equation}
\left(\eta^{mn}-\sum_{P}I_{P}^{mn}e^{\mathrm{i}k_{P}\cdot x}\right)\left(\eta_{np}+\sum_{Q}H_{Qnp}e^{\mathrm{i}k_{Q}\cdot x}\right)=\delta_{p}^{m},
\end{equation}
which yields
\begin{align}
I_{P}^{mn} & =\eta^{mp}\eta^{nq}H_{Ppq}-\eta^{nq}\sum_{P=Q\cup R}I_{Q}^{mp}H_{Rpq},\nonumber \\
 & =\eta^{mp}\eta^{nq}H_{Ppq}-\eta^{mp}\sum_{P=Q\cup R}H_{Qpq}I_{R}^{nq}.\label{eq:gravity-I-H-map}
\end{align}
The equivalence between the first and second lines is straightforward
to demonstrate. In particular, one can show that
\begin{equation}
I_{P}^{mn}=\eta^{mp}\eta^{nq}\sum_{i=1}^{|P|}(-1)^{i}\sum_{P=P_{1}\cup\ldots\cup P_{i}}(H_{P_{1}}\cdots H_{P_{i}})_{pq},\label{eq:graviton-multi-inversion}
\end{equation}
with $(H_{P_{1}}\cdots H_{P_{i}})_{mn}$ defined via the natural contractions
with the flat metric, for example 
\begin{equation}
(H_{P_{1}}H_{P_{2}}H_{P_{3}})_{mn}=H_{P_{1}mm_{1}}\eta^{m_{1}n_{1}}H_{P_{2}n_{1}m_{2}}\eta^{m_{2}n_{2}}H_{P_{3}n_{2}n}.
\end{equation}
This expansion may be more easily understood if we simply forget about
the spacetime indices for a moment. We then have
\begin{align}
1-I & =\left(1+H\right)^{-1}\nonumber \\
I & =\sum_{n=1}^{\infty}(-1)^{n+1}H^{n},
\end{align}
which is of course reminiscent of the NLSM construction. As usual,
the series expansion on the right hand side is truncated because the
single-particle labels are taken to be nilpotent. Equation \eqref{eq:graviton-multi-inversion}
manifestly displays an arbitrary number of interaction vertices. The
sum
\[
\sum_{P=P_{1}\cup\ldots\cup P_{i}}
\]
defined in equation \eqref{eq:shorthand-ordered}, is essentially
connected to an $(i+1)$-point interaction vertex in the perturbiner
framework.

\subsubsection{General coordinate transformations}

In an infinitesimal form, the transformation law of the metric leads
to
\begin{equation}
\delta g_{mn}=g_{mp}\partial_{n}\lambda^{p}+g_{np}\partial_{m}\lambda^{p}+\lambda^{p}\partial_{p}g_{mn},\label{eq:GCT-linearized}
\end{equation}
where $\lambda^{m}=x^{m}-\tilde{x}^{m}$ and $\delta g_{mn}=\tilde{g}_{mn}-g_{mn}$.
If we assume a multi-particle expansion of the form
\begin{equation}
\lambda^{m}(x)=\sum_{P}\Lambda_{P}^{m}e^{\mathrm{i}k_{P}\cdot x},
\end{equation}
it is direct to show that the currents $H_{Pmn}$ transform as
\begin{equation}
\delta H_{Pmn}=\mathrm{i}\left(k_{Pm}\Lambda_{Pn}+k_{Pn}\Lambda_{Pm}+\sum_{P=Q\cup R}(k_{Rn}H_{Qmp}+k_{Rm}H_{Qnp}+k_{Qp}H_{Qmn})\Lambda_{R}^{p}\right).\label{eq:graviton-multi-gauge-transf}
\end{equation}

In terms of the multi-particle currents, the harmonic gauge reads
\begin{equation}
k_{P}^{n}H_{Pmn}-\frac{1}{2}\eta^{np}k_{Pm}H_{np}=\sum_{P=Q\cup R}I_{Q}^{np}\left(\frac{1}{2}k_{Rm}H_{Rnp}-k_{Rp}H_{Rmn}\right).\label{eq:harmonic-multi-particle}
\end{equation}
The residual gauge transformations are then straightforward to obtain,
and the transformation of equation \eqref{eq:harmonic-multi-particle}
leads to a recursive definition of $\Lambda_{P}$ in the form
\begin{equation}
\Lambda_{Pm}=\frac{1}{s_{P}}\sum_{P=Q\cup R}(\cdots)_{m},\label{eq:graviton-residual-multi-parameter}
\end{equation}
where the right hand side is given by products of the currents $H$,
$\Lambda$ and their momenta, labeled with sub-words of $P$.
The precise form is not really illuminating, and this schematic expression
will suffice here for the discussion on the gauge invariance of tree
level scattering amplitudes.

\subsubsection{Gravitational Berends--Giele currents and tree level amplitudes\label{subsec:Gravitational-Berends=002013Giele}}

In order to determine the recursion for the currents $H_{Pmn}$, we
substitute the multi-particle expansions \eqref{eq:multi-graviton}
and \eqref{eq:multi-graviton-inverse} in the gauge fixed equation
of motion \eqref{eq:EFE-gauge-fixed}. The result is
\begin{align}
s_{P}H_{Pmn} & =\sum_{P=Q\cup R}I_{Q}^{pq}\left(k_{Qm}k_{Rp}H_{Rnq}+k_{Qn}k_{Rp}H_{Rmq}+k_{Rp}k_{Rq}H_{Rmn}-\frac{1}{2}k_{Qm}k_{Rn}H_{Rpq}\right)\nonumber \\
 & +\sum_{P=Q\cup R}k_{Qp}H_{Qmr}(k_{Rq}H_{Rns}-k_{Rs}H_{Rnq})\eta^{pq}\eta^{rs}\nonumber \\
 & -\sum_{P=Q\cup R\cup S}k_{Qp}H_{Qmr}(k_{Rq}H_{Rns}-k_{Rs}H_{Rnq})(I_{S}^{pq}\eta^{rs}+\eta^{pq}I_{S}^{rs})\nonumber \\
 & +\sum_{P=Q\cup R\cup S\cup T}k_{Qp}H_{Qmr}(k_{Rq}H_{Rns}-k_{Rs}H_{Rnq})I_{S}^{pq}I_{T}^{rs}.\label{eq:graviton-multi-recursion}
\end{align}
Since there is no color ordering, $H_{Pmn}$ is automatically symmetric
in the exchange of any single-particle labels of the word $P$. The
above recursion embodies the generalization of the Berends--Giele
currents to gravity. A cubic version of pure-gravity Berends--Giele
currents was already presented in \cite{Cheung:2017kzx}. However,
the metric density formulation (first proposed in \cite{Landau:1975pou},
which will be discussed in subsection \ref{subsec:Pure-gravity})
ends up polluting the coupling to matter with arbitrary point interaction
vertices. The more general formulation within the perturbiner framework
came later in \cite{Gomez:2021shh}, which we are following here. 

Let us present some examples. The two-particle current is given by
\begin{multline}
s_{12}H_{12mn}=(k_{2}\cdot h_{1}\cdot h_{2})_{n}k_{1m}+(k_{1}\cdot h_{2}\cdot h_{1})_{n}k_{2m}-(k_{2}\cdot h_{1})_{m}(k_{1}\cdot h_{2})_{n}\\
+\frac{1}{2}(k_{2}\cdot h_{1}\cdot k_{2})h_{2mn}+\frac{1}{2}(k_{1}\cdot h_{2}\cdot k_{1})h_{1mn}-\frac{1}{2}k_{1m}k_{2n}(h_{1}\cdot h_{2})+(m\leftrightarrow n),\label{eq:two-graviton-current}
\end{multline}
with the dot product ($\cdot$) denoting the spacetime vector index
contractions, e.g., $(k_{1}\cdot h_{2})_{m}=k_{1}^{n}h_{2mn}$ and
$(h_{1}\cdot h_{2})=h_{1}^{mn}h_{2mn}$. The expression for the three-particle
current is already quite long, so we will leave it in terms of the
two-particle currents,
\begin{multline}
s_{123}H_{123mn}=(k_{1}\cdot k_{23})(h_{1}\cdot H_{23})_{mn}+(k_{2}\cdot k_{13})(h_{2}\cdot H_{13})_{mn}+(k_{3}\cdot k_{12})(h_{3}\cdot H_{12})_{mn}\\
+k_{1m}(k_{23}\cdot h_{1}\cdot H_{23})_{n}+k_{23m}(k_{1}\cdot H_{23}\cdot h_{1})_{n}-(k_{23}\cdot h_{1})_{m}(k_{1}\cdot H_{23})_{n}\\
+k_{2m}(k_{13}\cdot h_{2}\cdot H_{13})_{n}+k_{13m}(k_{2}\cdot H_{13}\cdot h_{2})_{n}-(k_{13}\cdot h_{2})_{m}(k_{2}\cdot H_{13})_{n}\\
+k_{3m}(k_{12}\cdot h_{3}\cdot H_{12})_{n}+k_{12m}(k_{3}\cdot H_{12}\cdot h_{3})_{n}-(k_{12}\cdot h_{3})_{m}(k_{3}\cdot H_{12})_{n}\\
+\frac{1}{2}(k_{23}\cdot h_{1}\cdot k_{23})H_{23mn}-\frac{1}{2}k_{1m}k_{23n}(h_{1}\cdot H_{23})+\frac{1}{2}(k_{1}\cdot H_{23}\cdot k_{1})h_{1mn}\\
+\frac{1}{2}(k_{13}\cdot h_{2}\cdot k_{13})H_{13mn}-\frac{1}{2}k_{2m}k_{13n}(h_{2}\cdot H_{13})+\frac{1}{2}(k_{2}\cdot H_{13}\cdot k_{2})h_{2mn}\\
+\frac{1}{2}(k_{12}\cdot h_{3}\cdot k_{12})H_{12mn}-\frac{1}{2}k_{3m}k_{12n}(h_{3}\cdot H_{12})+\frac{1}{2}(k_{3}\cdot H_{12}\cdot k_{3})h_{3mn}\\
-(h_{1}\cdot h_{2}\cdot h_{3})(k_{1m}k_{2n}+k_{1m}k_{3n}+k_{2m}k_{1n}+k_{2m}k_{3n}+k_{3m}k_{1n}+k_{3m}k_{2n})\\
-\frac{1}{2}s_{12}(h_{1}\cdot h_{3}\cdot h_{2})_{mn}-\frac{1}{2}s_{13}(h_{1}\cdot h_{2}\cdot h_{3})_{mn}-\frac{1}{2}s_{23}(h_{2}\cdot h_{1}\cdot h_{3})_{mn}\\
-(k_{1}\cdot h_{3}\cdot k_{2})(h_{1}\cdot h_{2})_{mn}-(k_{1}\cdot h_{2}\cdot k_{3})(h_{1}\cdot h_{3})_{mn}-(k_{2}\cdot h_{1}\cdot k_{3})(h_{2}\cdot h_{3})_{mn}\\
+(k_{2}\cdot h_{1})_{m}(k_{1}\cdot h_{3}\cdot h_{2})_{n}+(k_{1}\cdot h_{2})_{m}(k_{2}\cdot h_{3}\cdot h_{1})_{n}+(k_{3}\cdot h_{1})_{m}(k_{1}\cdot h_{2}\cdot h_{3})_{n}\\
+(k_{1}\cdot h_{3})_{m}(k_{3}\cdot h_{2}\cdot h_{1})_{n}+(k_{3}\cdot h_{2})_{m}(k_{2}\cdot h_{1}\cdot h_{3})_{n}+(k_{2}\cdot h_{3})_{m}(k_{3}\cdot h_{1}\cdot h_{2})_{n}\\
+k_{12m}[(k_{3}\cdot h_{1}\cdot h_{2}\cdot h_{3})_{n}+(k_{3}\cdot h_{2}\cdot h_{1}\cdot h_{3})_{n}]+(k_{3}\cdot h_{1}\cdot h_{2}\cdot k_{3})h_{3mn}\\
+k_{13m}[(k_{2}\cdot h_{1}\cdot h_{3}\cdot h_{2})_{n}+(k_{2}\cdot h_{3}\cdot h_{1}\cdot h_{2})_{n}]+(k_{2}\cdot h_{1}\cdot h_{3}\cdot k_{2})h_{2mn}\\
+k_{23m}[(k_{1}\cdot h_{2}\cdot h_{3}\cdot h_{1})_{n}+(k_{1}\cdot h_{3}\cdot h_{2}\cdot h_{1})_{n}]+(k_{1}\cdot h_{2}\cdot h_{3}\cdot k_{1})h_{1mn}+(m\leftrightarrow n).
\end{multline}
Within the field theories we have so far discussed, this three-particle
current is undeniably the longest one. Note, in particular, that roughly
half of the expression corresponds to a four-point contact interaction.

Finally, the tree-level scattering amplitudes are given by
\begin{align}
\mathcal{M}_{N+1} & =\frac{1}{2\kappa}\lim_{s_{1\ldots N}\to0}h_{N+1}^{mn}s_{1\ldots N}H_{1\ldots Nmn},\nonumber \\
 & =\frac{1}{2\kappa}\lim_{s_{1\ldots N}\to0}(h_{N+1})_{mn}s_{1\ldots N}I_{1\ldots N}^{mn},\label{eq:graviton-amplitude}
\end{align}
which mimics the usual expressions with which we have been working.
The equivalence between the first and the second lines trivially follows
from \eqref{eq:graviton-multi-inversion}, as only the highest ranking
current survives the limit.

As an example, the three-graviton amplitude can be easily computed
using the two-particle current of equation \eqref{eq:two-graviton-current}.
The result is
\begin{align}
2\kappa\mathcal{M}_{3} & =2(k_{2}\cdot h_{1}\cdot h_{2}\cdot h_{3}\cdot k_{1})+2(k_{1}\cdot h_{2}\cdot h_{1}\cdot h_{3}\cdot k_{2})-2(k_{2}\cdot h_{1}\cdot h_{3}\cdot h_{2}\cdot k_{1})\nonumber \\
 & +(k_{2}\cdot h_{1}\cdot k_{2})(h_{2}\cdot h_{3})+(k_{1}\cdot h_{2}\cdot k_{1})(h_{1}\cdot h_{3})-(k_{1}\cdot h_{3}\cdot k_{1})(h_{1}\cdot h_{2}),\nonumber \\
 & =\frac{1}{4}h_{1}^{mn}h_{2}^{pq}h_{3}^{rs}V_{mpr}V_{nqs}.
\end{align}
Momentum conservation and transversality of the graviton polarizations
were used in order to rewrite the amplitude symmetrically in the three
graviton legs, and to make explicit the cubic gluon vertex $V_{mnp}$
of equation \eqref{eq:three-gluon-vertex}.

The residual gauge invariance implied by the harmonic gauge choice
extends to the amplitude. For instance, if we take 
\begin{equation}
\delta h_{N+1}^{mn}=k_{N+1}^{m}\lambda^{n}+k_{N+1}^{n}\lambda^{m},
\end{equation}
we obtain
\begin{align}
\delta\mathcal{M}_{N+1} & =\frac{1}{2\kappa}\lim_{s_{1\ldots N}\to0}\delta h_{N+1}^{mn}s_{1\ldots N}H_{1\ldots Nmn},\nonumber \\
 & =\frac{1}{\kappa}\lim_{s_{1\ldots N}\to0}\lambda^{m}k_{N+1}^{n}s_{1\ldots N}H_{1\ldots Nmn},\nonumber \\
 & =-\frac{1}{\kappa}\lim_{s_{1\ldots N}\to0}\lambda^{m}s_{1\ldots N}(k_{1\ldots N}^{n}H_{1\ldots Nmn}),\nonumber \\
 & =-\frac{1}{2\kappa}\lim_{s_{1\ldots N}\to0}\lambda^{m}s_{1\ldots N}(k_{1\ldots N}^{m}\eta^{np}H_{1\ldots Nnp}),\nonumber \\
 & =\frac{1}{2\kappa}\lim_{s_{1\ldots N}\to0}(k_{N+1}\cdot\lambda)s_{1\ldots N}(\eta^{np}H_{1\ldots Nnp})\nonumber \\
 & =0,
\end{align}
where we have used momentum conservation $k_{N+1}=-k_{1\ldots N}$
and equation \eqref{eq:harmonic-multi-particle}. For the latter,
note that the right hand side of \eqref{eq:harmonic-multi-particle}
is suppressed in the limit. The condition $(k_{N+1}\cdot\lambda)=0$
follows from the discussion around equation \eqref{eq:graviton-residual-traceless}.
We can also demonstrate the gauge invariance of $\mathcal{M}_{N+1}$
with respect to any of the other graviton legs (i.e., with $\delta H_{1\ldots Nmn}$
as presented in \eqref{eq:graviton-multi-gauge-transf}). The invariance
of the amplitude follows similar arguments, we just have to keep in
mind the general form of the multi-particle parameter \eqref{eq:graviton-residual-multi-parameter}.

\subsubsection*{A word on normalization conventions}

The overall normalization of $\mathcal{M}$ may appear arbitrary,
but it follows from the fact that the Einstein--Hilbert action (see
equation \eqref{eq:EH-action}) is accompanied by a $\kappa^{-1}$
factor. Throughout this section we have suppressed factors of $\kappa$
in the propagators and intermediate currents. They are distributed
differently across the recursion depending on the conventions adopted.
For instance, it is common to have $\kappa$ absorbed into the field
redefinition $g_{mn}=\eta_{mn}+\sqrt{\kappa}h_{mn}$, in which case
propagators are $\kappa$-free and the coupling is carried entirely
by the vertices. In either case, the overall power of $\kappa$ at
any given order is fixed by dimensional analysis and the structure
of the action, so the reader can in principle reinstate the suppressed
factors by tracing them back to the action and the graviton propagator
derived from it.

\subsubsection{Soft limit\label{subsec:Soft-limit}}

We will conclude the pure gravity tree level discussion with a short
note on soft limits.

It has been long known that the soft limit of graviton amplitudes
has a universal behavior. When the momentum of one of the external
gravitons is taken to zero, the amplitude factorizes into a universal
soft factor multiplied by the lower-point amplitude, a statement that
holds to leading order in the soft momentum regardless of the specifics
of the theory. This universality, first established by Weinberg in
the context of low-energy theorems for graviton emission \cite{Weinberg:1964ew,Weinberg:1965nx},
reflects the long-range nature of the gravitational interaction. Unlike
the gauge theory case, the leading soft graviton factor involves a
sort of weighted sum over the ``hard'' momenta contracted with the
soft polarization tensor (see equation \eqref{eq:graviton-leading-soft}
below), reflecting the spin-2 nature of the graviton. In the perturbiner
framework, the structure of the multi-particle currents makes the
leading order soft limit analysis remarkably transparent already at
the level of the recursion relations, as we now discuss.

We will take $h_{N+1}^{mn}$ as the soft graviton of the amplitude
\eqref{eq:graviton-amplitude}, parametrizing its momentum as $k_{N+1}^{\mu}=\tau q^{\mu}$,
with $q^{2}=0$ and soft parameter $\tau$. In the soft limit ($\tau\to0$),
the dominant contributions in $H_{1\ldots Nmn}$ is straightforward
to identify. It comes from the multi-particle currents with $(N-1)$
particles, which carry the poles of the generalized Mandelstam variables
with $(N-1)$ momenta. For example,
\begin{align}
s_{2\ldots N} & =(k_{2}+\ldots+k_{N})^{2},\nonumber \\
 & =(k_{1}+k_{N+1})^{2},\nonumber \\
 & =2\tau(k_{1}\cdot q),
\end{align}
where we assume momentum conservation of a $(N+1)$-point amplitude.
In this case, the $N$-particle currents $H_{1\ldots Nmn}$ can then
be expressed as
\begin{equation}
s_{1\ldots N}H_{1\ldots Nmn}=k_{1m}k_{1n}(h_{1}\cdot H_{2\ldots N})+\textrm{sym}(1,\ldots,N)+\mathcal{O}(\tau^{0}),
\end{equation}
where the leading soft singularity coming from the $(N-1)$-particle
currents has been flagged, and the symmetrization over the $N$ single-particle
labels is explicit.

As for the amplitude, the leading order contribution can then be cast
as
\begin{align}
\mathcal{M}_{N+1} & =\frac{1}{2\kappa}\lim_{s_{1\ldots N}\to0}h_{N+1}^{mn}s_{1\ldots N}H_{1\ldots Nmn},\nonumber \\
 & =\frac{1}{2\kappa}\lim_{s_{1\ldots N}\to0}h_{N+1}^{mn}\left[k_{1m}k_{1n}(h_{1}\cdot H_{2\ldots N})+\textrm{sym}(1,\ldots,N)\right]+\mathcal{O}(\tau^{0}),\nonumber \\
 & =\lim_{s_{1\ldots N}\to0}\left[\frac{(k_{1}\cdot h_{N+1}\cdot k_{1})}{s_{2\ldots N}}\left(\frac{1}{2\kappa}s_{2\ldots N}(h_{1}\cdot H_{2\ldots N})\right)+\textrm{sym}(1,\ldots,N)\right]+\mathcal{O}(\tau^{0}).
\end{align}
Accompanied with the soft limit, the expression inside the parentheses
yields the $N$-point amplitude at leading order in $\tau$. Therefore,
we have
\begin{equation}
\left.\mathcal{M}_{N+1}\right|_{\textrm{soft}}=\frac{1}{\tau}\left(\sum_{i=1}^{N}\frac{(k_{i}\cdot h_{N+1}\cdot k_{i})}{2(k_{i}\cdot q)}\right)\mathcal{M}_{N}+\mathcal{O}(\tau^{0}),\label{eq:graviton-leading-soft}
\end{equation}
manifesting the universal Weinberg pole. Regarding sub-leading contributions
(see e.g. \cite{Cachazo:2014fwa,Chakrabarti:2017ltl,Chakrabarti:2017zmh}),
they are not as trivial to obtain explicitly. However, since the invariance
under general coordinate transformations (diffeomorphisms) is in-built
in the construction, sub-leading soft limits should also be reproduced.

This is a good point to introduce a disclaimer regarding the perturbiner
framework for pure gravity. We of course have a streamlined way to
obtain the scattering trees, but at the end of the day they should
coincide with traditional diagrammatic computations. On the other
hand, there are many scattering amplitude methods that can be used
to compute graviton amplitudes. The BCFW recursion relations \cite{Britto:2004ap,Britto:2005fq},
originally formulated for gauge theory and extended to gravity shortly
after \cite{Bedford:2005yy,Cachazo:2005ca}, provide an on-shell
recursive construction of graviton amplitudes based on complex momentum
shifts that avoids any reference to an action or off-shell quantities.
The CHY formalism \cite{Cachazo:2013gna,Cachazo:2013hca,Cachazo:2014nsa}
offers a strikingly compact representation of graviton amplitudes
as integrals over the moduli space of punctured Riemann spheres, with
the integrand built from solutions to the scattering equations. The
double copy \cite{Kawai:1985xq,Bern:2008qj,Bern:2010ue}, rooted
in the KLT relations between open and closed string amplitudes and
later recast in terms of color-kinematics duality, expresses graviton
amplitudes directly in terms of gauge theory building blocks, bypassing
gravitational Feynman rules altogether. These on-shell methods are
usually much more efficient than the perturbiner output for pure graviton
amplitudes. Interestingly, they do not extend as naturally to more
general theories when different types of fields (matter) are involved.
The BCFW recursion requires careful treatment of the boundary terms that
arise for non-trivial matter couplings (e.g. \cite{Arkani-Hamed:2008bsc}),
and the double copy, while applicable to a restricted class of matter-coupled
theories (e.g. \cite{Johansson:2015oia,Johansson:2019dnu}), is not
straightforwardly defined beyond them. The CHY formalism has been
extended to include certain matter fields \cite{Cachazo:2014nsa},
but the construction of the relevant integrands becomes increasingly
involved. As we will see next, these extra ingredients are seamlessly
accommodated in classical multi-particle solutions.

\subsection{Coupling to matter}

The field equations \eqref{eq:EFE} can be derived from the minimum
action principle applied to
\begin{equation}
S=S_{\textrm{E.H.}}+S_{\textrm{matter}},\label{eq:action-gravity-full}
\end{equation}
where $S_{\textrm{E.H.}}$ is the Einstein--Hilbert action with a
cosmological constant,
\begin{equation}
S_{\textrm{E.H.}}=\frac{1}{2\kappa}\int d^{D}x\,\sqrt{-g}(R-2\Lambda),\label{eq:EH-action}
\end{equation}
and $S_{\textrm{matter}}$ is some matter action. Here we have $g\equiv\det(g_{mn})$,
such that $\delta g=gg^{mn}\delta g_{mn}$, and
\begin{equation}
T_{mn}=-\frac{2}{\sqrt{-g}}\frac{\delta}{\delta g^{mn}}S_{\textrm{matter}}.
\end{equation}

Equation \eqref{eq:EFE} leads to
\begin{equation}
R=\frac{2}{(2-D)}(\kappa g^{mn}T_{mn}-\Lambda D),
\end{equation}
and can be written in a trace-reversed form as
\begin{equation}
R_{mn}+\left(\frac{2}{2-D}\right)\Lambda g_{mn}=\kappa\left(T_{mn}+\frac{1}{(2-D)}g_{mn}g^{pq}T_{pq}\right),\label{eq:EFE-tracerev-matter}
\end{equation}
which is in the most convenient form to determine multi-graviton solutions
in the presence of matter.

\subsubsection{Scalars}

In the case of scalars coupled to gravity, a typical equation of motion
would be
\begin{equation}
g^{mn}\nabla_{m}\nabla_{n}\phi-\mathrm{m}^{2}\phi=F(\phi),\label{eq:scalar-eom-curved}
\end{equation}
where the left hand side takes the covariant form of \eqref{eq:massless-free-scalar},
the kinetic operator, with
\begin{equation}
g^{mn}\nabla_{m}\nabla_{n}\phi=g^{mn}(\partial_{m}\partial_{n}\phi-\Gamma_{mn}^{p}\partial_{p}\phi),\label{eq:covariant-box}
\end{equation}
and the right hand side comes from a (self-interaction) potential
$V(\phi)$, with $F=\partial_{\phi}V$ . There are relevant models
in which the scalar couples to the spacetime curvature as well. We
will see them in the next section. For now, we keep working around
flat space.

The energy-momentum tensor of the scalar field is given by
\begin{equation}
T_{mn}=\partial_{m}\phi\partial_{n}\phi-\frac{1}{2}g_{mn}[g^{pq}\partial_{p}\phi\partial_{q}\phi+\mathrm{m}^{2}\phi^{2}+2V(\phi)],\label{eq:EMT-scalar}
\end{equation}
and it is straightforward to check it is covariantly conserved using
the equation of motion \eqref{eq:scalar-eom-curved},
\begin{equation}
\nabla^{n}T_{mn}=0.
\end{equation}

In terms of multi-particle expansions, we start with the ansatze \eqref{eq:scalar-multiparticle},
\eqref{eq:multi-graviton}, and \eqref{eq:multi-graviton-inverse}.
The equation of motion \eqref{eq:scalar-eom-curved} leads to the
recursion
\begin{equation}
(s_{P}+\mathrm{m}^{2})\Phi_{P}=\sum_{P=Q\cup R}(k_{R}\cdot I_{Q}\cdot k_{R})\Phi_{R}-\mathcal{F}_{P},
\end{equation}
where $F(\phi)=\sum_{P}\mathcal{F}_{P}e^{\mathrm{i}k_{P}\cdot x}$,
and $\mathcal{F}_{P}$ is at least quadratic in the scalar multi-particle
coefficients (i.e., $V(\phi)\sim\phi^{3}$). Note also that the second
term on the right hand side of equation \eqref{eq:covariant-box}
vanishes because of the harmonic gauge. For the graviton multi-particle
recursion, we just add
\[
\kappa\mathcal{T}_{Pmn}+\frac{\kappa}{(2-D)}\left(\eta_{mn}\eta^{pq}\mathcal{T}_{Ppq}+\sum_{P=Q\cup R}(H_{Qmn}\eta^{pq}-\eta_{mn}I_{Q}^{pq})\mathcal{T}_{Rpq}-\sum_{P=Q\cup R\cup S}H_{Qmn}I_{R}^{pq}\mathcal{T}_{Spq}\right)
\]
to the right hand side of equation \eqref{eq:graviton-multi-recursion},
where
\begin{equation}
T_{mn}=\sum_{P}\mathcal{T}_{Pmn}e^{\mathrm{i}k_{P}\cdot x},
\end{equation}
which is derived from the multi-particle expansion of the energy momentum
tensor \eqref{eq:EMT-scalar}.

Let us work out the three-point amplitude with one graviton ($1$)
and two scalars $(2$ and $3$). This is the simplest example with
mixed field content. The relevant currents are
\begin{align}
(s_{12}+\mathrm{m}^{2})\Phi_{12} & =(k_{2}\cdot h_{1}\cdot k_{2})\phi_{2}\\
\mathcal{T}_{23mn} & =(\frac{1}{2}\eta_{mn}s_{23}-k_{2m}k_{3n}-k_{2n}k_{3m})\phi_{2}\phi_{3}\\
s_{23}H_{23mn} & =-\frac{2\kappa}{(2-D)}\eta_{mn}\mathrm{m}^{2}\phi_{2}\phi_{3}-\kappa(k_{2m}k_{3n}+k_{2n}k_{3m})\phi_{2}\phi_{3}.
\end{align}
It is then straightforward to demonstrate that the prescriptions \eqref{eq:(n+1)-pointscalar}
and \eqref{eq:graviton-amplitude} agree, leading to the three-point
amplitude
\begin{align}
A_{3} & =(k_{2}\cdot h_{1}\cdot k_{2})\phi_{2}\phi_{3},\nonumber \\
 & =-(k_{2}\cdot h_{1}\cdot k_{3})\phi_{2}\phi_{3}.
\end{align}
Invariance under the residual gauge transformation $\delta h_{1mn}=k_{1m}\lambda_{n}+k_{1n}\lambda_{m}$
is easy to demonstrate using the on-shell conditions $k_{1}\cdot k_{2}=k_{1}\cdot k_{3}=0$.

\subsubsection{Yang--Mills theory}

The source-free Yang--Mills equation of motion in curved spacetime is given by $\nabla^n F_{mn}=0$, where $\nabla_m$ is the combined gauge- and spacetime-covariant derivative. The equation of motion can be expressed  as
\begin{equation}
g^{np}\partial_{p}F_{mn}=g^{np}\left(\mathrm{i}[A_{p},F_{mn}]+\Gamma_{mp}^{q}F_{qn}+\Gamma_{np}^{q}F_{mq}\right),\label{eq:YM-eom-curved}
\end{equation}
with the field-strength $F_{mn}$ given in equation \eqref{eq:YM-fieldstrength}.
We also need the energy-momentum tensor derived from the curved space
version of \eqref{eq:YM-action},
\begin{equation}
T_{mn}=\frac{1}{g_{\textrm{YM}}^{2}}\left(g^{pq}\textrm{Tr}(F_{mp}F_{nq})-\frac{1}{4}g_{mn}g^{pr}g^{qs}\textrm{Tr}(F_{pq}F_{rs})\right).
\end{equation}

In order to obtain the multi-particle solutions of the gluon and the
graviton fields, we fix the respective redundancies. For the latter
we will take the harmonic gauge \eqref{eq:harmonic-gauge}. For the
former, we will choose the curved space version of the Lorenz gauge,
\begin{equation}
g^{mn}(\partial_{m}A_{n}-\Gamma_{mn}^{p}A_{p})=0.\label{eq:Lorenz-gauge-curved}
\end{equation}
Because of the harmonic gauge, the second term inside the parenthesis
drops out, so we are effectively left with $g^{mn}\partial_{m}A_{n}=0$.
In terms of the multi-particle expansions \eqref{eq:YM-perturbiner-color-dressed}
and \eqref{eq:multi-graviton-inverse}, this gauge choice translates
to
\begin{equation}
\eta^{mn}k_{Pm}\mathcal{A}_{Pn}^{a}=\sum_{P=Q\cup R}I_{Q}^{mn}k_{Rm}\mathcal{A}_{Rn}^{a}.
\end{equation}

From equation \eqref{eq:YM-eom-curved} we can derive the recursion
for the currents $\mathcal{A}_{P}^{a}$, and the result is
\begin{multline}
s_{P}\mathcal{A}_{P}^{a}=-f_{bc}^{\phantom{bc}a}\sum_{P=Q\cup R}(\mathrm{i}k_{Q}^{n}\mathcal{A}_{Qm}^{b}\mathcal{A}_{Rn}^{c}+\eta^{np}\mathcal{A}_{Qp}^{b}\mathcal{F}_{Rmn}^{c})\\
+f_{bc}^{\phantom{bc}a}\sum_{P=Q\cup R\cup S}(\mathrm{i}k_{Qp}\mathcal{A}_{Qm}^{b}\mathcal{A}_{Rn}^{c}+\mathcal{A}_{Qp}^{b}\mathcal{F}_{Rmn}^{c})I_{S}^{np}\\
+\sum_{P=Q\cup R}k_{Rp}I_{Q}^{np}(k_{Qm}\mathcal{A}_{Rn}^{a}+k_{Rn}\mathcal{A}_{Rm}^{a})+\mathrm{i}\eta^{pq}k_{Q}^{n}H_{Qmp}\mathcal{F}_{Rqn}^{a}\\
-\mathrm{i}\sum_{P=Q\cup R\cup S}k_{Qp}H_{Qmr}\mathcal{F}_{Rqn}^{a}(\eta^{np}I_{S}^{qr}+I_{S}^{np}\eta^{qr})+\mathrm{i}\sum_{P=Q\cup R\cup S\cup T}k_{Qp}H_{Qmr}\mathcal{F}_{Rqn}^{a}I_{S}^{np}I_{T}^{qr},\label{eq:YM-perturbiner-curved}
\end{multline}
where we have introduced the multi-particle expansion of the field-strength,
\begin{align}
F_{mn}^{a} & =\sum_{P}\mathcal{F}_{Pmn}^{a}e^{\mathrm{i}k_{P}\cdot x},\\
\mathcal{F}_{Pmn}^{a} & =\mathrm{i}(k_{Pm}\mathcal{A}_{Pn}^{a}-k_{Pn}\mathcal{A}_{Pm}^{a})+f_{bc}^{\phantom{bc}a}\sum_{P=Q\cup R}\mathcal{A}_{Qm}^{b}\mathcal{A}_{Rn}^{c}.
\end{align}
In the absence of gravitons, the first line of equation \eqref{eq:YM-perturbiner-curved}
leads back to the recursion \eqref{eq:YM-perturbiner-color-dressed},
as expected.

As an example, we can compute the three-point amplitude with one graviton
($1$) and two gluons ($2$ and $3$). The relevant multi-particle
currents are
\begin{align}
s_{12}\mathcal{A}_{12m}^{a} & =(k_{2}\cdot h_{1}\cdot\varepsilon_{2}^{a})k_{1m}+(k_{2}\cdot h_{1}\cdot k_{2})\varepsilon_{2m}^{a}-(k_{1}\cdot\varepsilon_{2}^{a})(k_{2}\cdot h_{1})_{m}+\frac{1}{2}s_{12}(\varepsilon_{2}^{a}\cdot h_{1})_{m},\\
\mathcal{T}_{23mn} & =\frac{1}{g_{\textrm{YM}}^{2}}\textrm{Tr}(T_{a}T_{b})\left((k_{3m}\varepsilon_{2n}^{a}+k_{3n}\varepsilon_{2m}^{a})(k_{2}\cdot\varepsilon_{3}^{b})+(k_{2m}\varepsilon_{3n}^{b}+k_{2n}\varepsilon_{3m}^{b})(k_{3}\cdot\varepsilon_{2}^{a})\right)\nonumber \\
 & -\frac{1}{g_{\textrm{YM}}^{2}}\textrm{Tr}(T_{a}T_{b})\left((k_{2m}k_{3n}+k_{2n}k_{3m})(\varepsilon_{2}^{a}\cdot\varepsilon_{3}^{b})+\frac{1}{2}s_{23}(\varepsilon_{2m}^{a}\varepsilon_{3n}^{b}+\varepsilon_{2n}^{a}\varepsilon_{3m}^{b})\right)\nonumber \\
 & +\frac{1}{g_{\textrm{YM}}^{2}}\textrm{Tr}(T_{a}T_{b})\left(\frac{1}{2}s_{23}(\varepsilon_{2}^{a}\cdot\varepsilon_{3}^{b})-(k_{3}\cdot\varepsilon_{2}^{a})(k_{2}\cdot\varepsilon_{3}^{b})\right)\eta_{mn},\\
s_{23}H_{23mn} & =\frac{\kappa}{g_{\textrm{YM}}^{2}}\textrm{Tr}(T_{a}T_{b})\left((k_{3m}\varepsilon_{2n}^{a}+k_{3n}\varepsilon_{2m}^{a})(k_{2}\cdot\varepsilon_{3}^{b})+(k_{2m}\varepsilon_{3n}^{b}+k_{2n}\varepsilon_{3m}^{b})(k_{3}\cdot\varepsilon_{2}^{a})\right)\nonumber \\
 & -\frac{\kappa}{g_{\textrm{YM}}^{2}}\textrm{Tr}(T_{a}T_{b})\left((k_{2m}k_{3n}+k_{2n}k_{3m})(\varepsilon_{2}^{a}\cdot\varepsilon_{3}^{b})+\frac{1}{2}s_{23}(\varepsilon_{2m}^{a}\varepsilon_{3n}^{b}+\varepsilon_{2n}^{a}\varepsilon_{3m}^{b})\right)\nonumber \\
 & +\frac{2}{(2-D)}\frac{\kappa}{g_{\textrm{YM}}^{2}}\textrm{Tr}(T_{a}T_{b})\left((k_{3}\cdot\varepsilon_{2}^{a})(k_{2}\cdot\varepsilon_{3}^{b})-\frac{1}{2}s_{23}(\varepsilon_{2}^{a}\cdot\varepsilon_{3}^{b})\right)\eta_{mn}.
\end{align}
Now, using the prescriptions \eqref{eq:Nptgluonamplitude} and \eqref{eq:graviton-amplitude},
we obtain
\begin{align}
\mathcal{M}_{3} & =\frac{1}{g_{\textrm{YM}}^{2}}\textrm{Tr}(T_{a}T_{b})\left((k_{3}\cdot h_{1}\cdot\varepsilon_{2}^{a})(k_{2}\cdot\varepsilon_{3}^{b})+(k_{2}\cdot h_{1}\cdot\varepsilon_{3}^{b})(k_{3}\cdot\varepsilon_{2}^{a})-(k_{2}\cdot h_{1}\cdot k_{3})(\varepsilon_{2}^{a}\cdot\varepsilon_{3}^{b})\right),\nonumber \\
 & =\frac{1}{g_{\textrm{YM}}^{2}}A_{3}.
\end{align}
The difference in the normalization comes from the overall factor
in front of the Yang--Mills action \eqref{eq:YM-action}. 

\subsection{Vielbein formalism and coupling to spinors}

The vielbein formalism addresses the issue that spinorial representations
are not compatible with the group of general coordinate transformations.
By introducing a local Minkowski frame at each point, one reduces
the structure group to $SO(1,D-1)$, which does admit spinorial representations,
allowing a covariant coupling of fermions to gravity (see e.g. \cite{Freedman:2012zz}).

This map between the curved metric $g_{mn}$ and the local flat metric
$\eta_{ab}$ is done through the vielbein $e_{m}^{a}$. Note now there
is a distinction between the flat indices\footnote{There are only so many letters we can use for indices, just keep in
mind that this vector index notation should not clash with some spinor
indices in previous sections. In order to avoid confusion, spinor
indices are left implicit in this subsection. This also helps to maintain
a dimension agnostic character for their representations.} $a,b,\ldots$ and curved indices $m,n,\ldots$. The metric can then
be expressed as
\begin{equation}
g_{mn}=e_{m}^{a}e_{n}^{b}\eta_{ab}.\label{eq:metric-vielbein}
\end{equation}
We can also introduce the inverse vielbein, $e_{a}^{m}$, satisfying
$e_{a}^{m}e_{m}^{b}=\delta_{b}^{a}$ and $e_{a}^{m}e_{n}^{a}=\delta_{n}^{m}$,
such that $\eta_{ab}=e_{a}^{m}e_{b}^{n}g_{mn}$.

In this procedure, we essentially turn the local Lorentz group into
a gauge redundancy. This is expected because the metric has $\frac{1}{2}D(D+1)$
components (symmetric tensor), while the vielbein has $D^{2}$ components,
though the extra freedom cannot have physical relevance. Indeed, the
left hand side of equation \eqref{eq:metric-vielbein} is oblivious
to the local Lorentz transformations
\begin{equation}
\delta_{\textrm{LL}}\tilde{e}_{m}^{a}=\Lambda_{\phantom{a}b}^{a}e_{m}^{b},
\end{equation}
where $\Lambda_{\phantom{a}b}^{a}=\Lambda_{\phantom{a}b}^{a}(x)$
is the transformation parameter, with $\Lambda^{ab}=\Lambda_{\phantom{a}c}^{a}\eta^{bc}=-\Lambda^{ba}$.
This redundancy accounts for the number mismatch, $\frac{1}{2}D(D-1)$,
between the components of the metric and the vielbein.

The gauge field associated with the local Lorentz group is the spin
connection, $\omega_{m}^{ab}$. Its field-strength is given by
\begin{equation}
R_{mn}^{ab}\equiv\partial_{m}\omega_{n}^{ab}-\partial_{n}\omega_{m}^{ab}+\eta_{cd}\omega_{m}^{ac}\omega_{n}^{db}-\eta_{cd}\omega_{n}^{ac}\omega_{m}^{db},\label{eq:Riemann-flattened}
\end{equation}
transforming covariantly under
\begin{equation}
\delta_{\textrm{LL}}\omega_{m}^{ab}=-\partial_{m}\Lambda^{ab}+\Lambda_{\phantom{a}c}^{a}\omega_{m}^{cb}-\Lambda_{\phantom{a}c}^{b}\omega_{m}^{ca}.
\end{equation}
However, the spin connection is not an independent object, otherwise
we would be adding extra degrees of freedom. Just like the Christoffel
symbol (which resembles a gauge field, cf. equation \eqref{eq:Riemann})
is written in terms of $g_{mn}$ via the metric compatibility condition,
$\nabla_{m}g_{np}=0$, the spin connection can be written in terms
of $e_{m}^{a}$ via the vielbein postulate,
\begin{align}
\nabla_{n}e_{m}^{a} & =\partial_{n}e_{m}^{a}-\Gamma_{mn}^{p}e_{p}^{a}+\eta_{bc}\omega_{n}^{ab}e_{m}^{c},\nonumber \\
 & =0.\label{eq:vielbein-postulate}
\end{align}
The spin connection is then expressed as
\begin{equation}
\omega_{m}^{ab}=\frac{1}{2}\eta^{ac}e_{c}^{n}(\partial_{m}e_{n}^{b}-\partial_{n}e_{m}^{b}+\eta^{bd}g_{mp}\partial_{n}e_{d}^{p})-(a\leftrightarrow b).\label{eq:spin-connection-vielbein}
\end{equation}

Equation \eqref{eq:Riemann-flattened} is nothing but the flattened
Riemann tensor,
\begin{equation}
R_{mn}^{ab}\equiv R_{\phantom{q}pmn}^{q}e_{q}^{a}e_{c}^{p}\eta^{bc}.
\end{equation}
This result can be cleanly derived from
\begin{equation}
[\nabla_{p},\nabla_{n}]e_{m}^{a}=0,
\end{equation}
which is a consequence of the vielbein postulate.

Finally, spinor couplings to the curved background are implemented
by replacing spacetime derivatives by their Lorentz-covariant version.
Under local Lorentz transformations, a given spinor $\psi$ transforms
as 
\begin{equation}
\delta_{\textrm{LL}}\psi=\frac{1}{4}\Lambda^{ab}(\gamma_{ab}\psi),
\end{equation}
where $\gamma_{ab}=\tfrac{1}{2}[\gamma{}_{a},\gamma_{b}]$ and $\gamma_{a}$
denotes the usual gamma matrices satisfying $\{\gamma_{a},\gamma_{b}\}=2\eta_{ab}$.
Its covariant derivative is then defined as
\begin{equation}
\nabla_{m}\psi=\partial_{m}\psi+\frac{1}{4}\omega_{m}^{ab}(\gamma_{ab}\psi),
\end{equation}
such that $\nabla_{m}\psi$ also transforms as a spinor.

\subsubsection{Equations of motion}

In terms of the vielbein and spin connection, it is straightforward
to rewrite the action \eqref{eq:action-gravity-full}. The scalar
curvature is given by
\begin{equation}
R=R_{mn}^{ab}e_{a}^{m}e_{b}^{n},
\end{equation}
and we also have $\sqrt{-g}=e$, where $e=\det e_{m}^{a}$. Because of a possible coupling to spinors, the matter
action can also depend on the spin connection.

In this setting, the equations of motion are more easily derived if
we vary the action with respect to the vielbein and the spin connection
independently. This is known as the Palatini variation \cite{Palatini:1919} (see also \cite{Carroll:2004st} for a more modern introduction). In the absence
of a cosmological constant, we obtain
\begin{equation}
R_{m}^{a}-\frac{1}{2}e_{m}^{a}R=\kappa T_{m}^{a},\label{eq:EFE-vielbein}
\end{equation}
where $R_{m}^{a}=R_{mn}^{ab}e_{b}^{n}$, and
\begin{align}
T_{m}^{a} & =-\frac{1}{e}\frac{\delta}{\delta e_{a}^{m}}S_{\textrm{matter}},\nonumber \\
 & =\eta^{ab}e_{b}^{n}T_{mn}.
\end{align}
In a trace-reversed form, equation \eqref{eq:EFE-vielbein} leads
to
\begin{equation}
R_{m}^{a}=\kappa\left(T_{m}^{a}+\frac{1}{(2-D)}e_{b}^{n}T_{n}^{b}e_{m}^{a}\right).\label{eq:EFE-vielbein-tracerev}
\end{equation}
Now, varying the action with respect to the spin connection yields
\begin{equation}
\partial_{n}(ee_{a}^{m}e_{b}^{n}-ee_{b}^{m}e_{a}^{n})+e\omega_{n}^{cd}(P^{-1})_{ab,cd}^{mn}+2\kappa\frac{\delta}{\delta\omega_{m}^{ab}}S_{\textrm{matter}}=0,
\end{equation}
where we define
\begin{align}
(P^{-1})_{ab,cd}^{mn} & \equiv\frac{1}{2}(e_{a}^{m}e_{d}^{n}-e_{d}^{m}e_{a}^{n})\eta_{bc}-\frac{1}{2}(e_{b}^{m}e_{d}^{n}-e_{d}^{m}e_{b}^{n})\eta_{ac}\nonumber \\
 & +\frac{1}{2}(e_{b}^{m}e_{c}^{n}-e_{c}^{m}e_{b}^{n})\eta_{ad}-\frac{1}{2}(e_{a}^{m}e_{c}^{n}-e_{c}^{m}e_{a}^{n})\eta_{bd}.
\end{align}
Notice that
\begin{multline}
P_{mn}^{ab,cd}=\frac{1}{(2-D)}(\eta^{ac}e_{m}^{b}e_{n}^{d}-\eta^{bc}e_{m}^{a}e_{n}^{d}-\eta^{ad}e_{m}^{b}e_{n}^{c}+\eta^{bd}e_{m}^{a}e_{n}^{c})\\
+\frac{1}{2}(\eta^{ac}e_{n}^{b}e_{m}^{d}-\eta^{bc}e_{n}^{a}e_{m}^{d}-\eta^{ad}e_{n}^{b}e_{m}^{c}+\eta^{bd}e_{n}^{a}e_{m}^{c}+\eta^{ac}\eta^{bd}g_{mn}-\eta^{bc}\eta^{ad}g_{mn}),
\end{multline}
satisfies
\begin{equation}
(P^{-1})_{ab,ef}^{mp}P_{pn}^{ef,cd}=\delta_{n}^{m}(\delta_{a}^{c}\delta_{b}^{d}-\delta_{a}^{d}\delta_{b}^{c}).
\end{equation}
Therefore, the equation of motion can be used to determine the spin
connection in terms of the vielbein and the matter action. The result
is
\begin{equation}
\omega_{m}^{ab}=\eta^{bc}e_{c}^{n}(\Gamma_{mn}^{p}e_{p}^{a}-\partial_{m}e_{n}^{a})+\kappa W_{m}^{ab},\label{eq:spin-connection-full}
\end{equation}
where we identify the matter contribution to the spin connection to
be
\begin{equation}
W_{m}^{ab}=-\frac{1}{e}P_{nm}^{cd,ab}\frac{\delta}{\delta\omega_{n}^{cd}}S_{\textrm{matter}}.
\end{equation}

Finally, by substituting the spin connection \eqref{eq:spin-connection-full}
back into the field equation \eqref{eq:EFE-vielbein-tracerev}, we obtain
a second order equation of motion for the vielbein, with the matter
coupling described by $T_{m}^{a}$ and $W_{m}^{ab}$.

\subsubsection{Multi-particle solutions and tree level scattering}

Once we start expanding the equation of motion for the vielbein, it
becomes clear that it is considerably longer than the one in the metric
formulation (e.g. equation \eqref{eq:EFE-tracerev-matter}). So we
will try to pack the different multi-particle expansions in a more convenient
form.

Since we are looking at classical solutions around flat space, we
introduce the expansions
\begin{align}
e_{m}^{a} & =\delta_{m}^{a}+\sum_{P}E_{Pm}^{a}e^{\mathrm{i}k_{P}\cdot x},\\
e_{a}^{m} & =\delta_{a}^{m}-\sum_{P}F_{Pa}^{m}e^{\mathrm{i}k_{P}\cdot x},
\end{align}
where we have the mixed Kronecker deltas $\delta_{a}^{m}$ and $\delta_{m}^{a}$.
This might feel counter-intuitive at first, but it is not different
than what we did in the metric formulation. We are simply doing a
perturbation around flat space, so the curved indices ($m$, $n$,
...) will be treated as flat, and we will be raising/lowering them
using the flat metric. Note that the spacetime signature of $g_{mn}$
is captured by the tangent space metric $\eta_{ab}$, so there is
no issue here regarding the expansion around the identity for the
vielbein.

Since we have $e_{a}^{m}e_{n}^{a}=\delta_{n}^{m}$ and $e_{a}^{m}e_{m}^{b}=\delta_{a}^{b}$,
it is easy to show that\begin{subequations}\label{eq:inverse-vielbein-recursion}
\begin{align}
\delta_{m}^{b}F_{Pa}^{m} & =\delta_{a}^{m}E_{Pm}^{b}-\sum_{P=Q\cup R}F_{Qa}^{m}E_{Rm}^{b},\\
\delta_{n}^{a}F_{Pa}^{m} & =\delta_{a}^{m}E_{Pn}^{a}-\sum_{P=Q\cup R}F_{Qa}^{m}E_{Rn}^{a}.
\end{align}
\end{subequations}The two equations are equivalent, and the demonstration
follows the same logic in the multi-particle coefficients $H_{Pmn}$
and $I_{P}^{mn}$.

Regarding gauge fixing, we can start with the local Lorentz invariance.
In order to make the connection to the metric formulation more immediate,
we eliminate the antisymmetric part of the vielbein,
\begin{equation}
\eta_{ab}E_{Pm}^{b}-\eta_{bm}E_{Pa}^{b}=0.\label{eq:vielbein-symgauge}
\end{equation}
The vanishing of the antisymmetric part of $F_{Pa}^{m}$ follows from
its recursive definition \eqref{eq:inverse-vielbein-recursion}.

We then introduce the multi-particle expansions associated to the
matter action. They can be generically cast as
\begin{align}
T_{m}^{a} & =\sum\mathcal{T}_{Pm}^{a}e^{\mathrm{i}k_{P}\cdot x},\\
W_{m}^{ab} & =\sum\mathcal{W}_{Pm}^{ab}e^{\mathrm{i}k_{P}\cdot x}.
\end{align}
Their specific form depends on the matter content, and we will see
an example soon.

Finally, the expansion of the spin connection can be written as
\begin{equation}
\omega_{m}^{ab}=\sum_{P}\Omega_{Pm}^{ab}e^{\mathrm{i}k_{P}\cdot x},
\end{equation}
where
\begin{equation}
\Omega_{Pm}^{ab}=\mathrm{i}k_{P}^{b}E_{Pm}^{a}-\mathrm{i}k_{P}^{a}E_{Pm}^{b}+\varTheta_{Pm}^{ab}+\kappa\mathcal{W}_{Pm}^{ab},
\end{equation}
and
\begin{multline}
\varTheta_{Pm}^{ab}=\frac{\mathrm{i}}{2}\bigg\{\sum_{P=Q\cup R}\{2(k_{Q}\cdot F_{R})^{a}E_{Qm}^{b}+k_{Qm}(E_{Q}\cdot F_{R})^{ab}+[(E_{Q}\cdot E_{R})^{ac}-(F_{R}\cdot E_{Q})^{ac}]k_{Q}^{b}\eta_{cm}\}\\
+\sum_{P=Q\cup R\cup S}[(k_{Q}\cdot F_{S})^{b}(F_{R}\cdot E_{Q})^{ac}-(k_{Q}\cdot F_{S})^{b}(E_{Q}\cdot E_{R})^{ac}-k_{Q}^{b}(F_{R}\cdot E_{Q}\cdot E_{S})^{ac}]\eta_{cm}\\
+\sum_{P=Q\cup R\cup S\cup T}(k_{Q}\cdot F_{S})^{b}(F_{R}\cdot E_{Q}\cdot E_{T})^{ac}\eta_{cm}\bigg\}-(a\leftrightarrow b).
\end{multline}
Here we have used equation \eqref{eq:spin-connection-full} with the
multi-particle expansions of the vielbein and $W_{m}^{ab}$, grouping
the non-linear vielbein contributions in $\varTheta_{Pm}^{ab}$.

Equation \eqref{eq:EFE-vielbein-tracerev} then leads to
\begin{multline}
s_{P}E_{Pm}^{a}=k_{Pm}(k_{P}\cdot E_{P})^{a}+k_{P}^{a}(k_{P}\cdot E_{P})_{m}-k_{Pm}k_{P}^{a}E_{Pn}^{b}\delta_{b}^{n}+\kappa\mathcal{T}_{Pm}^{a}+\frac{\kappa}{(2-D)}\mathcal{T}_{Pn}^{b}\delta_{b}^{n}\delta_{m}^{a}\\
+\mathrm{i}k_{Pb}(\varTheta_{Pm}^{ab}+\kappa\mathcal{W}_{Pm}^{ab})-\mathrm{i}k_{Pm}(\varTheta_{Pn}^{ab}+\kappa\mathcal{W}_{Pn}^{ab})\delta_{b}^{n}+\sum_{P=Q\cup R}\bigg\{\eta_{cd}\delta_{b}^{n}\Omega_{Qn}^{ac}\Omega_{Rm}^{db}-\eta_{cd}\delta_{b}^{n}\Omega_{Qm}^{ac}\Omega_{Rn}^{db}\\
+(\mathrm{i}k_{Qm}\Omega_{Qn}^{ab}-\mathrm{i}k_{Qn}\Omega_{Qm}^{ab})F_{Rb}^{n}+\frac{\kappa}{(2-D)}\mathcal{T}_{Qn}^{b}(\delta_{b}^{n}E_{Rm}^{a}-\delta_{m}^{a}F_{Rb}^{n})\bigg\}\\
+\sum_{P=Q\cup R\cup S}\bigg\{\eta_{cd}\Omega_{Qm}^{ac}\Omega_{Rn}^{db}-\eta_{cd}\Omega_{Qn}^{ac}\Omega_{Rm}^{db}-\frac{\kappa}{(2-D)}\mathcal{T}_{Qn}^{b}E_{Rm}^{a}\bigg\} F_{Sb}^{n},
\end{multline}
which is almost in the form we have been using to define the recursion.
The final ingredient is the gauge fixing of diffeomorphisms, with
\begin{equation}
(k_{P}\cdot E_{P})_{m}=\frac{1}{2}k_{Pm}E_{Pn}^{a}\delta_{a}^{n},\label{eq:lin-harmonic-vielbein-multi}
\end{equation}
which is just the linearized form of the harmonic gauge,
\begin{equation}
(\eta^{mn}\eta_{ab}-\tfrac{1}{2}\delta_{a}^{m}\delta_{b}^{n})\partial_{m}e_{n}^{b}=0.\label{eq:lin-harmonic-vielbein}
\end{equation}
With this choice, we finally have
\begin{multline}
s_{P}E_{Pm}^{a}=\kappa\mathcal{T}_{Pm}^{a}+\frac{\kappa}{(2-D)}\mathcal{T}_{Pn}^{b}\delta_{b}^{n}\delta_{m}^{a}+\mathrm{i}k_{Pb}(\varTheta_{Pm}^{ab}+\kappa\mathcal{W}_{Pm}^{ab})-\mathrm{i}k_{Pm}(\varTheta_{Pn}^{ab}+\kappa\mathcal{W}_{Pn}^{ab})\delta_{b}^{n}\\
+\sum_{P=Q\cup R}\bigg\{\eta_{cd}(\delta_{b}^{n}\Omega_{Qn}^{ac}\Omega_{Rm}^{db}-\delta_{b}^{n}\Omega_{Qm}^{ac}\Omega_{Rn}^{db})+(\mathrm{i}k_{Qm}\Omega_{Qn}^{ab}-\mathrm{i}k_{Qn}\Omega_{Qm}^{ab})F_{Rb}^{n}\\
+\frac{\kappa}{(2-D)}\mathcal{T}_{Qn}^{b}(\delta_{b}^{n}E_{Rm}^{a}-\delta_{m}^{a}F_{Rb}^{n})\bigg\}\\
+\sum_{P=Q\cup R\cup S}\bigg\{\eta_{cd}(\Omega_{Qm}^{ac}\Omega_{Rn}^{db}-\Omega_{Qn}^{ac}\Omega_{Rm}^{db})-\frac{\kappa}{(2-D)}\mathcal{T}_{Qn}^{b}E_{Rm}^{a}\bigg\} F_{Sb}^{n}.\label{eq:vielbein-recursion}
\end{multline}
This equation establishes the recursive definition of the vielbein
multi-particle currents.

The single-particle polarizations, $E_{im}^{a}=\eta^{an}h_{imn}$,
are identified with the gravitons. In the gauge \eqref{eq:lin-harmonic-vielbein},
they satisfy
\begin{equation}
k_{i}^{n}h_{imn}=\frac{1}{2}k_{im}\eta^{np}h_{inp}.
\end{equation}
We can then use part of the residual gauge invariance to set $\eta^{mn}h_{imn}=0$,
such that the graviton polarizations are symmetric, cf. equation \eqref{eq:vielbein-symgauge},
traceless, and transversal ($k_{i}^{n}h_{imn}=0$). The residual gauge
transformation is given by
\begin{equation}
\delta h_{imn}=k_{im}\lambda_{in}+k_{in}\lambda_{im},\label{eq:residual-gauge-vielbein}
\end{equation}
with $k_{i}\cdot\lambda_{i}=0$.

Following the usual prescription, the $(N+1)$-point amplitude involving
gravitons can be defined as
\begin{align}
\mathcal{M}_{N+1} & =\frac{1}{\kappa}\lim_{s_{1\ldots N}\to0}\eta_{ab}(h_{N+1})_{a}^{m}s_{1\ldots N}E_{1\ldots Nm}^{b},\nonumber \\
 & =\frac{1}{\kappa}\lim_{s_{1\ldots N}\to0}(h_{N+1})_{m}^{a}s_{1\ldots N}F_{1\ldots Na}^{m},\label{eq:vielbein-Npoint}
\end{align}
which is similar to equation \eqref{eq:graviton-amplitude}. Just
like in subsection \ref{subsec:Gravitational-Berends=002013Giele},
we can demonstrate invariance under residual gauge transformations
and leading soft limit behavior. The steps are essentially the same,
so they will not be repeated here.

\subsubsection{Coupling to spinors}

In the final part of this section we will have a quick look at the
coupling of gravity to spin $1/2$ particles in four dimensions.

In flat space, a massless spinor action can be cast as 
\begin{equation}
S_{\psi}^{\textrm{flat}}=\mathrm{i}\int d^{4}x\,\psi^{\alpha}\sigma_{\alpha\dot{\alpha}}^{a}\partial_{a}\bar{\psi}^{\dot{\alpha}}.
\end{equation}
We have already seen this in subsection \ref{subsec:WZmodel}. The minimal
coupling to gravity is simply given by
\begin{align}
S_{\psi} & =\mathrm{i}\int d^{4}x\,e\psi^{\alpha}e_{a}^{m}\sigma_{\alpha\dot{\alpha}}^{a}\nabla_{m}\bar{\psi}^{\dot{\alpha}},\\
\nabla_{m}\bar{\psi}^{\dot{\alpha}}= & \partial_{m}\bar{\psi}^{\dot{\alpha}}+\frac{1}{4}\omega_{m}^{bc}(\sigma_{bc}\bar{\psi})^{\dot{\alpha}},
\end{align}
such that
\begin{align}
T_{m}^{a} & =\mathrm{i}\psi^{\alpha}\sigma_{\alpha\dot{\alpha}}^{b}\nabla_{n}\bar{\psi}^{\dot{\alpha}}(e_{m}^{a}e_{b}^{n}-\delta_{b}^{a}\delta_{m}^{n}),\\
W_{m}^{ab} & =\mathrm{i}e_{m}^{c}(\psi\sigma_{c}\sigma^{ab}\bar{\psi}),
\end{align}
and $\sigma^{ab}=(\sigma^{a}\sigma^{b}-\sigma^{b}\sigma^{a})/2$.

The spinor equation of motion is simply
\begin{equation}
e_{a}^{m}\sigma_{\alpha\dot{\alpha}}^{a}\nabla_{m}\bar{\psi}^{\dot{\alpha}}=0,
\end{equation}
with a similar form for $\psi$. After introducing the multi-particle
ansatze as in equation \eqref{eq:WZ-multi-ansatze}, we obtain
\begin{multline}
\mathrm{i}\sigma_{\alpha\dot{\alpha}}^{m}k_{Pm}\bar{\Psi}_{P}^{\dot{\alpha}}\\=\sum_{P=Q\cup R}[\mathrm{i}\sigma_{\alpha\dot{\alpha}}^{a}k_{Qm}\bar{\Psi}_{Q}^{\dot{\alpha}}F_{Ra}^{m}-\frac{1}{4}\Omega_{Rm}^{bc}(\sigma^{m}\sigma_{bc}\bar{\Psi}_{Q})_{\alpha}]
+\frac{1}{4}\sum_{P=Q\cup R\cup S}F_{Ra}^{m}\Omega_{Sm}^{bc}(\sigma^{a}\sigma_{bc}\bar{\Psi}_{Q})_{\alpha},
\end{multline}
which can then be used to compute the tree level scattering between
gravitons and spin $1/2$ fermions.

As an example, let us consider the three-point scattering between
two spinors ($\psi_{1}^{\alpha}$ and $\bar{\psi}_{2}^{\dot{\alpha}}$)
and one graviton ($h_{3mn}$). The relevant currents are
\begin{align}
\sigma_{\alpha\dot{\alpha}}^{m}k_{23m}\bar{\Psi}_{23}^{\dot{\alpha}} & =(k_{2}\cdot h_{3})_{m}(\sigma^{m}\bar{\psi}_{2})_{\alpha},\\
\mathcal{T}_{12m}^{a} & =(\psi_{1}\sigma^{a}\bar{\psi}_{2})k_{2m},\\
\mathcal{W}_{Pm}^{ab} & =\mathrm{i}(\psi_{1}\sigma_{m}\sigma^{ab}\bar{\psi}_{2}),\\
s_{12}E_{12m}^{a} & =\kappa(\psi_{1}\sigma^{a}\bar{\psi}_{2})k_{2m}-\kappa k_{12b}(\psi_{1}\sigma_{m}\sigma^{ab}\bar{\psi}_{2})-2\kappa k_{12m}(\psi_{1}\sigma^{a}\bar{\psi}_{2}).
\end{align}
Now, using either the prescription \eqref{eq:WZNpoint} (with $\mathcal{N}=1$)
or \eqref{eq:vielbein-Npoint}, we obtain
\begin{equation}
\mathcal{M}_{3}=\frac{1}{2}(\psi_{1}\sigma^{m}\bar{\psi}_{2})(k_{2}^{n}-k_{1}^{n})h_{3mn},
\end{equation}
which is invariant under the residual gauge transformations \eqref{eq:residual-gauge-vielbein}.

\

\noindent \hrulefill

\

We have seen that the perturbiner framework seamlessly handles the
tree level scattering of gravitons and matter particles. It introduces
a democratic, systematic methodology that is computationally more
efficient than the diagrammatic approach we learn in traditional field
theory courses. In the next section we will take a step further and
investigate classical multi-particle solutions beyond Minkowski space.

\section{Perturbiner beyond flat space\label{sec:AdS}}

Our focus so far was field theory in flat space. The simple reason
is that in this case the building blocks of the multi-particle solutions
are plane waves, $e^{\mathrm{i}k\cdot x}$. A conceptually important
distinction arises when moving away from flat space: the notion of
asymptotic states, and therefore of scattering amplitudes in the traditional
S-matrix sense, is no longer available. In flat space, the multi-particle
solutions encode on-shell scattering data precisely because free solutions
propagate to spatial infinity as well-separated plane waves. In the
presence of a boundary or a non-trivial background geometry, this
picture breaks down, and our ``free'' solutions most certainly become
complicated functions of the spacetime coordinates. In a few cases,
however, we can identify flat-space patterns that enable a simple
enough construction of multi-particle states, with a direct generalization
of the tree level scattering. The natural observables become instead
boundary correlation functions, or correlators, which encode the response
of bulk fields to boundary sources. This shift from amplitudes to
correlators is not merely a technical inconvenience but reflects a
genuine change in the physical question being asked.

The extension of perturbiner methods to curved backgrounds has a limited
history. Early work by Rosly and Selivanov explored perturbiner constructions
in the self-dual sector of gravity, which can be thought of as a curved
background in a restricted sense. More systematic attempts to go beyond
flat space remained scarce for a long time, largely because the plane-wave
structure that underlies the recursion is tied to the isometries of
Minkowski space. The backgrounds we will consider here (flat space
with a boundary \cite{Gomez:2026yno}, anti-de Sitter space \cite{Armstrong:2022mfr}),
are among the simplest non-trivial cases where enough symmetry survives
to make the construction tractable. They involve only a partial breaking
of spacetime translation invariance. In flat space, translation invariance
enforces momentum conservation, which is what allows the multi-particle
ansatz to organize so cleanly into momentum eigenstates. In anti de
Sitter space, translations along the boundary directions are preserved,
so a partial momentum-space description remains available, but the
radial direction breaks the full translational symmetry. This is reflected
in the structure of the free solutions, which are no longer pure plane
waves but rather products of boundary plane waves and functions of
the radial coordinate. In flat space with a boundary, the situation
is analogous: translations parallel to the boundary are preserved
while the perpendicular direction is not, and the free solutions must
satisfy boundary conditions that modify their form relative to the
unbounded case.

In the section we will understand how the multi-particle solutions
overcome part of the obstacles introduced by curved backgrounds. By
the end of the section, we will have the necessary tools to derive,
for instance, cosmological correlators.

\subsection{Flat space with a boundary\label{subsec:flat-boundary}}

In this subsection we are going to analyze classical multi-particle
solutions of massless theories in a spacetime with a boundary in one
spatial direction, which we will denote by $z\in[0,\infty)$. The
remaining directions will be denoted by $x^{\mu}$, with $\mu=0,\ldots,d-1$
and a Minkowski metric $\eta_{\mu\nu}$.

In this case, a free massless scalar satisfies the equation of motion
\begin{align}
\Box\phi & =(\eta^{\mu\nu}\partial_{\mu}\partial_{\nu}+\partial_{z}^{2})\phi,\nonumber \\
 & =0.
\end{align}
The physical solutions to this equation can be expanded in momentum
eigenfunctions given by
\begin{equation}
\phi(x,z)\propto e^{\mathrm{i}k_{\mu}x^{\mu}}e^{-kz},\label{eq:single-massless-boundary}
\end{equation}
where the on-shell condition is simply $k=\sqrt{k_{\mu}k^{\mu}}$.
We will assume $k_{\mu}k^{\mu}>0$, so the solution \eqref{eq:single-massless-boundary}
exponentially decays in the $z$ direction.

In the interacting theory, momentum is conserved only along the $x^{\mu}$
directions. For instance, in a three-particle process, we have
\begin{equation}
k_{1}^{\mu}+k_{2}^{\mu}+k_{3}^{\mu}=0,
\end{equation}
but $k_{1}+k_{2}+k_{3}>0$ . This is, of course, a boundary effect,
since momentum is not conserved along the $z$ direction. This special
feature has interesting consequences when describing the correlator
of single-particle states, as will be clear below.

\subsubsection{Massless $\phi^{3}$ theory\label{subsec:flat-boundar-scalar}}

Let us consider a simple $\phi^{3}$ theory with equation of motion
\eqref{eq:scalar-cubic} and $\mathrm{m}^{2}=0$. The analogous multi-particle
ansatz to \eqref{eq:scalar-multiparticle} is given by
\begin{equation}
\phi(x,z)=\sum_{P}\Phi_{P}(z)e^{\mathrm{i}k_{P\mu}x^{\mu}},\label{eq:scalar-multi-flatboundary}
\end{equation}
where the multi-particle coefficients depend on the coordinate $z$.
After replacing this ansatz in the equation of motion, we obtain
\begin{equation}
(\partial_{z}^{2}-k_{P}^{2})\Phi_{P}=-\frac{\lambda}{2}\sum_{P=Q\cup R}\Phi_{Q}\Phi_{R}.\label{eq:boundary-phi3-recursion}
\end{equation}
Inverting the kinetic operator on the left hand side is not as trivial
as what we have seen so far. But it is not hard either, we simply
need to find the appropriate Green's function $G_{P}(z,w)$ satisfying
\begin{equation}
(\partial_{z}^{2}-k_{P}^{2})G_{P}(z,w)=\delta(z-w).\label{eq:green-boundary}
\end{equation}

The non-homogeneous solution to this equation is given by the one-dimensional
Yukawa potential,
\begin{equation}
G_{P}(z,w)=-\frac{1}{2k_{P}}e^{-k_{P}|z-w|}.
\end{equation}
We then have to specify a boundary condition, which translates to
adding a homogeneous piece to the equation above. We will require
the Green's function to satisfy
\begin{equation}
G_{P}(0,w)=G_{P}(z,0)=0,
\end{equation}
which is nothing but a homogeneous Dirichlet condition. Physically,
it implies that the boundary information appears only through the
single-particle states in \eqref{eq:scalar-multi-flatboundary}, and
fields interact only in the bulk ($z>0$ region). In this case, the
solution to equation \eqref{eq:green-boundary} is given by
\begin{equation}
G_{P}(z,w)=\frac{1}{2k_{P}}(e^{-k_{P}(z+w)}-e^{-k_{P}|z-w|}).
\end{equation}

We can now define the recursion for the multi-particle coefficients
via the inversion of the kinetic operator. The result is
\begin{equation}
\Phi_{P}(z)=-\frac{\lambda}{2}\sum_{P=Q\cup R}\int_{0}^{\infty}dw\,G_{P}(z,w)\Phi_{Q}(w)\Phi_{R}(w),
\end{equation}
with $\Phi_{P}(0)=0$ for $|P|>1$. Recall that $e^{\mathrm{i}k_{p\mu}x^{\mu}}$
has already been factored out in the recursion. It is easy to check
that $\Phi_{P}(z)$ satisfies equation \eqref{eq:boundary-phi3-recursion}.
Its basic building blocks are the single-particle solutions $\phi=\phi_{p}e^{-k_{p}z}$,
with $k_{p}=\sqrt{k_{p}^{2}}$. 

A two-particle current $\Phi_{12}$ is given by
\begin{equation}
\Phi_{12}(z)=-\frac{\lambda}{2k_{12}}\phi_{1}\phi_{2}\int_{0}^{\infty}dw(e^{-k_{12}(z+w)}-e^{-k_{12}|z-w|})e^{-(k_{1}+k_{2})w}.
\end{equation}
By separating the integration domains ($w<z$ and $w>z$), it is straightforward
to show that
\begin{equation}
\Phi_{12}(z)=\frac{\lambda}{s_{1,2}}\phi_{1}\phi_{2}(e^{-(k_{1}+k_{2})z}-e^{-k_{12}z}),
\end{equation}
with $s_{1,2}=k_{12}^{2}-(k_{1}+k_{2})^{2}$.

For the three-particle coefficients, we obtain
\begin{multline}
\Phi_{123}(z)=\frac{\lambda^{2}}{s_{1,2,3}s_{1,2}}\phi_{1}\phi_{2}\phi_{3}(e^{-(k_{1}+k_{2}+k_{3})z}-e^{-k_{123}z})\\
-\frac{\lambda^{2}}{s_{12,3}s_{1,2}}\phi_{1}\phi_{2}\phi_{3}(e^{-(k_{12}+k_{3})z}-e^{-k_{123}z})+\textrm{cyclic }(1,2,3),\label{eq:Phi_123-boundary}
\end{multline}
in which we define
\begin{align}
s_{P_{1},\ldots,P_{i}} & =k_{P_{1},\ldots,P_{i}m}k_{P_{1},\ldots,P_{i}}^{m},\nonumber \\
 & =k_{P_{1}\ldots P_{n}}^{2}-(k_{P_{1}}+\ldots+k_{P_{n}})^{2}.
\end{align}
Note that in both $\Phi_{12}$ and $\Phi_{123}$, there is a contribution
that solves the free equation of motion (respectively $e^{-k_{12}z}$
and $e^{-k_{123}z}$). They precisely ensure the vanishing of the
multi-particle coefficients at the boundary
\begin{equation}
\Phi_{12}(0)=\Phi_{123}(0)=0.
\end{equation}

Now we are in a better position to define a more refined ansatz to
the multi-particle solution. It will be cast as
\begin{align}
\phi(x,z) & =\sum_{P}\sum_{i=1}^{|P|}\sum_{P=P_{1}\cup\ldots\cup P_{i}}\frac{1}{i!}\Phi^{(i)}(P_{1},\ldots,P_{a})e^{\mathrm{i}k_{P_{1},\ldots,P_{i}}\cdot x},\label{eq:scalar-multi-boundary}
\end{align}
with the scalar product defined in $D=d+1$ dimensions,
\begin{align}
k\cdot x & =k_{\mu}x^{\mu}+k_{z}z,\nonumber \\
 & =k_{\mu}x^{\mu}+\mathrm{i}kz.
\end{align}
Here we have
\begin{align}
k_{P_{1},\ldots,P_{i}}^{\mu} & =k_{P_{1}}^{\mu}+\ldots+k_{P_{i}}^{\mu},\\
k_{P_{1},\ldots,P_{i}}^{z} & =-\mathrm{i}(k_{P_{1}}+\ldots+k_{P_{i}}),
\end{align}
with $k_{P}=\sqrt{k_{P}^{\mu}k_{P\mu}}$. The factor $i!$ in equation
\eqref{eq:scalar-multi-boundary} simply accounts for the indistinguishability
of the permutations of $\{P_{1},\ldots,P_{i}\}$.

The new multi-particle coefficients are not $z$-dependent. They are
related to $\Phi_{P}(z)$ as
\begin{equation}
\Phi_{P}(z)=\sum_{i=1}^{|P|}\sum_{P=P_{1}\cup\ldots\cup P_{i}}\frac{1}{i!}\Phi^{(i)}(P_{1},\ldots,P_{i})e^{-(k_{P_{1}}+\ldots+k_{P_{i}})z}.\label{eq:Phi3-boundary-newexpansion}
\end{equation}
Note that the terms with $i=1$ in \eqref{eq:scalar-multi-boundary}
satisfy the free equation of motion, since $k_{P}=\sqrt{k_{P}^{\mu}k_{P\mu}}$.
For this reason, they cannot be recursively determined. However, they
are fixed by the boundary conditions we had initially set: $\Phi_{P}(0)=0$
for $|P|>1$, i.e.,
\begin{equation}
\Phi_{P}(0)=0\Rightarrow\Phi^{(1)}(P)=-\sum_{i=2}^{|P|}\sum_{P=P_{1}\cup\ldots\cup P_{i}}\frac{1}{i!}\Phi^{(i)}(P_{1},\ldots,P_{i}).\label{eq:Phi(1)-def}
\end{equation}
As usual, one-letter words are associated to single-particle states,
$\Phi^{(1)}(p)=\phi_{p}$.

Finally, we simply replace the new ansatz in the equation of motion
to obtain
\begin{multline}
\Phi^{(i)}(P_{1},\ldots,P_{i})=\frac{\lambda}{2s_{P_{1},\ldots,P_{i}}}\sum_{j=1}^{i-1}\frac{1}{j!(i-j)!}\Phi^{(j)}(P_{1},\ldots,P_{j})\Phi^{(i-j)}(P_{j+1},\ldots,P_{i})\\
+\textrm{permutations}(P_{1},\ldots,P_{i}).
\end{multline}
Let us see some examples. For $i=2$ we have
\begin{equation}
\Phi^{(2)}(P_{1},P_{2})=\frac{\lambda}{s_{P_{1},P_{2}}}\Phi^{(1)}(P_{1})\Phi^{(1)}(P_{2}),\label{eq:Phi(2)-def}
\end{equation}
while for $i=3$ we get to
\begin{equation}
\Phi^{(3)}(P_{1},P_{2},P_{3})=\frac{\lambda^{2}}{s_{P_{1},P_{2},P_{3}}}\left(\frac{1}{s_{P_{1},P_{2}}}+\frac{1}{s_{P_{1},P_{3}}}+\frac{1}{s_{P_{2},P_{3}}}\right)\Phi^{(1)}(P_{1})\Phi^{(1)}(P_{2})\Phi^{(1)}(P_{3}).\label{eq:Phi(3)-def}
\end{equation}
If we compare them with the analogous results in section \ref{sec:multiparticle solutions},
the similarity is clear. In particular, note that $\Phi^{(n)}(1,\ldots,n)$
has the exact same structure of $\Phi_{1\ldots n}$ in equation \eqref{eq:cubic-scalar-recursion},
with $s_{p_{1},\ldots,p_{i}}$ replaced by $s_{p_{1}\ldots p_{i}}$.
More generally, the words $P_{j}$ in the argument of the multi-particle
coefficients are analogous to letters in the usual recursion, an indivisible
unity. $\Phi^{(1)}(P_{j})$ then behaves as a single-particle scalar
polarization.

We can now reproduce, for instance, equation \eqref{eq:Phi_123-boundary},
which was computed via the inversion of the kinetic operator through
the Green's function. From equation \eqref{eq:Phi3-boundary-newexpansion}
we obtain
\begin{multline}
\Phi_{123}(z)=\Phi^{(1)}(123)e^{-k_{123}z}+\Phi^{(2)}(12,3)e^{-(k_{12}+k_{3})z}+\Phi^{(2)}(13,2)e^{-(k_{13}+k_{2})z}\\
+\Phi^{(2)}(23,1)e^{-(k_{23}+k_{1})z}+\Phi^{(3)}(1,2,3)e^{-(k_{1}+k_{2}+k_{3})z}.
\end{multline}
Now, using \eqref{eq:Phi(1)-def}, it is straightforward to show that
\begin{multline}
\Phi_{123}(z)=\Phi^{(2)}(1,23)\left(e^{-(k_{1}+k_{23})z}-e^{-k_{123}z}\right)+\Phi^{(2)}(2,13)\left(e^{-(k_{2}+k_{13})z}-e^{-k_{123}z}\right)\\
+\Phi^{(2)}(3,12)\left(e^{-(k_{3}+k_{12})z}-e^{-k_{123}z}\right)+\Phi^{(3)}(1,2,3)\left(e^{-(k_{1}+k_{2}+k_{3})z}-e^{-k_{123}z}\right),
\end{multline}
which manifestly satisfies $\Phi_{123}(0)=0$. Finally, using equations
\eqref{eq:Phi(2)-def} and \eqref{eq:Phi(3)-def}, we recover $\Phi_{123}(z)$
in \eqref{eq:Phi_123-boundary}.

The natural question at this point is how to compute the scattering
data in a theory with boundary. In order to do so, let us recall the
scalar prescription \eqref{eq:(n+1)-pointscalar}, which can be expressed
as
\begin{equation}
A_{n+1}\propto\int d^{D}x\lim_{s_{N}\to\mathrm{m}^{2}}(\phi_{n+1}e^{\mathrm{i}k_{n+1}\cdot x})\Box_{D}(\Phi_{1\ldots n}e^{\mathrm{i}k_{1\ldots n}\cdot x}).\label{eq:scalar-prescription-integralform}
\end{equation}
The kinetic operator ($\Box_{D}$ in $D$ dimensions) yields the Mandelstam
variable $s_{P}$, and the integration introduces the momentum conservation
delta $\delta^{D}(k_{1}+\ldots+k_{n+1})$. We can mimic this construction
for the theory with boundary. In momentum space, the correlator translates
to the following object:
\[
\int d^{d}x\int_{0}^{\infty}dz\,\phi_{n+1}(x,z)\underbrace{\left(\partial_{z}^{2}-k_{1\ldots n}^{2}\right)}_{\Box_{d+1}}\Phi_{1\ldots n}(z)e^{\mathrm{i}k_{1\ldots n\mu}x^{\mu}},
\]
which is a direct extension of \eqref{eq:scalar-prescription-integralform}.
The integration over the $x^{\mu}$ directions leads to
\[
\delta^{d}(k_{1}+\ldots+k_{n}+k_{n+1}),
\]
but the integral over $z$ introduces a new ingredient. We finally
define the $(n+1)$-point correlator as
\begin{multline}
\tilde{A}_{n+1}=\phi_{n+1}\int_{0}^{\infty}dz\,e^{-k_{n+1}z}\left(k_{1\ldots n}^{2}-\partial_{z}^{2}\right)\Phi_{1\ldots n}(z),\\
=\sum_{i=2}^{n}\sum_{1\ldots n=P_{1}\cup\ldots\cup P_{i}}\frac{1}{i!}\frac{1}{(k_{P_{1}}+\ldots+k_{P_{i}}+k_{n+1})}s_{P_{1},\ldots,P_{i}}\Phi^{(i)}(P_{1},\ldots,P_{i})\phi_{n+1}.\label{eq:phi3-boundary-correlator}
\end{multline}

It might not be obvious yet, but the correlator above has a simple
interpretation in terms of scattering amplitudes of the boundary-less
theory. This is rather surprising, as we now have a theory with a
boundary! The pattern might be easier to identify after some examples.
The three-point correlator is given by
\begin{equation}
\tilde{A}_{3}=\frac{1}{(k_{1}+k_{2}+k_{3})}\left[\lambda\phi_{1}\phi_{2}\phi_{3}\right].
\end{equation}
If we take the residue around $k_{1}+k_{2}+k_{3}=0$, we are simply
left with the usual three-point amplitude, with momentum conservation
in all $D=d+1$ directions. For the four-point correlator, we obtain
\begin{multline}
\tilde{A}_{4}=\frac{1}{(k_{12}+k_{3}+k_{4})}\left[\lambda\Phi^{(1)}(12)\phi_{3}\phi_{4}\right]\\
+\frac{1}{(k_{23}+k_{1}+k_{4})}\left[\lambda\Phi^{(1)}(23)\phi_{1}\phi_{4}\right]+\frac{1}{(k_{13}+k_{2}+k_{4})}\left[\lambda\Phi^{(1)}(13)\phi_{2}\phi_{4}\right]\\
+\frac{1}{(k_{1}+k_{2}+k_{3}+k_{4})}\left[\left(\frac{1}{s_{1,2}}+\frac{1}{s_{2,3}}+\frac{1}{s_{1,3}}\right)\lambda^{2}\phi_{1}\phi_{2}\phi_{3}\phi_{4}\right].
\end{multline}
Once more, by taking the residue around $k_{1}+k_{2}+k_{3}+k_{4}=0$,
we obtain the four-point amplitude of the $\phi^{3}$ theory without
boundary. However, we now have a richer pole structure. For instance,
we could take the residue around $k_{12}+k_{3}+k_{4}=0$ and obtain
a \emph{three-point} amplitude in which legs $1$ and $2$ are concatenated
into the generalized polarization $\Phi^{(1)}(12)$, the first term
of the above equation. One might be worried about the fact that leg
$4$ appears asymmetrically with respect to legs $1$, $2$, and $3$.
So let us take the residue of $A_{4}$ around $k_{34}+k_{1}+k_{2}=0$
and show that this asymmetry is only apparent\footnote{This asymmetry becomes relevant when the interaction vertices involve
$z$-derivatives, as we will see in the Yang--Mills discussion.}. The poles are hiding in
\begin{align}
\frac{1}{s_{1,2}} & =\frac{1}{(k_{12}+k_{1}+k_{2})(k_{12}-k_{1}-k_{2})},\nonumber \\
 & =\frac{1}{(k_{34}+k_{1}+k_{2})(k_{12}-k_{1}-k_{2})},
\end{align}
where we have used momentum conservation from the first to the second
line. Therefore,
\begin{align}
\left.\tilde{A}_{4}\right|_{k_{34}+k_{1}+k_{2}\to0} & =-\frac{1}{(k_{34}+k_{1}+k_{2})}\frac{1}{s_{3,4}}\lambda^{2}\phi_{1}\phi_{2}\phi_{3}\phi_{4},\nonumber \\
 & =\frac{1}{(k_{34}+k_{1}+k_{2})}\left[\lambda\phi_{1}\phi_{2}\Phi^{(1)}(34)\right].
\end{align}
The expression inside the square brackets is just the three-point
amplitude with external legs $1$, $2$, and the generalized polarization
with concatenated legs $3$ and $4$.

Finally, note that the correlator \eqref{eq:phi3-boundary-correlator}
can be expressed as
\begin{equation}
\tilde{A}_{n+1}=\sum_{i=2}^{n}\sum_{1\ldots N=P_{1}\cup\ldots\cup P_{i}}\frac{1}{i!}\frac{1}{(k_{P_{1}}+\ldots+k_{P_{i}}+k_{n+1})}A_{i+1}(P_{1},\ldots,P_{i};n+1),
\end{equation}
where $A_{i+1}$ coincides with the usual amplitude prescription \eqref{eq:(n+1)-pointscalar}
with generalized polarizations $\Phi^{(1)}(P_{i})$, as long as we
keep in mind that momentum conservation does not hold in the $z$
direction.

\subsubsection{Yang--Mills}

For the Yang--Mills theory, we are going to work with color-stripped
currents. The multi-particle ansatz is given by
\begin{align}
A_{m}(x,z) & =\sum_{P}\boldsymbol{A}_{Pm}(z)e^{\mathrm{i}k_{P\mu}x^{\mu}}T^{a_{P}},\nonumber \\
 & =\sum_{P}T^{a_{P}}\sum_{i=1}^{|P|}\sum_{P=P_{1}\ldots P_{i}}\boldsymbol{A}_{m}^{(i)}(P_{1},\ldots,P_{i})e^{\mathrm{i}k_{P_{1},\ldots,P_{i}}\cdot x}\label{eq:YM-boundary-ansatz}
\end{align}
which is the natural generalization of equation \eqref{eq:CS-YM-perturbiner}.

We would like to determine currents $\boldsymbol{A}_{m}^{(i)}(P_{1},\ldots,P_{i})$
in order for $A_{m}(x,z)$ to be a classical solution of the Yang--Mills
equation
\begin{equation}
\Box A_{m}=2\mathrm{i}[A^{n},\partial_{n}A_{m}]-\mathrm{i}[A^{n},\partial_{m}A_{n}]+[[A_{m},A_{n}],A^{n}],
\end{equation}
where the Lorenz gauge was chosen, i.e., $\partial^{\mu}A_{\mu}+\partial^{z}A_{z}=0$.
Solutions to the linearized equation (single-particle states) can
be cast as
\[
A_{m}(x,z)=\sum_{i}\varepsilon_{im}e^{\mathrm{i}k_{i\mu}x^{\mu}}e^{-k_{i}z},
\]
with $k_{i}\cdot\varepsilon_{i}=k_{i}\cdot k_{i}=0$. In order to
determine the multi-particle currents, we plug-in the ansatz in the
equation of motion and arrive at
\begin{multline}
s_{P_{1},\ldots,P_{i}}\boldsymbol{A}_{m}^{(i)}(P_{1},\ldots,P_{i})=\\
\sum_{j=1}^{i-1}\bigg\{[\boldsymbol{A}^{(j)}(P_{1},\ldots,P_{j})\cdot\boldsymbol{A}^{(i-j)}(P_{j+1},\ldots,P_{i})](k_{P_{1},\ldots,P_{j}m}-k_{P_{j+1},\ldots,P_{i}m})\\
-2[k_{P_{1},\ldots,P_{j}}\cdot\boldsymbol{A}^{(i-j)}(P_{j+1},\ldots,P_{i})]\boldsymbol{A}_{m}^{(j)}(P_{1},\ldots,P_{j})\\
+2[k_{P_{j+1},\ldots,P_{i}}\cdot\boldsymbol{A}^{(j)}(P_{1},\ldots,P_{j})]\boldsymbol{A}_{m}^{(i-j)}(P_{j+1},\ldots,P_{i})\bigg\}\\
+\sum_{j=1}^{i-2}\sum_{l=b+1}^{i-1}\bigg\{2[\boldsymbol{A}^{(j)}(P_{1},\ldots,P_{j})\cdot\boldsymbol{A}^{(i-l)}(P_{l+1},\ldots,P_{i})]\boldsymbol{A}_{m}^{(l-j)}(P_{j+1},\ldots,P_{l})\\
-[\boldsymbol{A}^{(l-j)}(P_{j+1},\ldots,P_{l})\cdot\boldsymbol{A}^{(i-l)}(P_{l+1},\ldots,P_{i})]\boldsymbol{A}_{m}^{(j)}(P_{1},\ldots,P_{j})\\
-[\boldsymbol{A}^{(j)}(P_{1},\ldots,P_{j})\cdot\boldsymbol{A}^{(l-j)}(P_{j+1},\ldots,P_{l})]\boldsymbol{A}_{m}^{(i-l)}(P_{l+1},\ldots,P_{i})\bigg\},
\end{multline}
which determines the recursion of $\boldsymbol{A}_{m}^{(i)}(P_{1},\ldots,P_{i})$
for $i\geq2$. This is just a rewriting of equation \eqref{eq:BG-YM}
in which we have the words $P_{i}$ behaving as single-particle labels.
For example, the two-particle current is given by
\begin{multline}
s_{P_{1},P_{2}}\boldsymbol{A}_{m}^{(2)}(P_{1},P_{2})=2[k_{P_{2}}\cdot\boldsymbol{A}^{(1)}(P_{1})]\boldsymbol{A}_{m}^{(1)}(P_{2})-2[k_{P_{1}}\cdot\boldsymbol{A}^{(1)}(P_{2})]\boldsymbol{A}_{m}^{(1)}(P_{1})\\
+[\boldsymbol{A}^{(1)}(P_{1})\cdot\boldsymbol{A}^{(1)}(P_{2})](k_{P_{1}m}-k_{P_{2}m}),
\end{multline}
which is similar to equation \eqref{eq:CS-YM-2particle}, while the
three-particle current can be expressed as
\begin{multline}
s_{P_{1},P_{2},P_{3}}\boldsymbol{A}_{m}^{(3)}(P_{1},P_{2},P_{3})=2[\boldsymbol{A}^{(1)}(P_{1})\cdot\boldsymbol{A}^{(1)}(P_{3})]\boldsymbol{A}_{m}^{(1)}(P_{2})\\
-[\boldsymbol{A}^{(1)}(P_{2})\cdot\boldsymbol{A}^{(1)}(P_{3})]\boldsymbol{A}_{m}^{(1)}(P_{1})-[\boldsymbol{A}^{(1)}(P_{1})\cdot\boldsymbol{A}^{(1)}(P_{2})]\boldsymbol{A}_{m}^{(1)}(P_{3})\\
+2[k_{P_{2},P_{3}}\cdot\boldsymbol{A}^{(1)}(P_{1})]\boldsymbol{A}_{m}^{(2)}(P_{2},P_{3})+2[k_{P_{3}}\cdot\boldsymbol{A}^{(2)}(P_{1},P_{2})]\boldsymbol{A}_{m}^{(1)}(P_{3})\\
-2[k_{P_{1}}\cdot\boldsymbol{A}^{(2)}(P_{2},P_{3})]\boldsymbol{A}_{m}^{(1)}(P_{1})-2[k_{P_{1},P_{2}}\cdot\boldsymbol{A}^{(1)}(P_{3})]\boldsymbol{A}_{m}^{(2)}(P_{1},P_{2})\\
+[\boldsymbol{A}^{(1)}(P_{1})\cdot\boldsymbol{A}^{(2)}(P_{2},P_{3})](k_{P_{1}m}-k_{P_{2},P_{3}m})+[\boldsymbol{A}^{(2)}(P_{1},P_{2})\cdot\boldsymbol{A}^{(1)}(P_{3})](k_{P_{1},P_{2}m}-k_{P_{3}m}),
\end{multline}
which matches equation \eqref{eq:CS-YM-3particle} when replacing
single-particle labels by words.

As in the $\phi^{3}$ theory, the currents $\boldsymbol{A}_{m}^{(1)}(P)$
satisfy the free equation ($s_{P}=0$), so they are not determined
by the above equation. One-letter words are associated to single-particle
polarizations, $\boldsymbol{A}_{m}^{(1)}(i)=\varepsilon_{im}$, but
we still need their definition for a generic word $P$. Like in the $\phi^{3}$
theory, we could simply impose the vanishing of the multi-particle
current at the boundary, $\boldsymbol{A}_{Pm}(0)=0$, which would
lead to
\begin{equation}
\boldsymbol{A}_{m}^{(1)}(P)=-\sum_{i=2}^{|P|}\sum_{P=P_{1}\ldots P_{i}}\boldsymbol{A}_{m}^{(i)}(P_{1},\ldots,P_{i}).
\end{equation}
However, this condition is not compatible with the Lorenz gauge. Instead,
we will simply demand that $\boldsymbol{A}_{Pm}(0)$ vanishes up to
pure gauge contributions. A simple solution is to insert a transversality
projector,
\begin{equation}
\boldsymbol{A}_{m}^{(1)}(P)=-\sum_{i=2}^{|P|}\sum_{P=P_{1}\ldots P_{i}}\left(\delta_{m}^{n}-\frac{k_{P}^{n}k_{P_{1},\ldots,P_{i}m}}{k_{P}\cdot k_{P_{1},\ldots,P_{i}}}\right)\boldsymbol{A}_{n}^{(i)}(P_{1},\ldots,P_{i}),\label{eq:YM-boundary-A1def}
\end{equation}
which leads to
\begin{align}
\boldsymbol{A}_{\mu}^{(1)}(P) & =-\sum_{i=2}^{|P|}\sum_{P=P_{1}\ldots P_{i}}\left(\boldsymbol{A}_{\mu}^{(i)}(P_{1},\ldots,P_{i})-\mathrm{i}\frac{k_{P\mu}}{k_{P}}\boldsymbol{A}_{z}^{(i)}(P_{1},\ldots,P_{i})\right)\\
\boldsymbol{A}_{z}^{(1)}(P) & =-\sum_{i=2}^{|P|}\sum_{P=P_{1}\ldots P_{i}}\left(\boldsymbol{A}_{z}^{(i)}(P_{1},\ldots,P_{i})+\frac{(k_{P_{1}}+\ldots+k_{P_{i}})}{k_{P}}\boldsymbol{A}_{z}^{(i)}(P_{1},\ldots,P_{i})\right).
\end{align}

In this case, we have
\begin{multline}
\boldsymbol{A}_{Pm}(z)=\sum_{i=2}^{|P|}\sum_{P=P_{1}\ldots P_{i}}\boldsymbol{A}_{m}^{(i)}(P_{1},\ldots,P_{a})\left(e^{-(k_{P_{1}}+\ldots+k_{P_{i}})z}-e^{-k_{P}z}\right)\\
+\mathrm{i}\frac{k_{P_{1},\ldots,P_{i}m}}{k_{P}}\boldsymbol{A}_{z}^{(i)}(P_{1},\ldots,P_{i})e^{-k_{P}z}.
\end{multline}
In particular, note that
\begin{equation}
\boldsymbol{A}_{Pm}(0)=\mathrm{i}\sum_{i=2}^{|P|}\sum_{P=P_{1}\ldots P_{i}}\frac{k_{P_{1},\ldots,P_{i}m}}{k_{P}}\boldsymbol{A}_{z}^{(i)}(P_{1},\ldots,P_{i}).
\end{equation}
In the ansatz \eqref{eq:YM-boundary-ansatz}, this result translates
to a pure gauge object at the boundary,
\begin{multline}
A_{m}(x,0)=\frac{\partial}{\partial x^{m}}\left(\sum_{P}\frac{1}{k_{P}}T^{a_{P}}\sum_{i=2}^{|P|}\sum_{P=P_{1}\ldots P_{i}}\boldsymbol{A}_{z}^{(i)}(P_{1},\ldots,P_{i})e^{\mathrm{i}k_{P_{1},\ldots,P_{i}}\cdot x}\right)_{z=0}\\
+\textrm{single-particle contributions},
\end{multline}
as we expected. The single-particle states are the only information
the field $A_{m}(x,z)$ has from the boundary, up to gauge transformations.

This is a good point to discuss gauge transformations of the Yang--Mills
theory with a boundary. In analogy with section \ref{sec:gauge-theories},
let us consider the a multi-particle expansion of the gauge parameter,
\begin{equation}
\lambda(x,z)=\sum_{P}T^{a_{P}}\sum_{i=1}^{|P|}\sum_{P=P_{1}\ldots P_{i}}\boldsymbol{\Lambda}^{(i)}(P_{1},\ldots,P_{i})e^{\mathrm{i}k_{P_{1},\ldots,P_{i}}\cdot x},
\end{equation}
and determine the residual gauge transformation that leaves the Lorenz
gauge invariant. Using equation \eqref{eq:YM-gaugetransf}, we find
that
\begin{multline}
\boldsymbol{\Lambda}^{(i)}(P_{1},\ldots,P_{i})=\frac{1}{s_{P_{1},\ldots,P_{i}}}\sum_{j=1}^{i-1}\bigg\{[k_{P_{1},\ldots,P_{i}}\cdot\boldsymbol{A}^{(i-j)}(P_{j+1},\ldots,P_{i})]\boldsymbol{\Lambda}^{(j)}(P_{1},\ldots,P_{j})\\
-[k_{P_{1},\ldots,P_{i}}\cdot\boldsymbol{A}^{(j)}(P_{1},\ldots,P_{j})]\boldsymbol{\Lambda}^{(i-j)}(P_{j+1},\ldots,P_{i})\bigg\},
\end{multline}
for $i\geq2$, which are ultimately written in terms of the free parameters
$\boldsymbol{\Lambda}^{(1)}(P)$. They comprise the usual transformation
of the single-particle polarizations, $\delta\varepsilon_{im}=k_{im}\lambda_{i}$,
but also of $\boldsymbol{A}_{m}^{(1)}(P)$, which is itself the coefficient
of a free solution. To all effects, $\boldsymbol{A}_{m}^{(1)}(P)$
behaves as a single-particle polarization, satisfying $k_{P}\cdot\boldsymbol{A}^{(1)}(P)=k_{P}\cdot k_{P}=0$.
Therefore, it is also defined up to a residual gauge transformation,
\begin{equation}
\delta\boldsymbol{A}_{m}^{(1)}(P)=k_{Pm}\boldsymbol{\Lambda}^{(1)}(P),\label{eq:YM-boundary-A1gauge}
\end{equation}
so we can keep in mind this freedom. For instance, the choices
\begin{align}
\boldsymbol{A}_{\mu}^{(1)}(P) & =-\sum_{i=2}^{|P|}\sum_{P=P_{1}\ldots P_{i}}\boldsymbol{A}_{\mu}^{(i)}(P_{1},\ldots,P_{i}),\\
\boldsymbol{A}_{z}^{(1)}(P) & =-\sum_{i=2}^{|P|}\sum_{P=P_{1}\ldots P_{i}}\frac{(k_{P_{1}}+\ldots+k_{P_{i}})}{k_{P}}\boldsymbol{A}_{z}^{(i)}(P_{1},\ldots,P_{i}),
\end{align}
and
\begin{align}
\boldsymbol{A}_{\mu}^{(1)}(P) & =-\left(\delta_{\mu}^{\nu}-\frac{k_{P}^{\nu}k_{P\mu}}{k_{P}^{2}}\right)\sum_{i=2}^{|P|}\sum_{P=P_{1}\ldots P_{i}}\boldsymbol{A}_{\nu}^{(i)}(P_{1},\ldots,P_{i}),\\
\boldsymbol{A}_{z}^{(1)}(P) & =0,
\end{align}
are related to \eqref{eq:YM-boundary-A1def} via residual gauge transformations,
cf. equation \eqref{eq:YM-boundary-A1gauge}. It is easy to check
they maintain the transversality condition $k_{P}\cdot\boldsymbol{A}^{(1)}(P)=0$.

Finally, we can define the color-ordered correlator as
\begin{multline}
\tilde{\mathscr{C}}(1,\ldots,N,N+1)\\
=\frac{1}{N+1}\int_{0}^{\infty}dz\,\varepsilon_{N+1}^{m}e^{-k_{N+1}z}\left(k_{P}^{2}-\partial_{z}^{2}\right)\boldsymbol{A}_{Pm}(z)+\textrm{cyclic}(1,\ldots,N,N+1),\\
=\frac{1}{N+1}\sum_{i=2}^{N}\sum_{1\ldots N=P_{1}\ldots P_{i}}\frac{s_{P_{1},\ldots,P_{i}}}{(k_{P_{1}}+\ldots+k_{P_{i}}+k_{N+1})}[\varepsilon_{N+1}\cdot\boldsymbol{A}^{(i)}(P_{1},\ldots,P_{i})]\\
+\textrm{cyclic}(1,\ldots,N,N+1).\label{eq:boundary-correlator-YM}
\end{multline}
Note that we are introducing a cyclic average in the definition. For
the theory without a boundary, cyclicity of the prescription \eqref{eq:YMpartialamplitude}
is a consequence of momentum conservation. Since the interaction vertices
of the Yang--Mills theory involve derivatives with respect to $z$,
the cyclicity of the partial amplitude would be destroyed with a straightforward
generalization of \eqref{eq:YMpartialamplitude}.

Because of the residual gauge transformation, one may ask whether
$\tilde{\mathscr{C}}$ is indeed invariant. It is not. For instance,
if we consider $\delta\varepsilon_{N+1}^{m}=k_{N+1}^{m}\lambda_{N+1}$,
one of the contributions to $\delta\tilde{\mathscr{C}}(1,\ldots,N,N+1)$
is proportional to
\begin{multline}
\sum_{i=2}^{N}\sum_{1\ldots N=P_{1}\ldots P_{i}}\frac{s_{P_{1},\ldots,P_{i}}}{(k_{P_{1}}+\ldots+k_{P_{i}}+k_{N+1})}\lambda_{N+1}[k_{N+1}\cdot\boldsymbol{A}^{(i)}(P_{1},\ldots,P_{i})]\\
=\sum_{i=2}^{N}\sum_{1\ldots N=P_{1}\ldots P_{i}}\frac{s_{P_{1},\ldots,P_{i}}}{(k_{P_{1}}+\ldots+k_{P_{i}}+k_{N+1})}\lambda_{N+1}[(k_{N+1}+k_{P_{1},\ldots,P_{i}})\cdot\boldsymbol{A}^{(i)}(P_{1},\ldots,P_{i})].
\end{multline}
Since we do not have momentum conservation along the $z$ direction,
we obtain
\begin{equation}
\delta\tilde{\mathscr{C}}(1,\ldots,N,N+1)=\frac{\mathrm{i}}{N+1}\sum_{i=2}^{N}\sum_{1\ldots N=P_{1}\ldots P_{i}}\lambda_{N+1}s_{P_{1},\ldots,P_{i}}\boldsymbol{A}_{z}^{(i)}(P_{1},\ldots,P_{i})+\ldots,
\end{equation}
where the ellipsis include additional contributions coming from the
variation of the multi-particle currents. This result can be traced
back to a sum of boundary contributions,
\begin{equation}
\delta\tilde{\mathscr{C}}(1,\ldots,N,N+1)=\frac{\mathrm{i}}{N+1}\int_{0}^{\infty}dz\,\partial_{z}\left[\lambda_{N+1}e^{-k_{N+1}z}\left(k_{P}^{2}-\partial_{z}^{2}\right)\boldsymbol{A}_{Pm}(z)+\ldots\right].
\end{equation}
Indeed, they are called boundary contact terms and, though at first
uncomfortable, they are necessary for the Ward identities to be satisfied.
This follows from the fact that an arbitrary (infinitesimal) variation
of the Yang--Mills action generates a boundary term
\begin{equation}
\delta S=\frac{1}{g_{\textrm{YM}}^{2}}\int d^{d}xdz\,\textrm{Tr}[\delta A^{n}(\nabla^{m}F_{mn})]+\frac{1}{g_{\textrm{YM}}^{2}}\int d^{d}x\,\textrm{Tr}\left.(F_{\mu z}\delta A^{\mu})\right|_{z=0}.
\end{equation}
The second term on the right hand side is the one relevant for this
discussion. It implies we have to set boundary conditions for the
fields (either Dirichlet-like, $\delta A^{\mu}=0$, or Neumann-like,
$F_{\mu z}=0$, at $z=0$). Therefore, gauge transformations that
do not vanish at the boundary must affect the boundary value of the
fields. This can be interpreted as a global symmetry for edge modes,
which are boundary degrees of freedom. This is all to say that the
non-invariance of the correlator $\tilde{\mathscr{C}}$ is not really
a problem, but a feature of our gauge choice. Alternatively, we could
have chosen a variation of the temporal gauge (see \ref{subsec:Temporal-gauge})
by fixing $A_{z}=0$. This is often referred to as axial gauge. In
this case, the recursion of the multi-particle currents is a bit more
cumbersome. It would be determined from the component $\mu$ of the
Yang--Mills equation of motion \eqref{eq:YMeom},
\begin{equation}
(\eta^{\mu\nu}\partial_{\mu}\partial_{\nu}+\partial_{z}^{2})A_{\mu}  =\partial_{\mu}(\partial^{\nu}A_{\nu})-\mathrm{i}[A_{\mu},\partial^{\nu}A_{\nu}] +2\mathrm{i}[A^{\nu},\partial_{\nu}A_{\mu}]-\mathrm{i}[A^{n},\partial_{\mu}A_{n}]+[[A_{\mu},A_{\nu}],A^{\nu}],
\end{equation}
while the $z$ component would lead to
\begin{multline}
k_{P_{1},\ldots,P_{i}}^{\mu}\boldsymbol{A}_{\mu}^{(i)}(P_{1},\ldots,P_{i})\\
=\frac{1}{k_{P_{1},\ldots,P_{i}}}\sum_{j=1}^{i-1}(k_{P_{j+1},\ldots,P_{i}}-k_{P_{1},\ldots,P_{j}})\boldsymbol{A}^{(j)\mu}(P_{1},\ldots,P_{j})\boldsymbol{A}_{\mu}^{(i-j)}(P_{j+1},\ldots,P_{i}).
\end{multline}
A possible advantage of this choice is that there is no residual gauge
transformation left. However, we will not extend this discussion on
alternative gauge choices, as the focus is to explore the connection
to flat space amplitudes that are usually presented in the Lorenz
gauge.

As in the $\phi^{3}$ theory, $\tilde{\mathscr{C}}(1,\ldots,N+1)$
can be expanded in terms of color-ordered flat space amplitudes computed
using the Berends--Giele currents with generalized (concatenated)
polarizations $\mathcal{A}_{m}^{(1)}(P)$. It can be expressed as
\begin{multline}
\tilde{\mathscr{C}}(1,\ldots,N+1)=\frac{1}{N+1}\sum_{i=2}^{N}\sum_{1\ldots N=P_{1}\ldots P_{i}}\frac{1}{(k_{P_{1}}+\ldots+k_{P_{i}}+k_{N+1})}\mathscr{C}(P_{1},\ldots,P_{i},N+1)\\
+\textrm{cyclic}(1,\ldots,N,N+1),
\end{multline}
where $\mathscr{C}(P_{1},\ldots,P_{i},N+1)$ matches the partial amplitudes
of equation \eqref{eq:YMpartialamplitude}.

As an example, we can compute the three-point correlator. The two-particle
currents essentially have the same form as the ones in flat space
without a boundary, and we obtain
\begin{equation}
\tilde{\mathscr{C}}(1,2,3)=\frac{1}{(k_{1}+k_{2}+k_{3})}(\varepsilon_{1}\cdot\varepsilon_{2})[(k_{1}-k_{2})\cdot\varepsilon_{3}]+\textrm{cyclic}(1,2,3).
\end{equation}
Following the discussion on residual gauge transformations, note that
$\delta\varepsilon_{3m}=k_{3m}\lambda$ leads to
\begin{equation}
\delta\tilde{\mathscr{C}}(1,2,3)=\lambda[(\varepsilon_{1}\cdot\varepsilon_{2})(k_{2}-k_{1})+\varepsilon_{1z}(k_{1}\cdot\varepsilon_{2})-\varepsilon_{2z}(k_{2}\cdot\varepsilon_{1})].
\end{equation}
As expected, the pole $(k_{1}+k_{2}+k_{3})^{-1}$ disappears, and
$\delta\tilde{\mathscr{C}}(1,2,3)$ is interpreted as a boundary contact
term.

In the next subsection, we are going to develop the perturbiner in
anti de Sitter. Although there are further technical ingredients,
we will see that it shares several features with the construction
on flat space with a boundary. In some cases (conformally coupled
theories), the correlators in AdS have exactly the same form.

\subsection{Anti de Sitter\label{subsec:perturbiner-AdS}}

The anti de Sitter (AdS) space is a vacuum solution of the field equations
\eqref{eq:EFE} with a cosmological constant. The global
AdS space is a hyperboloid embedded in a higher-dimensional flat space
($X^{M}$, with $M=0,\ldots,D$) with two time dimensions ($X^{0}$
and $X^{D}$), constrained as
\begin{equation}
\eta_{\mu\nu}X^{\mu}X^{\nu}+X_{D-1}^{2}-X_{D}^{2}=-L^{2},
\end{equation}
where $L$ is the AdS radius, and $\mu,\nu=0,\ldots D-2$. We will
work here in the Poincar\'e patch, so we define two light-cone directions,
$X^{\pm}=X^{D}\pm X^{D-1}$ and use the surface constraint to solve
for either $X^{+}$ or $X^{-}$. For example,
\begin{equation}
X^{-}=\frac{1}{X^{+}}(L^{2}+\eta_{\mu\nu}X^{\mu}X^{\nu}).
\end{equation}
Now we define
\begin{equation}
\begin{array}{rcl}
X^{\mu}=\frac{L}{z}x^{\mu}, &  & X^{+}=\frac{L^{2}}{z},\end{array}
\end{equation}
with $z>0$ denoting the radial AdS coordinate, such that
\begin{equation}
X^{-}=\frac{1}{z}(z^{2}+\eta_{\mu\nu}x^{\mu}x^{\nu}).
\end{equation}
Note that in this parametrization we cover only half of the AdS space
($X^{+}>0$). Finally, the square of the line element is given by
\begin{align}
\eta_{MN}dX^{M}dX^{M} & =\eta_{\mu\nu}dX^{\mu}dX^{\nu}+dX^{+}dX^{-},\nonumber \\
 & =\frac{L^{2}}{z^{2}}(\eta_{\mu\nu}dx^{\mu}dx^{\nu}+dz^{2}),\nonumber \\
 & \equiv g_{mn}^{\textrm{AdS}}dx^{m}dx^{n},
\end{align}
where $g_{mn}^{\textrm{AdS}}$ is the $\mathrm{AdS}_{d+1}$ metric
in the Poincar\'e patch. And this is the metric we are going to use
in order to build multi-particle solutions in AdS.

Since we are going to need them soon, let us spell out the relevant
quantities derived from $g_{mn}^{\textrm{AdS}}$. The non-vanishing
Christoffel symbols are
\begin{equation}
\begin{array}{ccccc}
(\Gamma^{\textrm{AdS}})_{\mu\nu}^{z}=\frac{1}{z}\eta_{\mu\nu}, &  & (\Gamma^{\textrm{AdS}})_{\nu z}^{\mu}=-\frac{1}{z}\delta_{\nu}^{\mu}, &  & (\Gamma^{\textrm{AdS}})_{zz}^{z}=-\frac{1}{z},\\
\Gamma_{z\mu\nu}^{\textrm{AdS}}=\frac{L^{2}}{z^{3}}\eta_{\mu\nu}, &  & \Gamma_{\mu\nu z}^{\textrm{AdS}}=-\frac{L^{2}}{z^{3}}\eta_{\mu\nu} &  & \Gamma_{zzz}^{\textrm{AdS}}=-\frac{L^{2}}{z^{3}}.
\end{array}
\end{equation}
Then we can show that the Ricci tensor and the Ricci scalar are respectively
given by
\begin{equation}
\begin{array}{ccc}
R_{mn}^{\textrm{AdS}}=-\frac{d}{L^{2}}g_{mn}^{\textrm{AdS}}, &  & R^{\textrm{AdS}}=-\frac{d(d+1)}{L^{2}},\end{array}
\end{equation}
Therefore, we have negative curvature. Finally, it is easy to check
that $g_{mn}^{\textrm{AdS}}$ is a solution of \eqref{eq:EFE} with
cosmological constant
\begin{equation}
\Lambda=-\frac{1}{2L^{2}}d(d-1).\label{eq:cosmo-constant}
\end{equation}

We are now ready to delve into the most common field theories studied
in AdS backgrounds.

\subsubsection{Scalars\label{subsec:Scalars-AdS}}

In a curved background, a typical scalar theory action is given by
\begin{equation}
S=-\int d^{D}x\sqrt{-g}\left(\frac{1}{2}g^{mn}\partial_{m}\phi\partial_{n}\phi-\frac{1}{2}\mathrm{m}^{2}\phi^{2}+V(\phi)+\frac{1}{2}\xi R\phi^{2}\right),
\end{equation}
where $V(\phi)$ is the scalar potential, and $\xi$ is the coupling
parameter between the scalar and the background curvature, $R$.

We will work with a fixed $\mathrm{AdS}_{d+1}$ background (no back
reaction from the scalar), so the action can be cast as
\begin{equation}
S=-\int d^{d}x\,dz\,\underbrace{\left(\frac{L}{z}\right)^{d+1}}_{\sqrt{-g}}\bigg\{\frac{z^{2}}{2L^{2}}\eta^{\mu\nu}\partial_{\mu}\phi\partial_{\nu}\phi+\frac{z^{2}}{2L^{2}}\partial_{z}\phi\partial_{z}\phi-\frac{1}{2}\mathrm{M}^{2}\phi^{2}+V(\phi)\bigg\},\label{eq:action-scalar-AdS}
\end{equation}
where we have the effective mass
\begin{equation}
\mathrm{M}^{2}=\mathrm{m}^{2}+\frac{\xi d(d+1)}{L^{2}}.\label{eq:scalar-AdS-effective-mass}
\end{equation}
The equation of motion derived from the action \eqref{eq:action-scalar-AdS}
is then given by
\begin{equation}
z^{2}\partial_{z}^{2}\phi+(1-d)z\partial_{z}\phi+z^{2}\eta^{\mu\nu}\partial_{\mu}\partial_{\nu}\phi+(L\mathrm{M})^{2}\phi=L^{2}\partial_{\phi}V.\label{eq:scalar-eom-AdS}
\end{equation}
In addition, the principle of minimum action yields a boundary contribution,
\[
\left[\left(\frac{L}{z}\right)^{d-1}\delta\phi\partial_{z}\phi\right]_{z=0},
\]
which is eliminated with the appropriate boundary conditions. We will
work with Dirichlet type boundary conditions, $\delta\phi=0$.

\noindent \hrulefill

\subsubsection*{A side comment: connection to flat space with a boundary}

Let us focus on the free equation of motion, so that the left hand
side of equation \eqref{eq:scalar-eom-AdS} vanishes. In the massless
case ($\mathrm{m}^{2}=0$), a field redefinition of the form $\phi\to z^{(d-1)/2}\phi$
leads to
\begin{equation}
z^{2}(\partial_{z}^{2}\phi+\eta^{\mu\nu}\partial_{\mu}\partial_{\nu}\phi)=d(d+1)\left(\frac{d-1}{4d}-\xi\right)\phi.
\end{equation}
Note there is a specific solution of the coupling parameter $\xi$
that leads to an equation of the exact same form we studied in subsection
\ref{subsec:flat-boundar-scalar}, given by
\begin{equation}
\xi_{c}=\frac{d-1}{4d}.\label{eq:conformal-coupling}
\end{equation}
These are called conformally coupled scalars, and their dynamics in
AdS can be completely captured by those in a flat space with a boundary.
One might be worried that the field redefinition would lead to $z$-dependent
coupling constants in the interaction terms, which is generally true.
In this case, however, we introduce Laplace transforms
\begin{equation}
z^{\alpha}\sim\int_{0}^{\infty}dEf(E,\alpha)e^{-Ez},
\end{equation}
so that the $z$-dependence of the coupling constant has the same
functional form of equation \eqref{eq:scalar-multi-boundary}, for
instance, so the multi-particle recursion is tractable.

\noindent \hrulefill

\

The single-particle states associated to equation \eqref{eq:scalar-eom-AdS}
are what we call bulk-to-boundary propagators. They are the objects
that carry the boundary data. In appendix \ref{sec:Free-solutions-AdS}
the free solutions are discussed in enough detail. Since we still
have translation symmetry along the boundary directions, the single-particle
solutions can be cast as
\begin{equation}
\phi(x,z)=\phi^{(\nu)}(z;k^{2})e^{\mathrm{i}k_{\mu}x^{\mu}},
\end{equation}
where $k^{2}=k^{\mu}k_{\mu}$, and $\phi^{(\nu)}(z;k^{2})$ satisfies
\begin{equation}
\mathcal{D}_{k}^{2}\phi^{(\nu)}=(\nu^{2}-\frac{1}{4}d^{2})\phi^{(\nu)},\label{eq:scalar-eom-AdS-nu}
\end{equation}
with
\begin{align}
\mathcal{D}_{k}^{2} & =z^{2}\partial_{z}^{2}+(1-d)z\partial_{z}-z^{2}k^{2},\\
\nu^{2} & =\frac{1}{4}d^{2}-L^{2}\mathrm{M}^{2}.\label{eq:solution-parameter}
\end{align}
We will also need to invert the kinetic operator of the free equation
of motion. 

\begin{equation}
\left(\mathcal{D}_{k}^{2}+\frac{1}{4}d^{2}-\nu^{2}\right)G^{(\nu)}(z,y;k^{2})=z^{d+1}\delta(z-y),\label{eq:bulk-to-bulk-eom}
\end{equation}
where $z^{d+1}$ comes from the measure $\sqrt{-g}$. Just like in
flat space with a boundary, we will work with the homogeneous Dirichlet
boundary condition, $G(0,y)=G(z,0)=0$, so the bulk-to-bulk propagator
carries no information about the boundary. A simple representation
of $G^{(\nu)}(z,y;k^{2})$ is given in equation \eqref{eq:bulk-to-bulk-wronskian}.

Now we have all the necessary input to build the classical multi-particle
solutions of \eqref{eq:scalar-eom-AdS}. The multi-particle ansatz
is exactly the same as in equation \eqref{eq:scalar-multi-flatboundary},
where the multi-particle function $\Phi_{P}(z)$ satisfies
\begin{equation}
\left(\mathcal{D}_{P}^{2}+\frac{1}{4}d^{2}-\nu^{2}\right)\Phi_{P}=\Phi_{P}^{(\textrm{int})}(z),\label{eq:scalar-recursion-AdS}
\end{equation}
where $\mathcal{D}_{P}^{2}=\mathcal{D}_{k_{P}}^{2}$. Here we have
$\Phi^{(\textrm{int})}(z)$ denoting the multi-particle expansion
of the right hand side of equation \eqref{eq:scalar-eom-AdS}. For
a cubic interaction, for instance, we have
\begin{equation}
\Phi_{P}^{(\textrm{int})}(z)\propto\sum_{P=Q\cup R}\Phi_{Q}(z)\Phi_{R}(z).
\end{equation}
We will leave it implicit, but its general construction is straightforward,
including the case with derivative interactions and involving other
fields (gluons, gravitons, etc...).

In order to determine the recursion for $\Phi_{P}(z)$, we simply
introduce the respective Green's function $G^{(\nu)}(z,y;k_{P}^{2})$.
We then have
\begin{equation}
\Phi_{P}(z)=\int_{0}^{\infty}\frac{dy}{y^{d+1}}G^{(\nu)}(z,y;k_{P}^{2})\Phi_{P}^{(\textrm{int})}(y),\label{eq:scalar-AdS-recursion-inversion}
\end{equation}
which we can show is a solution of equation \eqref{eq:scalar-recursion-AdS}.
Notice that we have essentially the same structures as in flat space
with a boundary. Now, however, the radial dependence of the multi-particle
functions $\Phi_{P}(z)$ is much more involved, and there does not
seem to exist a simple (recursion-friendly) basis to expand them, unlike the exponentials in equation \eqref{eq:Phi3-boundary-newexpansion},
for example.

\subsubsection{Gluons}

The construction of gluon multi-particle solutions in AdS starts with
equation \eqref{eq:YM-eom-curved}. In the Poincar\'e patch, it can
be recast as\begin{subequations}\label{eq:YM-eom-AdS}
\begin{multline}
\partial_{z}^{2}A_{\mu}+\frac{(3-d)}{z}\partial_{z}A_{\mu}+\eta^{\nu\rho}\partial_{\nu}\partial_{\rho}A_{\mu}=-\mathrm{i}[A_{z},\partial_{\mu}A_{z}]-\mathrm{i}[A^{\nu},\partial_{\mu}A_{\nu}]+2\mathrm{i}[A_{z},\partial_{z}A_{\mu}]\\-2\mathrm{i}[\partial^{\nu}A_{\mu},A_{\nu}]
+[A_{z},[A_{z},A_{\mu}]]-[A^{\nu},[A_{\mu},A_{\nu}]]+\partial_{\mu}G-\mathrm{i}[A_{\mu},G],
\end{multline}
in the boundary directions, and as
\begin{multline}
\partial_{z}^{2}A_{z}+\frac{(3-d)}{z}\partial_{z}A_{z}+\eta^{\mu\nu}\partial_{\mu}\partial_{\nu}A_{z}-\frac{(3-d)}{z^{2}}A_{z}=-\mathrm{i}[A^{\mu},\partial_{z}A_{\mu}]+2\mathrm{i}[A^{\mu},\partial_{\mu}A_{z}]+\mathrm{i}[A_{z},\partial_{z}A_{z}]\\
-[A^{\mu},[A_{z},A_{\mu}]]+\partial_{z}G-\mathrm{i}[A_{z},G],
\end{multline}
\end{subequations}in the radial direction. We will be lowering (raising)
boundary indices using the flat metric $\eta_{\mu\nu}$ ($\eta^{\mu\nu}$).
Here we have introduced
\begin{equation}
G=\partial^{\nu}A_{\nu}+\partial_{z}A_{z}+\frac{(3-d)}{z}A_{z},
\end{equation}
which we will set to zero moving forward ($G=0$), as our gauge choice.
For $d=3$, $G=0$ coincides with the Lorenz gauge, and the equations
of motion \eqref{eq:YM-eom-AdS} lead back to our discussion on Yang--Mills
in flat space with a boundary. This is not an accident, as the YM
action in four dimensions (remember, $D=d+1$) is invariant under
conformal transformations of the metric.

After a rescaling of the fields ($A_{m}\to A_{m}/z$), equation \eqref{eq:YM-eom-AdS}
can be put in a form more similar to the scalar models discussed in
subsection \ref{subsec:Scalars-AdS}, with\begin{subequations}\label{eq:YM-eom-AdS-recursion-ready}
\begin{multline}
\left(z^{2}\partial_{z}^{2}+(1-d)z\partial_{z}+z^{2}\eta^{\nu\rho}\partial_{\nu}\partial_{\rho}-(1-d)\right)A_{\mu}=-\mathrm{i}z[A_{z},\partial_{\mu}A_{z}]-\mathrm{i}z[A^{\nu},\partial_{\mu}A_{\nu}]\\
+2\mathrm{i}z[A_{z},\partial_{z}A_{\mu}]-2\mathrm{i}[A_{z},A_{\mu}]-2\mathrm{i}z[\partial^{\nu}A_{\mu},A_{\nu}]+[A_{z},[A_{z},A_{\mu}]]+[A^{\nu},[A_{\nu},A_{\mu}]],
\end{multline}
in the boundary directions, and as
\begin{multline}
\left(z^{2}\partial_{z}^{2}+(1-d)z\partial_{z}+z^{2}\eta^{\nu\rho}\partial_{\nu}\partial_{\rho}-(4-2d)\right)A_{z}=-\mathrm{i}z[A^{\mu},\partial_{z}A_{\mu}]+2\mathrm{i}z[A^{\mu},\partial_{\mu}A_{z}]\\
+\mathrm{i}z[A_{z},\partial_{z}A_{z}]+[A^{\mu},[A_{\mu},A_{z}]],
\end{multline}
\end{subequations}
in the radial direction. The gauge choice $G=0$ translates to
\begin{equation}
\partial^{\mu}A_{\mu}+\partial_{z}A_{z}+\frac{(2-d)}{z}A_{z}=0.\label{eq:gluon-gauge-AdS}
\end{equation}

The free solutions of equation \eqref{eq:YM-eom-AdS-recursion-ready}
are identified with the gluons, and are given by
\begin{align}
A_{\mu}(x,z) & =\varepsilon_{\mu}e^{\mathrm{i}k_{\nu}x^{\nu}}\phi^{(d/2-1)}(z),\\
A_{z}(x,z) & =\varepsilon_{z}e^{\mathrm{i}k_{\nu}x^{\nu}}\phi^{(d/2-2)}(z),
\end{align}
where $\varepsilon_{\mu}$ and $\varepsilon_{z}$ are the components
of the gluon polarization. The radial component has a different
effective mass, but this is consistent with the gauge choice, since\begin{subequations}\label{eq:raising-lowering-nu}
\begin{align}
\left(\partial_{z}-\frac{(\nu+d/2)}{z}\right)\phi^{(\nu)}(z;k^{2}) & =-2\nu\phi^{(\nu+1)}(z;k^{2}),\\
\left(\partial_{z}+\frac{\nu-d/2}{z}\right)\phi^{(\nu)}(z;k^{2}) & =\frac{2}{(1-\nu)}\left(\frac{k^{2}}{4}\right)\phi^{(\nu-1)}(z;k^{2}).
\end{align}
\end{subequations}The operators on the left hand side act like a
raising/lowering operators on $\nu$. In particular, this equation
implies a residual gauge transformation given by
\begin{equation}
\begin{array}{rcl}
\delta\varepsilon_{\mu}=k_{\mu}\lambda, &  & \delta\varepsilon_{z}=\frac{\mathrm{i}}{(d-4)}k^{2}\lambda\end{array},\label{eq:residual-gluon-AdS}
\end{equation}
which we will use to set $\varepsilon_{z}=0$.

For pure Yang--Mills in AdS, we can work with the color-stripped
multi-particle ansatz
\begin{equation}
A_{m}(x,z)=\sum_{P}\boldsymbol{A}_{Pm}(z)e^{\mathrm{i}k_{P\mu}x^{\mu}}T^{a_{P}}.
\end{equation}
The gauge fixed equations of motion \eqref{eq:YM-eom-AdS-recursion-ready}
then lead to\begin{subequations}
\begin{multline}
\left(\mathcal{D}_{P}^{2}+(d-1)\right)\boldsymbol{A}_{P\mu}\\
=\sum_{P=QR}\left(\boldsymbol{A}_{Qz}(zk_{R\mu}\boldsymbol{A}_{Rz}+2\mathrm{i}z\partial_{z}\boldsymbol{A}_{R\mu}-2\mathrm{i}\boldsymbol{A}_{R\mu})+\boldsymbol{A}_{Q}^{\nu}(zk_{R\mu}\boldsymbol{A}_{R\nu}-2k_{R\nu}\boldsymbol{A}_{R\mu})-(Q\leftrightarrow R)\right)\\
+\sum_{P=QRS}\left(\boldsymbol{A}_{Qz}\boldsymbol{A}_{Rz}\boldsymbol{A}_{S\mu}-\boldsymbol{A}_{Qz}\boldsymbol{A}_{R\mu}\boldsymbol{A}_{Sz}+\boldsymbol{A}_{Q}^{\nu}\boldsymbol{A}_{R\nu}\boldsymbol{A}_{S\mu}-\boldsymbol{A}_{Q}^{\nu}\boldsymbol{A}_{R\mu}\boldsymbol{A}_{S\nu}+(Q\leftrightarrow S)\right),
\end{multline}
and
\begin{multline}
\left(\mathcal{D}_{P}^{2}+(2d-4)\right)\boldsymbol{A}_{Pz}=\sum_{P=QR}\left(\mathrm{i}z\boldsymbol{A}_{Qz}\partial_{z}\boldsymbol{A}_{Rz}-z\boldsymbol{A}_{Q}^{\mu}(\mathrm{i}\partial_{z}\boldsymbol{A}_{R\mu}+2k_{R\mu}\boldsymbol{A}_{Rz})-(Q\leftrightarrow R)\right)\\
+\sum_{P=QRS}\left(\boldsymbol{A}_{Q}^{\mu}\boldsymbol{A}_{R\mu}\boldsymbol{A}_{Sz}-\boldsymbol{A}_{Q}^{\mu}\boldsymbol{A}_{Rz}\boldsymbol{A}_{S\mu}+(Q\leftrightarrow S)\right).
\end{multline}
\end{subequations}In order to actually compute $\boldsymbol{A}_{P\mu}$
and $\boldsymbol{A}_{Pz}$, the procedure is similar to the one described
in equation \eqref{eq:scalar-AdS-recursion-inversion} with the appropriate
Green's functions, respectively $G^{(d/2-1)}(z,y;k_{P}^{2})$ and
$G^{(d/2-2)}(z,y;k_{P}^{2})$. Finally, we should keep in mind there
are other gauge choices, with different advantages and disadvantages.
From a diagrammatic computation point of view, the most common is
the axial gauge, $A_{z}=0$. In \cite{Armstrong:2022mfr} a boundary
transversal gauge was used ($\partial^{\mu}A_{\mu}=0$), in which
the recursion for $A_{Pz}$ is purely algebraic.

\subsubsection{Gravitons}

In the case of gravitons in anti-de Sitter, we essentially repeat
the procedure described in section \ref{sec:gravity}. Instead
of expanding around flat space ($g_{mn}=\eta_{mn}+H_{mn}$), we expand
the field equations \eqref{eq:EFE} in empty space around $g_{mn}^{\textrm{AdS}}$,
with cosmological constant \eqref{eq:cosmo-constant}. The expansion
is certainly longer, but nothing we cannot manage. The trace-reversed
equation is given by
\begin{equation}
R_{mn}=-\frac{d}{L^{2}}g_{mn},\label{eq:eom-AdS-graviton}
\end{equation}
where $g_{mn}=g_{mn}^{\textrm{AdS}}+\tilde{H}_{mn}$. The linearized
form of this equation takes a remarkably simple form, and can be cast
as\begin{subequations}\label{eq:graviton-linearized-AdS}
\begin{align}
[z^{2}\partial_{z}^{2}+(5-d)z\partial_{z}+z^{2}\eta^{\nu\rho}\partial_{\nu}\partial_{\rho}+(4-2d)]\tilde{H}_{\mu\nu} & =z^{2}(\partial_{\mu}G_{\nu}+\partial_{\nu}G_{\mu})-2z\eta_{\mu\nu}G,\label{eq:eom-AdS-graviton-munu}\\{}
[z^{2}\partial_{z}^{2}+(5-d)z\partial_{z}+z^{2}\eta^{\nu\rho}\partial_{\nu}\partial_{\rho}+(3-d)]\tilde{H}_{z\mu} & =z^{2}(\partial_{z}G_{\mu}+\partial_{\mu}G)+2zG_{\mu}+z\partial_{\mu}\tilde{H}_{zz},\label{eq:eom-AdS-graviton-zmu}\\{}
[z^{2}\partial_{z}^{2}+(3-d)z\partial_{z}+z^{2}\eta^{\nu\rho}\partial_{\nu}\partial_{\rho}]\tilde{H}_{zz} & =2z(z\partial_{z}+1)G,\label{eq:eom-AdS-graviton-zz}
\end{align}
\end{subequations}where $G_{\mu}$ and $G$ are given by
\begin{eqnarray}
G_{\mu} & \equiv & \partial^{\nu}\tilde{H}_{\mu\nu}-\frac{1}{2}\eta^{\nu\rho}\partial_{\mu}\tilde{H}_{\nu\rho}+\partial_{z}\tilde{H}_{z\mu}-\frac{1}{2}\partial_{\mu}\tilde{H}_{zz}+(3-d)\frac{1}{z}\tilde{H}_{z\mu}\label{eq:harmonic-AdS-boundary}\\
G & \equiv & \partial^{\mu}\tilde{H}_{z\mu}-\frac{1}{2}\eta^{\mu\nu}\partial_{z}\tilde{H}_{\mu\nu}+\frac{1}{2}\partial_{z}\tilde{H}_{zz}+\frac{1}{z}(1-d)\tilde{H}_{zz}-\frac{1}{z}\eta^{\mu\nu}\tilde{H}_{\mu\nu}.\label{eq:harmonic-AdS-radial}
\end{eqnarray}
It is easy to show that these objects are very similar to a linearized
form of the harmonic gauge in AdS, but each has one additional term\footnote{More precisely, they can be expressed as
\begin{align*}
z^{2}G & =\left.g^{mn}\Gamma_{zmn}\right|_{\textrm{linear}}-\frac{d}{2}\partial_{z}g^{zz}\tilde{H}_{zz},\\
z^{2}G_{\mu} & =\left.g^{mn}\Gamma_{\mu mn}\right|_{\textrm{linear}}-\frac{(d-1)}{2}\partial_{z}g^{zz}\tilde{H}_{z\mu}.
\end{align*}
}. Moving on, our gauge choice will be $G_{\mu}=G=0$.

Note that under the coordinate transformations of equation \eqref{eq:GCT-linearized},
given by\begin{subequations}
\begin{eqnarray}
\delta\tilde{H}_{\mu\nu} & = & \partial_{\nu}\lambda_{\mu}+\partial_{\mu}\lambda_{\nu}-\frac{2}{z}\eta_{\mu\nu}\lambda_{z},\\
\delta\tilde{H}_{z\mu} & = & \partial_{\mu}\lambda_{z}+\partial_{z}\lambda_{\mu}+\frac{2}{z}\lambda_{\mu},\\
\delta\tilde{H}_{zz} & = & 2\partial_{z}\lambda_{z}+\frac{2}{z}\lambda_{z},
\end{eqnarray}
\end{subequations}the harmonic gauge transforms as
\begin{align}
\delta G_{\mu} & =\frac{1}{z^{2}}[z^{2}\partial_{z}^{2}+(5-d)\partial_{z}+z^{2}\eta^{\nu\rho}\partial_{\nu}\partial_{\rho}+(4-2d)]\lambda_{\mu},\\
\delta G & =\frac{1}{z^{2}}[z^{2}\partial_{z}^{2}+(3-d)z\partial_{z}+z^{2}\eta^{\nu\rho}\partial_{\nu}\partial_{\rho}+(1-d)]\lambda_{z},
\end{align}
so we still have a residual gauge invariance as long as the parameters
$\lambda_{\mu}$ and $\lambda_{z}$ satisfy the equations 
\begin{align}
[z^{2}\partial_{z}^{2}+(5-d)\partial_{z}+z^{2}\partial^{\nu}\partial_{\nu}+(4-2d)]\lambda_{\mu} & =0,\\{}
[z^{2}\partial_{z}^{2}+(3-d)z\partial_{z}+z^{2}\partial^{\mu}\partial_{\mu}+(1-d)]\lambda_{z} & =0,
\end{align}
but how can we effectively use them?

First, let us recast $H_{mn}$ as
\begin{equation}
\tilde{H}_{mn}(x,z)=h_{mn}(z)e^{\mathrm{i}k_{\mu}x^{\mu}},
\end{equation}
so the plane-wave behavior in the boundary directions is explicit.
Next, consider the following transformation,
\begin{equation}
\delta\left(\partial_{z}h_{zz}+\frac{(2-d)}{z}h_{zz}\right)=2k^{\mu}k_{\mu}\lambda_{z}.
\end{equation}
Let us assume that the boundary momentum is not light-like\footnote{For light-like momentum ($k^{\mu}k_{\mu}=0$) all the physical solutions
to the equations of motion become a power law in $z$. In a sense,
the bulk-to-boundary propagator becomes universal, as it does not
depend on the momenta of the boundary states, and there is only a
rigid tensor structure left.}, i.e., $k^{\mu}k_{\mu}\neq0$. We can then use $\lambda_{z}$ to
set the expression in parenthesis to zero, which implies that
\begin{equation}
h_{zz}\propto z^{d-2}.
\end{equation}
In turn, equation \eqref{eq:eom-AdS-graviton-zz} yields $k^{\mu}k_{\mu}h_{zz}=0$.
The only possible solution is then $h_{zz}=0$, and there is no more
residual gauge transformation associated to the parameter $\lambda_{z}$.
We can do a similar analysis for the combination
\begin{equation}
\delta\left(\partial_{z}h_{z\mu}+\frac{(3-d)}{z}h_{z\mu}\right)=k^{\nu}k_{\nu}\lambda_{\mu},
\end{equation}
and use $\lambda_{\mu}$ to set the combination in parenthesis to
zero, leading to
\begin{equation}
h_{z\mu}\propto z^{d-3}.
\end{equation}
However, equation \eqref{eq:eom-AdS-graviton-zmu} implies that the
proportionally constant is zero. In this case, the physical degrees
of freedom of the graviton are all contained in $\tilde{H}_{\mu\nu}$.
After dealing with the residual gauge redundancy, the gauge fixing
condition $G=0$ leads to
\begin{equation}
\left(\partial_{z}+\frac{2}{z}\right)\eta^{\mu\nu}h_{\mu\nu}=0,
\end{equation}
which effectively implies $\eta^{\mu\nu}h_{\mu\nu}=0$, since it is
the only compatible solution with the equation of motion \eqref{eq:eom-AdS-graviton-munu}.
Finally, the gauge condition $G_{\mu}=0$ yields
\begin{equation}
k^{\nu}h_{\mu\nu}=0.
\end{equation}
Therefore, the graviton polarization $h_{\mu\nu}$ is (boundary) traceless
and transversal.

Before moving on to the multi-graviton solutions, note that the linearized
equations \eqref{eq:graviton-linearized-AdS} are not in the standard
scalar form of \eqref{eq:scalar-eom-AdS}. This can be achieved with
a simple scaling, though the components $\tilde{H}_{\mu\nu}$, $\tilde{H}_{z\mu}$,
and $\tilde{H}_{zz}$ do not scale uniformly. Observe that\begin{subequations}
\begin{align}
[z^{2}\partial_{z}^{2}+(5-d)z\partial_{z}+z^{2}\partial^{\mu}\partial_{\mu}+(4-2d)]\tilde{H}_{\mu\nu} & =\frac{1}{z^{2}}[z^{2}\partial_{z}^{2}+(1-d)z\partial_{z}+z^{2}\partial^{\mu}\partial_{\mu}](z^{2}\tilde{H}_{\mu\nu}),\\{}
[z^{2}\partial_{z}^{2}+(5-d)z\partial_{z}+z^{2}\partial^{\mu}\partial_{\mu}+(3-d)]\tilde{H}_{z\mu} & =\frac{1}{z^{2}}[z^{2}\partial_{z}^{2}+(1-d)z\partial_{z}+z^{2}\partial^{\mu}\partial_{\mu}+(d-1)](z^{2}\tilde{H}_{z\mu}),\\{}
[z^{2}\partial_{z}^{2}+(3-d)z\partial_{z}+z^{2}\partial^{\mu}\partial_{\mu}]\tilde{H}_{zz} & =\frac{1}{z}[z^{2}\partial_{z}^{2}+(1-d)z\partial_{z}+z^{2}\partial^{\mu}\partial_{\mu}+(d-1)](z\tilde{H}_{zz}).
\end{align}
\end{subequations}Therefore, the ansatze for the multi-particle expansion
of $g_{mn}$ and $g^{mn}$,
\begin{align}
g_{mn} & =g_{mn}^{\textrm{AdS}}+\sum_{P}\tilde{H}_{Pmn}(z)e^{\mathrm{i}k_{P\mu}x^{\mu}},\label{eq:multi-graviton-AdS}\\
g^{mn} & =(g^{\textrm{AdS}})^{mn}-\sum_{P}\tilde{I}_{P}^{mn}(z)e^{\mathrm{i}k_{P\mu}x^{\mu}},\label{eq:multi-graviton-inverse-AdS}
\end{align}
can be more conveniently expressed as\begin{subequations}\label{eq:multi-graviton-AdS-components}
\begin{align}
g_{\mu\nu} & =\frac{L^{2}}{z^{2}}\eta_{\mu\nu}+\frac{L^{2}}{z^{2}}\sum_{P}H_{P\mu\nu}(z)e^{\mathrm{i}k_{P\mu}x^{\mu}},\\
g_{z\mu} & =\frac{L^{2}}{z^{2}}\sum_{P}H_{Pz\mu}(z)e^{\mathrm{i}k_{P\mu}x^{\mu}},\\
g_{zz} & =\frac{L^{2}}{z^{2}}+\frac{L}{z}\sum_{P}H_{Pzz}(z)e^{\mathrm{i}k_{P\mu}x^{\mu}},
\end{align}
\end{subequations}with inverse\begin{subequations}\label{eq:multi-graviton-inverse-AdS-components}
\begin{align}
g^{\mu\nu} & =\frac{z^{2}}{L^{2}}\eta^{\mu\nu}-\frac{z^{2}}{L^{2}}\sum_{P}I_{P}^{\mu\nu}(z)e^{\mathrm{i}k_{P\mu}x^{\mu}},\\
g^{z\mu} & =-\frac{z^{2}}{L^{2}}\sum_{P}I_{P}^{z\mu}(z)e^{\mathrm{i}k_{P\mu}x^{\mu}},\\
g^{zz} & =\frac{z^{2}}{L^{2}}-\frac{z}{L}\sum_{P}I_{P}^{zz}(z)e^{\mathrm{i}k_{P\mu}x^{\mu}}.
\end{align}
\end{subequations}Since we require that $g_{mp}g^{pn}=\delta_{m}^{n}$,
the currents $I_{P}^{mn}$ are recursively determined as\begin{subequations}
\begin{align}
\eta_{\mu\rho}I_{P}^{\rho\nu} & =\eta^{\rho\nu}H_{P\mu\rho}-\sum_{P=Q\cup R}(H_{Q\mu\rho}I_{R}^{\rho\nu}+H_{Qz\mu}I_{R}^{z\nu}),\\
I_{P}^{z\mu} & =\eta^{\mu\nu}H_{P\nu z}-\eta^{\mu\nu}\sum_{P=Q\cup R}(H_{Q\nu\rho}I_{R}^{\rho z}-\frac{L}{z}H_{Q\nu z}I_{R}^{zz}),\nonumber \\
 & =\eta^{\mu\nu}H_{P\nu z}-\sum_{P=Q\cup R}(H_{Qz\nu}I_{R}^{\mu\nu}+\frac{z}{L}H_{Pzz}I_{P}^{z\mu}),\\
I_{P}^{zz} & =\frac{z^{2}}{L^{2}}H_{Pzz}-\frac{z}{L}\sum_{P=Q\cup R}(H_{Qz\mu}I_{R}^{\mu z}+H_{Qzz}I_{R}^{zz}).
\end{align}
\end{subequations}The asymmetric scaling of $H_{Pzz}$ introduces
some extra powers of $z$ here and there, but this is consistent with
the overall field redefinition necessary to have a standard form for
the bulk-to-bulk propagators. The currents with one-letter words correspond
to the bulk-to-boundary propagators (our single-particle states, as
usual), expressed as
\begin{equation}
H_{i\mu\nu}=h_{i\mu\nu}\phi^{(d/2)}(z;k_{i}^{2}),
\end{equation}
and $H_{iz\mu}=H_{izz}=0$. Note that $H_{i\mu\nu}$ satisfies the
free equation
\begin{equation}
\mathcal{D}_{k_{i}}^{2}H_{i\mu\nu}=0,
\end{equation}
with $k_{i}^{\nu}h_{i\mu\nu}=\eta^{\mu\nu}h_{i\mu\nu}=0$.

In order to determine the recursion for the multi-particle currents,
we replace the ansatze above in equation \eqref{eq:eom-AdS-graviton}.
The interaction part is quite long, but the full result can be expressed
as\begin{subequations}\label{eq:multi-graviton-AdS-recursion}
\begin{align}
\mathcal{D}_{P}^{2}H_{P\mu\nu} & =4\sum_{P=Q\cup R}(H_{Q\mu\rho}I_{R}^{\rho\sigma}+H_{Qz\mu}I_{R}^{z\sigma})\eta_{\nu\sigma}+2z^{2}G_{P\mu\nu}\nonumber \\
 & +2\eta_{\mu\nu}\sum_{P=Q\cup R}[(1-d)(H_{Qz\rho}I_{R}^{\rho z}+H_{Qzz}I_{R}^{zz})-H_{Q\rho\sigma}I_{R}^{\rho\sigma}-H_{Qz\rho}I_{R}^{z\rho}],\\{}
[\mathcal{D}_{P}^{2}+(d-1)]H_{Pz\mu} & =2d\sum_{P=Q\cup R}(H_{Qz\rho}I_{R}^{\rho\nu}+\frac{z}{L}H_{Pzz}I_{P}^{z\nu})\eta_{\mu\nu}+2z^{2}G_{Pz\mu}+\mathrm{i}\frac{z^{2}}{L}k_{P\mu}H_{Pzz},\\
\frac{z}{L}[\mathcal{D}_{P}^{2}+(d-1)]H_{Pzz} & =2d\sum_{P=Q\cup R}(H_{Qz\mu}I_{R}^{\mu z}+H_{Qzz}I_{R}^{zz})+2z^{2}G_{Pzz},
\end{align}
\end{subequations}It has a relatively compact form, since most of
the non-linear contributions are hidden in the object $G_{Pmn}$, whose explicit expression is considerably more involved. It can be cast as
\begin{multline}
G_{Pmn}=\sum_{P=Q\cup R}\Big\{\tilde{I}_{Q}^{pq}(\partial_{n}\tilde{\Gamma}_{Rpmq}-\partial_{p}\tilde{\Gamma}_{Rqmn})\\
+\vphantom{\sum_{P=Q\cup R}}(g^{\textrm{AdS}})^{pq}(g^{\textrm{AdS}})^{rs}(\tilde{\Gamma}_{Qpmr}\tilde{\Gamma}_{Rqns}-\tilde{\Gamma}_{Qpmn}\tilde{\Gamma}_{Rqrs})+\tilde{I}_{Q}^{pq}\tilde{I}_{R}^{rs}(\Gamma_{pmr}^{\textrm{AdS}}\Gamma_{qns}^{\textrm{AdS}}-\Gamma_{pmn}^{\textrm{AdS}}\Gamma_{qrs}^{\textrm{AdS}})\\
\vphantom{\sum_{P=Q\cup R}}-[\tilde{I}_{Q}^{pq}(g^{\textrm{AdS}})^{rs}+(g^{\textrm{AdS}})^{pq}\tilde{I}_{Q}^{rs}](\Gamma_{pmr}^{\textrm{AdS}}\tilde{\Gamma}_{Rqns}+\tilde{\Gamma}_{Rpmr}\Gamma_{qns}^{\textrm{AdS}}-\Gamma_{pmn}^{\textrm{AdS}}\tilde{\Gamma}_{Rqrs}-\tilde{\Gamma}_{Rpmn}\Gamma_{qrs}^{\textrm{AdS}})\Big\}\\
+\sum_{P=Q\cup R\cup S}\Big\{[\tilde{I}_{Q}^{pq}(g^{\textrm{AdS}})^{RS}+(g^{\textrm{AdS}})^{PQ}\tilde{I_{Q}}^{rs}](\tilde{\Gamma}_{Rpmr}\tilde{\Gamma}_{Sqns}-\tilde{\Gamma}_{Rpmn}\tilde{\Gamma}_{Sqrs})\\
\vphantom{\sum_{P=Q\cup R}}-\tilde{I}_{Q}^{pq}\tilde{I}_{R}^{rs}(\Gamma_{pmr}^{\textrm{AdS}}\tilde{\Gamma}_{Sqns}+\tilde{\Gamma}_{Spmr}\Gamma_{qns}^{\textrm{AdS}}-\Gamma_{pmn}^{\textrm{AdS}}\tilde{\Gamma}_{Sqrs}-\tilde{\Gamma}_{Spmn}\Gamma_{qrs}^{\textrm{AdS}})\Big\}\\
-\sum_{P=Q\cup R\cup S\cup T}\tilde{I}_{Q}^{pq}\tilde{I}_{R}^{rs}(\tilde{\Gamma}_{Spmr}\tilde{\Gamma}_{Tqns}-\tilde{\Gamma}_{Spmn}\tilde{\Gamma}_{Tqrs}).
\end{multline}
Here we are using the covariant form of the expansions \eqref{eq:multi-graviton-AdS}
and \eqref{eq:multi-graviton-inverse-AdS}, and the specific components
can be read via a direct comparison with equations \eqref{eq:multi-graviton-AdS-components}
and \eqref{eq:multi-graviton-inverse-AdS-components}, in which the
scalings have been introduced. We also define
\begin{equation}
\tilde{\Gamma}_{Pqmn}=\frac{1}{2}(\partial_{m}\tilde{H}_{Pnq}+\partial_{n}\tilde{H}_{Pmq}-\partial_{q}\tilde{H}_{Pmn}),
\end{equation}
where the derivative is simply $\partial_{z}$ in the radial direction
or $\mathrm{i}k_{P\mu}$ in the boundary directions. The recursions
displayed in equation \eqref{eq:multi-graviton-AdS-recursion} were
obtained using the gauge choices \eqref{eq:harmonic-AdS-boundary}
and \eqref{eq:harmonic-AdS-radial}. Their multi-particle version
are respectively given by
\begin{align}
k_{P}^{\nu}H_{P\mu\nu}-\frac{1}{2}k_{P\mu}\eta^{\nu\rho}H_{P\nu\rho} & =\frac{\mathrm{i}}{z}[z\partial_{z}+(1-d)]H_{Pz\mu}+\frac{1}{2}\frac{z}{L}k_{P\mu}H_{Pzz},\\
k_{P}^{\mu}H_{Pz\mu} & =\frac{\mathrm{i}}{2L}[z\partial_{z}+(2d-1)]H_{Pzz}-\frac{\mathrm{i}}{2}\eta^{\mu\nu}\partial_{z}H_{P\mu\nu}.
\end{align}
As in the case of gluons, the computation of $H_{P\mu\nu}$, $H_{Pz\mu}$,
and $H_{Pzz}$ follows the steps described in equation \eqref{eq:scalar-AdS-recursion-inversion}
with the appropriate Green's functions, respectively $G^{(d/2)}(z,y;k_{P}^{2})$
for $H_{P\mu\nu}$, and $G^{(d/2-2)}(z,y;k_{P}^{2})$ for $H_{Pz\mu}$
and $H_{Pzz}$.

For mixed theories (for instance, including gravitons, gluons, and
scalars), the construction is completely analogous to what we have
seen so far. The corresponding equations of course get a bit longer,
but the intermediate steps generalize quite naturally. In \cite{Armstrong:2022mfr}
the reader can find a couple of explicit computations in a different
gauge. In the next subsection we will introduce the correlator expressions
computed from the multi-particle currents in AdS.

\subsubsection{Correlators and Witten diagrams}

Inspired by our construction in flat space with a boundary, we can
easily propose the formulae for correlators in AdS using the multi-particle
currents introduced above.

Let us start with scalars. The $(n+1)$-point correlator in AdS is
given by
\begin{equation}
\tilde{A}_{n+1}=-\int_{0}^{\infty}\frac{dz}{z^{d+1}}\,\phi_{n+1}^{(\nu)}\left(\mathcal{D}_{1\ldots n}^{2}+\frac{1}{4}d^{2}-\nu^{2}\right)\Phi_{1\ldots n}(z),\label{eq:boundary-scalar-correlator-AdS}
\end{equation}
where $\phi_{n+1}^{(\nu)}=\phi^{(\nu)}(z;k_{n+1}^{2})$, the invariant
measure $\sqrt{-g}$ in the Poincar\'e patch is expressed as $z^{d+1}$,
and the term in parenthesis is the usual kinetic operator. This is
a straightforward generalization of the first line in equation \eqref{eq:phi3-boundary-correlator}.
Indeed, they coincide for conformally coupled scalars with a $z$-dependent
coupling constant (see the discussion around equation \eqref{eq:conformal-coupling}).
For polynomial self-interactions, equation \eqref{eq:boundary-scalar-correlator-AdS}
perfectly matches the diagrammatic computation. As an example, let
us consider a $\phi^{3}$ interaction. The three-point correlator
is given by
\begin{equation}
\tilde{A}_{3}\propto\int_{0}^{\infty}\frac{dz}{z^{d+1}}\,\phi_{1}^{(\nu)}\phi_{2}^{(\nu)}\phi_{3}^{(\nu)},
\end{equation}
where momentum conservation along the boundary directions is left
implicit. This is exactly the same output of the three-point Witten
diagram displayed below. In the diagram, the boundary ($z=0$) is
represented by the dashed line, while the solid lines are the bulk-to-boundary
propagators.\begin{equation}
  \tilde{A}_{3} 
  \qquad \longleftrightarrow \qquad
  \raisebox{-0.4\height}{%
    \begin{tikzpicture}[scale=0.7]
 
  \draw[thick,dashed] (0,0) circle (2cm);
 
 
  \coordinate (z) at (0,0);
  \fill (z) circle (2pt);
  \node[right, font=\small] at (0.08,0.1) {$z$};
 
  \coordinate (p1) at (210:2cm);
  \coordinate (p2) at (330:2cm);
  \coordinate (p3) at (90:2cm);
 
  \fill (p1) circle (2pt);
  \fill (p2) circle (2pt);
  \fill (p3) circle (2pt);
 
  \draw[thick] (z) -- (p1);
  \draw[thick] (z) -- (p2);
  \draw[thick] (z) -- (p3);
 
  \node[below left,  font=\normalsize] at (p1) {$1$};
  \node[below right, font=\normalsize] at (p2) {$2$};
  \node[above,yshift=3pt,       font=\normalsize] at (p3) {$3$};

\end{tikzpicture}}
\end{equation}

\

\noindent The interaction happens in the bulk (coordinate $z$),
which is integrated over the whole space. Note, in particular, this
is a natural source of divergences. Depending on the value of $\nu$,
the integrand might not converge fast enough as $z\to0$. This is
usually resolved via the introduction of an integration cut-off at
$z=\epsilon$, with the divergent pieces as $\epsilon\to0$ being
canceled by boundary counter terms (see e.g. \cite{Henningson:1998gx,deHaro:2000vlm}).

In this case, the interaction does not depend on $z$-derivatives.
This is relevant because the recursive construction of $\Phi_{P}(z)$
does not have access to the external leg that will be attached to
it. Effectively, this means that the bulk-to-boundary propagators
present in $\Phi_{P}$ have a special role. This asymmetry disappears,
for instance, when we have momentum conservation. Since momentum is
not conserved along the radial direction, the expression \eqref{eq:boundary-scalar-correlator-AdS}
would match the respective Witten diagrams computations only up to
boundary contributions (total derivatives in $z$) if derivative interactions
were to be considered. One way to fix this is to explicitly introduce
an averaged formula with respect to the single-particle labels. This
is necessary, for instance, in the Yang--Mills theory.

Similarly to equation \eqref{eq:boundary-correlator-YM}, the color-ordered
correlator of gluons in AdS is given by
\begin{multline}
\tilde{\mathscr{C}}(1,\ldots,N,N+1)=-\frac{1}{N+1}\varepsilon_{N+1}^{\mu}\int_{0}^{\infty}\frac{dz}{z^{d+1}}\,\phi_{N+1}^{(d/2-1)}\left(\mathcal{D}_{1\ldots N}^{2}+(d-1)\right)\boldsymbol{A}_{1\ldots N\mu}\\
-\frac{1}{N+1}\varepsilon_{N+1}^{z}\int_{0}^{\infty}\frac{dz}{z^{d+1}}\,\phi_{N+1}^{(d/2-2)}\left(\mathcal{D}_{1\ldots N}^{2}+(2d-4)\right)\boldsymbol{A}_{1\ldots Nz}\\
+\textrm{cyclic}(1,\ldots,N,N+1),
\end{multline}
where the cyclic averaging is explicitly added. Note here that we
have not fixed the residual gauge redundancy (for instance by setting
$\varepsilon^{z}=0$). It is straightforward to check the integrands
in $\tilde{\mathscr{C}}(1,\ldots,N,N+1)$ change by total derivatives
with respect to the residual transformations. This is consistent with
the Ward identities of the theory. In order to see this, let us consider
the specific contribution
\begin{multline}
\delta\tilde{\mathscr{C}}\propto\delta\varepsilon_{N+1}^{\mu}\int_{0}^{\infty}\frac{dz}{z^{d+1}}\,\phi_{N+1}^{(d/2-1)}\left(\mathcal{D}_{1\ldots N}^{2}+(d-1)\right)\boldsymbol{A}_{1\ldots N\mu}\\
+\delta\varepsilon_{N+1}^{z}\int_{0}^{\infty}\frac{dz}{z^{d+1}}\,\phi_{N+1}^{(d/2-2)}\left(\mathcal{D}_{1\ldots N}^{2}+(2d-4)\right)\boldsymbol{A}_{1\ldots Nz}+\ldots,
\end{multline}
where $\delta\varepsilon_{N+1}^{\mu}$ and $\delta\varepsilon_{N+1}^{z}$
are given in equation \eqref{eq:residual-gluon-AdS}. Note that the
gauge condition \eqref{eq:gluon-gauge-AdS} is expressed as
\begin{equation}
\mathrm{i}k_{P}^{\mu}\boldsymbol{A}_{P\mu}+\partial_{z}\boldsymbol{A}_{Pz}+\frac{(2-d)}{z}\boldsymbol{A}_{Pz}=0,
\end{equation}
in the terms of the multi-particle currents. Now, using momentum conservation
along the boundary and recalling equation \eqref{eq:raising-lowering-nu},
the contribution above can be recast as
\begin{multline}
\delta\tilde{\mathscr{C}}\propto-\mathrm{i}\int_{0}^{\infty}\frac{dz}{z^{d+1}}\,\phi_{N+1}^{(d/2-1)}\left(\mathcal{D}_{1\ldots N}^{2}+(d-1)\right)\left(\partial_{z}+\frac{(2-d)}{z}\right)\boldsymbol{A}_{1\ldots Nz}\\
-\mathrm{i}\int_{0}^{\infty}\frac{dz}{z^{d+1}}\,\left(\partial_{z}-\frac{1}{z}\right)\phi_{N+1}^{(d/2-1)}\left(\mathcal{D}_{1\ldots N}^{2}+(2d-4)\right)\boldsymbol{A}_{1\ldots Nz}+\ldots,
\end{multline}
Finally, we use the property
\begin{equation}
\left(\mathcal{D}_{P}^{2}+(d-1)\right)\left(\partial_{z}+\frac{(2-d)}{z}\right)=\left(\partial_{z}-\frac{d}{z}\right)\left(\mathcal{D}_{P}^{2}+(2d-4)\right),
\end{equation}
which leads to
\begin{equation}
\delta\tilde{\mathscr{C}}\propto-\mathrm{i}\int_{0}^{\infty}dz\,\partial_{z}\left(\frac{1}{z^{d+1}}\phi_{N+1}^{(d/2-1)}\left(\mathcal{D}_{1\ldots N}^{2}+(2d-4)\right)\boldsymbol{A}_{1\ldots Nz}\right)+\ldots,
\end{equation}
as we expected. A similar analysis holds when we consider the residual
gauge transformations of the multi-particle currents. This is akin
to what we have discussed in flat space (section \ref{sec:gauge-theories}),
and we can show that the color-ordered correlator $\tilde{\mathscr{C}}$
is invariant up to boundary contributions.

Finally, the expression for graviton correlators in AdS is given by
\begin{align}
\tilde{\mathcal{M}}_{N+1} & =-\frac{1}{N+1}\int_{0}^{\infty}\frac{dz}{z^{d+1}}I_{N+1}^{zz}\left(\mathcal{D}_{1\ldots N}^{2}+(d-1)\right)H_{1\ldots Nzz}\nonumber \\
 & -\frac{2}{N+1}\int_{0}^{\infty}\frac{dz}{z^{d+1}}I_{N+1}^{z\mu}\left(\mathcal{D}_{1\ldots N}^{2}+(d-1)\right)H_{1\ldots Nz\mu}\nonumber \\
 & -\frac{1}{N+1}\int_{0}^{\infty}\frac{dz}{z^{d+1}}I_{N+1}^{\mu\nu}\left(\mathcal{D}_{1\ldots N}^{2}\right)H_{1\ldots N\mu\nu}+\textrm{perm}\{N+1|1,\ldots,N\}.\label{eq:graviton-correlator-AdS}
\end{align}
where $\textrm{perm}\{N+1|1,\ldots,N\}$ denotes the $N$ additional
terms obtained by cycling the single-particle labels $1,\ldots,N+1$.
The factor of $2$ on the second line comes from the symmetric contributions
$I_{N+1}^{z\mu}(\ldots)H_{1\ldots Nz\mu}$ and $I_{N+1}^{\mu z}(\ldots)H_{1\ldots N\mu z}$.
The transformation of $\tilde{\mathcal{M}}_{N+1}$ under residual
gauge transformations follows the same logic of the gluon correlators,
and it is possible to show that it leads to a boundary contribution.

As an example, let us compute the three-point graviton correlator.
We will fix the residual gauge redundancy by setting $H_{izz}=H_{iz\mu}=0$.
In this case, we need to compute
\begin{equation}
\mathcal{D}_{12}^{2}H_{12\mu\nu}=4[(h_{1}\cdot h_{2})_{\mu\nu}+(h_{1}\cdot h_{2})_{\nu\mu}]+2z^{2}G_{P\mu\nu}-4\eta_{\mu\nu}(h_{1}\cdot h_{2}),
\end{equation}
where $G_{12\mu\nu}$ is given by
\begin{multline}
G_{12\mu\nu}=\frac{1}{2}k_{1\mu}(k_{1}\cdot h_{2}\cdot h_{1})_{\nu}+\frac{1}{2}k_{1\nu}(k_{1}\cdot h_{2}\cdot h_{1})_{\mu}-\frac{1}{2}h_{1\mu\nu}(k_{1}\cdot h_{2}\cdot k_{1})+\frac{1}{2}(k_{1}\cdot h_{2})_{\mu}(k_{2}\cdot h_{1})_{\nu}\\
+\frac{1}{2}k_{2\mu}(k_{2}\cdot h_{1}\cdot h_{2})_{\nu}+\frac{1}{2}k_{2\nu}(k_{2}\cdot h_{1}\cdot h_{2})_{\mu}-\frac{1}{2}h_{2\mu\nu}(k_{2}\cdot h_{1}\cdot k_{2})+\frac{1}{2}(k_{1}\cdot h_{2})_{\nu}(k_{2}\cdot h_{1})_{\mu}\\
-\frac{1}{2}(k_{1\mu}k_{1\nu}+k_{2\mu}k_{2\nu})(h_{1}\cdot h_{2})-\frac{1}{4}(k_{1\mu}k_{2\nu}+k_{2\mu}k_{1\nu})(h_{1}\cdot h_{2})\\
-\frac{1}{2}(k_{1}\cdot k_{2})(h_{1}\cdot h_{2})_{\mu\nu}-\frac{2}{z^{2}}(h_{1}\cdot h_{2})_{\mu\nu}-\frac{1}{2}(k_{1}\cdot k_{2})(h_{1}\cdot h_{2})_{\nu\mu}-\frac{2}{z^{2}}(h_{1}\cdot h_{2})_{\nu\mu}\\
+\frac{1}{2z}\eta_{\mu\nu}\partial_{z}(h_{1}\cdot h_{2})-\frac{2}{z^{2}}\eta_{\mu\nu}(h_{1}\cdot h_{2})+\frac{1}{2}(\partial_{z}h_{1}\cdot\partial_{z}h_{2})_{\mu\nu}+\frac{1}{2}(\partial_{z}h_{1}\cdot\partial_{z}h_{2})_{\nu\mu}.
\end{multline}
The vector contractions here are only along the boundary directions.
For example, $(h_{1}\cdot h_{2})_{\mu\nu}=\eta^{\rho\sigma}h_{1\mu\rho}h_{2\nu\sigma}$,
and $(k_{2}\cdot h_{1})_{\mu}=k_{2}^{\nu}h_{1\mu\nu}$. Now, after
substituting these results in \eqref{eq:graviton-correlator-AdS},
we obtain
\begin{align}
\tilde{\mathcal{M}}_{3} & =\frac{1}{4}h_{1}^{\mu\nu}h_{2}^{\rho\sigma}h_{3}^{\gamma\lambda}V_{\mu\rho\gamma}V_{\nu\sigma\lambda}\int_{0}^{\infty}\frac{dz}{z^{d-1}}\phi_{1}^{(d/2)}(z)\phi_{2}^{(d/2)}(z)\phi_{3}^{(d/2)}(z)\nonumber \\
 & -\frac{1}{3}(h_{1}\cdot h_{2}\cdot h_{3})\int_{0}^{\infty}dz\,\partial_{z}\left(\frac{1}{z^{d-1}}\partial_{z}\left(\phi_{1}^{(d/2)}\phi_{2}^{(d/2)}\phi_{3}^{(d/2)}\right)\right).\label{eq:graviton-3pt-AdS}
\end{align}
where we have used $k_{123}^{\mu}=0$, and the equations of motion
\begin{equation}
\mathcal{D}_{i}^{2}\phi_{i}^{(d/2)}=0.
\end{equation}
The object $V_{\mu\rho\gamma}$ is the three-point gluon vertex of
equation \eqref{eq:three-gluon-vertex}, so $\tilde{\mathcal{M}}_{3}$
naturally hints at some double-copy structure. The total derivative
in the second line of \eqref{eq:graviton-3pt-AdS} is called a boundary
contact term. It diverges as $z\to0$, as we had already anticipated,
but its renormalization is well understood. Since this is a bit out
of our scope here, we refer the reader to \cite{Armstrong:2022mfr}
for a short discussion on this and further references.

\

\noindent \hrulefill

\

The construction developed in this section for anti de Sitter can
be extended, with appropriate modifications, to other curved backgrounds.
A natural candidate is de Sitter space, which shares with AdS the
property of being a maximally symmetric solution of the Einstein field
equations with a cosmological constant, but with opposite sign. At
the level of the Poincar\'e patch metric, de Sitter can be reached
from AdS via a Wick rotation of the radial coordinate ($z\to\mathrm{i}\eta$).
This operation, for instance, maps the exponential decay of bulk-to-boundary
propagators into oscillatory behavior\footnote{A recent review \cite{Hinterbichler:2026xqf} presents in detail
the representations of the algebra of isometries of de Sitter space.
These representations are connected to the various types of fields
that can propagate there, and constitute a sound material to base
the construction of multi-particle solutions in de Sitter.}. This analytic continuation carries over to the multi-particle ansatz
and allows one to import results from the AdS construction directly.
The physical interpretation changes accordingly: the natural observables
in de Sitter are not boundary correlators, but rather the so-called
in-in correlators relevant for cosmological perturbation theory, which
encode the statistics of primordial fluctuations (see e.g. \cite{Maldacena:2002vr,Weinberg:2005vy}).
The perturbiner framework is well-suited to organize these computations
recursively, as long as we deal with the subtleties that do not arise
in AdS (such as the time-like boundary, which is the cosmological
horizon).

In the next section, we are going to show that the perturbiner framework
can go beyond tree level computations via a systematic construction
of one-loop integrands via algebraic recursions. 

\section{Loop-level extension\label{sec:loop}}

We have seen over and over that the perturbiner is a reliable framework
to build scattering trees. This is only possible because the multi-particle
coefficients can be interpreted as scattering amplitudes with one
off-shell leg, the one we finally attach to our amplitude/correlator
formulas in their final computation.

One might wonder whether we can have more than this one off-shell
leg. And this questioning has an immediate ramification. Let us consider
a one-loop diagram. In simple terms, it can be built from a scattering
tree with two off-shell legs, which are then connected with the respective
propagator and integrated over. 

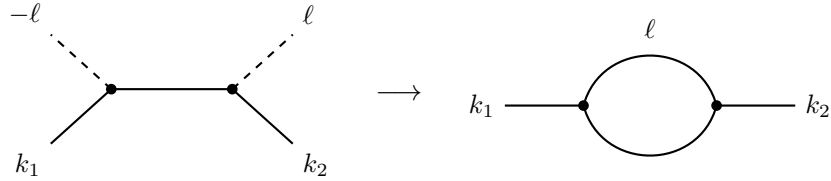
\begin{equation*}
  \raisebox{-0.45\height}{%
    \begin{tikzpicture}[scale=0.8]
 
 
      \fill (-1, 0) circle (2.5pt);
      \fill (1, 0) circle (2.5pt);
 
      \draw[thick] (-1,0) -- (1,0);
 
      \draw[thick, dashed] (-1,0) -- (-2, 0.9);
      \node[above left, font=\small] at (-2,0.9) {$-\ell$};
 
      \draw[thick] (-1,0) -- (-2,-0.9);
      \node[below left, font=\small] at (-2,-0.9) {$k_1$};
 
      \draw[thick, dashed] (1,0) -- (2, 0.9);
      \node[above right, font=\small] at (2,0.9) {$\ell$};
 
      \draw[thick] (1,0) -- (2,-0.9);
      \node[below right, font=\small] at (2,-0.9) {$k_2$};
 
    \end{tikzpicture}}
  \quad \longrightarrow \quad
  \raisebox{-0.45\height}{%
    \begin{tikzpicture}[scale=0.8]
 
 
      \fill (-1.1, 0) circle (2.5pt);
      \fill (1.1, 0) circle (2.5pt);
 
      \draw[thick] (-1.1,0) .. controls (-0.8, 1.1) and (0.8, 1.1) .. (1.1,0);
 
      \draw[thick] (-1.1,0) .. controls (-0.8,-1.1) and (0.8,-1.1) .. (1.1,0);
 
      \draw[thick] (-1.1,0) -- (-2.4,0);
      \node[left, font=\small] at (-2.4,0) {$k_1$};
 
      \draw[thick] (1.1,0) -- (2.4,0);
      \node[right, font=\small] at (2.4,0) {$k_2$};
 
      \node[above, font=\small] at (0, 0.95) {$\ell$};
 
    \end{tikzpicture}}
\end{equation*}This procedure is a fundamental physical interpretation of a loop
diagram, with a virtual (off-shell) particle in the loop. In this
section, we are going to implement this construction step by step,
establishing the recursions for one-loop integrands starting from
the tree-level framework of multi-particle solutions. Subsection \ref{subsec:scalar-oneloop}
involves a more detailed combinatorial analysis, but it builds the
core understanding of the one-loop implementation of the perturbiner.
We will then work with the metric density formulation of gravity and
its one-loop construction. More generally, we will develop the necessary
tools to build algebraic recursions for one-loop integrands in any
of the field theories we have seen so far, formulated in Minkowski
space.

\subsection{Scalars with polynomial self-interactions\label{subsec:scalar-oneloop}}

We will first discuss scalar models. They are rich enough to accommodate
arbitrary-point interactions but without being encumbered by gauge
redundancies.

\subsubsection{Back to the beginning: $\phi^{3}$ theory}

In section \ref{sec:multiparticle solutions} we have introduced classical
multi-particle solutions using the massless $\phi^{3}$ theory. Let
us reconsider the expansion \eqref{eq:scalar-multiparticle}, now
putting the single-particle states off the mass-shell. Therefore,
the ansatz
\begin{equation}
\phi(x)=\sum_{P}\Phi_{P}e^{\mathrm{i}k_{P}\cdot x},
\end{equation}
with
\begin{equation}
\Phi_{P}=\frac{1}{s_{P}}\frac{\lambda}{2}\sum_{P=Q\cup R}\Phi_{Q}\Phi_{R},\label{eq:scalar-multi-off}
\end{equation}
no longer satisfies the equation of motion, since we now have
\begin{equation}
\Box\phi+\frac{\lambda}{2}\phi^{2}=-\sum_{p}k_{p}^{2}\phi_{p}e^{\mathrm{i}k_{P}\cdot x}.\label{eq:scalar-loop-off-shell}
\end{equation}
Here the sum on the right hand side goes over single-particle states
only. In other words, the interacting (non-linear) part of the equation
of motion is satisfied. For this theory, in particular, the interaction
vertices do not distinguish on-shell and off-shell states. 

Now, let us single out one single-particle label ($\ell$) and define
$\Phi_{\ell P}=\Lambda_{P}(\ell)$, with $\Phi_{\ell\emptyset}=1$.
Equation \eqref{eq:scalar-multi-off} then becomes
\begin{align}
\Lambda_{P}(\ell) & =\frac{1}{s_{\ell P}}\frac{\lambda}{2}\sum_{\ell P=Q\cup R}\Phi_{Q}\Phi_{R},\nonumber \\
 & =\frac{\lambda}{s_{\ell P}}\Phi_{P}+\frac{\lambda}{s_{\ell P}}\sum_{P=Q\cup R}\Lambda_{Q}\Phi_{R},\label{eq:scalar-loop-motivation}
\end{align}
where we denote by $\ell^{m}$ the momentum of the particle $\ell$.
Ultimately, $\Lambda_{P}$ is a function of scattering trees $\Phi_{P}$,
recursively defined with the above equation. Does $\Lambda_{P}$ have
a physical interpretation in terms of scattering amplitudes? Almost.
In order to understand why, it is easier to analyze some lower multiplicity
cases.

Starting with $P=1$, we have\begin{equation}
  \Lambda_{1}(\ell) = \frac{\lambda}{(\ell+k_{1})^{2}}\Phi_{1}
  \qquad \longleftrightarrow \qquad
  \raisebox{-0.4\height}{%
    \begin{tikzpicture}[scale=0.7]
      \draw[thick] (-1.4,0) -- (0,0);
      \node[left, font=\small] at (-1.4,0) {$k_1$};
      \fill (0,0) circle (2pt);
      \draw[thick] (0.55,0) circle (0.55cm);
      \node[right, font=\small] at (1.15,0) {$\ell$};
    \end{tikzpicture}}
\end{equation}When $k_{1}=0$, this object looks like the \emph{integrand} of a
one-particle tadpole. However, the numerical factor is off by a factor
of $\frac{1}{2}$. This is because the tadpole has an automorphism
$\ell\to-\ell$, so when integrating over $\ell$ we are overcounting
the diagram contribution.

For $P=12$ we can do the same analysis, with\begin{equation}
  \Lambda_{12}(\ell) = \frac{\lambda}{(\ell+k_{12})^{2}}\Phi_{12}
  + \frac{\lambda}{(\ell+k_{12})^{2}}(\Lambda_{1}\Phi_{2}+\Lambda_{2}\Phi_{1})
\end{equation}
\[
  \longleftrightarrow \quad
  \raisebox{-0.4\height}{%
    \begin{tikzpicture}[scale=0.7]
      \draw[thick] (-1.4,0) -- (0,0);
      \node[left, font=\small] at (-1.4,0) {$k_{12}$};
      \fill (0,0) circle (2pt);
      \draw[thick] (0.55,0) circle (0.55cm);
      \node[right, font=\small] at (1.15,0) {$\ell$};
    \end{tikzpicture}}
  \quad + \quad
  \raisebox{-0.4\height}{%
    \begin{tikzpicture}[scale=0.7]
      \draw[thick] (-1.8,0) -- (-0.6,0);
      \node[left, font=\small] at (-1.8,0) {$k_1$};
      \fill (-0.55,0) circle (2pt);
      \draw[thick] (-0.55,0) .. controls (-0.3,0.7) and (0.3,0.7) .. (0.55,0);
      \draw[thick] (-0.55,0) .. controls (-0.3,-0.7) and (0.3,-0.7) .. (0.55,0);
      \fill (0.55,0) circle (2pt);
      \draw[thick] (0.55,0) -- (1.55,0);
      \node[right, font=\small] at (1.55,0) {$k_2$};
      \node[font=\small] at (0,0.8) {$\ell$};
    \end{tikzpicture}}
\]Once more we see the overcounting. When $k_{12}=0$, the first term
is identified with the integrand of a two-particle tadpole, again
off by a $\frac{1}{2}$ factor. The second term in the equation is
associated to the two-particle bubble diagram. The two terms inside
the parenthesis contribute equally, as they are related by $\ell\to-\ell$.
So for what would be a bubble diagram with symmetry factor $\frac{1}{2}$
we get instead a factor of $2$, so we are off by a factor of $\frac{1}{4}$
.

For $P=123$, with $k_{123}=0$, we have
\begin{equation}
\Lambda_{123}(\ell)=\frac{\lambda}{\ell^{2}}\Phi_{123}+\frac{\lambda}{\ell^{2}}[\Lambda_{12}\Phi_{3}+\Lambda_{3}\Phi_{12}+\textrm{perm}(1,2,3)].
\end{equation}
The three-particle tadpole integrand, associated to the first term,
is again off by a $\frac{1}{2}$ factor. We also have
\begin{equation}
\Lambda_{12}\Phi_{3}+\Lambda_{3}\Phi_{12}=\frac{\lambda}{(\ell+k_{12})^{2}}\Phi_{12}\Phi_{3}+\frac{\lambda}{(\ell+k_{3})^{2}}\Phi_{3}\Phi_{12}+\frac{\lambda}{(\ell+k_{12})^{2}}(\Lambda_{1}\Phi_{2}+\Lambda_{2}\Phi_{1})\Phi_{3}.\label{eq:loop-phi3-3particle}
\end{equation}
The first two terms on the right-hand side are related by $\ell\to-\ell$,
associated to a bubble diagram with an external three $\Phi_{12}$
and external leg $\Phi_{3}$. Following the same arguments as in the
$P=12$ case, we have a factor of 2 instead of $\frac{1}{2}$, so
we are off by a factor of $\frac{1}{4}$ . The last term in equation
\eqref{eq:loop-phi3-3particle} corresponds to two copies of the triangle
diagram with external legs $1$, $2$, and $3$. Together with the
remaining permutations of $(1,2,3)$ we have six times the expected
contribution.

From this discussion, the overcounting of one-loop diagram contributions
in $\Lambda_{P}$ is clear. The question then is whether it is possible
to redefine it in such a way that each diagram contribution is correctly
reproduced (with the expected symmetry factor) up to integration over
the loop momentum $\ell$. In order for it to fit a recursion, a modified
$\Lambda_{P}$ could at most be defined as
\begin{equation}
\tilde{\Lambda}_{P}=\frac{\lambda}{s_{\ell P}}\left(\frac{1}{2}\Phi_{P}+\sum_{P=Q\cup R}g_{Q,R}\tilde{\Lambda}_{Q}\Phi_{R}\right).
\end{equation}
The $\frac{1}{2}$ factor is unique, as it can be easily traced back
to tadpole diagrams. The coefficients $g_{Q,R}\leq1$, if they exist,
should balance the diagram overcounting. Note that $g_{Q,R}$ cannot
depend on the momenta nor, more generally, on the particle polarizations.
This would introduce alien dynamics in the one-loop amplitude. 

The integrand of the one-loop amplitude would be simply defined as
\begin{equation}
I_{|P|}=\lim_{k_{P}\to0}\tilde{\Lambda}_{P}(\ell).\label{eq:phi3-loopintegrand}
\end{equation}
We can then look at specific diagram contributions and try to find
a solution for $g_{Q,R}$. For instance, there are different contributions
to the bubble diagram with (fixed) external trees $\Phi_{Q}$ and
$\Phi_{R}$. In $I_{|P|}$, they come from
\begin{equation}
\frac{\lambda}{\ell^{2}}(g_{Q,R}\tilde{\Lambda}_{Q}\Phi_{R}+g_{R,Q}\tilde{\Lambda}_{R}\Phi_{Q})=\frac{1}{2}\frac{\lambda^{2}}{\ell^{2}}\left(\frac{1}{(\ell+k_{Q})^{2}}g_{Q,R}+\frac{1}{(\ell+k_{R})^{2}}g_{R,Q}\right)\Phi_{Q}\Phi_{R}+\ldots,
\end{equation}
where on the left hand side we singled out only the flagged contributions.
Since we have $k_{P}=0$, the respective bubble integrand is given
by
\begin{equation}
I_{\Circle}(Q,R)=\frac{1}{2}\frac{\lambda^{2}}{\ell^{2}(\ell+k_{Q})^{2}}(g_{Q,R}+g_{R,Q})\Phi_{Q}\Phi_{R}.
\end{equation}
We know that this integrand has a symmetry factor of $\frac{1}{2}$,
so we obtain the first condition on the coefficients,
\begin{equation}
g_{Q,R}+g_{R,Q}=1.\label{eq:phi3-loopcondition1}
\end{equation}

Now, if we focus on the integrand of the triangle diagram with external
trees $\Phi_{Q}$, $\Phi_{R}$, and $\Phi_{S}$, its contributions
come from permutations of
\begin{align}
g_{Q\cup R,S}\tilde{\Lambda}_{Q\cup R}\Phi_{S} & =\frac{\lambda^{2}}{\ell^{2}(\ell+k_{QR})^{2}}g_{Q\cup R,S}(g_{Q,R}\tilde{\Lambda}_{Q}\Phi_{R}+g_{R,Q}\tilde{\Lambda}_{R}\Phi_{Q})\Phi_{S}+\ldots,\\
 & =\frac{1}{2}\frac{\lambda^{3}}{\ell^{2}(\ell+k_{QR})^{2}}g_{Q\cup R,S}\left(\frac{1}{(\ell+k_{Q})^{2}}g_{Q,R}+\frac{1}{(\ell+k_{R})^{2}}g_{R,Q}\right)\Phi_{Q}\Phi_{R}\Phi_{S}+\ldots.
\end{align}
By adding all permutations of $Q$, $R$, and $S$, and using momentum
conservation ($k_{Q}+k_{R}+k_{S}=0$), the one-loop integrand is given
by
\begin{multline}
I_{\vartriangle}(Q,R,S)=\frac{\lambda^{3}}{2}\Phi_{Q}\Phi_{R}\Phi_{S}\bigg\{\frac{g_{Q\cup R,S}}{\ell^{2}(\ell+k_{QR})^{2}}\bigg(\frac{g_{Q,R}}{(\ell+k_{Q})^{2}}+\frac{g_{R,Q}}{(\ell+k_{R})^{2}}\bigg)\\
+\frac{g_{Q\cup S,R}}{\ell^{2}(\ell+k_{QS})^{2}}\bigg(\frac{g_{Q,S}}{(\ell+k_{Q})^{2}}+\frac{g_{S,Q}}{(\ell+k_{S})^{2}}\bigg)+\frac{g_{R\cup S,Q}}{\ell^{2}(\ell+k_{RS})^{2}}\bigg(\frac{g_{R,S}}{(\ell+k_{R})^{2}}+\frac{g_{S,R}}{(\ell+k_{S})^{2}}\bigg)\bigg\}.
\end{multline}
All terms inside the curly brackets can be related to one another
via a change of the integration variable (momentum) $\ell$. For example,
\begin{align}
\left.\frac{1}{\ell^{2}(\ell+k_{RS})^{2}(\ell+k_{S})^{2}}\right|_{\ell\to-\ell} & =\frac{1}{\ell^{2}(\ell+k_{QR})^{2}(\ell+k_{Q})^{2}},\\
\left.\frac{1}{\ell^{2}(\ell+k_{QR})^{2}(\ell+k_{R})^{2}}\right|_{\ell\to-\ell-k_{QR}} & =\frac{1}{\ell^{2}(\ell+k_{QR})^{2}(\ell+k_{Q})^{2}}.
\end{align}
Given that the correct integrand for this triangle diagram is
\begin{equation}
I_{\vartriangle}(Q,R,S)=\frac{\lambda^{3}}{\ell^{2}(\ell+k_{Q})^{2}(\ell+k_{QR})^{2}}\Phi_{Q}\Phi_{R}\Phi_{S},
\end{equation}
we obtain a second condition on the coefficients $g_{Q,R}$, given
by
\begin{equation}
g_{Q\cup R,S}(g_{Q,R}+g_{R,Q})+g_{Q\cup S,R}(g_{Q,S}+g_{S,Q})+g_{R\cup S,Q}(g_{R,S}+g_{S,R})=2.
\end{equation}
Using equation \eqref{eq:phi3-loopcondition1}, we simply have
\begin{equation}
g_{Q\cup R,S}+g_{R\cup S,Q}+g_{Q\cup S,R}=2,
\end{equation}
which also implies that
\begin{equation}
g_{Q,R\cup S}+g_{R,Q\cup S}+g_{S,Q\cup R}=1.
\end{equation}

It is interesting to observe that the above relations already suggest
a solution. The coefficients $g_{Q,R}$ are combinatoric in nature
and there is natural ingredient that fits this expectation: the length
of the subwords $|Q|$ and $|R|$. Having this is mind, there is a
simple proposal that solves the above conditions, given by
\begin{equation}
g_{Q,R}=\frac{|Q|}{|Q\cup R|}.\label{eq:phi3-loopcombinatoric}
\end{equation}
But what about higher multiplicity trees? Let us consider the so-called
$n$-gon diagram, which consists of $n$ trees $\Phi_{P_{i}}$, with
$i=1,\ldots,n$, each one attached to a different vertex of the $n$-gon
(recall that we have only cubic interactions so far). The associated
one-loop integrand is simply
\begin{equation}
I_{n}(P_{1},\ldots,P_{n})=\frac{\lambda^{n}}{(\ell+k_{P_{1}})^{2}(\ell+k_{P_{1}P_{2}})^{2}\cdots(\ell+k_{P_{1}\ldots P_{n}})^{2}}\Phi_{P_{1}}\ldots\Phi_{P_{n}}.\label{eq:phi3-ngon}
\end{equation}
Such a contribution can be easily identified in equation \eqref{eq:phi3-loopintegrand},
and it comes from
\begin{multline}
I_{n}(P_{1},\ldots,P_{n})=\frac{\lambda^{n}}{2(\ell+k_{P_{1}})^{2}(\ell+k_{P_{1}P_{2}})^{2}\cdots(\ell+k_{P_{1}\ldots P_{n}})^{2}}\\
\times\Phi_{P_{1}}\ldots\Phi_{P_{n}}(g_{P_{1}\cup\ldots\cup P_{n-1},P_{n}}\times g_{P_{1}\cup\ldots\cup P_{n-2},P_{n-1}}\times\ldots\times g_{P_{1},P_{2}})\\
+\frac{\lambda^{n}}{2(\ell+k_{P_{n}})^{2}(\ell+k_{P_{n-1}P_{n}})^{2}\cdots(\ell+k_{P_{1}\ldots P_{n}})^{2}}\\
\times\Phi_{P_{1}}\ldots\Phi_{P_{n}}(g_{P_{n}\cup\ldots\cup P_{2},P_{1}}\times g_{P_{n}\cup\ldots\cup P_{3},P_{2}}\times\ldots\times g_{P_{n},P_{n-1}})\\
+\textrm{cyclic}(P_{1},\ldots,P_{n}).\label{eq:phi3-ngon-new}
\end{multline}
Unlike the bubble and triangle diagrams, the order of the external
trees leads to inequivalent one-loop contributions in $n$-gons with
$n\geq4$. Therefore, the equivalent contributions here are the cyclic
permutations of the trees and their inverse order. Now notice that
\begin{multline}
g_{P_{1}\cup\ldots\cup P_{n-1},P_{n}}\times g_{P_{1}\cup\ldots\cup P_{n-2},P_{n-1}}\times\ldots\times g_{P_{1},P_{2}}\\
=\left(\frac{|P_{1}\cup\ldots\cup P_{n-1}|}{|P_{1}\cup\ldots\cup P_{n}|}\right)\times\left(\frac{|P_{1}\cup\ldots\cup P_{n-2}|}{|P_{1}\cup\ldots\cup P_{n-1}|}\right)\times\ldots\times\left(\frac{|P_{1}|}{|P_{1}\cup P_{2}|}\right)\\
=\frac{|P_{1}|}{|P_{1}\cup\ldots\cup P_{n}|}.
\end{multline}
With the addition of cyclic permutations, we simply have
\begin{equation}
\frac{|P_{1}|}{|P_{1}\cup\ldots\cup P_{n}|}+\textrm{cyclic}(P_{1},\ldots,P_{n})=1.
\end{equation}
Therefore, equations \eqref{eq:phi3-ngon} and \eqref{eq:phi3-ngon-new}
exactly match. This demonstrates that \eqref{eq:phi3-loopintegrand}
generates the complete one-loop integrand of the $\phi^{3}$ theory.
Let us know see how this construction extends to higher-point vertices.

\subsubsection{General case}

We can consider a more general scalar theory with polynomial self-interactions,
and a recursive definition of the currents given by
\begin{equation}
\Phi_{P}=\frac{1}{s_{P}}\sum_{n=2}\sum_{P=P_{1}\cup\ldots\cup P_{n}}\frac{\lambda^{(n+1)}}{n!}\Phi_{P_{1}}\ldots\Phi_{P_{n}},
\end{equation}
where we introduce $\lambda^{(n)}$ to denote the coupling constant
for the $n$-point interaction vertex.

Again, by defining $\Phi_{\ell P}=\Lambda_{P}(\ell)$, with $\Phi_{\ell\emptyset}=1$,
we obtain
\begin{align}
\Phi_{\ell P} & =\frac{1}{s_{\ell P}}\sum_{n=2}\sum_{\ell P=P_{1}\cup\ldots\cup P_{n}}\frac{\lambda^{(n+1)}}{n!}\Phi_{P_{1}}\ldots\Phi_{P_{n}},\\
\Lambda_{P}(\ell) & =\frac{1}{s_{\ell P}}\sum_{n=1}\sum_{P=P_{1}\cup\ldots\cup P_{n}}\frac{\lambda^{(n+2)}}{n!}\Phi_{P_{1}}\ldots\Phi_{P_{n}}\nonumber \\
 & +\frac{1}{s_{\ell P}}\sum_{n=2}\sum_{P=P_{1}\cup\ldots\cup P_{n}}\frac{\lambda^{(n+1)}}{(n-1)!}\Lambda_{P_{1}}\ldots\Phi_{P_{n}}.\label{eq:scalar-loop-overcounted}
\end{align}
We know that trying to interpret $\Lambda_{P}$ as a one-loop integrand
will lead to the overcounting of some diagrams, so we can define instead
\begin{align}
\tilde{\Lambda}_{P}(\ell) & =\frac{1}{2s_{\ell P}}\sum_{n=1}\sum_{P=P_{1}\cup\ldots\cup P_{n}}\frac{\lambda^{(n+2)}}{n!}\Phi_{P_{1}}\ldots\Phi_{P_{n}}\nonumber \\
 & +\frac{1}{s_{\ell P}}\sum_{n=2}\sum_{P=P_{1}\cup\ldots\cup P_{n}}\frac{\lambda^{(n+1)}}{(n-1)!}g_{P_{1},P_{2},\ldots,P_{n}}^{(n+1)}\tilde{\Lambda}_{P_{1}}\ldots\Phi_{P_{n}}.\label{eq:phin-loopintegrand}
\end{align}
The factor of $\frac{1}{2}$ in the first line comes directly from
the tadpole diagrams. The task now is to determine $g_{P_{1},P_{2},\ldots,P_{n}}^{(n)}<1$,
as we have a similar setup to the $\phi^{3}$ theory. Notice that,
by construction, $g_{P_{1},P_{2},\ldots,P_{n}}^{(n)}$ is symmetric
in the exchange of any two words $P_{i}$ with $i\geq2$, with $P_{1}$
being the asymmetric argument.

Let us start with the one-loop integrands with external trees $\Phi_{Q}$
and $\Phi_{R}$. The respective contribution coming from \eqref{eq:phin-loopintegrand}
is given by
\begin{equation}
I(Q,R)=\frac{1}{2\ell^{2}}\left(\lambda^{(4)}+\frac{\lambda^{(3)}}{(\ell+k_{Q})^{2}}(g_{Q,R}^{(3)}+g_{R,Q}^{(3)})\right)\Phi_{Q}\Phi_{R}.
\end{equation}
The first term inside the parenthesis is the tadpole diagram with
a quartic vertex, already with the correct symmetry factor. The second
term is the bubble diagram, which we have already seen. Therefore,
we require that
\begin{equation}
g_{Q,R}^{(3)}+g_{R,Q}^{(3)}=1.
\end{equation}

The one-loop integrand with three external trees ($\Phi_{Q}$, $\Phi_{R}$,
and $\Phi_{S}$), already after taking into account changes in the
loop momentum variable, is given by
\begin{multline}
I(Q,R,S)=\frac{1}{2\ell^{2}}\lambda^{(5)}\Phi_{Q}\Phi_{R}\Phi_{S}+\frac{\lambda^{(3)}\lambda^{(4)}}{2\ell^{2}(\ell+k_{Q})^{2}}(g_{R\cup S,Q}^{(3)}+g_{Q,R,S}^{(4)})\Phi_{Q}\Phi_{R}\Phi_{S}\\
+\frac{\lambda^{(3)}\lambda^{(4)}}{2\ell^{2}(\ell+k_{R})^{2}}(g_{Q\cup S,R}^{(3)}+g_{R,Q,S}^{(4)})\Phi_{Q}\Phi_{R}\Phi_{S}+\frac{\lambda^{(3)}\lambda^{(4)}}{2\ell^{2}(\ell+k_{S})^{2}}(g_{Q\cup R,S}^{(3)}+g_{S,Q,R}^{(4)})\Phi_{Q}\Phi_{R}\Phi_{S}\\
+\frac{\lambda^{(3)}}{2\ell^{2}(\ell+k_{Q})^{2}(\ell+k_{QR})^{2}}\\
\times[g_{Q\cup R,S}^{(3)}(g_{Q,R}^{(3)}+g_{R,Q}^{(3)})+g_{Q\cup S,R}^{(3)}(g_{Q,S}^{(3)}+g_{S,Q}^{(3)})+g_{R\cup S,Q}^{(3)}(g_{R,S}^{(3)}+g_{S,R}^{(3)})]\Phi_{Q}\Phi_{R}\Phi_{S}.
\end{multline}
We promptly recognize the associated diagrams. There is a tadpole with
a quintic vertex (coupling constant $\lambda^{(5)}$), three bubble
diagrams with a cubic and a quartic vertex, and one triangle diagram,
with three cubic vertices. In this case, we must impose
\begin{equation}
g_{Q\cup R,S}^{(3)}+g_{S,Q,R}^{(4)}=1,
\end{equation}
and
\begin{equation}
g_{Q\cup R,S}^{(3)}(g_{Q,R}^{(3)}+g_{R,Q}^{(3)})+g_{Q\cup S,R}^{(3)}(g_{Q,S}^{(3)}+g_{S,Q}^{(3)})+g_{R\cup S,Q}^{(3)}(g_{R,S}^{(3)}+g_{S,R}^{(3)})=2.
\end{equation}

Notice that all the conditions that we have so far derived are satisfied
when
\begin{align}
g_{P_{1},P_{2}}^{(3)} & =\frac{|P_{1}|}{|P_{1}\cup P_{2}|},\\
g_{P_{1},P_{2},P_{3}}^{(4)} & =\frac{|P_{1}|}{|P_{1}\cup P_{2}\cup P_{3}|}.
\end{align}
Therefore, we will simply propose that
\begin{equation}
g_{P_{1},P_{2},\ldots,P_{n}}^{(n+1)}=\frac{|P_{1}|}{|P_{1}\cup\ldots\cup P_{n}|},\label{eq:phin-loopcoefficient}
\end{equation}
and demonstrate this result for an arbitrary configuration. Let us
consider a one-loop diagram formed with $N$ vertices, such that each
vertex with coupling constant $\lambda^{(n_{i}+1)}$ has $(n_{i}-1)$
external trees. We will denote by $P_{\{n_{i}\}}$ the union of the
$(n_{i}-1)$ words $P_{j}$ connected to the $i$-th vertex, and by
$\Phi_{\{n_{i}\}}$ the product of the respective trees. In this case,
the contributions to this diagram derived from \eqref{eq:phin-loopintegrand}
are given by
\begin{multline}
I(\{n_{1}\},\ldots,\{n_{N}\})\\
=\frac{g_{P_{\{n_{2}\}}\cup\ldots\cup P_{\{n_{N}\}},P_{\{n_{1}\}}}^{(n_{1}+1)}g_{P_{\{n_{3}\}}\cup\ldots\cup P_{\{n_{N}\}},P_{\{n_{2}\}}}^{(n_{2}+1)}\ldots g_{P_{\{n_{N-1}\}},P_{\{n_{N}\}}}^{(n_{N}+1)}}{2\ell^{2}(\ell+k_{\{n_{2},\ldots,n_{N}\}})^{2}\ldots(\ell+k_{\{n_{N}\}})^{2}}\prod_{i=1}^{N}\lambda^{(n_{i}+1)}\Phi_{\{n_{i}\}}\\
+\frac{g_{P_{\{n_{1}\}}\cup\ldots\cup P_{\{n_{N-1}\}},P_{\{n_{N}\}}}^{(n_{N}+1)}g_{P_{\{n_{1}\}}\cup\ldots\cup P_{\{n_{N-2}\}},P_{\{n_{N-1}\}}}^{(n_{N-1}+1)}\ldots g_{P_{\{n_{1}\}},P_{\{n_{2}\}}}^{(n_{1}+1)}}{2\ell^{2}(\ell+k_{\{n_{1},\ldots,n_{N-1}\}})^{2}\ldots(\ell+k_{\{n_{1}\}})^{2}}\prod_{i=1}^{N}\lambda^{(n_{i}+1)}\Phi_{\{n_{i}\}}\\
+\textrm{cyclic}(\{n_{1}\},\ldots,\{n_{N}\}).
\end{multline}
Similarly to equation \eqref{eq:phi3-ngon-new}, we have to add the
cyclic combination of tree sets $\{n_{i}\}$ and also their inverse
order, which correspond to identical integrand contributions. Using
\eqref{eq:phin-loopcoefficient}, it is then straightforward to show
that
\begin{equation}
g_{P_{\{n_{2}\}}\cup\ldots\cup P_{\{n_{N}\}},P_{\{n_{1}\}}}^{(n_{1}+1)}g_{P_{\{n_{3}\}}\cup\ldots\cup P_{\{n_{N}\}},P_{\{n_{2}\}}}^{(n_{2}+1)}\ldots g_{P_{\{n_{N-1}\}},P_{\{n_{N}\}}}^{(n_{N}+1)}+\textrm{cyclic}(\{n_{1}\},\ldots,\{n_{N}\})=1,
\end{equation}
which leads to the correct one-loop integrand of the configuration
analyzed.

Therefore, the one-loop integrand of a scalar with polynomial self-interactions
can be cast as
\begin{multline}
I_{1\textrm{-loop}}(P)=\frac{1}{2\ell^{2}}\sum_{n=1}\sum_{P=P_{1}\cup\ldots\cup P_{n}}\frac{\lambda^{(n+2)}}{n!}\Phi_{P_{1}}\ldots\Phi_{P_{n}}\\
+\frac{1}{\ell^{2}}\sum_{n=2}\sum_{P=P_{1}\cup\ldots\cup P_{n}}\frac{\lambda^{(n+1)}}{(n-1)!}\frac{|P_{1}|}{|P|}\tilde{\Lambda}_{P_{1}}\ldots\Phi_{P_{n}},\label{eq:scalar-polynomial-oneloop}
\end{multline}
where momentum conservation $k_{P}^{m}=0$ is implicit. The one-loop
amplitude, which still has to be regularized, is given by
\begin{equation}
A_{1\textrm{-loop}}(P)\propto\int d^{D}\ell I_{1\textrm{-loop}}(P),
\end{equation}
where the normalization factor can be fixed by some lower-point computation.
We have demonstrated here that the integrand \eqref{eq:scalar-polynomial-oneloop}
correctly generates the diagrammatic expansion of the one-loop amplitude
$A_{1\textrm{-loop}}(P)$.

\noindent \hrulefill

\subsubsection*{Carefully going off the mass-shell}

There is one point we should discuss before tackling the more interesting
case of gravity. It concerns the connection between the off-shell
scattering trees and the equations of motion, in particular regarding
our premise of going off the mass-shell with the single-particle states,
like in equation \eqref{eq:scalar-loop-off-shell}.

Let us try to understand this point in a more elaborate context, and
consider the group-valued equation of the NLSM \eqref{eq:NLSM-group-eom},
\begin{equation}
U^{-1}\partial^{m}\partial_{m}U=-\partial^{m}U^{-1}\partial_{m}U.\label{eq:NLSM-off-1}
\end{equation}
If we multiply both sides of the equation by $U$ on the left, we obtain
\begin{equation}
\partial^{m}\partial_{m}U=-U\partial^{m}U^{-1}\partial_{m}U,\label{eq:NLSM-off-2}
\end{equation}
and the two versions are of course equivalent, as we can easily invert
the operation. Now we consider off-shell single-particle states and
the multi-particle expansions of equations \eqref{eq:NLSM-group-Uexpansion}
and \eqref{eq:NLSM-group-Ubarexpansion}. For simplicity, we take
only two external states:
\begin{equation}
U(x)=\mathds{1}+U_{1}e^{\mathrm{i}k_{1}\cdot x}T^{a_{1}}+U_{2}e^{\mathrm{i}k_{2}\cdot x}T^{a_{2}}+U_{12}e^{\mathrm{i}k_{12}\cdot x}T^{a_{1}}T^{a_{2}}+U_{21}e^{\mathrm{i}k_{12}\cdot x}T^{a_{2}}T^{a_{1}},
\end{equation}
\begin{multline}
U^{-1}(x)=\mathds{1}-U_{1}e^{\mathrm{i}k_{1}\cdot x}T^{a_{1}}-U_{2}e^{\mathrm{i}k_{2}\cdot x}T^{a_{2}}\\
-(U_{12}-\mathrm{i}U_{1}U_{2})e^{\mathrm{i}k_{12}\cdot x}T^{a_{1}}T^{a_{2}}-(U_{21}-\mathrm{i}U_{2}U_{1})e^{\mathrm{i}k_{12}\cdot x}T^{a_{2}}T^{a_{1}},
\end{multline}
with $k_{i}^{2}\neq0$. Equation \eqref{eq:NLSM-off-1} yields the
following solution for $U_{12}$,
\begin{equation}
U_{12}=\frac{1}{s_{12}}[k_{2}^{2}-(k_{1}\cdot k_{2})]U_{1}U_{2},
\end{equation}
while equation \eqref{eq:NLSM-off-2} leads to
\begin{equation}
U_{12}=-\frac{1}{s_{12}}(k_{1}\cdot k_{2})U_{1}U_{2}.
\end{equation}
As long as the single-particle states are on-shell, $U_{12}$ computed
using \eqref{eq:NLSM-off-1} or \eqref{eq:NLSM-off-2} give the same
result. With the single-particle states off the mass-shell, the scattering
trees cease to be equivalent. Recall that we are now only solving
the interaction part of the equation of motion, just like in \eqref{eq:scalar-loop-off-shell}.
If we want to match the diagrammatic approach, our equations of motion
have to be taken directly from the variation of the action, which
can be generically expressed for a set of fields $\Phi_{I}$ as
\begin{equation}
\delta S\propto\int d^{D}x\,\delta\Phi_{I}\times F^{I}(\Phi)+\partial_{m}F^{m}(\Phi),
\end{equation}
where $F^{I}=0$ denotes the equations of motion and $F^{m}$ denotes
possible boundary terms. We should not resort to any operation that
introduces/removes interactions involving the off-shell single particle
states. As it turns out, this subtlety is clearly manifested in gravity,
and this is largely related to the measure factor $\sqrt{-g}$.

\subsection{Pure gravity\label{subsec:Pure-gravity}}

If we try to naively extend the one-loop construction of scalars to
gravity, we will immediately run into ambiguities. The first and more
obvious is the fact that we are dealing with a gauge theory. Therefore,
in order to be able to quantize the model one has to address its gauge
redundancies, something that is usually done with the introduction
of ghost fields. We will come back to this soon. The second obstacle
boils down to the invariant spacetime measure, given by $d^{D}x$$\sqrt{-g}$.
Since we would like to consider at least one additional off-shell
state in order to close the loop, it is clear that the following equations
will lead to different recursions for the multi-particle currents:
\begin{equation}
\begin{array}{c}
R_{mn}=0,\\
\sqrt{-g}(R_{mn}-\frac{1}{2}g_{mn}R)=0.
\end{array}\label{eq:EFE-inequivalent-off-shell}
\end{equation}
In the second equation, off-shell single-particle states contain additional
interaction terms that are simply not present in the first one. And
it is possible to show that the resulting one-loop integrands are
physically inequivalent, just like what we saw for the NLSM in the
previous subsection.

In order for our one-loop construction to match the diagrammatic approach,
the second form of the equation of motion should be used (plus ghost
interactions). In the general case of gravity coupled to matter, we
must take into account the contributions that $\sqrt{-g}$ brings
along. Its multi-particle expansion can be introduced as follows.
Consider the function
\begin{equation}
f(\epsilon)=\sqrt{-\det(\mathds{1}+\epsilon H)},
\end{equation}
where $\epsilon$ is a parameter and $H$ is some matrix. The first
derivative with respect to $\epsilon$ can be expressed as
\begin{equation}
\frac{\partial f}{\partial\epsilon}=\tfrac{1}{2}f(\epsilon)\textrm{tr}\mathtt{h},
\end{equation}
and here we define the matrix product $\mathtt{h}\equiv(\mathds{1}+\epsilon H)^{-1}\cdot H$.
Note that it satisfies
\begin{align}
\frac{\partial\mathtt{h}}{\partial\epsilon} & =\left[\frac{\partial}{\partial\epsilon}(\mathds{1}+\epsilon H)^{-1}\right]\cdot H,\nonumber \\
 & =-(\mathds{1}+\epsilon H)^{-1}\cdot H\cdot(\mathds{1}+\epsilon H)^{-1}\cdot H,\nonumber \\
 & =-\mathtt{h}\cdot\mathtt{h},
\end{align}
Therefore, higher derivatives of $f$ with respect to $\epsilon$
lead to
\begin{align}
\frac{\partial^{2}f}{\partial\epsilon^{2}} & =\tfrac{1}{4}f(\epsilon)(\textrm{tr}\mathtt{h})^{2}-\tfrac{1}{2}f(\epsilon)\textrm{tr}(\mathtt{h}\cdot\mathtt{h}),\\
\frac{\partial^{3}f}{\partial\epsilon^{3}} & =\tfrac{1}{8}f(\epsilon)(\textrm{tr}\mathtt{h})^{3}-\tfrac{3}{4}f(\epsilon)(\textrm{tr}\mathtt{h})\textrm{tr}(\mathtt{h}\cdot\mathtt{h})+f(\epsilon)\textrm{tr}(\mathtt{h}\cdot\mathtt{h}\cdot\mathtt{h}),\\
\frac{\partial^{4}f}{\partial\epsilon^{4}} & =\tfrac{1}{16}f(\epsilon)(\textrm{tr}\mathtt{h})^{4}-\tfrac{3}{4}f(\epsilon)(\textrm{tr}\mathtt{h})^{2}\textrm{tr}(\mathtt{h}\cdot\mathtt{h})+\tfrac{3}{4}f(\epsilon)\textrm{tr}(\mathtt{h}\cdot\mathtt{h})\textrm{tr}(\mathtt{h}\cdot\mathtt{h})\nonumber \\
 & +2f(\epsilon)(\textrm{tr}\mathtt{h})\textrm{tr}(\mathtt{h}\cdot\mathtt{h}\cdot\mathtt{h})-3f(\epsilon)\textrm{tr}(\mathtt{h}\cdot\mathtt{h}\cdot\mathtt{h}\cdot\mathtt{h}),
\end{align}
and so on. Now we can recall the perturbiner expansion for the graviton
in equation \eqref{eq:multi-graviton}. In this case, the matrix $\mathtt{h}$
can be expressed as
\begin{equation}
\mathtt{h}_{mn}(x)\Bigg\vert_{\epsilon=0}=\sum_{P}H_{Pmn}e^{\mathrm{i}k_{P}\cdot x}.
\end{equation}
Therefore, the derivatives of $f$ can be written as a multi-particle
expansion of the form
\begin{equation}
\frac{\partial^{i}f_{P}}{\partial\epsilon^{i}}\Bigg\vert_{\epsilon=0}=\sum_{P=Q_{1}\cup Q_{2}\cup\cdots\cup Q_{i}}G^{m_{1}n_{1}m_{2}n_{2}\dots m_{i}n_{i}}H_{Q_{1}m_{1}n_{1}}H_{Q_{2}m_{2}n_{2}}\cdots H_{Q_{i}m_{i}n_{i}},
\end{equation}
where the tensor $G$ can be read from the previous equations. The
first three orders are given by\begin{subequations}
\begin{align}
G^{mn} & =\tfrac{1}{2}\eta^{mn},\\
G^{mnpq} & =\tfrac{1}{4}(\eta^{mn}\eta^{pq}-2\eta^{mp}\eta^{nq}),\\
G^{mnpqrs} & =(\tfrac{1}{8}\eta^{mn}\eta^{pq}\eta^{rs}-\tfrac{3}{4}\eta^{mn}\eta^{pr}\eta^{qs}+\eta^{mr}\eta^{np}\eta^{qs}).
\end{align}
\end{subequations}More generally, we can think of $f(\epsilon)$
as the generator of the multi-particle expansion of $\sqrt{-g}$,
with
\begin{equation}
\sqrt{-\mathrm{det}g}=1+\sum_{P}E_{P}e^{\mathrm{i}k_{P}\cdot x},
\end{equation}
and
\begin{equation}
E_{P}=\sum_{i=1}^{|P|}\tfrac{1}{i!}\sum_{P=Q_{1}\cup Q_{2}\cup\cdots\cup Q_{i}}G^{m_{1}n_{1}m_{2}n_{2}\dots m_{i}n_{i}}H_{Q_{1}m_{1}n_{1}}H_{Q_{2}m_{2}n_{2}}\cdots H_{Q_{i}m_{i}n_{i}}.
\end{equation}

The above construction serves to illustrate the fact that we can work
with $\sqrt{-g}$ within multi-particle expansions, though the formulas
involved are not as simple to obtain. For pure gravity, however, we
can perform a field redefinition on the metric in order to absorb
the measure factor. This is known as the metric density formulation
and was introduced by Landau and Lifshitz in \cite{Landau:1975pou}.

\subsubsection{Metric density formulation}

Let us define the metric density
\begin{equation}
\mathfrak{g}^{mn}=\sqrt{-g}g^{mn},
\end{equation}
with inverse $\mathfrak{g}_{mn}$. The Einstein--Hilbert action (with
$\Lambda=0$) can then be expressed as
\begin{multline}
S_{\textrm{E.H.}}=\frac{1}{8\kappa}\int d^{D}x\,\bigg\{\mathfrak{g}^{mn}(\partial_{m}\mathfrak{g}^{pq}\partial_{n}\mathfrak{g}_{pq}-2\partial_{m}\mathfrak{g}^{pq}\partial_{p}\mathfrak{g}_{nq})\\
-\frac{1}{(2-D)}\mathfrak{g}^{mn}\partial_{m}\mathfrak{g}^{pq}\mathfrak{g}_{pq}\partial_{n}\mathfrak{g}^{rs}\mathfrak{g}_{rs}-4\partial_{m}\mathfrak{B}^{m}\bigg\},\label{eq:EH-action-density}
\end{multline}
where $\mathfrak{B}^{m}$ is given by
\begin{equation}
\mathfrak{B}^{m}=\partial_{n}\mathfrak{g}^{mn}+\left(\frac{1}{2-D}\right)\mathfrak{g}^{mn}\mathfrak{g}^{pq}\partial_{n}\mathfrak{g}_{pq}.
\end{equation}
Since we will consider perturbations around flat space, we can disregard
the boundary term in \eqref{eq:EH-action-density}.

We know $S_{\textrm{E.H.}}$ is invariant under general coordinate
transformations. Under their linearized version \ref{eq:GCT-linearized},
the metric density transforms as
\begin{align}
\delta\mathfrak{g}_{mn} & =\mathfrak{g}_{mp}\partial_{n}\lambda^{p}+\mathfrak{g}_{np}\partial_{m}\lambda^{p}+\lambda^{p}\partial_{p}\mathfrak{g}_{mn}-\mathfrak{g}_{mn}\partial_{p}\lambda^{p},\\
\delta\mathfrak{g}^{mn} & =\partial_{p}(\lambda^{p}\mathfrak{g}^{mn})-\mathfrak{g}^{np}\partial_{p}\lambda^{m}-\mathfrak{g}^{mp}\partial_{p}\lambda^{n}.
\end{align}

In order to proceed with the loop construction, we need to properly
gauge fix the action. This is the only way we can obtain the correct
off-shell currents for loop integrands. One of the cleanest approaches
to do so is through the introduction of ghost fields\footnote{A good introductory reference to the Batalin--Vilkovisky (BV) formulation
can be found in \cite{Fuster:2005eg}.}. In this case, we promote the gauge parameters $\lambda^{m}$ to
a ghost field $c^{m}$, with fermionic statistics. We will work with
the harmonic gauge \eqref{eq:harmonic-gauge}, which in terms of the
metric density is nicely expressed as
\begin{equation}
\partial_{m}\mathfrak{g}^{mp}=0.\label{eq:harmonic-density}
\end{equation}
After introducing the gauge fixing term, the gauge fixed action is
given by
\begin{multline}
S=\frac{1}{8\kappa}\int d^{D}x\,\bigg\{\mathfrak{g}^{mn}\partial_{m}\mathfrak{g}^{pq}\partial_{n}\mathfrak{g}_{pq}-2\mathfrak{g}^{mn}\partial_{m}\mathfrak{g}^{pq}\partial_{p}\mathfrak{g}_{nq}-2\eta_{mn}\partial_{p}\mathfrak{g}^{mp}\partial_{q}\mathfrak{g}^{nq}\\
-\frac{1}{(2-D)}\mathfrak{g}^{mn}\partial_{m}\mathfrak{g}^{pq}\mathfrak{g}_{pq}\partial_{n}\mathfrak{g}^{rs}\mathfrak{g}_{rs}+2[\mathfrak{g}^{np}\partial_{p}c^{m}+\mathfrak{g}^{mp}\partial_{p}c^{n}-\partial_{p}(c^{p}\mathfrak{g}^{mn})]\partial_{m}b_{n}\bigg\},\label{eq:gauge-fixed-action-metric-density}
\end{multline}
where $b_{m}$ is the antighost field $c^{m}$, and $\eta_{mn}$ is
again the flat metric. This action is invariant under the nilpotent
BRST transformations\begin{subequations}
\begin{align}
\delta_{\textrm{BRST}}\mathfrak{g}^{mn} & =\xi[\partial_{p}(c^{p}\mathfrak{g}^{mn})-\mathfrak{g}^{np}\partial_{p}c^{m}-\mathfrak{g}^{mp}\partial_{p}c^{n}],\\
\delta_{\textrm{BRST}}c^{m} & =\xi c^{n}\partial_{n}c^{m},\\
\delta_{\textrm{BRST}}b_{m} & =-2\xi\eta_{mn}\partial_{p}g^{np},
\end{align}
\end{subequations}where $\xi$ is an anticommuting global parameter.
From this point on, the gauge fixing condition is imposed only on
the single-particle solutions (recall that the physical states have
to be annihilated by the BRST charge).

Finally, we can derive the equations of motion from the gauge fixed
action. Up to boundary terms, the variation of equation \eqref{eq:gauge-fixed-action-metric-density}
can be written as
\begin{equation}
\delta S=-\frac{1}{4\kappa}\int d^{D}x\,(\delta\mathfrak{g}^{mn}\mathcal{G}_{mn}+\delta c^{m}\mathcal{C}_{m}+\mathcal{B}^{m}\delta b_{m}),\label{eq:pure-gravity-variation}
\end{equation}
where we have
\begin{align}
\mathcal{B}^{m} & =\partial_{p}(\mathfrak{g}^{np}\partial_{n}c^{m})-\partial_{p}(c^{p}\partial_{n}\mathfrak{g}^{mn}),\\
\mathcal{C}_{m} & =\partial_{p}(\mathfrak{g}^{np}\partial_{n}b_{m})+\partial_{p}\mathfrak{g}^{np}\partial_{m}b_{n},
\end{align}
and
\begin{multline}
\mathcal{G}_{mn}=\partial_{p}(\mathfrak{g}^{pq}\partial_{q}\mathfrak{g}_{mn})-\frac{1}{2}\partial_{m}\mathfrak{g}^{pq}\partial_{n}\mathfrak{g}_{pq}\\
-\eta_{mp}\partial_{n}\partial_{q}\mathfrak{g}^{pq}+\partial_{q}(\mathfrak{g}_{mp}\partial_{n}\mathfrak{g}^{pq})-\eta_{np}\partial_{m}\partial_{q}\mathfrak{g}^{pq}+\partial_{q}(\mathfrak{g}_{np}\partial_{m}\mathfrak{g}^{pq})\\
+\frac{1}{2}\mathfrak{g}^{pq}\mathfrak{g}^{rs}\partial_{p}\mathfrak{g}_{mr}(\partial_{s}\mathfrak{g}_{nq}-\partial_{q}\mathfrak{g}_{ns})+\frac{1}{2}\mathfrak{g}^{pq}\mathfrak{g}^{rs}\partial_{p}\mathfrak{g}_{nr}(\partial_{s}\mathfrak{g}_{mq}-\partial_{q}\mathfrak{g}_{ms})\\
+\frac{1}{2}\frac{1}{(2-D)}[\mathfrak{g}^{pq}\mathfrak{g}^{rs}\partial_{m}\mathfrak{g}_{pq}\partial_{n}\mathfrak{g}_{rs}+2\mathfrak{g}_{mn}\partial_{p}(\mathfrak{g}^{pq}\mathfrak{g}^{rs}\partial_{q}\mathfrak{g}_{rs})]\\
-\partial_{m}c^{p}\partial_{n}b_{p}-\partial_{m}(c^{p}\partial_{p}b_{n})-\partial_{n}c^{p}\partial_{m}b_{p}-\partial_{n}(c^{p}\partial_{p}b_{m}).\label{eq:graviton-eom-off-shell}
\end{multline}
The equations of motion for the $c$ ghost, the $b$ ghost, and the
metric density are respectively given by
\begin{equation}
\mathcal{B}^{m}=\mathcal{C}_{m}=\mathcal{G}_{mn}=0.
\end{equation}

Since we are interested in promoting additional external legs off
the mass-shell, we must work with the equations of motion in the form
above. For example, the reader might be tempted to take the trace-reversed
form of equation $\mathcal{G}_{mn}=0$. However, as we have already
discussed, this would introduce extra interactions compared to the
diagrammatic approach. 

We are now ready to construct the off-shell scattering trees using
the gauge-fixed equations of motion in pure gravity, including ghost
contributions.

\subsubsection{Off-shell multi-particle expansions}

Our first step here is to build the multi-particle currents associated
to the classical solutions of the interacting theory, our scattering
trees. Consider the following ansatze,
\begin{align}
b_{m}(x) & =\sum_{P}B_{Pm}e^{\mathrm{i}k_{P}\cdot x},\\
c^{m}(x) & =\sum_{P}C_{P}^{m}e^{\mathrm{i}k_{P}\cdot x},\\
\mathfrak{g}_{mn}(x) & =\eta_{mn}+\sum_{P}H_{Pmn}e^{\mathrm{i}k_{P}\cdot x},\\
\mathfrak{g}^{mn}(x) & =\eta^{mn}-\sum_{P}I_{P}^{mn}e^{\mathrm{i}k_{P}\cdot x}.
\end{align}
Observe that we denote the currents $H_{Pmn}$ and $I_{P}^{mn}$ as
in section \ref{sec:gravity}. It follows from $\mathfrak{g}_{np}\mathfrak{g}^{mp}=\delta_{n}^{m}$
that they can be recursively mapped to one another via equation \eqref{eq:gravity-I-H-map}.
We just have to keep in mind they are different objects, as they now
parametrize the multi-particle expansion of the metric density.

For our purposes, single-particle states can also be associated to
ghost fields. Their linearized equations of motion are simply
\begin{equation}
\Box b_{m}=\Box c^{m}=0.
\end{equation}
However, we are only interested in their loop contributions, so their
momenta are always off-shell. Their full equations of motion lead to
\begin{align}
C_{P}^{m} & =\frac{1}{s_{P}}\sum_{P=Q\cup R}I_{Q}^{np}C_{R}^{q}\mathcal{V}_{npq}^{(g)m}(Q,P),\label{eq:ghost-recursion-off-shell}\\
B_{Pm} & =\frac{1}{s_{P}}\sum_{P=Q\cup R}I_{Q}^{np}B_{Rq}\bar{\mathcal{V}}_{mnp}^{(g)q}(P,R),
\end{align}
where we define
\begin{align}
\mathcal{V}_{npq}^{(g)m}(Q,P) & =\frac{1}{2}(\delta_{q}^{m}k_{Pp}k_{Pn}-\delta_{q}^{m}k_{Pp}k_{Qn}-\delta_{p}^{m}k_{Pq}k_{Qn})+(n\leftrightarrow p),\\
\bar{\mathcal{V}}_{mnp}^{(g)q}(P,R) & =\frac{1}{2}(\delta_{m}^{q}k_{Pp}k_{Rn}+\delta_{n}^{q}k_{Pp}k_{Rm}-\delta_{n}^{q}k_{Rp}k_{Rm})+(n\leftrightarrow p),
\end{align}
which are identified with the three-point vertices with one $b$ ghost,
one $c$ ghost, and one graviton. For example, $\mathcal{V}_{npq}^{(g)m}(P,Q)$
can be depicted as\begin{equation}
\mathcal{V}_{npq}^{(g)m}(P,Q) \;=\;
\raisebox{-0.45\height}{%
\begin{tikzpicture}[scale=0.9]

  \fill (0,0) circle (2.5pt);

  \draw[decoration={coil, aspect=0, segment length=4pt, amplitude=3pt},
        decorate, thick] (0,0) -- (-1.4, 1.0);
  \node[above left, font=\small] at (-1.4,1.0) {$I_{Q}^{np}$};

  \draw[ dashed] (0,0) -- (1.4, 1.0);
  \node[above right, font=\small] at (1.4,1.0) {$C^q_R$};

  \draw[dashed] (0,0) -- (0,-1.4);
  \node[below, font=\small] at (0,-1.4) {$k_P,\, m$};

  \node[right, font=\footnotesize, gray] at (0.1,-0.7) {$\downarrow$};

\end{tikzpicture}}
\end{equation}where $k_{Q}$ and $k_{R}$ are the incoming momenta of the graviton
and the $c$ ghost, respectively, and $k_{P}$ is the outgoing momentum
of the $b$ ghost. It is easy to show that $\mathcal{V}_{npq}^{(g)m}(P,Q)$
and $\bar{\mathcal{V}}_{mnp}^{(g)q}(P,R)$ are equivalent with the
proper relabeling. For example, consider particle $1$ to be a $c$
ghost, particle $2$ to be a graviton, and particle $3$ to be a $b$
ghost. Then we can show that
\begin{equation}
s_{12}C_{12}^{m} b_{3m}=s_{23}c_{1}^{m}B_{23m}.
\end{equation}
This is an off-shell statement, and only momentum conservation is
being used.

The graviton multi-particle recursion comes out after a lengthy computation,
and the vanishing of equation \eqref{eq:graviton-eom-off-shell} (up
to single-particle off-shell states) leads to
\begin{multline}
s_{P}\mathbb{P}_{mnpq}I_{P}^{pq}=\sum_{P=Q\cup R}2B_{Qq}C_{R}^{p}\mathcal{V}_{mnp}^{(g)q}(P,Q)+\frac{1}{2}I_{Q}^{pq}I_{R}^{rs}\mathcal{V}_{mnpqrs}^{(3)}(Q,R)\\
+\sum_{P=Q\cup R\cup S}I_{Q}^{pq}I_{R}^{rs}H_{S}^{tu}V_{mnpqrstu}^{(4)}(Q,R,S)+\sum_{P=Q\cup R\cup S\cup T}I_{Q}^{pq}I_{R}^{rs}H_{S}^{tu}H_{T}^{vz}V_{mnpqrstuvz}^{(5)}(Q,R,S),\label{eq:graviton-recursion-off-shell}
\end{multline}
where we have introduced
\begin{equation}
\mathbb{P}_{mnpq}=\frac{1}{2}\eta_{mp}\eta_{nq}+\frac{1}{2}\eta_{mq}\eta_{np}+\frac{1}{(2-D)}\eta_{mn}\eta_{pq},
\end{equation}
with inverse
\begin{equation}
(\mathbb{P}^{-1})^{mnpq}=\frac{1}{2}(\eta^{mp}\eta^{nq}+\eta^{mq}\eta^{np}-\eta^{mn}\eta^{pq}),
\end{equation}
such that
\begin{equation}
\mathbb{P}_{mnrs}(\mathbb{P}^{-1})^{rspq}=\frac{1}{2}(\delta_{m}^{p}\delta_{n}^{q}+\delta_{m}^{q}\delta_{n}^{p}).
\end{equation}

The tensor structures appearing in equation \eqref{eq:graviton-recursion-off-shell}
are given by
\begin{multline}
\mathcal{V}_{mnpqrs}^{(3)}(Q,R)=k_{QRr}k_{Qm}\eta_{np}\eta_{qs}+k_{QRp}k_{Rm}\eta_{nr}\eta_{qs}-k_{Qs}k_{Rp}\eta_{mr}\eta_{nq}+\frac{1}{2}k_{QRp}k_{Rq}P_{mnrs}\\
+\frac{1}{2}k_{QRr}k_{Qs}P_{mnpq}-\frac{1}{2}k_{Qm}k_{Rn}P_{pqrs}-\frac{1}{2}s_{QR}P_{mnpr}\eta_{qs}-\frac{1}{2}s_{Q}P_{mrpq}\eta_{ns}-\frac{1}{2}s_{R}P_{mprs}\eta_{nq}+(m\leftrightarrow n),\label{eq:graviton-off-3ptvertex}
\end{multline}
which is connected to the three-graviton vertex, 
\begin{multline}
V_{mnpqrstu}^{(4)}(Q,R,S)=k_{QRSp}k_{Qn}\eta_{ms}\delta_{r}^{t}\delta_{q}^{u}-k_{Qs}k_{Rq}\eta_{mp}\eta_{rt}\eta_{nu}-\frac{1}{4}k_{Qm}k_{RSn}\eta_{ps}\eta_{rt}\eta_{qu}\\
+\frac{1}{2}(k_{QRSp}k_{RSq}-k_{Rp}k_{Sq})\eta_{ms}\eta_{rt}\eta_{nu}+\frac{1}{2}[(k_{Q}\cdot k_{R})-s_{QRS}]\eta_{mp}\eta_{qs}\eta_{rt}\eta_{nu}+\frac{1}{2}(k_{Q}\cdot k_{RS})\eta_{mp}\eta_{ns}\eta_{rt}\eta_{qu}\\
+\frac{1}{2}\frac{1}{(2-D)}[k_{QRSp}k_{Rq}\eta_{mn}\eta_{rt}\eta_{su}-k_{Qm}k_{Rn}\eta_{pq}\eta_{rt}\eta_{su}-(k_{QR}\cdot k_{Q})\eta_{pr}\eta_{qs}\eta_{mt}\eta_{nu}]\\
+\frac{1}{2}\frac{1}{(2-D)}[k_{Qr}k_{QRs}\eta_{pq}\eta_{mt}\eta_{nu}-(k_{QRS}\cdot k_{Q})\eta_{mn}\eta_{ps}\eta_{rt}\eta_{qu}-s_{Q}\eta_{ms}\eta_{pq}\eta_{rt}\eta_{nu}]\\
+(m\leftrightarrow n),\label{eq:graviton-off-V4}
\end{multline}
and
\begin{multline}
V_{mnpqrstuvz}^{(5)}(Q,R,S)=\frac{1}{2}\eta_{mv}\eta_{pz}\eta_{qu}[(k_{Q}\cdot k_{RS})\eta_{nr}\eta_{st}-k_{Qr}k_{Ss}\eta_{nt}]-\frac{1}{2}k_{Qr}k_{Rq}\eta_{mt}\eta_{nv}\eta_{pu}\eta_{sz}\\
+\frac{1}{2}\frac{1}{(2-D)}\eta_{mv}\eta_{nz}\eta_{rt}[k_{QRSp}k_{Rq}\eta_{su}-(k_{QRS}\cdot k_{Q})\eta_{ps}\eta_{qu}]-\frac{1}{4}\frac{1}{(2-D)}k_{Qm}k_{Rn}\eta_{pt}\eta_{qu}\eta_{rv}\eta_{sz}\\
+(m\leftrightarrow n).\label{eq:graviton-off-V5}
\end{multline}
Before presenting the general vertex structure, we should note that
the expressions for $V^{(4)}$ and $V^{(5)}$ are not unique. For
example, let us take the following expression,
\begin{align}
\sum_{P=Q\cup R\cup S}k_{Qm}I_{Q}^{pq}H_{Rpq}k_{Sn}H_{Srs} & =\sum_{P=Q\cup R\cup S}k_{Qm}I_{Q}^{pq}\eta_{pt}\eta_{qu}I_{R}^{tu}k_{Sn}H_{Srs}\nonumber \\
 & +\sum_{P=Q\cup R\cup S\cup T}k_{Qm}I_{Q}^{pq}\eta_{pt}I_{R}^{tu}H_{Tqu}k_{Sn}H_{Srs},\label{eq:off-V4-ambiguity-1}\\
 & =\sum_{P=Q\cup R\cup S}k_{Qm}I_{Q}^{pq}H_{Rpq}k_{Sn}\eta_{rt}\eta_{su}I_{S}^{tu}\nonumber \\
 & +\sum_{P=Q\cup R\cup S\cup T}k_{Qm}I_{Q}^{pq}H_{Rpq}k_{Sn}\eta_{rt}I_{S}^{tu}H_{Tsu},\label{eq:off-V4-ambiguity-2}
\end{align}
The difference between \eqref{eq:off-V4-ambiguity-1} and \eqref{eq:off-V4-ambiguity-2}
comes from which current $H$ on the left hand side has been rewritten
in terms of the current $I$ using equation \eqref{eq:gravity-I-H-map}.
Of course, the underlying physics does not depend on this choice.
Ultimately, this is connected to the fact that the recursion is not
expressed solely in terms of $H$ or $I$. Since the left hand side
of equation \eqref{eq:graviton-recursion-off-shell} is already expressed
in terms of the current $I$, we can use equation \eqref{eq:gravity-I-H-map}
and write
\begin{equation}
H_{Pmn}=\eta_{mp}\eta_{nq}\sum_{i=1}^{|P|}\sum_{P=P_{1}\cup\ldots\cup P_{i}}(I_{P_{1}}\cdots I_{P_{i}})^{pq}.\label{eq:graviton-multi-inversion-HtoI}
\end{equation}
This is the inverse of equation \eqref{eq:graviton-multi-inversion},
and we will use it next to rewrite the recursion in a more democratic
form.

By substituting equation \eqref{eq:graviton-multi-inversion-HtoI}
into our main recursion, we will unavoidably introduce arbitrary point
vertices. This is not a surprise, just the unpacking of something
we knew all along: the Einstein--Hilbert action has an infinite number
of interaction vertices. Indeed, equation \eqref{eq:graviton-recursion-off-shell}
can be rewritten as
\begin{multline}
s_{P}\mathbb{P}_{mnpq}I_{P}^{pq}=2\sum_{P=Q\cup R}B_{Qq}C_{R}^{p}\mathcal{V}_{mnp}^{(g)q}(P,Q)\\
+\sum_{i=2}^{\infty}\left(\frac{1}{i!}\right)\sum_{P=P_{1}\cup\ldots\cup P_{i}}\mathcal{V}_{mnm_{1}n_{1}\ldots m_{i}n_{i}}^{(i+1)}(P_{1},\ldots,P_{i})\prod_{j=1}^{i}I_{P_{j}}^{m_{j}n_{j}}.\label{eq:graviton-off-full-recursion}
\end{multline}
The vertex $\mathcal{V}^{(3)}$ is given in equation \eqref{eq:graviton-off-3ptvertex},
which is already symmetric in the two attached currents. For $i\geq3$,
the pattern is easy to identify. The four point vertex comes directly
from $\eqref{eq:graviton-off-V4}$, and it is given by
\begin{equation}
\mathcal{V}_{mnm_{1}n_{1}m_{2}n_{2}m_{3}n_{3}}^{(4)}(P_{1},P_{2},P_{3})=V_{mnm_{1}n_{1}m_{2}n_{2}m_{3}n_{3}}^{(4)}(P_{1},P_{2},P_{3})+S(1,2,3),
\end{equation}
where $S(1,2,3)$ adds the permutations of the labels $1$, $2$,
and $3$. So there are $3!$ terms in total, which precisely cancel
the respective $i!$ factor in equation \eqref{eq:graviton-off-full-recursion}.
This permutation would have been automatically implemented in the
multi-particle sum, but the idea is to make them explicit in order
to identify $\mathcal{V}^{(4)}$ with the four-graviton vertex. For
$\mathcal{V}^{(5)}$ we have contributions from $\eqref{eq:graviton-off-V4}$
and from $\eqref{eq:graviton-off-V5}$. It can be cast as
\begin{multline}
\mathcal{V}_{mnm_{1}n_{1}m_{2}n_{2}m_{3}n_{3}m_{4}n_{4}}^{(5)}(P_{1},P_{2},P_{3},P_{4})=V_{mnm_{1}n_{1}m_{2}n_{2}m_{3}n_{3}m_{4}n_{4}}^{(5)}(P_{1},P_{2},P_{3})\\
+V_{mnm_{1}n_{1}m_{2}n_{2}m_{3}n_{4}}^{(4)}(P_{1},P_{2},P_{3}\cup P_{4})\eta_{n_{3}m_{4}}+S(1,2,3,4),
\end{multline}
where $S(1,2,3,4)$ introduces the remaining $(4!-1)$ permutations
of the labels $1$, $2$, $3$, and $4$. Both $\mathcal{V}^{(4)}$
and $\mathcal{V}^{(5)}$ match available results in the literature
(e.g. \cite{Capper:1973pv,Brandt:1992dk}). By carefully analyzing
the emerging pattern, the general construction for $\mathcal{V}^{(i)}$
can be written as
\begin{multline}
\mathcal{V}_{mnm_{1}n_{1}\ldots m_{i}n_{i}}^{(i+1)}(P_{1},\ldots,P_{i})=V_{mnm_{1}n_{1}m_{2}n_{2}m_{3}n_{i}}^{(4)}(P_{1},P_{2},P_{3}\cup\ldots\cup P_{i})\prod_{j=3}^{i-1}\eta_{n_{j}m_{j+1}}\\
+\sum_{l=4}^{i}V_{mnm_{1}n_{1}m_{2}n_{2}m_{3}n_{l-1}m_{l}n_{i}}^{(5)}(P_{1},P_{2},P_{3}\cup\ldots\cup P_{l-1})\prod_{j=4}^{l-1}\eta_{n_{j-1}m_{j}}\prod_{k=l}^{i-1}\eta_{n_{k}m_{k+1}}\\
+S(1,\ldots,i),
\end{multline}
which, after the symmetrization of the index pairs $\{m_{i},n_{i}\}$,
provides a closed formula for the $(i+1)$-point graviton vertex derived
from the action \eqref{eq:EH-action-density}.

Finally, using the standard perturbiner approach, we can build off-shell
scattering trees for gravitons and ghosts. They are given by
\begin{equation}
\mathcal{M}_{N+1}=\frac{1}{2\kappa}s_{1\ldots N}h_{N+1}^{mn}\mathbb{P}_{mnpq}I_{1\ldots N}^{pq},
\end{equation}
in which momentum conservation is left implicit. Notice that the usual
limit $s_{1\ldots N}\to0$ is absent, as this would put the graviton
$N+1$ on the mass-shell. From this equation it is straightforward
to compute tree level scattering amplitudes. The ghosts currents are
taken to zero, while the single-particle states associated to the
graviton should be made on-shell. Physical states have to be annihilated
by the BRST charge, and this is where the traceless, transversal conditions
come from in the BRST construction.

In the next subsection we are going to mimic the scalar construction
and build the one-loop integrands in pure gravity, both with gravitons
and ghosts running in the loop.

\subsubsection{One-loop integrands}

In the graviton recursion \eqref{eq:graviton-off-full-recursion},
the off-shell scattering trees are automatically symmetrized. This
would have been achieved by the sum over subwords, but the analogous
construction to \eqref{eq:scalar-loop-motivation} in the scalar theory
demands a symmetric vertex. The reason is simple, as we are going
to be singling out an external leg $\ell$. If we do this before having
a symmetric vertex, the combinatoric analysis of subsection \ref{subsec:scalar-oneloop}
gets mangled, and we lose the systematic control of the diagram overcounting.

Our first step is to establish a recursion for $I_{\ell P}^{mn}$,
where $\ell$ is the special leg with polarization $h_{\ell}^{mn}$.
Using equation \eqref{eq:graviton-off-full-recursion}, we obtain
\begin{multline}
s_{\ell P}\mathbb{P}_{mnpq}I_{\ell P}^{pq}=\sum_{i=2}^{\infty}\sum_{\ell P=P_{1}\cup\ldots\cup P_{i}}\frac{1}{i!}\mathcal{V}_{mnm_{1}n_{1}\ldots m_{i}n_{i}}^{(i+1)}(P_{1},\ldots,P_{i})\prod_{j=1}^{i}I_{P_{j}}^{m_{j}n_{j}}+\textrm{ghosts},\\
=\sum_{i=2}^{\infty}\sum_{P=P_{1}\cup\ldots\cup P_{i-1}}\frac{1}{(i-1)!}h_{\ell}^{m_{i}n_{i}}\mathcal{V}_{mnm_{1}n_{1}\ldots m_{i}n_{i}}^{(i+1)}(P_{1},\ldots,P_{i-1},\ell)\prod_{j=1}^{i-1}I_{P_{j}}^{m_{j}n_{j}}\\
+\sum_{i=2}^{\infty}\sum_{P=P_{1}\cup\ldots\cup P_{i}}\frac{1}{(i-1)!}I_{\ell P_{i}}^{m_{i}n_{i}}\mathcal{V}_{mnm_{1}n_{1}\ldots m_{i}n_{i}}^{(i+1)}(P_{1},\ldots,P_{i-1},\ell P_{i})\prod_{j=1}^{i-1}I_{P_{j}}^{m_{j}n_{j}}+\textrm{ghosts}.
\end{multline}
From the first to the second line we have just separated the contribution
from the single-letter currents $I_{\ell}^{mn}=h_{\ell}^{mn}$. We
can drop the ghosts for now, as the primary objective is to consider
one-loop integrands with external gravitons. If we identify $I_{\ell P}^{mn}$
with $h_{\ell}^{rs}K_{Prs}^{mn}(\ell)$, the equation above can be
recast as
\begin{multline}
K_{Prs}^{mn}(\ell)=D_{\ell P}^{mnpq}\bigg\{\sum_{i=2}^{\infty}\sum_{P=P_{1}\cup\ldots\cup P_{i-1}}\frac{1}{(i-1)!}\mathcal{V}_{pqm_{1}n_{1}\ldots m_{i-1}n_{i-1}rs}^{(i+1)}(P_{1},\ldots,P_{i-1},\ell)\prod_{j=1}^{i-1}I_{P_{j}}^{m_{j}n_{j}}\\
+\sum_{i=2}^{\infty}\sum_{P=P_{1}\cup\ldots\cup P_{i}}\frac{1}{(i-1)!}K_{Prs}^{m_{i}n_{i}}(\ell)\mathcal{V}_{pqm_{1}n_{1}\ldots m_{i}n_{i}}^{(i+1)}(P_{1},\ldots,P_{i-1},\ell P_{i})\prod_{j=1}^{i-1}I_{P_{j}}^{m_{j}n_{j}}\bigg\}.
\end{multline}
This object is the graviton analogue of the scalar $\Lambda_{P}(\ell)$
in equation \eqref{eq:scalar-loop-overcounted}. Apart from the extra
spacetime indices, they have the same structure. Furthermore, we recognize
here the graviton propagator,
\begin{equation}
D_{P}^{mnpq}=\frac{1}{s_{P}}(\mathbb{P}^{-1})^{mnpq},
\end{equation}
which naturally fits the sewing procedure used to build a graviton
loop.

By extending the analogy with the scalar model with polynomial interactions,
we finally define
\begin{multline}
\tilde{K}_{Prs}^{mn}(\ell)=D_{\ell P}^{mnpq}\bigg\{\sum_{i=2}^{\infty}\sum_{P=P_{1}\cup\ldots\cup P_{i-1}}\left(\frac{1}{2}\right)\frac{1}{(i-1)!}\mathcal{V}_{pqm_{1}n_{1}\ldots m_{i-1}n_{i-1}rs}^{(i+1)}(P_{1},\ldots,P_{i-1},\ell)\prod_{j=1}^{i-1}I_{P_{j}}^{m_{j}n_{j}}\\
+\sum_{i=2}^{\infty}\sum_{P=P_{1}\cup\ldots\cup P_{i}}\frac{1}{(i-1)!}\left(\frac{|P_{i}|}{|P|}\right)\tilde{K}_{P_{i}rs}^{m_{i}n_{i}}(\ell)\mathcal{V}_{pqm_{1}n_{1}\ldots m_{i}n_{i}}^{(i+1)}(P_{1},\ldots,P_{i-1},\ell P_{i})\prod_{j=1}^{i-1}I_{P_{j}}^{m_{j}n_{j}}\bigg\}.
\end{multline}
The diagram overcounting is taken care by the factor of $1/2$ in
the first line, and by the ratio $|P_{i}|/|P|$ in the second line.

The one-loop graviton integrand is defined as the trace $\tilde{K}_{Prs}^{mn}(\ell)$
with $k_{P}^{m}=0$ (momentum conservation of the external legs),
with
\begin{equation}
\mathcal{I}_{P}^{\textrm{graviton}}(\ell)=\tilde{K}_{Pmn}^{mn}(\ell).\label{eq:graviton-loop}
\end{equation}

As for the contribution with ghosts in the loop, we simply extend
the sewing procedure to the ghost currents. Let us take, for example,
the current $C_{\ell P}^{m}\equiv c_{\ell}^{n}J_{Pn}^{m}(\ell)$,
in which we single out an external ghost leg $\ell$, with polarization
$c_{\ell}^{m}$. Using equation \eqref{eq:ghost-recursion-off-shell},
we obtain
\begin{align}
C_{\ell P}^{m} & =\frac{1}{s_{\ell P}}\sum_{\ell P=Q\cup R}I_{Q}^{np}C_{R}^{q}\mathcal{V}_{npq}^{(g)m}(Q,\ell P),\\
J_{Pn}^{m}(\ell) & =\frac{1}{s_{\ell P}}I_{P}^{rs}\mathcal{V}_{rsn}^{(g)m}(P,\ell P)+\frac{1}{s_{\ell P}}\sum_{\ell P=Q\cup R}I_{Q}^{rs}J_{Rn}^{p}(\ell)\mathcal{V}_{rsp}^{(g)m}(Q,\ell P).
\end{align}
From the first to the second line we have separated the contribution
from the one-letter current $C_{\ell}^{q}$. Note that $I_{\ell}^{np}=0$,
as the leg $\ell$ is now identified with the ghost in order to close
the loop. The ghost propagator is just like the scalar theory here.
We can then define a modified current $\tilde{J}_{Pn}^{m}(\ell)$
that suppresses the diagram overcounting from the ghost loop. It is
straightforward to show the answer is
\begin{equation}
\tilde{J}_{Pn}^{m}(\ell)=\frac{1}{2}\frac{1}{s_{\ell P}}I_{P}^{rs}\mathcal{V}_{rsn}^{(g)m}(P,\ell P)+\frac{1}{s_{\ell P}}\sum_{\ell P=Q\cup R}\frac{|Q|}{|P|}I_{Q}^{rs}J_{Rn}^{p}(\ell)\mathcal{V}_{rsp}^{(g)m}(Q,\ell P).
\end{equation}
The ghost contribution to the one-loop integrand is then given by
\begin{equation}
\mathcal{I}_{P}^{\textrm{ghost}}(\ell)=\tilde{J}_{Pm}^{m}(\ell).
\end{equation}

Therefore, the complete one-loop integrand for the correlator of $N$
gravitons is given by
\begin{equation}
\mathcal{I}_{N}(\ell)=\mathcal{I}_{1\ldots N}^{\textrm{graviton}}(\ell)-\mathcal{I}_{1\ldots N}^{\textrm{ghost}}(\ell).
\end{equation}
As usual, the ghost loop comes with a negative sign. This sign is
not visible from the equations of motion alone, as it is encoded in
the Grassmann statistics of the ghost fields in the gauge-fixed action.
It is a realization of the familiar minus sign for fermion
loops in the path integral formulation.

The $N$-graviton amputated correlator $\mathcal{M}_{N}^{1\textrm{-loop}}$
is finally given by
\begin{equation}
\mathcal{M}_{N}^{1\textrm{-loop}}=\int\frac{d^{d}\ell}{(2\pi)^{d}}\,\mathcal{I}_{N}(\ell).
\end{equation}
We refer the reader to reference \cite{Gomez:2024xec} for additional
details and the explicit computation of the graviton self-energy in
the perturbiner framework. Note that the construction of the one-loop integrand is a well-defined problem, and the perturbiner framework is able to produce it. Renormalizability of the theory is a different question altogether. In particular, we know that gravity is not renormalizable \cite{tHooft:1974toh}. Questions about the ultraviolet behavior of the theory, regularization, and renormalization concern the (integrated) amplitudes, and lie beyond the scope of this review.

\subsection{Higher loops\label{subsec:Higher-loops}}

The one-loop construction we have so far discussed is based on a direct
algebraic extension of the tree-level perturbiner. It was obtained
by singling out one additional off-shell leg and carefully tracking
the diagram overcounting through combinatorial coefficients. This
approach is systematic and self-contained: the one-loop integrand
is derived entirely from the classical equations of motion. There
is no reference to path integrals or quantum field theory beyond the
tree-level framework, though we still have a minor input from the
path integral formulation (like the negative sign for the ghost loop).

The possibility of extending the perturbiner framework to loop level
via a quantum equation of motion was already noted in \cite{Gomez:2021shh},
though a concrete implementation was not pursued there. This idea
was nicely developed in full in \cite{Lee:2022aiu}, where the quantum
effective action formalism is used to derive a quantum perturbiner
expansion. The starting point is the Dyson--Schwinger equation, the
quantum analogue of the classical equation of motion, which is obtained
from the classical one by a systematic deformation involving functional
derivatives. This naturally generates a hierarchy of descendant fields
(functional derivatives of the classical field with respect to the
external source) whose perturbiner expansions encode loop integrals
explicitly. Each order in $\hbar$ then yields a closed set of recursion
relations for the quantum off-shell currents, which can be solved
iteratively. This formalism has been applied to $\phi^{4}$ theory
and pure Yang--Mills, and in principle extends to arbitrary loop
order and any QFT with an action principle.

The two approaches are complementary. The quantum effective action
framework is more structured and better suited for pushing to higher
loop orders, since it provides a clean organizational principle through
the $\hbar$ expansion and the hierarchy of descendant equations.
On the other hand, the algebraic construction of this section is considerably
simpler to implement at one loop. It requires only the tree-level
recursion as input and produces the integrand directly, without the
need to introduce descendant fields or solve coupled systems of equations.
For practical one-loop computations, particularly in gauge theories
and gravity, where the recursive structure is already complex at tree
level, the algebraic approach offers a more direct route.

The extension to two loops and beyond within the algebraic framework
remains an open question\footnote{There were some recent developments focusing on planar loop integrands \cite{Tao:2025fch,Tao:2026irg}, which are combinatorially easier to deal with.}. The combinatorial analysis of diagram overcounting
becomes more involved at higher loop orders, as the singling out of
multiple off-shell legs introduces additional identifications. Whether
the simple ratio $|P_{1}|/|P|$ that governs the one-loop coefficients
generalizes to a clean closed-form expression at two loops is an interesting
open problem, and one that the connection to the quantum effective
action framework of \cite{Lee:2022aiu} may help to illuminate.

\section{Summary of recent applications\label{sec:summary}}

In this final section we will quickly go through some of the recent
research involving perturbiner methods. This is by no means an exhaustive
account of the many interesting modern applications. Each subsection
is focused on a different topic, more or less relevant depending on
the taste of the reader, with open directions to be investigated.
We then conclude with some final remarks regarding the status of the
perturbiner framework and other possible future applications.

\subsection{Non-Lagrangian equations of motion\label{subsec:Non-Lagrangian-eom}}

An interesting application of the perturbiner method concerns the
question of whether a consistent perturbative S-matrix can exist for
field theories that lack an action principle. This is a fundamental
question in quantum field theory. While on-shell methods such as the BCFW
recursion and unitarity cuts do not strictly require a Lagrangian,
it is far from obvious whether arbitrary equations of motion can define
a sensible quantum theory. The perturbiner method provides a natural
testing ground for this question, since it extracts scattering amplitudes
directly from the classical equations of motion, without any reference
to an action.

This has been recently explored in reference \cite{Escudero:2022zdz}
and in reference \cite{Chakrabarti:2024crx}. The former is based
on the non-Abelian Navier--Stokes equation proposed in \cite{Cheung:2020djz},
and introduces a realization of the double copy at the level of perturbiner
expansions. The latter explores the interacting $p$-form theories
introduced in \cite{Broccoli:2021pvv}. These theories became known
as third-way consistent. Even though their equation of motion, generically
expressed as a flatness condition $G=0$, cannot be derived from an
action principle, the integrability condition $dG\propto G$ is satisfied
on the mass-shell (i.e., $G=0$). The perturbiner method operates
without any reference to a Lagrangian, as the associated multi-particle
currents are constructed recursively from the equations of motion
alone. Tree-level partial amplitudes are extracted via the standard
prescription, with a computation that is entirely parallel to what
one would do in a Lagrangian theory.

The result is intriguing: all third-way theories of this type fail
to produce unitary scattering amplitudes, already at tree level. This
failure manifests as a breakdown of factorization, since the four-point
amplitude does not correctly decompose into products of three-point
amplitudes in the expected channels. The perturbiner framework makes
this diagnosis transparent, since factorization is directly visible
in the pole structure of the multi-particle currents. Interestingly,
there is a natural way to circumvent this obstruction via a non-trivial
realization of the field-strength $G$, parametrized by a constant
$\alpha$. Precisely at the value of $\alpha$ in which factorization
is restored, the theory is identified with the higher-dimensional
Freedman-Townsend model \cite{Freedman:1980us}, which does follow
from an action. In this sense, the perturbiner method does not just
compute amplitudes: it can be used as a diagnostic tool to determine
whether a given set of equations of motion is compatible with a consistent
perturbative quantum theory, and to reverse-engineer the Lagrangian
when one exists.

\subsection{Chiral string theories\label{subsec:Chiral-string-theories}}

Among other applications, the perturbiner has been used as a consistency
check tool in chiral string models \cite{Siegel:2015axg}. Chiral
strings can be seen as the tensionful version of ambitwistor strings
\cite{Mason:2013sva}. The latter has an exclusively massless spectrum,
while chiral strings may contain a finite number of massive resonances.
Because of the finite spectrum, these theories are more amenable to
a Lagrangian description, since a finite number of fields can in principle
be accommodated in a local action. Interestingly, the computation
of string amplitudes can be done without any knowledge of the respective
field theory action.

In \cite{Guillen:2021mwp}, such a Lagrangian has been proposed to
describe the interactions between massless and one massive state in
the heterotic chiral string. In order to confirm the consistency of
the proposal, scattering amplitudes computed directly from the string
model, and using the perturbiner, were matched against each other.

In \cite{Carabine:2023yxv}, the chiral bosonic string was analyzed
in detail. A series of consistency conditions led to the complete
set of equations of motion of the model, including all corrections
parametrized by the string tension. Once more, (string) scattering
amplitudes computed using string theory techniques and the perturbiner
were used to confirm the validity of the derived equations of motion.

Together, these results establish the perturbiner as an efficient
consistency check in the construction of string-inspired field theories,
complementing and cross-validating amplitude computations from the
world-sheet.

\subsection{Reconstruction of the Schwarzschild metric from flat space}

An interesting application of the perturbiner method other than the
computation of scattering amplitudes was recently presented in \cite{Damgaard:2024fqj},
where the authors use a perturbiner-like recursion to construct the
Schwarzschild solution in harmonic gauge order by order in the post-Minkowskian
expansion. Starting from the Einstein equations with a point-particle
source, they compute the one-point function of the metric perturbation
around flat space. This quantity is organized as a formal expansion
in powers of the source current. Although the recursive construction
involves momentum-space integrals resembling loop integrals (similar
to the ideas introduced in \cite{Lee:2022aiu}, which we have briefly
discussed in subsection \ref{subsec:Higher-loops}), these arise entirely
within the classical perturbative solution. Substituting this expansion
into the equations of motion generates recursive off-shell currents
for the gravitational field, allowing the metric perturbation to be
computed systematically to arbitrary order.

The resulting expansion converges in its expected domain and resums
to the exact Schwarzschild metric in harmonic gauge, providing one
of the first examples in which a classical solution of Einstein gravity
is recovered from an all-order perturbiner construction. Although
the calculation is still formulated as a perturbation around Minkowski
space, it demonstrates that perturbiner methods naturally extend beyond
the computation of S-matrix elements and can be used to reconstruct
non-trivial classical backgrounds. This perspective provides useful
motivation for extending perturbiner techniques to genuinely curved
geometries, where the recursive construction is formulated directly
on the background spacetime rather than as a perturbation of flat
space, and going beyond what was presented in section \ref{sec:AdS}.

\subsection{Gluon correlators in $\textrm{AdS}_{4}$}

The perturbiner construction of \ref{subsec:flat-boundary} and \ref{subsec:perturbiner-AdS}
naturally raises a question: how much flat-space amplitude data is
encoded in the full AdS correlator? This has been recently answered
for Yang--Mills theory in $\textrm{AdS}_{4}$ in \cite{Gomez:2026yno}.
The main result is that every tree-level gluon correlator admits an
exact decomposition into flat-space scattering amplitudes at all multiplicities,
with the energy poles providing a clear organizational structure to the expansion.

The key observation is that the merged polarizations $\mathcal{A}_{m}^{(1)}(P)$
introduced in subsection \ref{subsec:flat-boundary}, which arise
from Dirichlet boundary conditions and parametrize free solutions
of the equation of motion, behave as genuine single-particle polarizations
in the Lorenz gauge. The Berends-Giele recursion for the multi-particle
currents then takes precisely the same form as in flat space, with
the subwords playing the role of external particle labels. This allows
one to define flat-space amplitudes with merged external data at any
multiplicity, which become the residues of energy poles in the correlator.
The leading energy pole encodes the flat-space limit, while sub-leading
poles carry curvature corrections through lower-point amplitudes with
concatenated legs.

A non-trivial feature of this result is that the amplitude expansion
structure only becomes manifest at the level of full correlators,
i.e., diagram by diagram it is invisible. The perturbiner framework
makes it explicit from the outset, providing a more transparent and
efficient route to AdS observables than conventional diagrammatic
methods, and establishing flat-space amplitudes as an organizing principle
of the full correlator at finite curvature.

\subsection{Tree level string scattering\label{subsec:Tree-level-string}}

String theory contains an infinite number of physical states (string
resonances). Therefore, any computation beyond three points, even
at tree level, introduces an exchange of virtual states spanning the
full tower of string resonances. This behavior is elegantly embodied
by the Veneziano amplitude, which expresses the tree level scattering
of four open string tachyons. It has an infinite number of poles associated
to each mass level of resonances. Recently, a new field-theory-like
expansion of the Veneziano amplitude has been introduced in \cite{Saha:2024qpt}.
The new series are analytic in all the poles, which manifestly appear
in the expression, and are amenable to level truncation. A first principles
derivation of such an expansion appears to require string field theory
(SFT), a second quantized version of string theory.

In SFT, the string field, $\Psi$, describes all the physical degrees
of freedom of the string spectrum. For the specific case of the open
bosonic string, the equation of motion of the string field is given
by \cite{Witten:1985cc}
\begin{equation*}
Q\Psi=\Psi\star\Psi.\label{eq:SFT-eom}
\end{equation*}
Here we have $Q$ denoting the string BRST charge, which can be thought
of as the kinetic operator, and the star product $\star$, which non-trivially
combines two string fields into one. This equation is very similar
to the standard $\phi^{3}$ theory we used to introduce many concepts
regarding the perturbiner. It would then be tempting to apply the
perturbiner method to the equation of motion \eqref{eq:SFT-eom}.
However, this approach has not yet been fruitful.

String theory amplitudes are usually computed via two-dimensional
conformal field theory (CFT) techniques. So one might attempt a perturbiner
recursion embedded in the CFT approach. This is exactly what has been proposed in \cite{Gomez:2026exq}. We have
been able to compute tree level open string scattering via a perturbiner
approach inspired by equation \eqref{eq:SFT-eom}. Besides producing
a series expansion for the Veneziano amplitude, much like the results
of \cite{Saha:2024qpt}, we are able to generate the series expansion
of arbitrary point amplitudes as long as the external states are specified.
These exciting results may finally establish the perturbiner as a
tool within string theory, with many natural open directions to explore,
including closed strings, superstrings, and extensions beyond tree
level.

\subsection{Final remarks}

It would be fair to say that the perturbiner is a mature and largely
complete framework for tree-level amplitudes in flat space. However,
the most interesting applications may lie ahead rather than behind, in particular regarding curved backgrounds, off-shell data for quantum
field theory, and string theory, where the recursive off-shell structure
provides access to objects that on-shell methods cannot easily reach.
It appears that the method has found its natural domain, which is
much broader than its origins suggested.

Taken together, the material presented in this review showcases the
evolution of the perturbiner from a tool for organizing tree-level
amplitudes into a general recursive framework of considerable scope,
applicable to an ever increasing and diverse set of scenarios. They range from
gauge theories, gravity, supersymmetric field theories, and the non-linear
sigma model and its relatives, to loop integrands, correlators in curved
backgrounds, non-Lagrangian theories, and now the first steps into
string theory. This breadth is not just a coincidence, and the core
idea of the method (classical multi-particle solutions as generating
functions for scattering data) is robust enough to survive the passage
from flat space to curved backgrounds, from tree level to loop level,
and from Lagrangian to non-Lagrangian settings. We often have to introduce
non-trivial adaptations, but the underlying recursive logic remains
intact throughout.

Several directions remain open and deserve further investigation.
The loop construction of section \ref{sec:loop} establishes a clean
algebraic framework at one loop, but its extension to two loops and
beyond within the same algebraic spirit remains an open problem. The
combinatorial structure governing diagram overcounting at one loop,
captured by the simple ratios of the form $|Q|/|P|$, may admit a
generalization, and the connection to the quantum effective action
approach of \cite{Lee:2022aiu} suggests one possible route. In curved
backgrounds, the perturbiner has been formulated for anti-de Sitter
space and flat space with a boundary. However, the de Sitter case
(physically relevant for cosmology and closely related to AdS by analytic
continuation) has only been discussed briefly and deserves a more
complete treatment. The double copy at the level of classical solutions
in curved backgrounds is another natural direction: while the perturbiner
in flat space can accommodate color-kinematics duality and
KLT relations at the level of currents, extending these structures
to AdS or de Sitter would be a significant development. Subsection
\ref{subsec:Non-Lagrangian-eom} raises a broader structural question:
to what extent can the perturbiner serve as a probe of the space of
consistent quantum theories, going beyond the specific case of third-way
theories? Finally, subsection \ref{subsec:Tree-level-string} opens
the door for a completely unexplored realm within the perturbiner framework.
It offers a tangible field theory dressing of tree level string theory
amplitudes, and a potentially fruitful path to explore from the perspective
of string field theory. This is potentially the most significant open
application, and string theory is where the perturbiner has the most
room to add something genuinely new. If these results generalize
cleanly to superstrings and closed strings, that would be a substantial
contribution.

From its unpretentious beginnings, the perturbiner now occupies a
distinctive position within the modern landscape of amplitude methods.
Unlike purely on-shell approaches such as BCFW recursion or the CHY
formalism, it operates at the level of classical field configurations
and retains an explicit off-shell leg throughout the construction.
This makes it less efficient than on-shell methods for computing pure
amplitude data in simple theories (see discussion at the end of subsection
\ref{subsec:Soft-limit}), but considerably more versatile when many
field ``flavours'' are involved or when the goal is to understand
the recursive structure of the theory itself. This includes probing
its symmetries and gauge structure, understanding its behavior under
deformations, and extending the framework to settings where asymptotic
states are not available. In this sense the perturbiner is not a competitor
to on-shell methods but a complement to them, one whose full potential
is still being explored.

\section*{Acknowledgments}

I would like to thank my collaborators and colleagues whose work and
insights have shaped this review: Connor Armstrong, Nicholas Carabine,
Subhroneel Chakrabarti, Humberto Gomez, Max Guillen, Henrik Johansson,
Sitender Pratap Kashyap, Kanghoon Lee, Arthur Lipstein, Cristhiam
Lopez-Arcos, Jiajie Mei, Sebastian Mizera, Joris Raeymaekers, Oliver
Schlotterer, Alexander Quintero Velez. In particular, I thank those
who introduced me to the subject and participated in innumerable essential
discussions that underpinned the results presented here. I would also like
to thank Chandrasekhar Bhamidipati for comments on the draft.
Finally, this work was supported by the Czech Science Foundation through
grant GA\v CR 25-16244S.

\printbibliography

\appendix

\section{Appendix: Free solutions in anti-de Sitter\label{sec:Free-solutions-AdS}}

In this appendix we will discuss the free solutions of the equation
of motion \eqref{eq:scalar-eom-AdS}. In other words, we will explore
the space of solutions of
\begin{equation}
z^{2}\partial_{z}^{2}\phi+(1-d)z\partial_{z}\phi+z^{2}\eta^{\mu\nu}\partial_{\mu}\partial_{\nu}\phi+(L\mathrm{M})^{2}\phi=0,
\end{equation}
which we have seen take the form
\begin{equation}
\phi(x,z)=\phi(z;k^{2})e^{\mathrm{i}k_{\mu}x^{\mu}}.\label{eq:free-scalar-AdS}
\end{equation}

After replacing $\phi(x,z)$ in the equation of motion, we obtain
an ordinary differential equation for the radial part, given by
\begin{equation}
z^{2}\partial_{z}^{2}\phi+(1-d)z\partial_{z}\phi-z^{2}k^{2}\phi=-(L\mathrm{M})^{2}\phi,\label{eq:bulk-to-boundary-eom}
\end{equation}
now with $\phi=\phi(z;k^{2})$, and $k^{2}=k_{\mu}k^{\mu}$. This
is a Bessel equation in disguise.

First of all, in order to simplify the analysis, we introduce
\begin{equation}
\phi=z^{(d-1)/2} \varphi(z;k^{2}),
\end{equation}
such that the differential equation for $\varphi(z;k)$ is given by
\begin{equation}
z^{2}(\partial_{z}^{2}\varphi-k^{2}\varphi)=(\nu^{2}-\frac{1}{4})\varphi,\label{eq:simplified-scalar-AdS}
\end{equation}
where $\nu$ is defined in equation \eqref{eq:solution-parameter}.

Before the general solution, we can look at the asymptotic ($z\to\infty$)
behavior of the equation, in which case the dominant piece comes from
the left hand side of \eqref{eq:simplified-scalar-AdS}, regardless
of the value of $\nu$. For $k^{2}>0$, we have
\begin{equation}
\varphi\sim A_{+}e^{+\bar{k}z}+B_{+}e^{-\bar{k}z},
\end{equation}
where $\bar{k}=\sqrt{|k_{\mu}k^{\mu}|}$. For $\bar{k}=0$ we simply
have
\begin{equation}
\varphi\sim A_{0}z+B_{0},
\end{equation}
while for $k^{2}<0$ we obtain
\begin{equation}
\varphi\sim A_{-}e^{+\mathrm{i}\bar{k}z}+B_{-}e^{-\mathrm{i}\bar{k}z},
\end{equation}
where $A_{i}$ and $B_{i}$ are numerical constants. Therefore, the
value of $k^{2}$ dictates the asymptotic behavior of the solution
(exponential growth/decay, oscillatory, linear/constant). The solutions
above are exact for $\nu^{2}=1/4$.

For the general solution of equation \eqref{eq:simplified-scalar-AdS},
we can use the Frobenius method, and introduce a series expansion
for $\varphi$, given by
\begin{equation}
\varphi=z^{\alpha}\sum_{n=0}^{\infty}c_{n}z^{2n},\label{eq:Frobenius-ansatz}
\end{equation}
where $\alpha$ and $c_{n}$ are coefficients to be determined. It
is easy to see that the odd power series is mapped to the even one
according to the two solutions of $\alpha$ (remember, this is a second
order equation), so it is enough to have only the even powers appearing
in the ansatz \eqref{eq:Frobenius-ansatz}. After substituting it
in the differential equation, we obtain the following conditions,
\begin{equation}
\alpha(\alpha-1)=\nu^{2}-\frac{1}{4},
\end{equation}
and
\begin{equation}
[(\alpha+2n+2)(\alpha+2n+1)-\nu^{2}+\frac{1}{4}]c_{n+1}=k^{2}c_{n},
\end{equation}
for $n\geq0$. Therefore, $\alpha$ can take two possible values,
\begin{equation}
\alpha_{\pm}=\frac{1}{2}\pm\nu,
\end{equation}
and the two solutions of equation \eqref{eq:simplified-scalar-AdS}
are
\begin{equation}
\varphi_{\pm}^{\nu}(z;k^{2})=z^{\alpha_{\pm}}\sum_{n=0}^{\infty}c_{n}^{\pm}z^{2n},
\end{equation}
with
\begin{equation}
c_{n+1}^{\pm}=\frac{k^{2}}{4(n+1)(n+1\pm\nu)}c_{n}^{\pm}
\end{equation}
for $n\geq0$. The solution of this recursion is given by
\begin{equation}
c_{n}^{\pm}=\left(\frac{k^{2}}{4}\right)^{n}\frac{c_{0}^{\pm}}{n!\Gamma(n\pm\nu+1)},
\end{equation}
where $c_{0}^{\pm}$ is just the overall normalization.

Now with so many possible solutions, how do we pick the ``physical''
ones to be used as bulk-to-boundary propagators? First we demand that
the exponents $\alpha_{\pm}$ are real, therefore
\begin{equation}
\frac{1}{4}d^{2}\geq L^{2}\mathrm{M}^{2},
\end{equation}
which translates to $\nu\in\mathbb{R}.$ This is known as the Breitenl\"ohner--Freedman
(BF) bound. There is a whole story behind this, but the bottom line
is that below the BF bound the system is unstable \cite{Breitenlohner:1982bm,Breitenlohner:1982jf}
(see also \cite{Witten:1998qj}). Precisely at the bound, a logarithmic
dependence is introduced in the solution. Again, there is a lot of
physics to discuss when the bound is saturated (e.g. \cite{Henneaux:2004zi}),
but we will assume we are above it, with $\nu>0$. In this case, as
we approach the boundary, the solution $\varphi_{+}$ decays faster
than $\varphi_{-}$. However, as $z\to\infty$ both solutions exponentially
grow. There is a specific linear combination of them that provides
the physically required behavior for a bulk-to-boundary propagator
as $z$ goes to infinity: exponential decay for $k^{2}>0$, and the 
in-falling condition ($e^{+\mathrm{i}\bar{k}z}$), for $k^{2}<0$,
which corresponds to the retarded Green's function in Lorentzian signature.
This combination corresponds to the so-called non-normalizable mode,
and is identified with the bulk-to-boundary propagator,
\begin{multline}
\phi^{(\nu)}(z;k^{2})=z^{d/2}\left(\frac{k^{2}}{4}\right)^{\nu/2}\Gamma(1-\nu)\sum_{n=0}^{\infty}\frac{1}{n!}\frac{1}{\Gamma(n-\nu+1)}\left(\frac{k^{2}z^{2}}{4}\right)^{n-\nu/2}\\
-z^{d/2}\left(\frac{k^{2}}{4}\right)^{\nu/2}\Gamma(1-\nu)\sum_{n=0}^{\infty}\frac{1}{n!}\frac{1}{\Gamma(n+\nu+1)}\left(\frac{k^{2}z^{2}}{4}\right)^{n+\nu/2},\label{eq:bulk-to-boundary-propagator}
\end{multline}
where we assume a non-integer $\nu$. The normalization is such that
\begin{equation}
\lim_{z\to0}\phi^{(\nu)}(z;k^{2})=z^{d/2-\nu}.
\end{equation}
The expression inside the square brackets is proportional to the (modified)
Bessel function of the second kind. As usual, for integer $\nu$ the
two solutions ($\varphi_{+}$ and $\varphi_{-}$) are not independent
and a logarithmic term must be introduced. This is standard textbook
material (see e.g. \cite{NIST:DLMF}). One simple approach is to
derive equation \eqref{eq:scalar-eom-AdS-nu} with respect to the
parameter, and evaluate it at $\nu=0$:
\begin{equation}
\left(\mathcal{D}_{k}^{2}+\frac{1}{4}d^{2}\right)\partial_{\nu}\phi^{(0)}=0.
\end{equation}
Therefore, we compute $\partial_{\nu}\phi^{(\nu)}$ from equation
\eqref{eq:bulk-to-boundary-propagator}, set $\nu=0$, and obtain
the desired bulk-to-boundary propagator. Other integer values of $\nu$
can be obtained using the raising/lowering operators displayed in
equation \eqref{eq:raising-lowering-nu}.

As for the so-called normalizable mode, it is expressed in terms of
$\varphi_{+}$ as
\begin{equation}
\tilde{\phi}^{(\nu)}(z;k^{2})=z^{d/2}\Gamma(1+\nu)\sum_{n=0}^{\infty}\frac{1}{n!\Gamma(n+\nu+1)}\left(\frac{k^{2}z^{2}}{4}\right)^{n+\nu/2},
\end{equation}
such that
\begin{equation}
\lim_{z\to0}\tilde{\phi}^{(\nu)}(z;k^{2})=z^{d/2+\nu}.
\end{equation}
 $\phi=z^{d/2} \varphi_{+}$ alone is the normalizable mode.

We also need to invert the operator
\[
z^{2}\partial_{z}^{2}+(1-d)z\partial_{z}-z^{2}k^{2}+(L\mathrm{M})^{2}
\]
in order to build the classical multi-particle solutions in AdS. In
other words, we need the Green's function satisfying equation \eqref{eq:bulk-to-bulk-eom}.
With the homogeneous Dirichlet boundary condition, the simplest representation
of the Green's function is given by
\begin{equation}
G^{(\nu)}(z,y;k^{2})=\frac{y^{d-1}}{W^{(\nu)}(y;k^{2})}\begin{cases}
\tilde{\phi}^{(\nu)}(z;k^{2})\phi^{(\nu)}(y;k^{2}) & z<y,\\
\phi^{(\nu)}(z;k^{2})\tilde{\phi}^{(\nu)}(y;k^{2}) & z>y,
\end{cases}\label{eq:bulk-to-bulk-wronskian}
\end{equation}
where $W^{(\nu)}(y)$ is the Wronskian,
\begin{equation}
W^{(\nu)}=\tilde{\phi}^{(\nu)}\partial_{z}\phi^{(\nu)}-\phi^{(\nu)}\partial_{z}\tilde{\phi}^{(\nu)}.
\end{equation}
It is straightforward to show that the bulk-to-bulk propagator \eqref{eq:bulk-to-bulk-wronskian}
satisfies \eqref{eq:bulk-to-bulk-eom} by integrating both sides of
the equation with respect to $z$. For actual computations within
the perturbiner framework, it is more convenient to use different
representations of the Green's function. They are widely available
in the literature (e.g. \cite{Freedman:1998tz}, but there are many
more for all different tastes).
\end{document}